\documentclass[
 amsmath,amssymb,
 aps,
 prb,
]{revtex4-2}

\usepackage{amsmath, amssymb}
\usepackage{comment}
\usepackage{bm}
\usepackage{siunitx}
\usepackage[italicdiff]{physics}
\usepackage{mathtools}
\usepackage{wrapfig}
\usepackage{color}
\usepackage{hyperref}
\usepackage{xurl}
\usepackage{array}
\hypersetup{unicode, hidelinks, colorlinks,linkcolor=blue, urlcolor=blue, citecolor=blue}
\mathtoolsset{showonlyrefs,showmanualtags}
\allowdisplaybreaks

\DeclareMathOperator{\sign}{sign}
\newcommand{\tspace}{{\lower2.5ex\hbox{\rule{0pt}{6ex}}}}
\newcommand{\clength}{\xi_{\text{\scalebox{0.7}{$\mathrm{SC}$}}}}
\renewcommand{\order}[1]{O(#1)}
\begin{document}
\title{Fluctuation contribution to Spin Hall Effect in Superconductors}
\author{Akimitsu Watanabe}
 \email{watanabeakimitsu@alumni.u-tokyo.ac.jp}
  \affiliation{
 Department of Physics, The University of Tokyo, Bunkyo, Tokyo 113-0033, Japan}
\author{Hiroto Adachi}
\email{hiroto.adachi@okayama-u.ac.jp}
\affiliation{Research Institute for Interdisciplinary Science, Okayama University, Okayama 700-8530, Japan}
\author{Yusuke~Kato}
 \email{yusuke@phys.c.u-tokyo.ac.jp}
\affiliation{Department of Basic Science, The University of Tokyo, Meguro-ku, Tokyo 153-8902, Japan}
%
\date{\today}
%
\begin{abstract}
 We theoretically study the contribution of superconducting fluctuation to extrinsic spin Hall effects in two- and three-dimensional electron gas and intrinsic spin Hall effects in two-dimensional electron gas with Rashba-type spin-orbit interaction. The Aslamazov-Larkin, Density-of-States, Maki-Thompson terms have logarithmic divergence $\ln\epsilon$ in the limit $\epsilon=(T- T_{\mathrm{c}})/T_{\mathrm{c}} \rightarrow +0$ in two-dimensional systems for both extrinsic and intrinsic spin Hall effects except the Maki-Thompson terms in extrinsic effect, which are proportional to $(\epsilon-\gamma_\varphi)^{-1}\ln\epsilon$ with a cutoff $\gamma_\varphi$ in two-dimensional systems. We found that the fluctuation effects on the extrinsic spin Hall effect have an opposite sign to that in the normal state, while those on the intrinsic spin Hall effect have the same sign.
\end{abstract}
\maketitle
\section{Introduction} \label{introduction-section1}
Spin current has opened a new venue to manipulate condensed matter systems. Spin transport experiments were firstly conducted by Tedrow and Meservey\cite{Tedrow1971}, who demonstrated that current flow across a ferromagnet-superconductor interface was spin-polarized. Subsequently, Aronov discussed that spin injection from ferromagnet to nonmagnetic metals could be used to amplify the ESR signals\cite{*[{}][{ [JETP Lett. $\bm{24}$, 32 (1976)].}] Aronov1976}. Johnson and Silsbee demonstrated that nonlocal response against the local charge current injection from ferromagnetic metal to nonmagnetic metal could be utilized to measure the spin relaxation time\cite{Johnson1985}. In the nonlocal response, a major role is played by the propagation of non-conserved spin current over a mesoscopic scale termed a spin diffusion length. In addition to finding efficient ways for spin injection and study of nonlocal response due to spin diffusion, spin-charge conversion (spin Hall effect \cite{Dyakonov1971,Hirsch1999,Zhang2000,Takahashi2002,Engel2005,Tse2006,Murakami2003,Sinova2004,Kato2004,Wunderlich2005,valenzuela2006,Saitoh2006,
Sinova2006,Sinova2015} and spin galvanic effect\cite{Edelstein1990,Ganichev2002}) is also an important issue in physics of spin transport. The spin Hall effect is categorized into two groups according to the origin, viz, extrinsic spin Hall effect \cite{Dyakonov1971, Hirsch1999, Zhang2000} caused by the spin-orbit interaction in the disorder potential and intrinsic spin Hall effect \cite{Murakami2003, Sinova2004} that occurs in a perfect crystal with the electric band structure split by the spin-orbit interaction.    

Those issues in spin transport have been addressed not only in normal metals but in superconductors\cite{Yang2021}. 
A theory of spin current injected into superconductors by taking account of charge imbalance and spin imbalance of quasiparticles was developed in 1995\cite{Zhao1995}. In 2012, H\"{u}bler et al. reported spin transport in superconducting Al, over distances of several microns, exceeding the normal-state spin-diffusion length and the charge-imbalance length\cite{Hubler2012}. Wakamura et al. observed spin-relaxation times in superconducting Nb, which is four times longer than that in the normal state\cite{Wakamura2014}. Wakamura et al. \cite{Wakamura2015} reported, in another paper, inverse spin Hall effect (ISHE), conversion from spin current to charge current in superconductors NbN. 
Recently, several efficient ways of injection of spin current into superconductors near the transition temperature $T_{\rm c}$ have been discussed theoretically in refs~\cite{Inoue2017, Kato2019} in terms of spin-pumping, spin-Seebeck effect, and strong coupling between spin and energy in spin-splitting quasiparticles\cite{Ojajarvi2021}. Jeon et al. reported that the conversion efficiency of magnon spin to quasiparticle charge in superconducting Nb via ISHE is enhanced compared with that in the normal state near $T_{\rm c}$ 
\cite{Jeon2018,Jeon2020}. Enhancement of the ISHE signal was observed even at the temperatures up to twice $T_{\rm c}$\cite{Jeon2020}. Those experimental results imply the importance of superconducting fluctuation effects on spin transport near $T_{\rm c}$. We note that earlier theoretical studies\cite{Takahashi2002,Kontani2009,Takahashi2012} but one \cite{Taira2021} have focused on spin Hall effect in superconductors below $T_{\rm c}$.

Fluctuation effects on transport properties above the superconducting transition temperature $T_{\rm c}$ were firstly studied by Aslamazov and Larkin\cite{Aslamazov1968}, Maki\cite{Maki1968a}, and Thompson\cite{Thompson1970} on electric conductivity\cite{Larkinlate2005}. Dominant fluctuation processes contributing to electric conductivity are the charge transport by the fluctuating Cooper pairs \cite{Aslamazov1968}, the reduction of the density of states by the presence of the fluctuating Cooper pairs, and the scattering of electrons by the fluctuation of Cooper pairs. These three processes are represented by the Feynman diagrams, each of which is called the Aslamazov-Larkin(AL) terms, DOS terms, and Maki-Thompson (MT) terms, respectively. The MT terms for electric conductivity contain the anomalous part, which diverges for all temperatures above $T_{\rm c}$ in one- and two-dimensional systems. This anomalous part is cut off by the phase-breaking parameter with various origins such as the paramagnetic impurities, magnetic fields, inelastic phonon scattering, and nonlinear fluctuation effects (See Secs. 8.3.3 and 8.3.4 in \cite{Larkinlate2005}).  
Depending on the phase-breaking parameter, either the AL terms or the MT terms dominantly contribute to electric conductivity. The superconducting fluctuation effect of extrinsic anomalous Hall effects\cite{Nagaosa2010}, which is closely related to the spin Hall effect, was studied by Li and Levchenko\cite{Li2020a}. In the ref.~\cite{Taira2021}, the spin Hall effect in the presence of a magnetic field was studied by consideration of the AL terms cooperating with the Hartree approximation.  

In this paper, we discuss the fluctuation effects on the spin Hall effect in superconductors above $T_{\rm c}$ in the absence of magnetic fields with the lowest order processes of the fluctuation propagator. We study the extrinsic spin Hall effect in the two and three-dimensional electron gas by incorporating the superconducting fluctuations in the model used by Tse and Das Sarma\cite{Tse2006}. We also investigate the intrinsic spin Hall effect by taking account of the superconducting fluctuations in the model used by Sinova et al. \cite{Sinova2004} viz, the two-dimensional electron gas with the Rashba spin-orbit interaction.  

Our main results are summarized as follows:
\begin{itemize}
    \item Extrinsic spin Hall effect in two dimensional electron gas [Sec.~\ref{extrinsic-section1}]: 
    In the presence of the side jump process, the DOS terms contribute dominantly and suppress the spin Hall conductivity in the normal state. They have a logarithmic divergence $\ln(1/\epsilon)$ in the limit $\epsilon=(T-T_{\mathrm{c}})/T_{\mathrm{c}} \rightarrow +0$. In the presence of the skew scattering process, either the DOS terms or the MT terms contribute dominantly and suppress the spin Hall conductivity in the normal state, depending on the relative ratio between three dimensionless parameters: $\epsilon$, the phase-breaking parameter $\gamma_\varphi$, and $T \tau / \hbar$ with the impurity scattering time $\tau$. The DOS terms have the logarithmic divergence $\ln(1/\epsilon)$ while the MT terms are proportional to $(\epsilon-\gamma_\varphi)^{-1}\ln(1/\epsilon)$. The results are summarized in Tables~\ref{discussion-table1} and \ref{discussion-table2} in Sec.~\ref{discussion-section0}.  
    \item Extrinsic spin Hall effect in three-dimensional electron gas [Sec.~\ref{extrinsic-section1}]: In the presence of the side jump process, the AL, DOS, and MT terms have no singularity concerning $\epsilon$. 
    The DOS terms contribute dominantly and suppress the spin Hall conductivity in the normal state. In the presence of the skew scattering process, the MT terms are proportional to 
    $(\epsilon^{1/2}+\gamma_\varphi^{1/2})^{-1}$ while the AL and DOS terms have no singularity. The MT terms thus contribute dominantly and suppress the spin Hall conductivity in the normal state. 

    \item Intrinsic spin Hall effect in Rashba model [Sec.~\ref{intrinsic-section1}]:  
The AL, DOS, and MT terms have logarithmic divergence $\ln(1/\epsilon)$ with the coefficients of the same order. The sum of these contributions has the same sign as the spin Hall conductivity in the normal state and enhances the spin Hall conductivity. The results are summarized in Tables~\ref{discussion-table3} in Sec.~\ref{discussion-section0}.
\end{itemize}

The rest of the present paper is organized as follows. The next section summarizes the extrinsic and intrinsic spin Hall effect in the normal state and fluctuation propagator. 
 In Sec.~\ref{extrinsic-section1}, we address the fluctuation effects on extrinsic spin Hall effects in the presence of side jump and skew scattering processes in two- and three-dimensional electron gas.  
 In Sec.~\ref{intrinsic-section1}, we discuss the fluctuation effects on intrinsic spin Hall effects in two-dimensional electron gas with Rashba spin-orbit interaction. In Sec.~\ref{discussion-section0}, we discuss singularity near $T_{\rm c}$ and magnitude of fluctuation contribution in AL, DOS, and MT terms.  We also raise several issues to be addressed in the future. In Sec.~\ref{conclusion-section1}, we conclude the present study. We defer the details of derivation in Secs.~\ref{extrinsic-section1} and \ref{intrinsic-section1} to supplemental materials. We list the symbols used in the main text and supplemental materials in the appendix. 
 
 Throughout this paper, we set the Boltzmann constant to be unity (i.e., $k_{\rm B}=1$) and take the electric charge of the carriers to be negative ($-e<0$). 

\section{Preliminaries: spin Hall effect in the normal state and fluctuation propagator}
This section aims to define the models, fix the notations, and summarize earlier results in the form seamless to the calculation presented in the following sections. 

\subsection{Extrinsic spin Hall effect}
In this section, we summarize the extrinsic spin Hall effect in the two-dimensional and three-dimensional electron gas following \cite{Tse2006}.
\subsubsection{Hamiltonian, Spin current density, and Charge current density}
We start with the following single-particle Hamiltonian:
\begin{align}
	H 
	=&
	\frac{(\bm{p} + e\bm{A})^2}{2m} - \frac{\lambda_0^2}{4\hbar} \qty[\bm{\sigma} \cp \grad \mathcal{V}(\bm{r})] \vdot (\bm{p} + e\bm{A})	 + \mathcal{V}(\bm{r}) \\
    \equiv&
	H_0 + H_{\mathrm{SO}} + \mathcal{V},\label{eq: H-normal-extrinsic}
\end{align}
which consists of the kinetic energy $H_{0}$, the spin-orbit interaction $H_{\mathrm{SO}}$ and the potential energy due to impurities $\mathcal{V}$.
$\mathcal{V}(\bm{r})$ can be written as $\mathcal{V}(\bm{r}) = \sum_{i} \mathcal{V}_{\mathrm{single}}(\bm{r} - \bm{R}_i)$, where $\mathcal{V}_{\mathrm{single}}(\bm{r})$ is a potential energy of single impurity and $\bm{R}_{i}$ is the position of the $i$-th impurity.
In Eq.~\eqref{eq: H-normal-extrinsic}, $\bm{A}$ denotes the spatially uniform vector potential. The symbol $\bm{\sigma} = {}^t(\sigma^x, \sigma^y, \sigma^z)$ denotes the Pauli matrices. The symbol $\lambda_0$, which has the dimension of length, is the coupling strength of spin-orbit interaction. 

The random average of the impurity potential is given by
\begin{align}
    \begin{aligned}
    	\expval{\mathcal{V}(\bm{r}_1)\mathcal{V}(\bm{r}_2)} =& n_{\rm i} v_0^2 \delta(\bm{r}_1 - \bm{r}_2)\\
    	\expval{\mathcal{V}(\bm{r}_1)\mathcal{V}(\bm{r}_2)\mathcal{V}(\bm{r}_3)} =& n_{\rm i} v_0^3 \delta(\bm{r}_1 - \bm{r}_2)\delta(\bm{r}_2 - \bm{r}_3),
    \end{aligned}\label{eq: randon-average}
\end{align}
where $n_{\rm i}$ is the density of impurities and $v_0$ is the uniform component of the Fourier transform of $\mathcal{V}_{\mathrm{single}}(\bm{r})$
\begin{align}
	v_0 = \int d\bm{r}\; \mathcal{V}_{\mathrm{single}}(\bm{r}).
\end{align}
The single-particle charge current is given by 
\begin{align}
	\bm{J}_{\rm c} = -e\bm{u}=-e\left(\frac{\bm{p}+ e\bm{A}}{m} - \frac{\lambda_0^2}{4\hbar} \bm{\sigma}\cp \grad \mathcal{V}\right),
\end{align}
in terms of the velocity $\bm{u}$
\begin{align}
	\bm{u}
	=&
	\frac{i}{\hbar} \comm{H}{\bm{r}} =
	\frac{\bm{p}+ e\bm{A}}{m} - \frac{\lambda_0^2}{4\hbar} \bm{\sigma}\cp \grad \mathcal{V}. 
\end{align}
We define the single-particle spin current by 
\begin{align}
	\bm{J}_{\rm s} = -\frac{e}{4} \pb{\sigma^z}{\bm{u}}=-e\qty(\frac{\bm{p}+ e\bm{A}}{2m} \sigma^z - \frac{\lambda_0^2}{8\hbar} \hat{\bm{z}} \cp \grad{\mathcal{V}} I) 
	 \qquad (\text{$I$ is the $2\times 2$ unit matrix}). \label{def-of-Js}
\end{align}
Each term in the Hamiltonian Eq.~\eqref{eq: H-normal-extrinsic} is rewritten in the second quantized form  as
\begin{align}
	H_0
	=&
	\sum_{\alpha \bm{k} } \psi_{\bm{k}\alpha}^\dag \frac{1}{2m} \qty(\hbar \bm{k} + e \bm{A}) ^2 \psi_{\bm{k} \alpha} 
\end{align}
\begin{align}
	H_{\mathrm{SO}}
		=&
	-\frac{i\lambda_0^2}{ 4V} \sum_{\alpha \beta} \sum_{\bm{k} \bm{k}'} \psi_{\bm{k}\alpha}^\dag (\bm{k}\cp \bm{k}') \vdot \bm{\sigma}_{\alpha \beta} \mathcal{V}_{\bm{k}-\bm{k}'} \psi_{\bm{k}'\beta }
	-
	\frac{ie\lambda_0^2}{4\hbar V} \sum_{\alpha \beta} \sum_{\bm{k}\bm{k}'} ((\bm{k}' - \bm{k}) \cp \bm{\sigma}_{\alpha \beta})\vdot \bm{A}\mathcal{V}_{\bm{k} - \bm{k}'} \psi_{\bm{k} \alpha}^\dag \psi_{\bm{k}' \beta}
	\label{eq: HSOC-second}
\end{align}
\begin{align}
	\mathcal{V} 
	=&
	\frac{1}{V} \sum_\alpha \sum_{\bm{k}\bm{k}'} \mathcal{V}_{\bm{k}-\bm{k}'} \psi_{\bm{k} \alpha}^\dag \psi_{\bm{k}' \alpha}, 
\end{align}
in terms of the creation $\psi^\dagger_{\bm{k}\alpha}$  and annihilation $\psi_{\bm{k}\alpha}$ operators of one-particle state with the wavevector $\bm{k}$ and $z$-component of spin $\alpha=\uparrow,\downarrow$. The symbol $V$ denotes the volume of the system. 

The spin current density is written in terms of the field operators
$\psi_\alpha	^\dag (\bm{r}),\psi_\alpha	(\bm{r})$ in the second quantized form as 
\begin{align}
	\bm{j}_{\rm s}(\bm{r}) 
	=&
	-e \sum_{\alpha \beta} \psi_\alpha	^\dag (\bm{r}) \qty(\frac{\bm{p} + e \bm{A}}{2m}\sigma^z_{\alpha \beta} - \frac{\lambda_0^2}{8\hbar} \hat{\bm{z}} \cp \grad V \delta_{\alpha \beta}) \psi_\beta (\bm{r})\label{eq: js-def}
\end{align}
and the uniform component of the Fourier transform of Eq.~\eqref{eq: js-def}, which we denote by $\bm{j}_{\rm s}$, is written as
\begin{align}
	\bm{j}_{\mathrm{s}}
	=&
-\frac{e}{2} \sum_{\alpha \beta} \sum_{\bm{k}\bm{k}'} \psi_{\bm{k} \alpha}^\dag \qty[\frac{\hbar \bm{k}}{m} \sigma_{\alpha \beta}^z \delta_{\bm{k}\bm{k}'} - \frac{i\lambda_0^2}{4\hbar V} \hat{\bm{z}} \delta_{\alpha \beta} \cp (\bm{k}-\bm{k}') \mathcal{V}_{\bm{k}-\bm{k}'} + \frac{e}{m} \bm{A} \sigma_{\alpha \beta}^z \delta_{\bm{k}\bm{k}'}] \psi_{\bm{k}' \beta}\label{eq: spin_current}\\
	\equiv&
	\bm{j}_{\mathrm{s}}^{(1)} + \bm{j}_{\mathrm{s}}^{(2)} + \bm{j}_{\mathrm{s}}^{(3)}.  
\end{align}
The charge current density is written in the second quantized form as 
\begin{equation}
\bm{j}_{\rm c}(\bm{r})=-e\sum_{\alpha}\psi_\alpha ^\dag(\bm{r}) \left(\frac{\bm{p}+e\bm{A}}{m}\right)\psi_\alpha (\bm{r})
+\frac{e\lambda_0^2}{4\hbar} \sum_{\alpha \beta} \bm{\sigma}_{\alpha \beta} \cp \grad \mathcal{V}(\bm{r}) \psi_\alpha ^\dag(\bm{r}) \psi_\beta (\bm{r}).
\end{equation}
The uniform component of the Fourier transform of charge current density, which we denote by $	\bm{j}_{\rm c}
$, is given by
\begin{align}
\bm{j}_{\mathrm{c}}
	=&
	-e\sum_{\alpha \beta} \sum_{\bm{k}\bm{k}'} \psi_{\bm{k} \alpha}^\dag \qty[\frac{\hbar \bm{k}}{m} \delta_{\alpha \beta} \delta_{\bm{k}\bm{k}'} - \frac{i\lambda_0^2}{4\hbar V} \bm{\sigma}_{\alpha \beta} \cp (\bm{k}-\bm{k}') \mathcal{V}_{\bm{k}-\bm{k}'} + \frac{e}{m} \bm{A} \delta_{\alpha \beta} \delta_{\bm{k}\bm{k}'}] \psi_{\bm{k}' \beta}\label{eq: charge_current}\\
	\equiv&
	\bm{j}_{\mathrm{c}}^{(1)} + \bm{j}_{\mathrm{c}}^{(2)} + \bm{j}_{\mathrm{c}}^{(3)}.
\end{align}
\subsubsection{Spin Hall conductivity}
We consider the spin current density against the uniform electric field with the use of the Kubo formula. The random average of the impurity potential Eq.~\eqref{eq: randon-average} becomes
\begin{align}
	\expval{\mathcal{V}_{\bm{k}}\mathcal{V}_{\bm{k}'}}
	=&
	n_{\rm i} v_0^2 V \delta_{\bm{k} + \bm{k}', 0}
\end{align}
and \begin{align}
	\expval{\mathcal{V}_{\bm{k}}\mathcal{V}_{\bm{k}'}\mathcal{V}_{\bm{k}''}} 
	=&
	n_{\rm i} v_0^3 V \delta_{\bm{k} + \bm{k}' + \bm{k}'', 0},
\end{align}
in the Fourier space.

The spin Hall coefficient (spin Hall conductivity) $\sigma_{xy}$ defined through $j_{\mathrm{s}\mu} = \sum_{\nu} \sigma_{\mu\nu}E_\nu$ with electric field $\bm{E}$ can be obtained as
\begin{align}
    \sigma_{\mu\nu} (\bm{q}, \omega) 
	=&
	\frac{\Phi_{\mu\nu}(\bm{q}, \hbar \omega+i\delta) - \Phi_{\mu\nu}(\bm{q}, i\delta)}{i(\omega+i\delta)} \label{general-sigma}\\
    	\Phi_{\mu\nu}(\bm{q}, i\omega_\nu)
	=&
	\frac{1}{V} \int_0^{\beta}\expval{T_u[j_{\mathrm{s},\mu,\bm{q}}(u) j_{\mathrm{c},\nu,-\bm{q}}(0)]} e^{i\omega_\nu u} du \label{general-phi},
\end{align}
where $u$ is the imaginary time and $\omega_\lambda = 2\pi T \lambda \;(\lambda \in \mathbb{Z})$ is the Bosonic  Matsubara frequency. 

\begin{figure}[htbp]
	\centering
\includegraphics[width=0.8\textwidth,page=1]{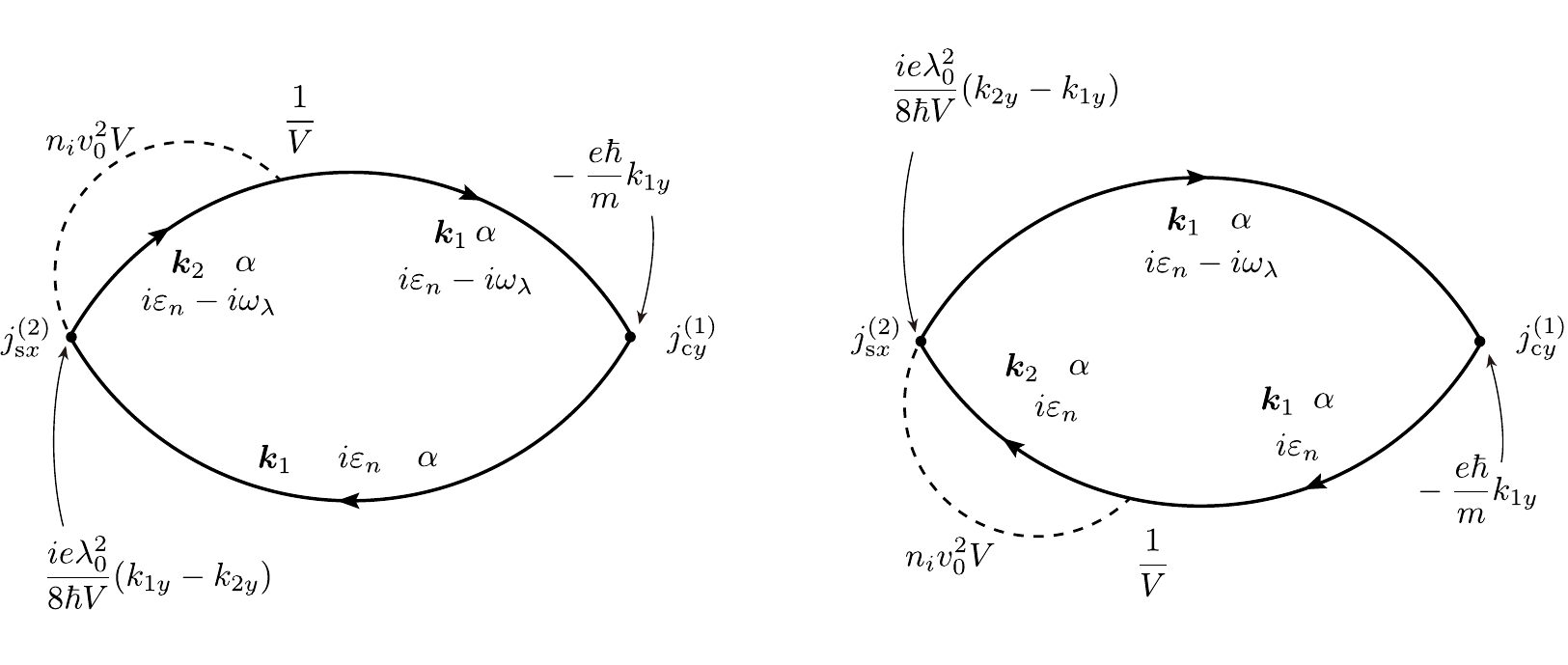}
	\caption{Feynman diagram of a contribution from the side jump to $\sigma_{xy}$ in the normal state.  The spin current vertex and charge current vertex come, respectively, $\bm{j}_{{\rm s}x}^{(2)}$ and $\bm{j}_{{\rm c}y}^{(1)}$.}
	\label{das-sarma-fig1}
\end{figure}

\begin{figure}[htbp]
	\centering
\includegraphics[width=0.8\textwidth,page=2]{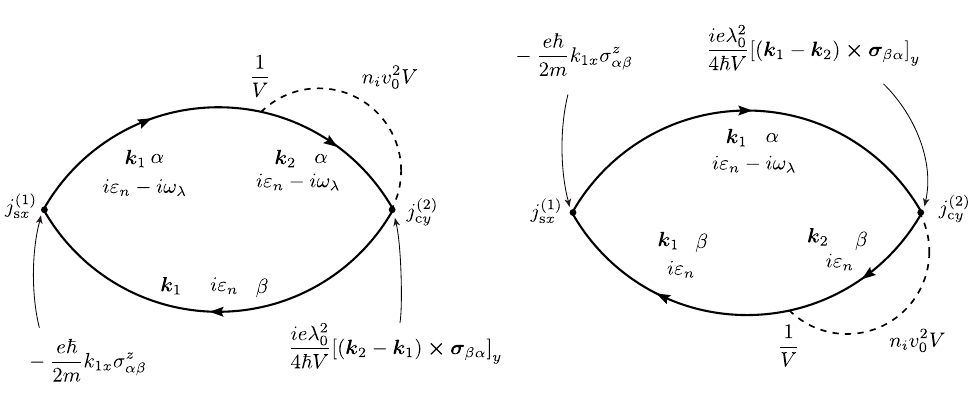}
	\caption{Feynman diagram of a contribution from the side jump to $\sigma_{xy}$ in the normal state.  The spin current vertex and charge current vertex, respectively,  come from $\bm{j}_{{\rm s},x}^{(1)}$ and $\bm{j}_{{\rm c},y}^{(2)}$.}
	\label{das-sarma-fig4}
\end{figure}

\begin{figure}[htbp]
	\centering
\includegraphics[width=0.8\textwidth,page=3]{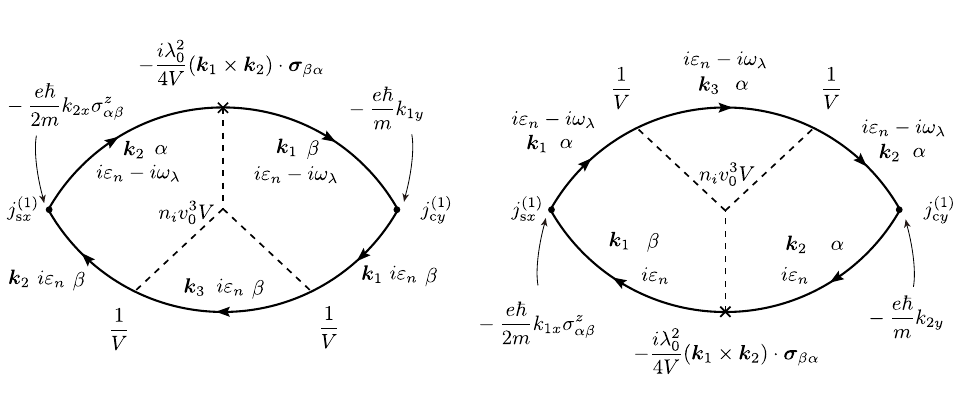}
	\caption{Feynman diagram of a contribution from the skew scattering to $\sigma_{xy}$ in the normal state. }
	\label{das-sarma-fig2}
\end{figure}

In the Feynman diagrams contributing to $\sigma_{xy}$ in the normal state, the spin-orbit interaction enters predominantly through either side jump (Figs.~\ref{das-sarma-fig1} and \ref{das-sarma-fig4}) or the skew scattering (Fig.~\ref{das-sarma-fig2})\cite{Tse2006}. 

The quantities assigned to the vertices and the dotted line are inscribed in the Figures. 
The solid line in the Figures corresponds to the free electron propagator 
\begin{equation}
	\mathcal{G}_\alpha (\bm{k}, i\varepsilon_n) 
	=
	\frac{1}{i\varepsilon_n - \varepsilon_{\bm{k}} + \mu + \frac{i\hbar}{2\tau} \sign(\varepsilon_n)}.\label{das-sarma-eq10}
\end{equation}
Here $\varepsilon_n=2\pi T(n+1/2)$ is the fermionic Matsubara frequency, $\varepsilon_{\bm{k}}=\hbar^2 k^2/(2m)$ and $\mu$, respectively, the kinetic energy and chemical potential of free electron. The mean free time $\tau$ is given by
\begin{equation}
\tau = \frac{\hbar}{2\pi N (0) n_{\rm i} v_0^2}, \label{das-sarma-tau}
\end{equation}
with the density of states $N(0)$ per spin at the Fermi surface in the normal state.
The dotted lines represent the random average of the impurity potentials, and the cross marks represent spin-orbit interactions.

Following the standard procedures, we can obtain spin Hall conductivity in the normal state
\begin{align}
	\sigma_{xy}^{\mathrm{SJ(normal)}} 
	=&
	\frac{e^2 \hbar}{2D m} \lambda_0^2 k_{\mathrm{F}}^2 N(0) \label{das-sarma-eq3}\\
	\sigma_{xy}^{\mathrm{SS(normal)}}
	=&
	\frac{\pi e^2 \hbar^2}{2D^2 m^2} \lambda_0^2 v_0 k_{\mathrm{F}}^4 N(0)^2 \tau, \label{das-sarma-eq4}
\end{align}
with $D=2,3$.  
\subsection{Intrinsic spin Hall effect}
We summarize the earlier results on intrinsic spin Hall effect developed by Sinova et al. \cite{Sinova2004} for the two-dimensional electron gas with Rashba spin-orbit interactions.

\subsubsection{Hamitonian, Spin current density, and Charge current density}
The Hamiltonian of the two-dimensional electron gas (2DEG) with Rashba-type spin-orbit interaction is written as 
\begin{align}
	H
	=&
	\frac{\bm{p}^2}{2m} I - \frac{\lambda}{\hbar} \bm{\sigma}\vdot (\hat{\bm{z}} \cp \bm{p})\\
	=&
	\mqty(\frac{\bm{p}^2}{2m} & \frac{\lambda}{\hbar} (p_y + ip_x) \\ \frac{\lambda}{\hbar} (p_y - ip_x) & \frac{\bm{p}^2}{2m}), 
\end{align}
where the region of 2DEG is taken to be $xy$ plane, which is perpendicular to the unit vector $\hat{\bm{z}}$ parallel to $z$-axis. $I$, $\bm{\sigma}={}^t(\sigma^x,\sigma^y,\sigma^z)$ are, respectively, the 2 by 2 unit matrix and the Pauli matrices in the spin space. 
The symbol $\lambda$ represents the coupling parameter of spin-orbit interaction, which is different from $\lambda_0$ used in extrinsic Hall effect.
The eigenvalues of this Hamiltonian $H$ are given by
\begin{equation}
	\varepsilon_{\pm}(\bm{p})
	=
	\frac{p^2}{2m} \pm \frac{\lambda}{\hbar} p.
\end{equation}
The eigenvectors are given by
\begin{align}
    \ket{\chi_\pm(\bm{p})}
    \equiv
	\frac{1}{\sqrt{2}} \mqty(\pm \frac{1}{p} (p_y + ip_x) \\ 1)
\end{align}
in spin space.
Let $\psi_{\bm{k}\sigma}$ be the Fourier transform of the field operator $\psi_\sigma(\bm{r})$ for annihilation of particle at the position $\bm{r}$ with $z$-component of spin $\sigma$. This operator $\psi_{\bm{k}\sigma}$ is expanded in terms of the annihilation operator $c_{\bm{p}\alpha}$ of the eigenstate of the Hamiltonian with spin state $\ket{\chi_{\alpha}(\bm{p})}$ for $\alpha=\pm$ as

\begin{align}
	\psi_{\bm{k}\sigma}
	=&
	\sum_{\alpha = \pm} \frac{1}{\sqrt{2}} \qty[\alpha \delta_{\sigma \uparrow} \frac{1}{k} (k_y + ik_x) + \delta_{\sigma\downarrow}] c_{\bm{k}\alpha} \label{sinova-eq2}.
\end{align}
We define the spin current density as $\bm{J}_{\mathrm{s}} = - (e / 4) \acomm{\sigma^z}{\bm{u}}$
in the same way as \eqref{def-of-Js} but differently from Sinova et al.(2004)  \cite{Sinova2004}.  The velocity operator $\bm{u}$ is given by 
\begin{equation}
    \bm{u}=\frac{i}{\hbar}\comm{H}{\bm{r}}
    =
    \frac{\bm{p}}{m} - \frac{\lambda}{\hbar} \bm{\sigma} \cp \hat{\bm{z}},
\end{equation}
which is followed by 
\begin{equation}
	\bm{J}_{\mathrm{s}}
    =
	-e \frac{\bm{p}}{2m} \sigma^z. 
	\label{eq: js-intrinsic-r}
\end{equation}
The uniform component of the Fourier transform of Eq.~\eqref{eq: js-intrinsic-r} is given by  \begin{align}
	\bm{j}_{\mathrm{s}}
	=&
	-e \sum_{\sigma \bm{k}} \frac{\hbar \bm{k}}{2m} \sigma_{\sigma \sigma}^{z} \psi_{\bm{k} \sigma}^\dag \psi_{\bm{k}\sigma},
\end{align}
which is rewritten as
\begin{equation}
	\bm{j}_{\mathrm{s}}
	=
	e\sum_{\bm{k}\alpha}\frac{\hbar \bm{k}}{2m} c_{\bm{k}\alpha}^\dag c_{\bm{k}-\alpha}. 
\end{equation}
The charge current density $\bm{j}_{\mathrm{c}}$ is given by the sum of the free electron part due to the kinetic energy and that due to the spin-orbit interaction $H_{\mathrm{SO}}$. We focus on the latter, which we denote by $\bm{j}_{\mathrm{c}}^{\rm SO}(\bm{r})$. The spin orbit interaction $H_{\mathrm{SO}}$ is written in the form of the second quantization
\begin{align}
	H_{\mathrm{SO}}
	=&
	-\frac{\lambda}{\hbar} \sum_{\sigma \sigma'}\int d\bm{r} \; \psi_\sigma^\dag(\bm{r}) \bm{\sigma}_{\sigma \sigma'} \vdot \qty(\hat{\bm{z}} \cp \frac{\hbar}{i}\grad)\psi_{\sigma'}(\bm{r}).
\end{align}
From the relation $\div \bm{j}_{\mathrm{c}}^{\rm SO}(\bm{r}) =  -(ie / \hbar) \comm{\sum_{\sigma} \psi^\dag_{\sigma}(\bm{r}) \psi_{\sigma}(\bm{r})}{H_{\mathrm{SO}}}$ and
\begin{equation}
-\frac{ie}{\hbar}\comm{\sum_{\sigma} \psi^\dag_{\sigma}(\bm{r}) \psi_{\sigma}(\bm{r})}{H_{\mathrm{SO}}}
=
\div \qty[\frac{e\lambda}{\hbar} \sum_{\sigma \sigma'} \qty(\bm{\sigma}_{\sigma \sigma'} \cp \hat{\bm{z}}) \psi_\sigma^\dag(\bm{r})\psi_{\sigma'}(\bm{r})],
\end{equation}
we obtain
\begin{align}
	\bm{j}^{\mathrm{SO}}_{\mathrm{c}} (\bm{r})
	=&
	\frac{e\lambda}{\hbar} \sum_{\sigma \sigma'} (\bm{\sigma}_{\sigma \sigma'} \cp \hat{\bm{z}}) \psi_\sigma^\dag(\bm{r}) \psi_{\sigma'}(\bm{r}). 
\end{align}
The uniform component of the Fourier transform of $\bm{j}^{\mathrm{SO}}_{\mathrm{c}} (\bm{r})$ is given by
\begin{align}
	\bm{j}_{\mathrm{c}}^{\mathrm{SO}}
	=&
	\frac{e\lambda}{\hbar} \sum_{\sigma \sigma'} (\bm{\sigma}_{\sigma \sigma'} \cp \hat{\bm{z}}) \sum_{\bm{k}} \psi_{\bm{k}\sigma}^\dag \psi_{\bm{k} \sigma'}. 
\end{align}
By adding the part $\bm{j}_{\mathrm{c}}^{(1)}$ due to kinetic energy, the uniform component of the Fourier transform of $\bm{j}_{\mathrm{c}} (\bm{r})$ is given by 
\[
\bm{j}_{\mathrm{c}} = -\sum_{\bm{k}\sigma} e \frac{\hbar \bm{k}}{m} \psi_{\bm{k} \sigma}^\dag \psi_{\bm{k} \sigma}
+
	\frac{e\lambda}{\hbar} \sum_{\sigma \sigma'} (\bm{\sigma}_{\sigma \sigma'} \cp \hat{\bm{z}}) \sum_{\bm{k}} \psi_{\bm{k}\sigma}^\dag \psi_{\bm{k} \sigma'}
.
\]
The $y$-component of  $\bm{j}_{\mathrm{c}}$  is expressed in terms of creation/annihilation operator of the energy eigenstates as
\begin{align}
	j_{\mathrm{c}y}
	=&
	\sum_{\bm{k}\alpha} \qty[-\frac{e\hbar}{m} k_y c_{\bm{k}\alpha}^\dag c_{\bm{k}\alpha} + \frac{e\lambda}{\hbar}\qty(-\frac{k_y}{k} \alpha c_{\bm{k}\alpha}^\dag c_{\bm{k}\alpha} + i \frac{k_x}{k} \alpha c_{\bm{k}\alpha}^\dag c_{\bm{k}-\alpha})].
\end{align}
\subsubsection{Spin Hall conductivity}
\begin{figure}[tbp]
	\centering
	\includegraphics[width=0.5\textwidth, page=2]{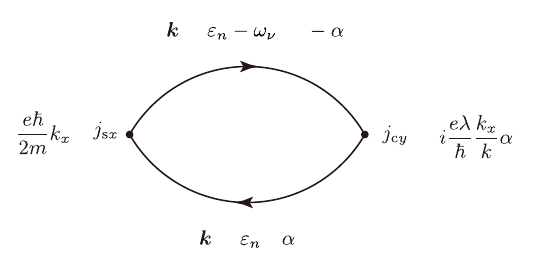}
\caption{Feynman diagram for the lowest order process for intrinsic spin Hall effect.}
	\label{sinova-fig1}
\end{figure}

Figure \ref{sinova-fig1} shows the diagram representing the lowest order process for spin Hall effect. 
The response function $\Phi_{xy}(\bm{q}, i\omega_\nu)$ and the spin Hall conductivity $\sigma_{xy}(\bm{q}, \omega)$ are obtained by using Eq.~\eqref{general-sigma} and Eq.~\eqref{general-phi} as in the extrinsic case.
It then suffices to calculate the response function, which is written as 
\begin{align}
	\Phi_{xy}(0,i\omega_\nu)
	=&
	-\frac{1}{V} \sum_{\alpha} \sum_{\bm{k}}T\sum_{n} \qty(\frac{e\hbar}{2m} k_x) \qty(i\frac{e\lambda}{\hbar}  \frac{k_x}{k} \alpha)
	\mathcal{G}_{\alpha}(\bm{k}, i\varepsilon_n) \mathcal{G}_{-\alpha}(\bm{k},i\varepsilon_n-i\omega_\nu)\\
	=&
	-\frac{1}{V}\sum_{\alpha} \alpha \sum_{\bm{k}} \frac{ie^2\lambda}{2m} \frac{k_x^2}{k} 
	\frac{f\bm{(}\varepsilon_\alpha(\bm{k})\bm{)} - f\bm{(}\varepsilon_{-\alpha}(\bm{k})\bm{)}}{\varepsilon_{\alpha}(\bm{k}) - \varepsilon_{-\alpha}(\bm{k}) - i\omega_\nu}
	\end{align}
in terms of the green function
\begin{equation}
	\mathcal{G}_{\alpha}(\bm{k}, i\varepsilon_n)
	=
	\frac{1}{i\varepsilon_n - \qty(\frac{\hbar^2 k^2}{2m} + \alpha \lambda k) + \mu}. \label{sinova-eq3}
\end{equation}
With use of Eq.~\eqref{general-sigma},
%
%
%
 $\sigma_{xy}(0, \omega\to 0)$ becomes \cite{Sinova2004}
\begin{align}
	\sigma_{xy}(0, \omega\to 0)
	\approx&
	-\frac{\hbar e^2}{16\pi m \lambda} \int_0^{\infty} dk \; 2\lambda k f'\qty(\frac{\hbar^2k^2}{2m})
	=
	\frac{e^2}{8\pi\hbar} f(0)
	\approx
	\frac{e^2}{8\pi\hbar} \label{sinova-eq4}
\end{align}
in the case where $\lambda k_{\mathrm{F}} \ll \varepsilon_{\mathrm{F}}$.
Note that $\sigma_{xy}(0, \omega\to 0)
$ does not depend on $\lambda$.

Exceptional simplicity of this model, viz, combination of parabolic band dispersion and linear momentum dependence of the spin-orbit interaction makes the spin Hall conductivity vanish in the presence of spin-conserving impurities, even in the limit of weak scatterers\cite{Inoue2004,Schliemann2004,Mishchenko2004,Rashba2004,Dimitrova2005,Murakami2006,Sinova2006,Raimondi2012}. 
However, this model is of importance to address the two-body interaction effect on the spin Hall effect\cite{Dimitrova2005,Shekhter2005}. We will thus adopt the Rashba model with an attractive short-range interaction in sec.~\ref{intrinsic-section1} to consider the superconducting fluctuation contribution to the spin Hall effect.  The superconducting property of this model was discussed in \cite{Gorkov2001,[{}][{ [Sov. Phys. JETP $\bm{68}$ 1244 (1989)].}]Edelstein1989}.  

\subsection{Cooperon and fluctuation propagator}
In this subsection, we give an overview of Cooperon and superconducting fluctuations, following the description in reference \cite{Larkinlate2005}.
We add the BCS-type two-body attractive interaction to the Hamiltonian $H_0$ in the normal state Eq.~\eqref{eq: H-normal-extrinsic}
\begin{equation}
H_{\rm int}=-\frac{g}{V} \sum_{\bm{k} \bm{k}' \bm{q}} 
	\psi_{\bm{k} + \bm{q},\uparrow}^\dag \psi_{-\bm{k}, \downarrow}^\dag \psi_{-\bm{k}',\downarrow} \psi_{\bm{k}' + \bm{q}, \uparrow}\label{eq: Hint}  
\end{equation}
in order to take account of the three types of the superconducting fluctuation effect. 
We define the Cooperon and fluctuation propagator of superconductivity in this subsection. 

\begin{figure}[htbp]
	\centering
	\includegraphics[width=1\textwidth, page=2]{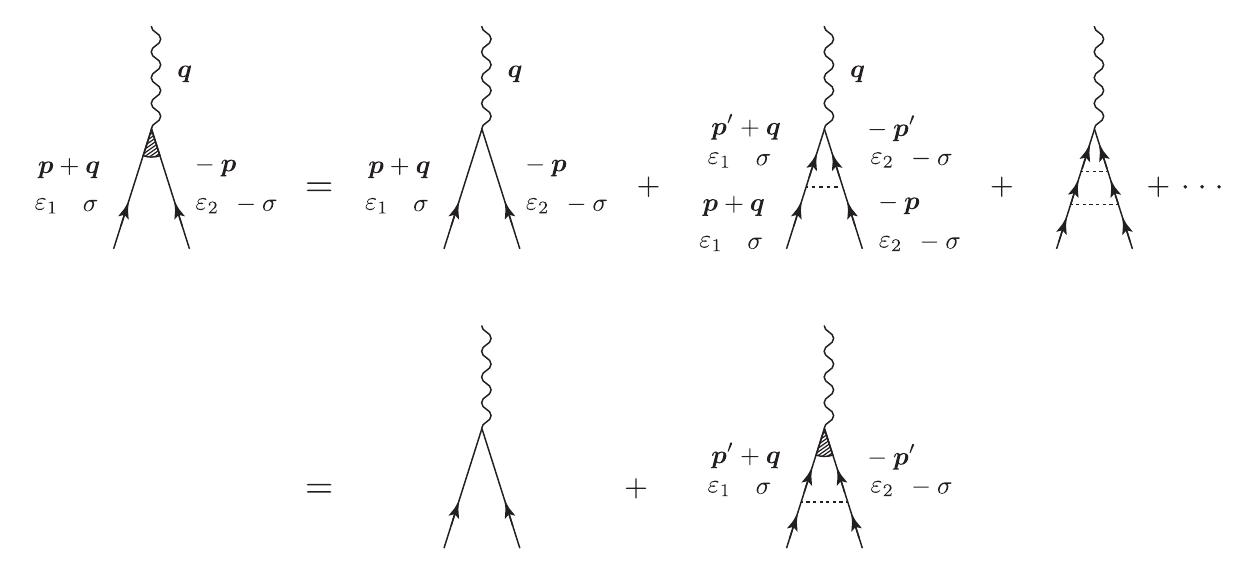}
	\caption{$C(\bm{q}, \varepsilon_1, \varepsilon_2)$.}
	\label{fluctuation-fig2}
\end{figure}

The shaded region represents the Cooperon $C(\bm{q},\varepsilon_1,\varepsilon_2)$, which is defined through the recursive relation
\begin{align}
	C(\bm{q}, \varepsilon_1, \varepsilon_2)
	=&
	1+ \frac{1}{V^2} n_i v_0^2 V \sum_{\bm{p}} \mathcal{G}(\bm{p}+\bm{q}, i\varepsilon_1) \mathcal{G}(-\bm{p}, i\varepsilon_2) C(\bm{q}, \varepsilon_1, \varepsilon_2)
\end{align}
or graphically by Fig.~\ref{fluctuation-fig2}. 
Explicit expression for $C(\bm{q},\varepsilon_1,\varepsilon_2)$ is give by
\begin{align}
	C(\bm{q}, \varepsilon_1, \varepsilon_2)
	=&
	\frac{\abs{\widetilde{\varepsilon}_1 - \widetilde{\varepsilon}_2}}{\abs{\varepsilon_1 - \varepsilon_2} + \frac{\hbar}{\tau} \frac{\expval{(\hbar \bm{v}_{\mathrm{F}} \vdot \bm{q})^2}_{\mathrm{F.S.}}}{\abs{\widetilde{\varepsilon}_1 - \widetilde{\varepsilon}_2}^2}\theta(-\varepsilon_1 \varepsilon_2)}.
\end{align}
From this expression, we see that $C(\bm{q}, \varepsilon_1, \varepsilon_2)$ becomes small unless  
\begin{equation}
\frac{\expval{(\hbar \bm{v}_{\mathrm{F}} \vdot \bm{q})^2}_{\mathrm{F.S.}}}{\tau / \hbar \abs{\widetilde{\varepsilon}_1 - \widetilde{\varepsilon}_2}^2} \lesssim \abs{\varepsilon_1 - \varepsilon_2} \label{eq: condition-for-Cooperon}   
\end{equation}
 is satisfied. The condition \eqref{eq: condition-for-Cooperon} reduces to 
\begin{align}
	q
	\ll&
	\frac{1}{\hbar v_{\mathrm{F}}} \abs{\widetilde{\varepsilon}_1 - \widetilde{\varepsilon}_2} \sqrt{\frac{T \tau}{\hbar}}
	\sim
	\begin{dcases}
		\frac{T}{\hbar v_{\mathrm{F}}} \sqrt{\frac{T \tau}{\hbar}} & (\text{when }T \tau / \hbar \gg 1 )\\
		\frac{1}{\tau v_{\mathrm{F}}} \sqrt{\frac{T \tau}{\hbar}} & (\text{when }T \tau / \hbar \ll 1).
	\end{dcases} \label{fluctuation-eq5}
\end{align}
This implies that the integral region with $\bm{q}$ is practically reduced to the region where the condition \eqref{fluctuation-eq5} is satisfied. 

\begin{figure}[htbp]
	\centering
	\includegraphics[width=1\textwidth, page=1]{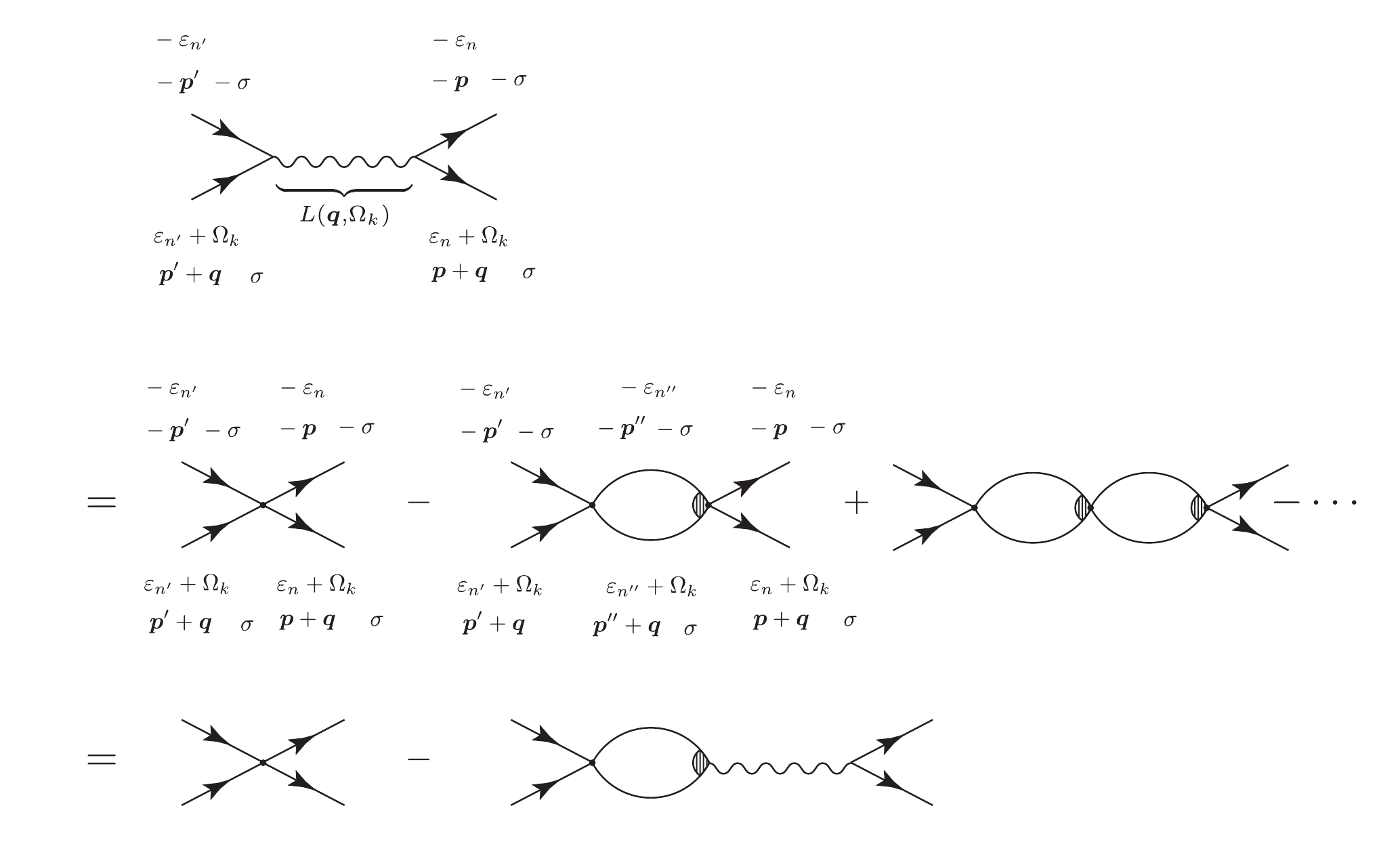}
	\caption{Diagramatical definition of $L(\bm{q}, i\Omega_k)$. }
	\label{fluctuation-fig1}
\end{figure}

The wavy lines represent the propagator of superconducting fluctuation, 
$L(\bm{q}, i\Omega_k)$, which is defined by Eq.~\eqref{eq: LqOmega-def} and graphically by Fig.~\ref{fluctuation-fig1}.
\begin{equation}
	L(\bm{q}, i\Omega_k)
	=
	-\frac{g}{V} + \frac{g}{V} T \sum_{n'', p''} \mathcal{G}(\bm{p}'' + \bm{q}, i\varepsilon_{n''+k}) \mathcal{G}(-\bm{p}'', -i\varepsilon_{n''})  C(\bm{q}, \varepsilon_{n''+k} , -\varepsilon_{n''}) L(\bm{q}, i\Omega_k). \label{eq: LqOmega-def}
\end{equation}
Explicit expression for $L(\bm{q}, i\Omega_k)$ is given by
\begin{equation}
	L(\bm{q}, i\Omega_k)
	=
	-\frac{1}{V N(0)} \frac{1}{\epsilon + \psi\qty(\frac{1}{2} + \frac{\abs{\Omega_k}}{4\pi T}) - \psi\qty(\frac{1}{2}) + \clength^2\bm{q}^2}
	\approx 
	-\frac{1}{V N(0)} \frac{1}{\epsilon + \frac{\pi \abs{\Omega_k}}{8 T} + \clength^2\bm{q}^2},
	\label{fluctuation-eq10}
\end{equation}
where we have introduced the coherence length $\clength$ and diffusion constant 
$\mathcal{D}$
\begin{align}
	\clength^2
	=
	\frac{\pi \hbar}{8T} \mathcal{D}
	=
	-\frac{v_{\mathrm{F}}^2 \tau^2}{D}\qty[\psi\qty(\frac{1}{2} + \frac{\hbar}{4\pi T \tau}) - \psi\qty(\frac{1}{2}) - \frac{\hbar}{4\pi T \tau} \psi^{(1)}\qty(\frac{1}{2})]. \label{fluctuation-eq12}
\end{align}
Here the di-Gamma function $\psi(z) \equiv \dv{z} \ln \Gamma(z) = \lim_{n_{\mathrm{c}} \to \infty} [- \sum_{n=0}^{n_{\mathrm{c}}-1} \frac{1}{n+z} + \ln n_{\mathrm{c}}]$ and $n$-th order derivative of $\psi(z)$, which we denote by $\psi^{(n)}(z)$ (polyGamma function)   are introduced. 
The coherence length reduces to 
\begin{align}
	\clength
	=&
	\begin{dcases}
		\sqrt{\frac{7\zeta(3)}{16\pi^2 D}} \frac{\hbar v_{\mathrm{F}}}{T} & (T \tau / \hbar \gg 1)\\
		\sqrt{\frac{\pi}{8D}} v_{\mathrm{F}}\tau \sqrt{\frac{\hbar}{T \tau}}& (T \tau / \hbar \ll 1)
	\end{dcases} \label{fluctuation-eq6}
\end{align}
in the dirty and clean limits, respectively. 
\section{Fluctuation effects on extrinsic spin Hall conductivity} \label{extrinsic-section1}
Near and above the superconducting transition temperature $T_{\rm c}$, we take account of the superconducting fluctuation via three types of the process; Aslamazov-Larkin (AL) term, the Maki-Thompson (MT) term, and the density of states (DOS) term. Those are known to be the most diverging in the electric conductivity when $T\rightarrow T_{\rm c}+0$.  

\subsection{Aslamazov-Larkin terms}

\begin{figure}[tbp]
	\begin{minipage}[b]{0.9\linewidth}
		\centering
		\includegraphics[scale=0.7, page=10]{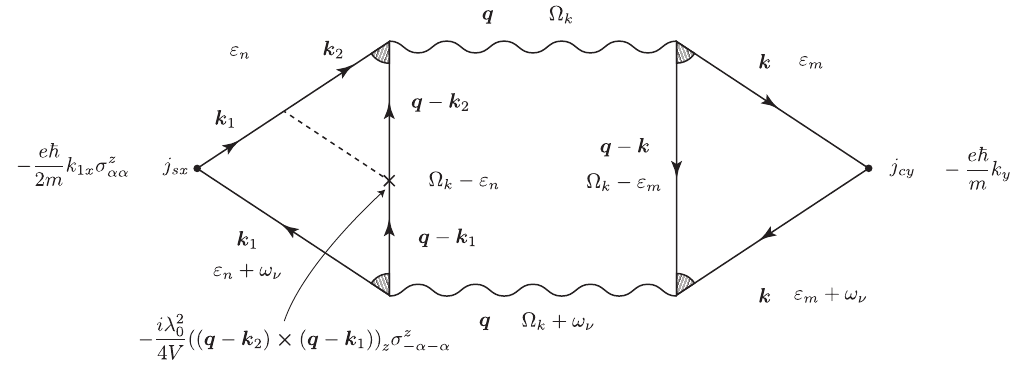}
		\\(a)
		\label{al-fig4a}
	\end{minipage}\\
	\begin{minipage}[b]{0.45\linewidth}
		\centering
		\includegraphics[scale=0.7, page=8]{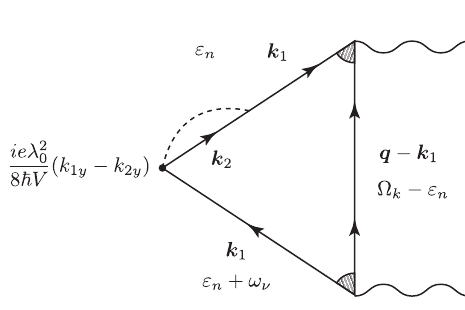}
		\\(b)
		\label{al-fig4b}
	\end{minipage}
	\begin{minipage}[b]{0.45\linewidth}
		\centering
		\includegraphics[scale=0.7, page=9]{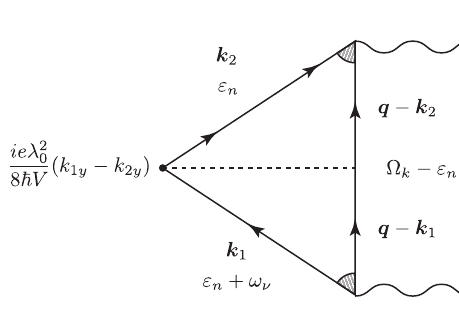}
		\\(c)
		\label{al-fig4c}
	\end{minipage}
	\caption{The diagram for the AL term in the presence of side jump.}
	\label{al-fig4}
\end{figure}
\begin{figure}[tbp]
	\begin{minipage}[b]{0.8\linewidth}
		\centering
		\includegraphics[width=\textwidth, page=1]{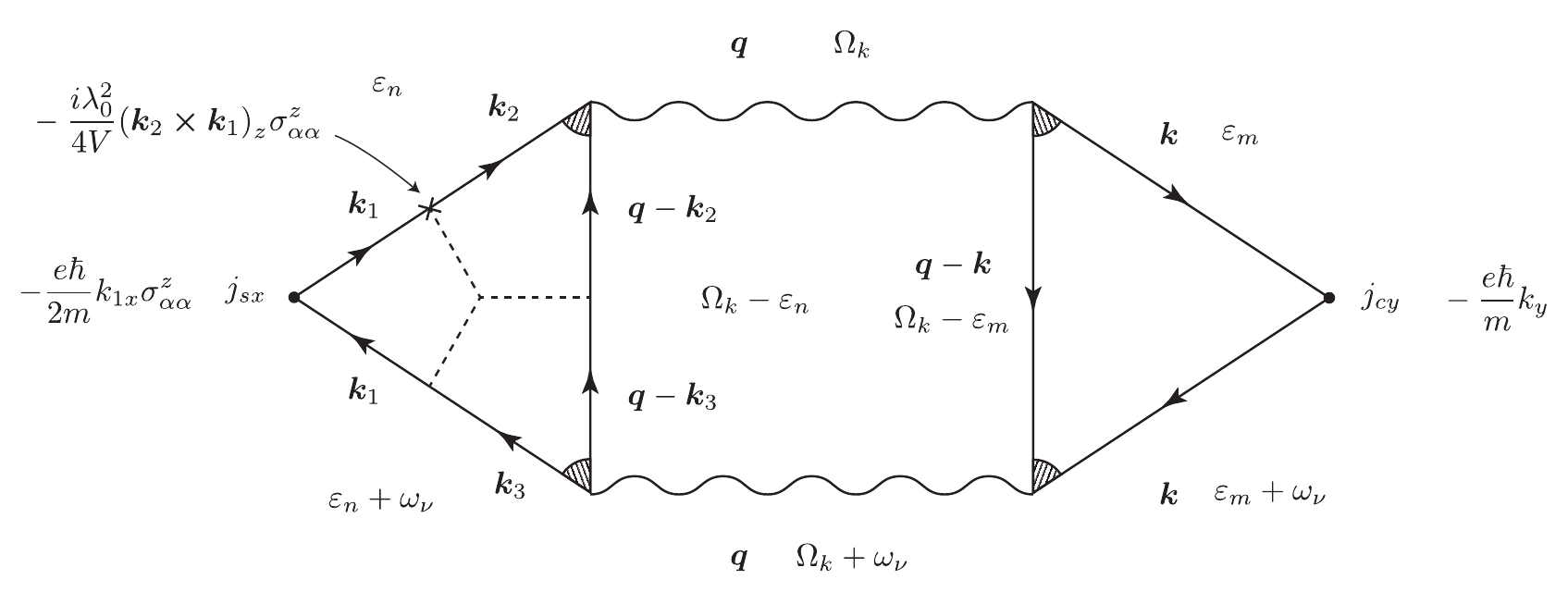}
		\\(a)
		\label{al-fig1a}
	\end{minipage}\\
	\begin{minipage}[b]{0.4\linewidth}
		\centering
		\includegraphics[scale=0.7, page=2]{extrinsic_AL.pdf}
		\\(b)
		\label{al-fig1b}
	\end{minipage}
	\begin{minipage}[b]{0.4\linewidth}
		\centering
		\includegraphics[scale=0.7, page=3]{extrinsic_AL.pdf}
		\\(c)
		\label{al-fig1c}
	\end{minipage}\\
	\begin{minipage}[b]{0.4\linewidth}
		\centering
		\includegraphics[scale=0.7, page=4]{extrinsic_AL.pdf}
		\\(d)
		\label{al-fig1d}
	\end{minipage}
	\caption{The diagram for the AL term in the presence of skew scattering.}
	\label{al-fig1}
\end{figure}

In this section, we consider the Hamiltonian $H = H_0 + H_{\mathrm{SO}} + \mathcal{V} + H_{\mathrm{int}}$, which is the sum of Eq.~\eqref{eq: H-normal-extrinsic} and Eq.~\eqref{eq: Hint}.
The Feynman diagrams [of the charge current-spin current correlation function $\Phi_{xy}^{\mathrm{AL}}(0, i\omega_\nu)$] for the AL terms with side jump are shown in Fig.~\ref{al-fig4} and those for AL terms with skew scattering are shown in Fig.~\ref{al-fig1}. We denote $\mathcal{B}_{\mathrm{c}}(\bm{q},\omega,\Omega)$ by the triangular part containing the charge current vertex (i.e. renormalized charge current vertex) and $\mathcal{B}_{\mathrm{s}}(\bm{q},\omega,\Omega)$ by the renormalized spin current vertex, respectively. The response function is then given by 
\begin{align}
	\Phi_{xy}(0, i\omega_\nu)
	=&
	\frac{1}{V}\sum_{\bm{q}} T \sum_{\Omega_k} L(\bm{q}, i\Omega_k) L(\bm{q}, i\Omega_k + i\omega_\nu)
	\mathcal{B}_{\mathrm{s}}(\bm{q},i\Omega_k, i\omega_\nu) \mathcal{B}_{\mathrm{c}}(\bm{q},i\Omega_k, i\omega_\nu).
	\label{eq: al-eq9-}
\end{align}
The $\order{\omega}$ contributions in Eq.~\eqref{eq: al-eq9-} yields the spin Hall conductivity.  In the electric conductivity, we can deduce the main contribution in the AL term by setting $\omega=0$ and $\Omega=0$ in the charge current vertices $\mathcal{B}_{\rm c}(\bm{q},\omega,\Omega)$ and retaining $\order{\omega}$ part in the fluctuation propagators. 
In contrast, in the spin Hall conductivity, the contributions with $\mathcal{B}_{\rm s}(\bm{q},0,0)$ vanishes and thus we have to maintain the frequencies $\Omega,\omega$ to be finite. Accordingly, the procedure to calculate the spin Hall conductivity, which is given below, is slightly different from that for AL term in electric conductivity,
\begin{enumerate}
	\item List the relevant diagrams (Fig.~\ref{al-fig4}
	for AL+ side jump and Fig.~\ref{al-fig1} for AL + skew scattering) and write down the expressions for the charge current-spin current correlation function $\Phi_{xy}^{\mathrm{AL}}(0, i\omega_\nu)$. 
	\item Expand $\mathcal{B}_{\mathrm{c}}(\bm{q},\omega,\Omega)$ and $\mathcal{B}_{\mathrm{s}}(\bm{q},\omega,\Omega)$ with respect of $\bm{q}$ up to the first order as $\mathcal{B}_{\mathrm{s}}(\bm{q},\omega,\Omega)\approx -i\zeta_{\rm s}V q_y B_{\mathrm{s}}(\omega,\Omega)$ and $\mathcal{B}_{\mathrm{c}}(\bm{q},\omega,\Omega)\approx \zeta_{\rm c}q_y V B_{\mathrm{c}}(\omega,\Omega)$ with coefficients $\zeta_{\rm s}$ and $\zeta_{\rm c}$. 
	\item Perform integrals in the expressions for $B_{\mathrm{c}}$ and $B_{\mathrm{s}}$ with respect to internal wave vectors $\bm{k}_1,\bm{k}_2\cdots$. 
	\item Transform the sum with $\Omega_k$ to contour integral.
	\item Expand the resultant expression with respect to $\omega$ after analytic continuation $i\omega_\nu \to \hbar \omega$. 
	\item  Retain the most singular terms in the limit of $\epsilon \to 0$. 
	\item Perform summation in $B_{\mathrm{c}}(\bm{q},\omega,\Omega)$, $B_{\mathrm{s}}(\bm{q},\omega,\Omega)$ with respect to internal frequencies $\varepsilon_n$ and $\varepsilon_m$. 
	\item Integrate the resultant expression with respect to $\bm{q}$. 
\end{enumerate}
The details of calculation along these procedures are given in Supplemental materials. Here we outline the flow of calculations. After Step 3, 
the response function reduces to 
\begin{align}
	\Phi_{xy}(0, i\omega_\nu)
		\approx&
	-i\zeta_{\rm s}\zeta_{\rm c}
	V\sum_{\bm{q}} \frac{\bm{q}^2}{D} T \sum_{\Omega_k} L(\bm{q}, i\Omega_k) L(\bm{q}, i\Omega_k + i\omega_\nu)
	B_{\mathrm{s}}(i\Omega_k, i\omega_\nu) B_{\mathrm{c}}(i\Omega_k, i\omega_\nu).
	\label{al-eq9}
\end{align}

After Steps 4 and 5, we find that
\begin{align}
	&
	\qty[T \sum_{\Omega_k} L(\bm{q}, i\Omega_k) L(\bm{q}, i\Omega_k + i\omega_\nu) B_{\mathrm{c}}(i\Omega_k, i\omega_\nu) B_{\mathrm{s}}(i\Omega_k, i\omega_\nu)]_{i\omega_\nu \to \hbar \omega \approx 0}\\
	=&
	\hbar \omega \frac{T}{2} L(\bm{q}, 0)^2 B_{\mathrm{c}2}(0, 0) \dv{x} \qty[B_{\mathrm{s}2}(-x, x) + B_{\mathrm{s}2}(0, x)]_{x=0} \label{al-eq3+}
    +\hbar\omega(\mbox{regular terms in the limit $\epsilon \rightarrow +0$})+\order{\omega^2}.
\end{align}
Here $B_{\mathrm{c,s2}}(Z, z)$ 
is analytic continuation $(i\Omega_k\rightarrow Z,i\omega_\nu\rightarrow z) $ of $B_{\mathrm{c,s}}(i\Omega_k, i\omega_\nu)$ under the condition $-{\rm Im}z<{\rm Im} Z<0$. 

After Step 7, we obtain the expressions for $\dv{x} \qty[B_{\mathrm{s}2}(-x, x) + B_{\mathrm{s}2}(0, x)]_{x=0}$ and
$B_{\mathrm{c}2}(0, 0)$ in terms of $\hbar/\tau$ and $T$. 
After Step 8, we obtain 
\begin{align}
	V \sum_{\bm{q}} \bm{q}^2 L(\bm{q}, 0)^2
	\approx&
	\frac{1}{4\pi N(0)^2 \clength^4} \ln \frac{1}{\epsilon} \label{al-eq14}
\end{align}
for $D=2$ and
\begin{align}
	V\sum_{\bm{q}}\bm{q}^2 L(\bm{q}, 0)^2
	\approx&
	\frac{1}{2\pi^2 N(0)^2 \clength^5}\label{eq: al-eq14++}
\end{align}
for $D=3$.

Finally, the resultant expression for the AL term of the spin Hall conductivity in the side jump process is given by

\begin{align}
	\sigma_{xy}^{\mathrm{AL-SJ}}
	\approx&
	\frac{2e^2 \hbar}{Dm} \lambda_0^2 k_{\mathrm{F}}^2 \frac{T}{\varepsilon_{\mathrm{F}}} N(0)
	\begin{dcases}
		\frac{1}{2} \ln \frac{1}{\epsilon} & (D=2)\\
		\frac{1}{\pi \clength} & (D=3).
	\end{dcases}
\end{align}
For $D=2$, the ratio of fluctuation conductivity to that in the normal state is given by 
\begin{align}
	\sigma_{xy}^{\mathrm{AL-SJ}} / \sigma_{xy}^{\mathrm{SJ(normal)}}
=&
	2 \frac{T}{\varepsilon_{\mathrm{F}}} \ln \frac{1}{\epsilon}
\end{align}
with use of Eq.~\eqref{das-sarma-eq3}.

The expression for the AL term of the spin Hall conductivity in the skew scattering process is given by
\begin{align}
	\sigma_{xy}^{\mathrm{AL-SS}}
	=&
	\frac{8\pi^2 e^2 \hbar^7 \lambda_0^2 n_{\mathrm{i}} v_0^3 k_{\mathrm{F}}^6 N(0)^2 T}{D^4 m^4} \qty(\frac{\tau}{\hbar})^6 \nonumber\\
	&\times
	\qty[\frac{\hbar}{4\pi T \tau} \psi^{(1)}\qty(\frac{1}{2}) + \psi\qty(\frac{1}{2}) - \psi\qty(\frac{1}{2} + \frac{\hbar}{4\pi T \tau})] \nonumber\\
	&\times
	\qty[2 \frac{\hbar}{4\pi T \tau}\psi^{(1)}\qty(\frac{1}{2}) + \frac{\hbar}{4\pi T \tau} \psi^{(1)}\qty(\frac{1}{2} + \frac{\hbar}{4\pi T \tau}) + 3 \psi\qty(\frac{1}{2}) - 3\psi\qty(\frac{1}{2} + \frac{\hbar}{4\pi T \tau})]\nonumber\\
	&\times
	\begin{dcases}
		\frac{1}{4\pi \clength^4} \ln \frac{1}{\epsilon} & (D=2)\\
		\frac{1}{2\pi^2 \clength^5} & (D=3).
	\end{dcases}
	\label{eq: sigma-AL-SS}
\end{align}
In the dirty limit$(T\tau \ll 1)$ in $D=2$ systems, Eq.~\eqref{eq: sigma-AL-SS} reduces to 
\begin{align}
	\sigma_{xy}^{\mathrm{AL-SS}} / \sigma_{xy}^{\mathrm{SS(normal)}}
	=&
	4 \frac{T}{\varepsilon_{\mathrm{F}}} \ln \frac{1}{\epsilon}
\end{align}
with use of Eqs.~\eqref{das-sarma-tau}, \eqref{das-sarma-eq4}, \eqref{fluctuation-eq12}.
In the clean limit ($T\tau / \hbar \gg 1$) in the two dimensional systems $D=2$, Eq.~\eqref{eq: sigma-AL-SS} reduces to
\begin{align}
	&
	\sigma_{xy}^{\mathrm{AL-SS}} / \sigma_{xy}^{\mathrm{SS(normal)}}\nonumber
		=
	2 \frac{T}{\varepsilon_{\mathrm{F}}}\ln \frac{1}{\epsilon}.
\end{align}

\subsection{DOS terms}\label{subsec: Dos-Extrinsic}
\begin{figure}[tbp]
\begin{tabular}{cc}
		\begin{minipage}[b]{0.45\linewidth}
			\centering
			\includegraphics[width=\linewidth,page=1]{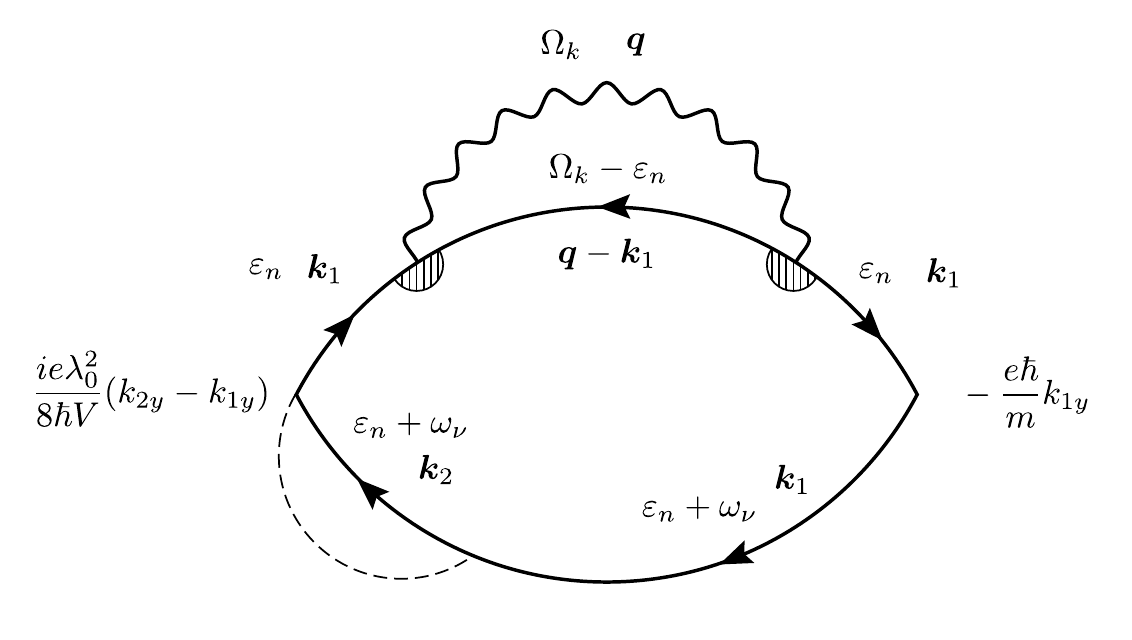}
		\\(a)
		\end{minipage}
		&
		\begin{minipage}[b]{0.45\linewidth}
			\centering
			\includegraphics[width=\linewidth,page=3]{extrinsic_DOS_sj.pdf}
		\\(b)
		\end{minipage}
		\\ & \\
		\begin{minipage}[b]{0.45\linewidth}
			\centering
			\includegraphics[width=\linewidth,page=5]{extrinsic_DOS_sj.pdf}
		\\(c)
		\end{minipage}
		&
		\begin{minipage}[b]{0.45\linewidth}
			\centering
			\includegraphics[width=\linewidth,page=7]{extrinsic_DOS_sj.pdf}
		\\(d)
		\end{minipage}
		\\ & 
\end{tabular}
	\caption{Diagrams for DOS term in the presence of side jump process.}
	\label{DOS-sj-figures}
\end{figure}

\begin{figure}[tbp]
	\begin{minipage}[b]{0.4\linewidth}
		\centering
		\includegraphics[width=\textwidth]{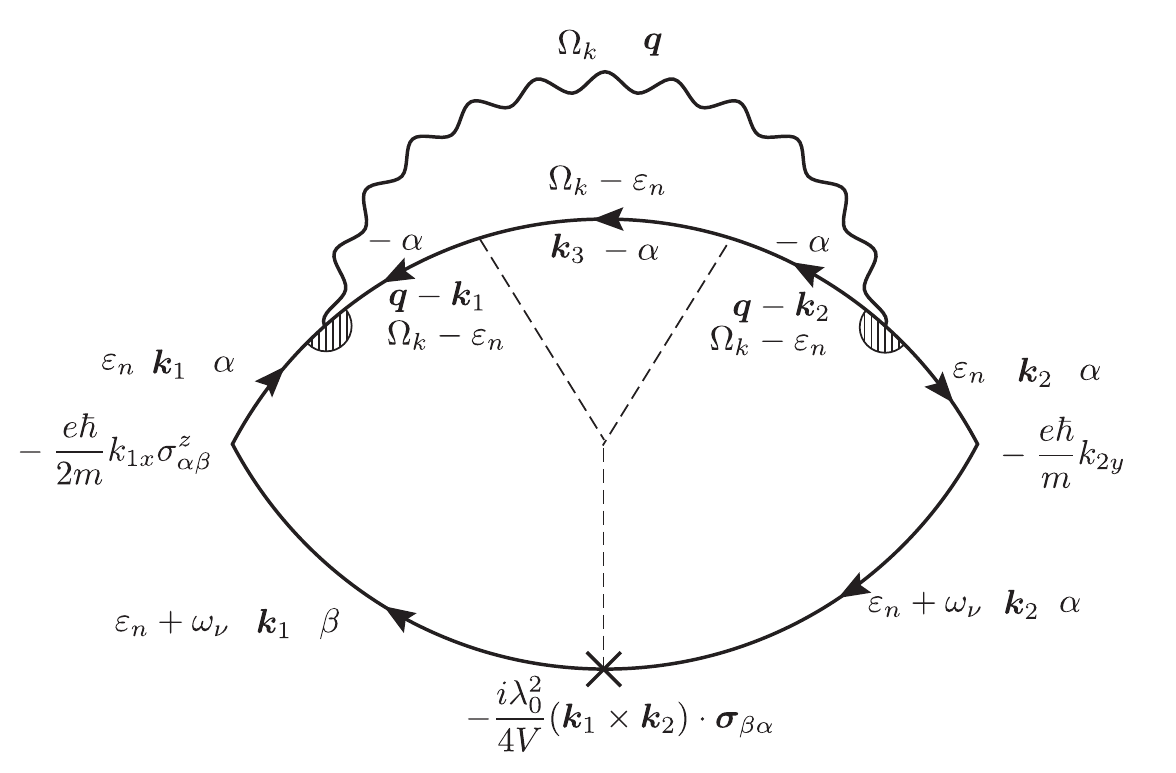}
		\\(a)
	\end{minipage}
	\begin{minipage}[b]{0.4\linewidth}
		\centering
		\includegraphics[width=\textwidth]{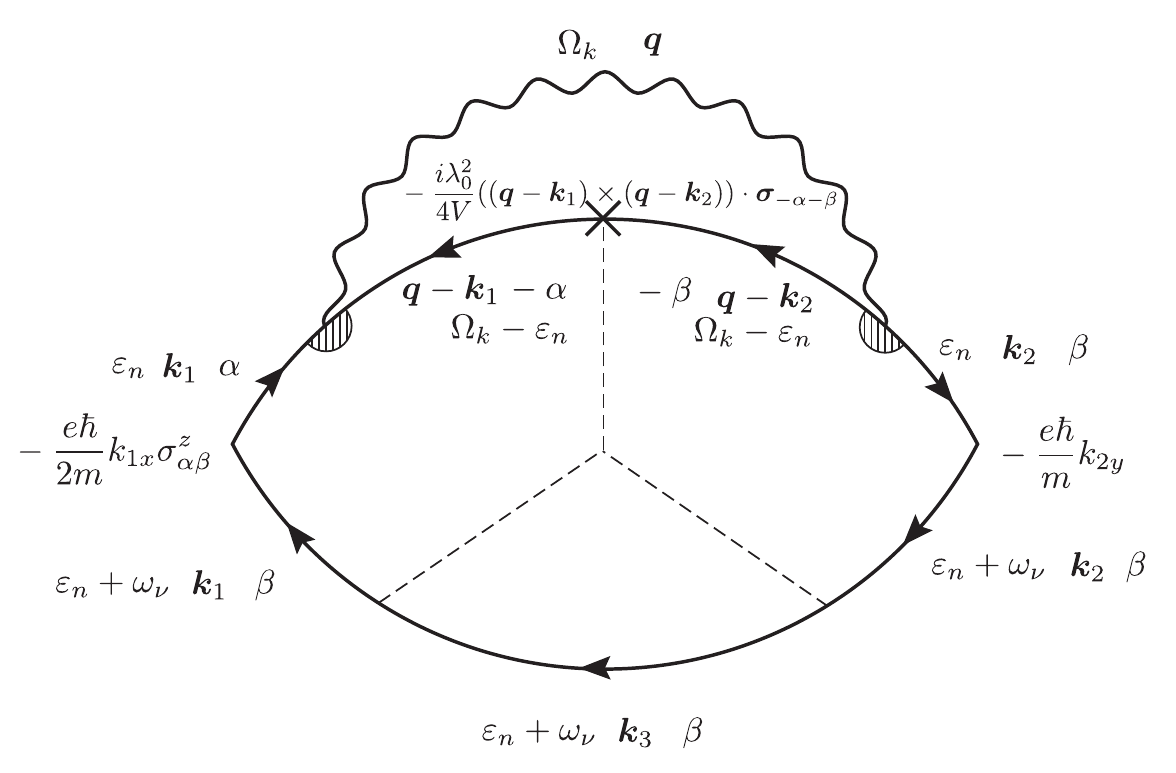}
		\\(b)
	\end{minipage}\\
	\hspace{5mm} \\
	\begin{minipage}[b]{0.4\linewidth}
		\centering
		\includegraphics[width=\textwidth]{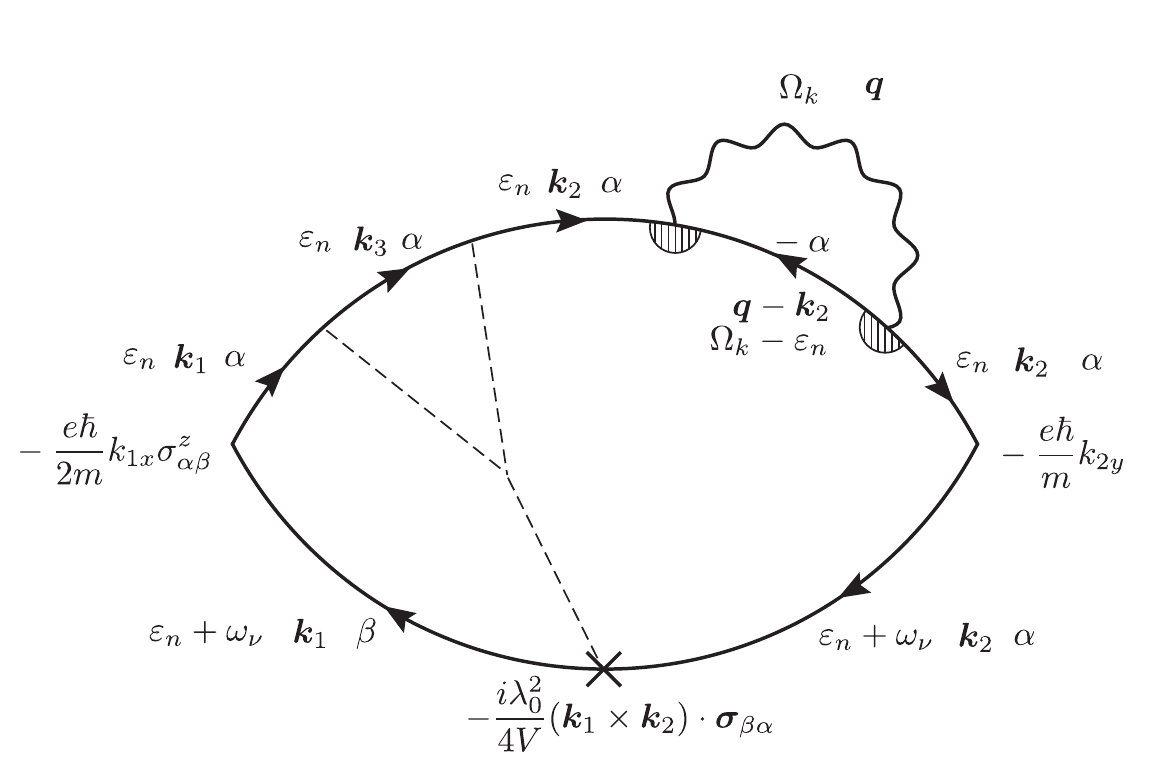}
		\\(c)
	\end{minipage}
	\begin{minipage}[b]{0.4\linewidth}
		\centering
		\includegraphics[width=\textwidth]{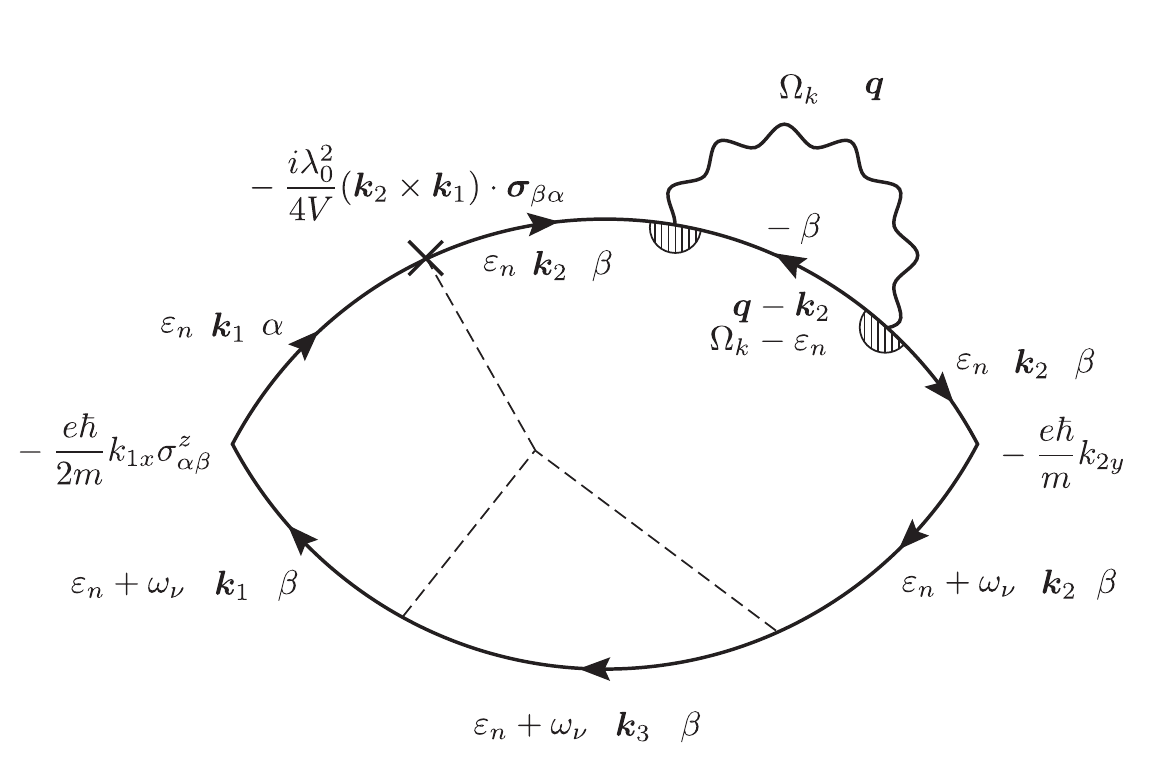}
		\\(d)
	\end{minipage}\\
	\hspace{5mm} \\
	\begin{minipage}[b]{0.4\linewidth}
		\centering
		\includegraphics[width=\textwidth]{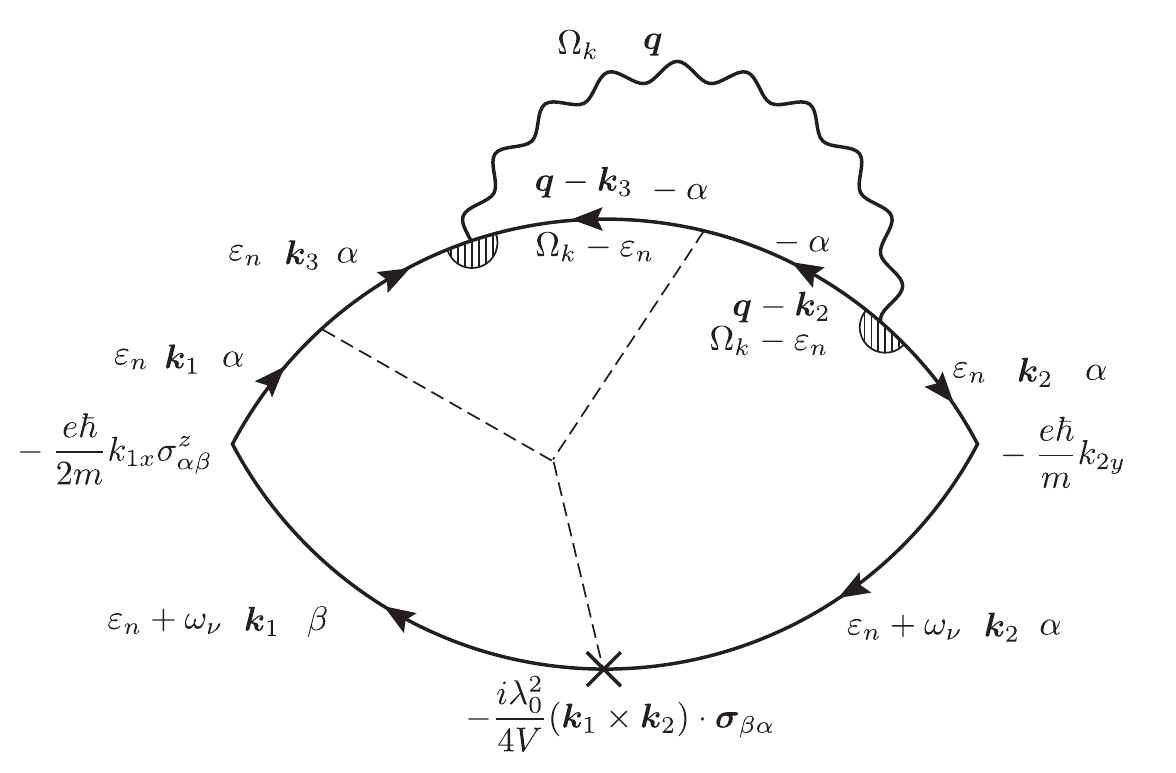}
		\\(e)
	\end{minipage}
	\begin{minipage}[b]{0.4\linewidth}
		\centering
		\includegraphics[width=\textwidth]{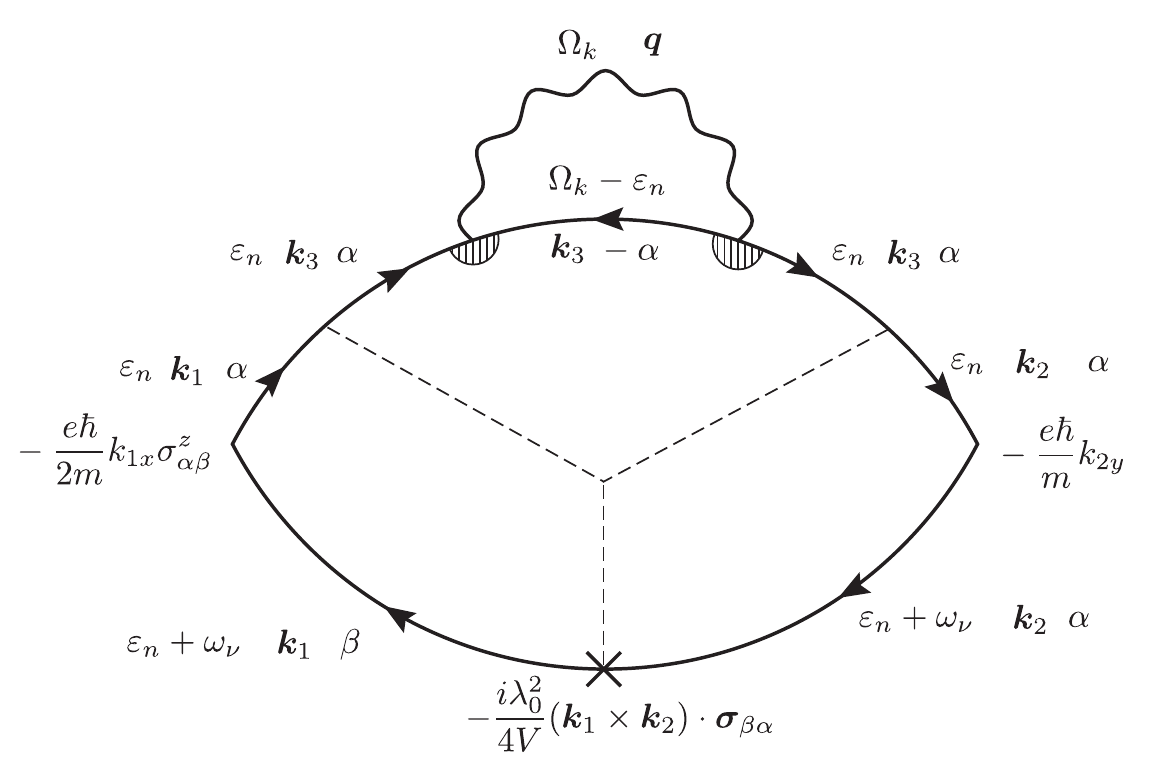}
		\\(f)
	\end{minipage}
	\caption{Diagrams of DOS terms in the presence of skew scattering.}
	\label{DOS-ss-figs}
\end{figure}

The DOS terms for the spin Hall conductivity are calculated in a way similar to those for electric conductivity. 
The procedure to calculate the spin Hall conductivity is given below. %
\begin{enumerate}
	\item List the relevant diagrams (Fig.~\ref{DOS-sj-figures} for DOS + side jump, and Fig.~\ref{DOS-ss-figs} for DOS + skew scattering) and write down the expressions for the charge current-spin current correlation function $\Phi_{xy}^{\mathrm{DOS}}(0, i\omega_\nu)$. 
\item Put $\Omega_k = 0$ in all quantities and $\bm{q}=0$ in all but $L(\bm{q}, i\Omega_k)$.
	\item Perform integration with respect to internal wave vectors $\bm{k}_1,\bm{k}_2\cdots$ with use of residue theorem. 
	\item Reduce the sum with $\varepsilon_n$ in the polygamma functions. 
	\item Expand the resultant expression with respect to $\omega$ after analytic continuation $i\omega_\nu \to \hbar \omega$. 
\item Perform integration in $L(\bm{q},0)$ with respect to $\bm{q}$.
\end{enumerate}

The details of calculations are given in Supplemental material. However, we here briefly outline the analysis. Figure~\ref{DOS-sj-figures} shows the diagrams of the DOS terms with the side jump process. The diagrams reflected vertically and horizontally give the same contributions as those shown here. 
The contributions from these diagrams to the spin Hall conductivity are put together as
\begin{align}
	\sigma_{xy}^{\mathrm{DOS-SJ}}
	=&
	\lim_{\omega\to 0} 
	\frac{e^2 \lambda_0^2 n_i v_0^2 k_{\mathrm{F}}^2 N(0)^2}{m D\omega} T\sum_{\Omega_k} \sum_{\bm{q}} L(\bm{q}, i\Omega_k) \Sigma_{xy}^{\mathrm{DOS-SJ}} (\bm{q}, i\Omega_k, \hbar\omega)\label{eq: sigmaxy-DOS-SJ-extrinsic}\\
	\Sigma_{xy}^{\mathrm{DOS-SJ}}(\bm{q}, i\Omega_k, i\omega_\nu)
	=&
	T \sum_{\varepsilon_n} C(\bm{q}, \varepsilon_n, \Omega_k-\varepsilon_n) ^2 (-I^{\rm a} + I^{\rm b} + I^{\rm c} + I^{\rm d}).
\end{align}
Here we introduce the notations $I^{j}(\bm{q}, \varepsilon_n, \Omega_k, \omega_\nu)$ with $j={\rm a,b,c,d}$, which we define as the integration of  the product of the Green functions obtained in the diagram $j$, where $\xi_{i} \equiv \frac{\hbar^2 \bm{k}_{i}^2}{2m} - \mu$.
For example, the expression for $I^{\rm a}(\bm{q}, \varepsilon_n, \Omega_k, \omega_\nu)$ is given by
\begin{equation}
I^a(\bm{q}, \varepsilon_n, \Omega_k, \omega_\nu) 
=
\int d\xi_1 \;\mathcal{G}(\bm{k}_1, i\varepsilon_n)^2 \mathcal{G}(\bm{k}_1, i\varepsilon_n+i\omega_\nu) \mathcal{G}(\bm{q}-\bm{k}_1, i\Omega_k-i\varepsilon_n) \int d\xi_2 \; \mathcal{G}(\bm{k}_2, i\varepsilon_n+i\omega_\nu).
\end{equation}
We retain the term with $\Omega_k=0$, which yields the most diverging contribution with respect to $\epsilon$. We further retain $\bm{q}$-dependence only in $L(\bm{q}, 0)$. 
With these simplification, Eq.~\eqref{eq: sigmaxy-DOS-SJ-extrinsic} becomes
\begin{align}
	\sigma_{xy}^{\mathrm{DOS-SJ}}
	\approx&
	\lim_{\omega\to 0} 
	\frac{e^2 \lambda_0^2 n_i v_0^2 k_{\mathrm{F}}^2 N(0)^2T}{m D \omega}\Sigma_{xy}^{\mathrm{DOS-SJ}} (0, 0, \hbar\omega)\left(\sum_{\bm{q}} L(\bm{q}, 0)\right).\label{eq: sigmaxy-DOS-SJ-extrinsic-reduced}
\end{align}
The part $\Sigma_{xy}^{\mathrm{DOS-SJ}} (0, 0, \hbar\omega)$ yields $\omega$-linear term 
\begin{align}
	\Sigma^{\mathrm{DOS-SJ}}_{xy} (0, 0, \omega)
	=&
	\hbar \omega \cdot 2\pi \qty(\frac{\tau}{\hbar})^3 
	\qty[
	2\psi\qty(\frac{1}{2}) - 2\psi \qty(\frac{1}{2} + \frac{\hbar}{4\pi T \tau})
		+ 3\frac{\hbar}{4\pi T \tau} \psi^{(1)} \qty(\frac{1}{2}) 
	-\qty(\frac{\hbar}{4\pi T \tau})^2 \psi^{(2)} \qty(\frac{1}{2} )
	] 
	+
	\order{\omega^2},\label{DOS-sj-eq2}
\end{align}
%
while the part $\left(\sum_{\bm{q}} L(\bm{q}, 0)\right)$ in Eq.~\eqref{eq: sigmaxy-DOS-SJ-extrinsic-reduced} contributes to singularity in the limit $\epsilon \rightarrow +0$, i. e .,   
\begin{align}
	\sum_{\bm{q}} L(\bm{q}, 0)
	\approx&
	-\frac{1}{N(0) \clength^D}
	\begin{dcases}
    	\frac{1}{4\pi} \ln \frac{1}{\epsilon} & (D=2)\\
    	\frac{1}{2\pi^2} & (D=3).
	\end{dcases}
\end{align}
Putting the above results, we arrive at the expression for the DOS terms for extrinsic spin Hall conductivity with the side jump process,
\begin{align}
	\sigma_{xy}^{\mathrm{DOS-SJ}}
	=&
	-\frac{2\pi e^2 \lambda_0^2n_{\mathrm{i}} v_0^2 k_F^2 N(0) T \hbar}{Dm \clength^D} \qty(\frac{\tau}{\hbar})^3\\
	&\times
	\qty[
	2\psi\qty(\frac{1}{2}) - 2\psi \qty(\frac{1}{2} + \frac{\hbar}{4\pi T \tau})
		+ 3\frac{\hbar}{4\pi T \tau} \psi^{(1)} \qty(\frac{1}{2}) 
	-\qty(\frac{\hbar}{4\pi T \tau})^2 \psi^{(2)} \qty(\frac{1}{2} )
	]\\ 
	&\times
	\begin{dcases}
		\frac{1}{4\pi} \ln \frac{1}{\epsilon} & (D=2)\\
		\frac{1}{2\pi^2} & (D=3).
	\end{dcases} \label{DOS-sj-eq8}
\end{align}
%
%
For $D=2$, the ratio of the spin Hall conductivity to that in the normal state is
\begin{align}
		\sigma_{xy}^{\mathrm{DOS-SJ}} / \sigma_{xy}^{\mathrm{SJ(normal)}}
	=&
	-\frac{7\zeta(3)}{\pi^3} \frac{\hbar / \tau}{\varepsilon_{\mathrm{F}}} \ln \frac{1}{\epsilon}\label{DOS-sj-normal-dirty}
\end{align}
in the dirty limit and 
\begin{align}
		\sigma_{xy}^{\mathrm{DOS-SJ}} / \sigma_{xy}^{\mathrm{SJ(normal)}}
		 =-\frac{2\pi^3}{7\zeta(3)} \frac{T}{\varepsilon_{\mathrm{F}}} \frac{T \tau}{\hbar} \ln \frac{1}{\epsilon}\label{DOS-sj-normal-clean}
\end{align}
in the clean limit. The opposite sign between the DOS terms and the spin Hall conductivity in the normal state is consistent with the suppression of the density of state by the superconducting fluctuation. 

The expression for the DOS term of the spin Hall conductivity in the skew scattering process is given by
\begin{align}
	\sigma^{\mathrm{DOS-SS}}_{xy}
	=&-\frac{e^2 \hbar^3 \lambda_0^2 n_{\mathrm{i}} v_0^3 N(0)^2 T}{2m^2 \clength^D} 
	\qty(\frac{k_{\mathrm{F}}^2}{D})^2 \\
	&\times
	8\pi^2 \qty(\frac{\tau}{\hbar})^4 \Biggl\{
	   \begin{aligned}[t]
		   -&
			3\qty[\psi\qty(\frac{1}{2} + \frac{\hbar}{4 \pi T \tau}) - \psi\qty(\frac{1}{2})]
			+
			\frac{\hbar}{4\pi T \tau} \qty[\psi^{(1)}\qty(\frac{1}{2} + \frac{\hbar}{4 \pi T \tau}) + 3\psi^{(1)}\qty(\frac{1}{2})]\\
			-&
			\qty(\frac{\hbar}{4\pi T \tau})^2 \psi^{(2)}\qty(\frac{1}{2})
			\Biggr\}
		\end{aligned}\\
	&\times
	\begin{dcases}
		\frac{1}{4\pi} \ln \frac{1}{\epsilon} & (D=2)\\
		\frac{1}{2\pi^2} & (D=3).
	\end{dcases}
\end{align}
For $D=2$, the ratio of the spin Hall conductivity to that in the normal state is
\begin{align}
	\sigma^{\mathrm{DOS-SS}}_{xy} / \sigma_{xy}^{\mathrm{SS(normal)}} 
	=&
	-\frac{14\zeta(3)}{\pi^3} \frac{\hbar / \tau}{\varepsilon_{\mathrm{F}}} \ln \frac{1}{\epsilon}.
\end{align}
in the dirty limit and  
\begin{align}
	\sigma_{xy}^{\mathrm{DOS-SS}} / \sigma_{xy}^{\mathrm{SJ(normal)}}
	=&
	 -\frac{4\pi^3}{7\zeta(3)} \frac{T}{\varepsilon_{\mathrm{F}}} \frac{T \tau}{\hbar} \ln \frac{1}{\epsilon}
\end{align}
in the clean limit. We notice again the opposite sign between the DOS terms and the spin Hall conductivity in the normal state. 

\subsection{Maki-Thompson terms}\label{subsec: Maki-Thompson}
\begin{figure}[tbp]
	\begin{minipage}[b]{0.45\linewidth}
		\centering
		\includegraphics[width=\textwidth]{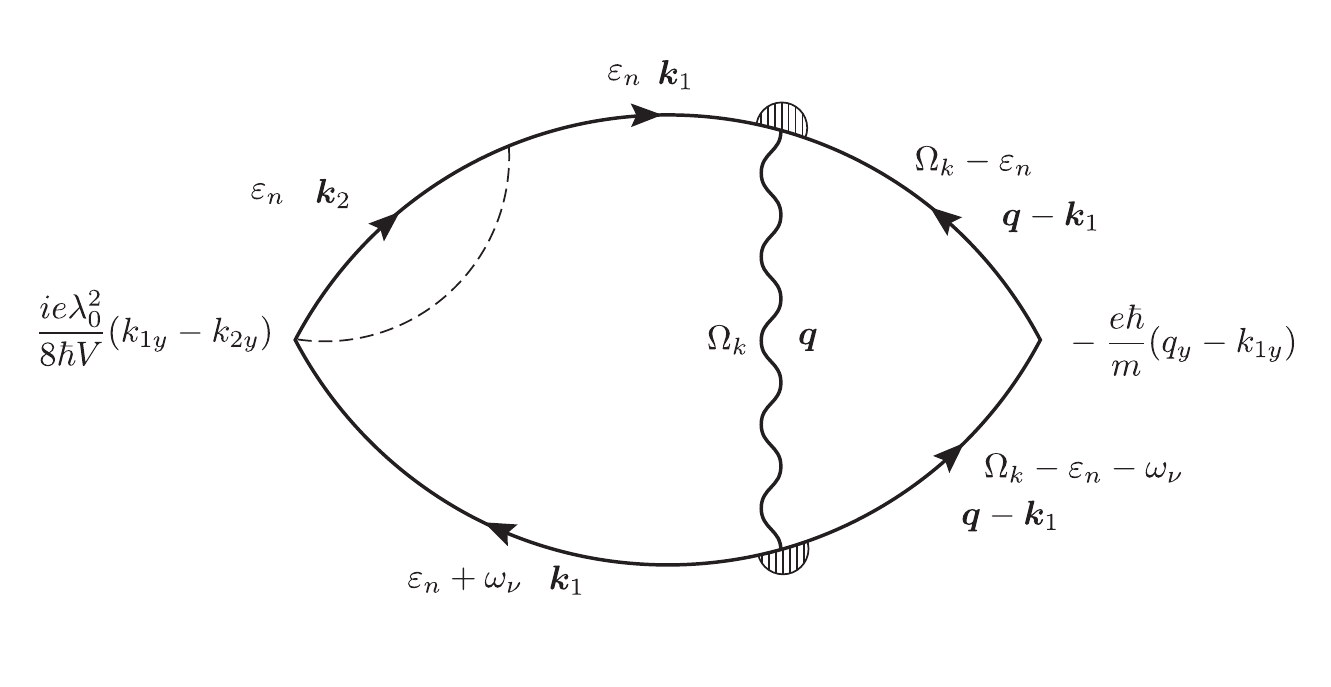}
		\\(a)
	\end{minipage}
	\begin{minipage}[b]{0.45\linewidth}
		\centering
		\includegraphics[width=\textwidth]{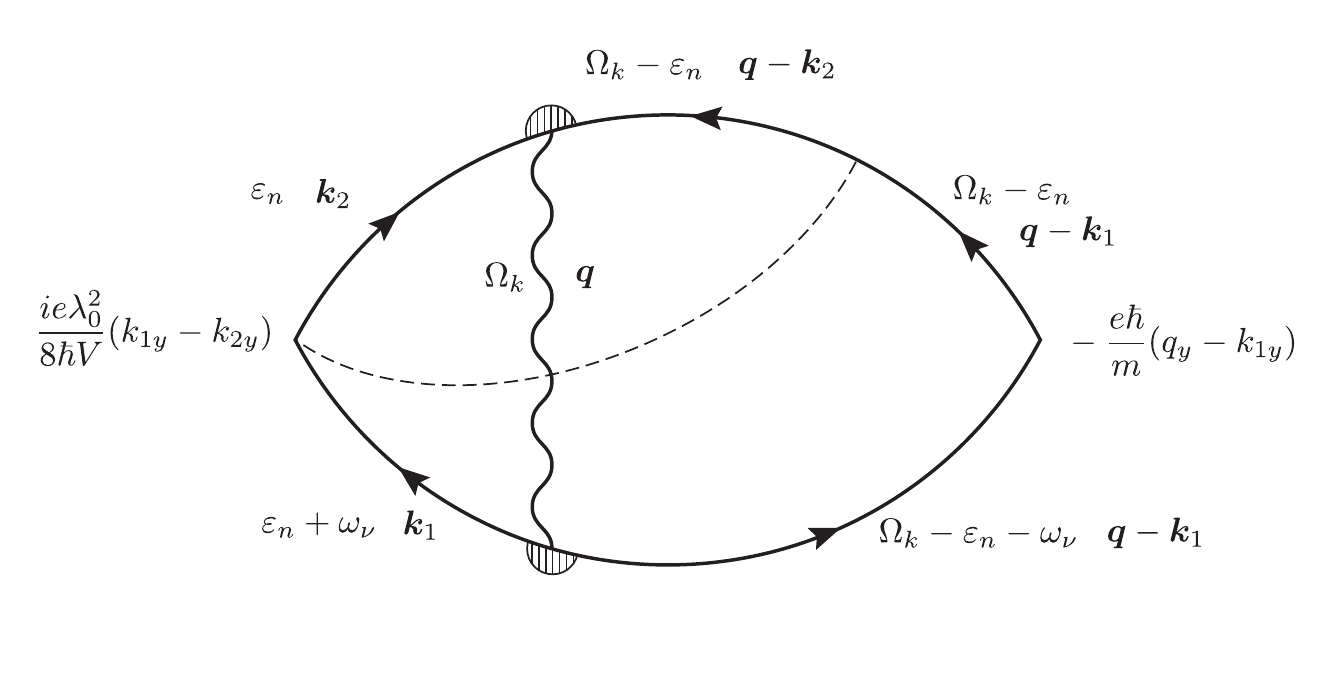}
		\\(b)
	\end{minipage}\\
	\caption{Diagrams of MT terms in the presence of side jump.}
	\label{mt-sj-figs}
\end{figure}

\begin{figure}[tbp]
	\begin{minipage}[b]{0.45\linewidth}
		\centering
		\includegraphics[width=\textwidth]{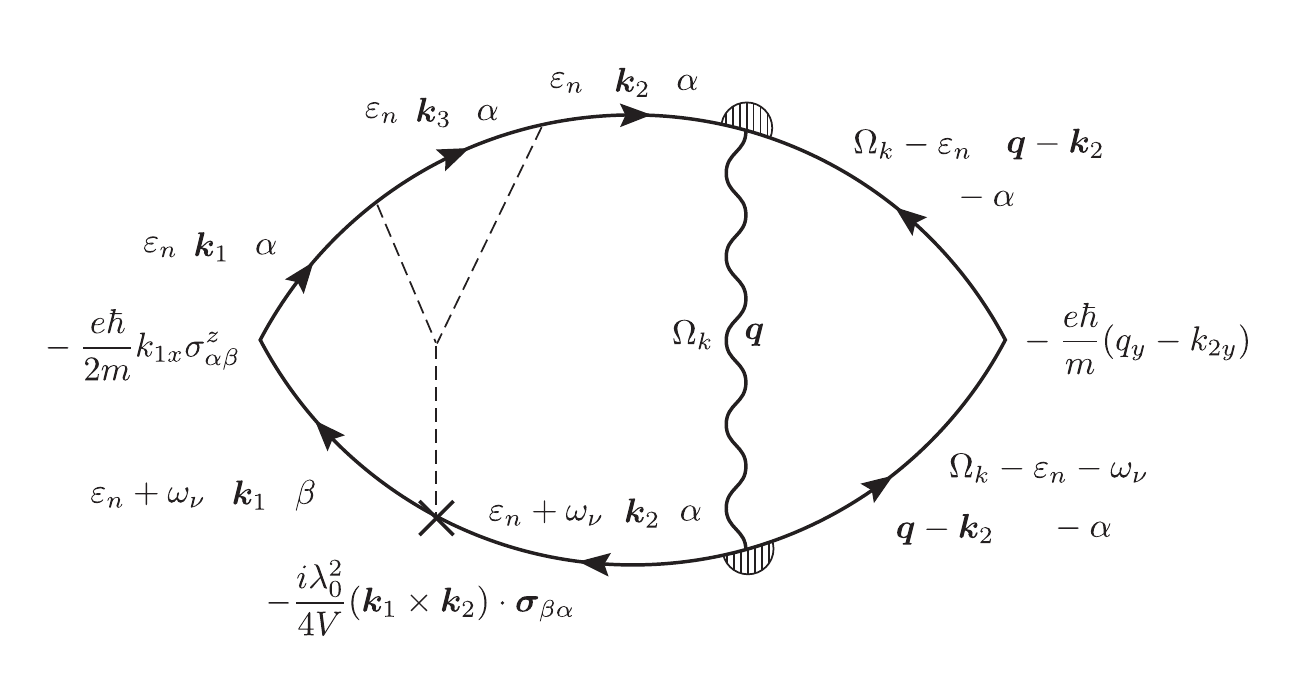}
		\\(a)
	\end{minipage}
	\begin{minipage}[b]{0.45\linewidth}
		\centering
		\includegraphics[width=\textwidth]{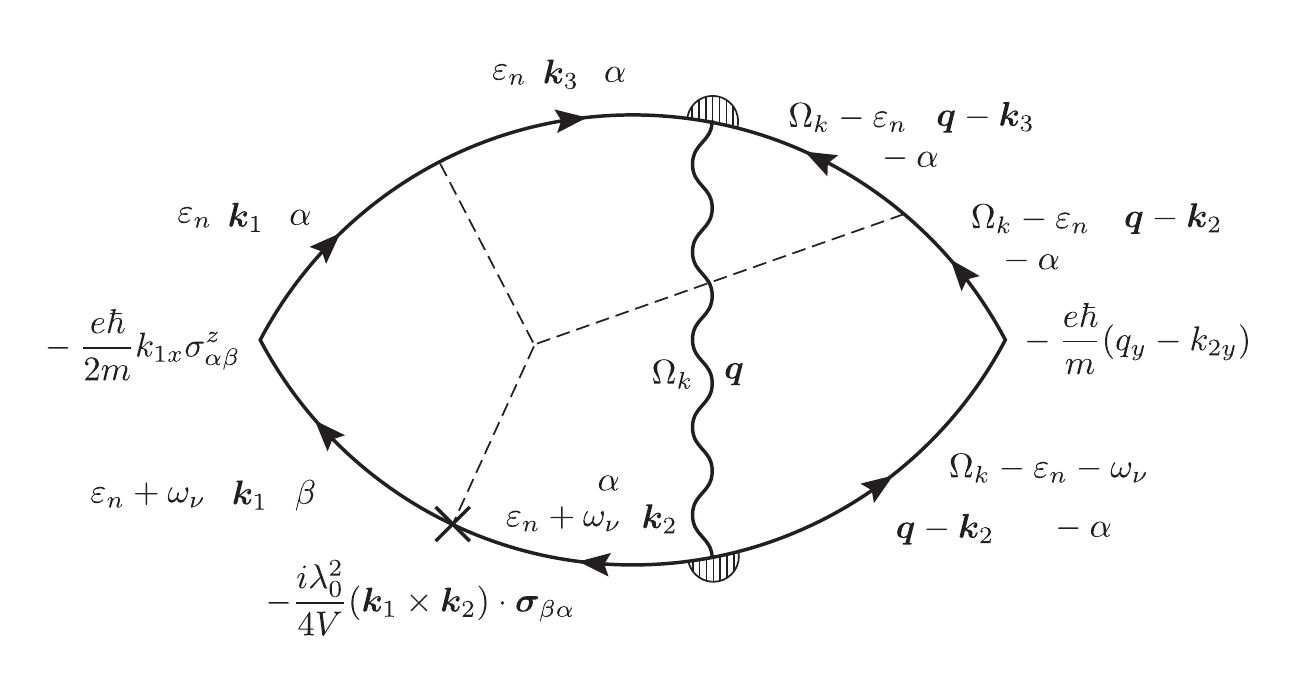}
		\\(b)
	\end{minipage}\\
	\begin{minipage}[b]{0.45\linewidth}
		\centering
		\includegraphics[width=\textwidth]{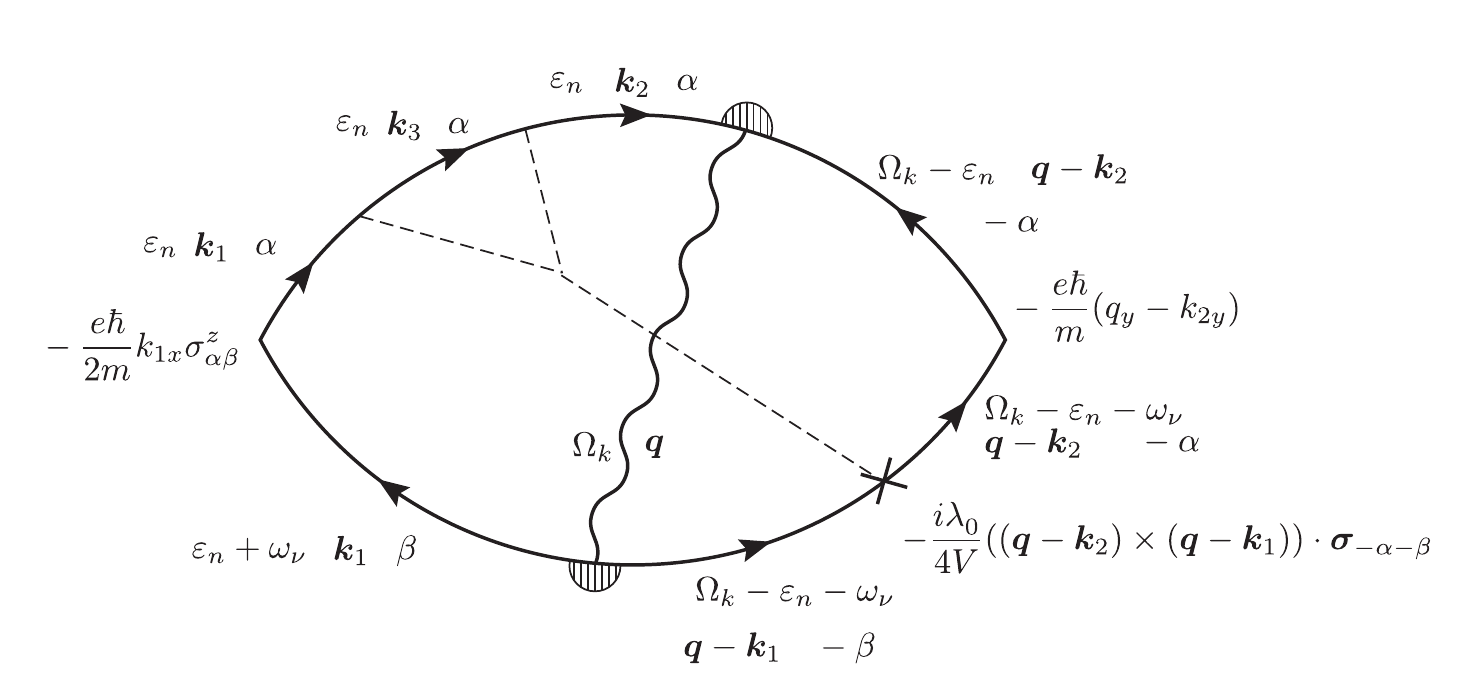}
		\\(c)
	\end{minipage}
	\caption{Diagrams of MT terms in the presence of skew scattering.}
	\label{mt-ss-figs}
\end{figure}

The MT terms are calculated similarly to those for DOS terms. The procedure to calculate the MT terms is given below. We note that the MT terms with the side jump process turn out to vanish in a way similar to that for the MT terms in the extrinsic anomalous Hall effect\cite{Li2020a}. 
\begin{enumerate}
	\item List the relevant diagrams (Fig.~\ref{mt-sj-figs} for MT + side jump and Fig.~\ref{mt-ss-figs} for MT + skew scattering)  and write down the expressions for the charge current-spin current correlation function $\Phi_{xy}^{\mathrm{MT}}(0, i\omega_\nu)$. 
	\item Put $i\Omega_k = 0$ in all quantities. 
	\item Perform integration with respect to internal wave vectors $\bm{k}_1,\bm{k}_2\cdots$ with use of the residue theorem. 
	\item Perform summation over $\varepsilon_n$. 
	\item Expand the resultant expression with respect to $\omega$ after analytic continuation $i\omega_\nu \to \hbar \omega$. 
	\item Separate {\it regular part} and {\it anomalous part}. All factors in the former is regular but $L(\bm{q}, 0)$ is regular in the limit of $\bm{q} \to 0$ while the anomalous part contains singular factor in addition to $L(\bm{q}, 0)$. 
	\item Integrate the regular part with $\bm{q}$ after setting $\bm{q} \to 0$ in all quantities but $L(\bm{q}, 0)$. 
	\item Integrate the anomalous part with $\bm{q}$ after introducing a phase-breaking relaxation time $\tau_\varphi$ to cutoff IR divergence.
\end{enumerate}
The details of the calculation is available in the supplemental material. 
The diagrams of the MT terms with skew scattering are shown in Fig.~\ref{mt-ss-figs}. The spin Hall conductivity is given  in the form of  
\begin{align}
	\sigma_{xy}^{\mathrm{MT-SS}}
	=&
	\lim_{\omega\to 0}
	\qty(-\frac{e^2 \hbar^2 \lambda_0^2 n_i v_0^3 k_{\mathrm{F}}^4 N(0)^3}{D^2m^2\omega}) T\sum_{\Omega_k} \sum_{\bm{q}} L(\bm{q}, i\Omega_k) \Sigma_{xy}^{\mathrm{MT-SS}} (\bm{q}, i\Omega_k, \hbar \omega)\\
	\Sigma_{xy}^{\mathrm{MT-SS}}(\bm{q}, i\Omega_k, i\omega_\nu)
	=&
	T \sum_{\varepsilon_n} C(\bm{q}, \varepsilon_n, \Omega_k-\varepsilon_n) C(\bm{q}, \varepsilon_n+\omega_\nu, \Omega_k - \varepsilon_n-\omega_\nu) \times\qty(\begin{minipage}{0.25\textwidth}the sum of the integral of product of  Green functions\end{minipage}).\\
\end{align}
The calculation is similar to that in the DOS terms but we have to separate $\Sigma_{xy}^{\mathrm{MT-SS}}$in the regular part and anomalous part
\begin{align}
	\Sigma_{xy}^{\mathrm{MT-SS}}(\bm{q}, 0, \omega_\nu)
	\xrightarrow{i \omega_\nu \to \hbar \omega} &
	\Sigma_{xy}^{\mathrm{MT-SS(reg)}}
	+
	\Sigma_{xy}^{\mathrm{MT-SS(an)}}(\bm{q}, 0, \omega),\\
	\Sigma_{xy}^{\mathrm{MT-SS(reg)}}
	=&
	-\hbar \omega \frac{\tau^2}{8 T^2 \hbar^2} \psi^{(2)}\qty(\frac{1}{2}),\\
	\Sigma_{xy}^{\mathrm{MT-SS(an)}}(\bm{q}, 0, \omega)
	=&
	-\hbar \omega \frac{\pi^3 \tau^2}{4T \hbar^2} \frac{1}{-i\hbar \omega + \hbar \mathcal{D} \bm{q}^2},
\end{align}
as for the MT terms in the electric conductivity.

As for the regular part,  we can proceed in a way similar to that in the DOS terms and obtain 
\begin{align}
	\sigma_{xy}^{\mathrm{MT-SS(reg)}}
	=&
	-\frac{e^2 \hbar^3 \lambda_0^2 n_{\mathrm{i}} v_0^3 k_{\mathrm{F}}^4 N(0)^2 \tau^2}{8D^2 m^2 T \hbar^2 \clength^D} \psi^{(2)}\qty(\frac{1}{2})
	\begin{dcases}
		\frac{1}{4\pi} \ln \frac{1}{\epsilon} & (D=2)\\
		\frac{1}{2\pi^2} & (D=3).
	\end{dcases}
\end{align}
As for the anomalous part, on the other hand, we introduce the phase breaking time $\tau_\varphi$ to cut-off the IR divergence as in the case of electric conductivity\cite{Thompson1970}.
We then obtain 
\begin{align}
	\sigma_{xy}^{\mathrm{MT-SS(an)}}
	=&
	-\frac{\pi^4 e^2 \hbar^3 \lambda_0^2 n_{\mathrm{i}} v_0^3 k_{\mathrm{F}}^4 N(0)^2 \tau^2}{32 D^2 m^2 T \hbar^2 \clength^D} 
	\begin{dcases}
		 \frac{1}{4\pi (\epsilon - \gamma_\varphi)} \ln \frac{\epsilon}{\gamma_\varphi} & (D=2)\\
		\frac{1}{4\pi (\sqrt{\epsilon}+ \sqrt{\gamma_\varphi})} & (D=3),
	\end{dcases}
\end{align}
with the dimensionless cutoff $\gamma_\varphi = \pi / 8T \tau_\varphi$. 
For $D=2$, the ratio of the spin Hall conductivity to that in the normal state is
\begin{align}
	\sigma^{\mathrm{MT-SS(reg)}}_{xy} / \sigma_{xy}^{\mathrm{SS(normal)}}
	=&
	\frac{7\zeta(3)}{\pi^3} \frac{\hbar / \tau}{\varepsilon_{\mathrm{F}}} \ln \frac{1}{\epsilon},\\
	\sigma^{\mathrm{MT-SS(an)}}_{xy} / \sigma_{xy}^{\mathrm{SS(normal)}}
	=&
	-\frac{\pi}{8} \frac{\hbar / \tau}{\varepsilon_{\mathrm{F}}} \frac{1}{\epsilon - \gamma_\varphi} \ln \frac{\epsilon}{\gamma_\varphi}
\end{align}
in the dirty limit 
and 
\begin{align}
	\sigma_{xy}^{\mathrm{MT-SS(reg)}} / \sigma_{xy}^{\mathrm{SS(normal)}}
	=&
		2 \frac{T}{\varepsilon_{\mathrm{F}}} \ln \frac{1}{\epsilon},\\
	\sigma_{xy}^{\mathrm{MT-SS(an)}} / \sigma_{xy}^{\mathrm{SS(normal)}}
	=&
		-\frac{\pi^4}{28\zeta(3)} \frac{T}{\varepsilon_{\mathrm{F}}} \frac{1}{\epsilon - \gamma_{\varphi}} \ln \frac{\epsilon}{\gamma_{\varphi}}
\end{align}
in the clean limit. 
\section{Fluctuation effects on intrinsic spin Hall conductivity in two-dimensional systems with Rashba-type spin-orbit interaction} \label{intrinsic-section1}
\subsection{Fluctuation propagator}
\begin{figure}[tbp]
	\centering
	\includegraphics[width=1\textwidth,page=1]{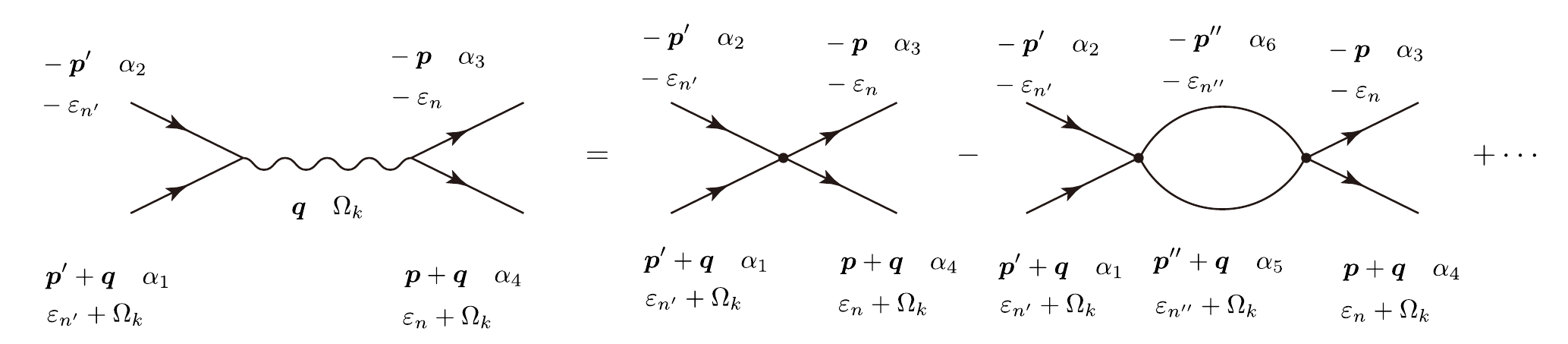}
	\caption{Feynman diagram for fluctuation propagator.}
	\label{intrinsic-fig1}
\end{figure}

\begin{figure}[tbp]
	\centering
	\includegraphics[width=0.5\textwidth,page=5]{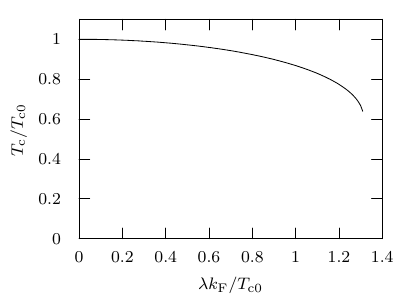}
	\caption{$\lambda$-dependence of $T_{\mathrm{c}}$.}
	\label{intrinsic-fig3}
\end{figure}

We first rewrite the two-body interaction with use of  Eq.~\eqref{sinova-eq2} as
\begin{align}
	H_{\mathrm{int}}
	=&
	-\frac{g}{V} \sum_{\bm{p} \bm{p}'\bm{q}} \psi_{\bm{p}+\bm{q}, \sigma}^\dag \psi_{-\bm{p}, -\sigma}^\dag \psi_{-\bm{p}', -\sigma} \psi_{\bm{p}' + \bm{q}, \sigma}\\
	=&
	-\frac{g}{4V} \sum_{\bm{p}\bm{p}'\bm{q}} \sum_{\alpha_1\sim \alpha_4} \alpha_2 \alpha_3 \frac{(-p_y + ip_x) (-p'_y - ip'_x)}{pp'} c_{\bm{p}+ \bm{q}, \alpha_1} ^\dag c_{-\bm{p}, \alpha_2}^\dag c_{-\bm{p}' ,\alpha_3} c_{\bm{p}' + \bm{q}, \alpha_4}.\\
\end{align}
Figure \ref{intrinsic-fig1} shows the diagram of fluctuation propagator.
As an inspection, we examine the diagram containing two $H_{\mathrm{int}}$ (the second term on the right hand side in Fig.~\ref{intrinsic-fig1}), which yields
\begin{align}
	&
	T \sum_{\alpha_5 \alpha_6} \sum_{\varepsilon_{n''} \bm{p}''} 
	\begin{aligned}[t]
		&
	\qty(-\frac{g}{4V}) \alpha_2 \alpha_6 \frac{(-p''_y + ip''_x)(-p'_y - ip'_x)}{p'' p'} \mathcal{G}_{\alpha_5} (\bm{p}''+\bm{q}, i\varepsilon_{n''} + i\Omega_k) \mathcal{G}_{\alpha_6} (-\bm{p}'', -i\varepsilon_{n''}) \\
		\times &
		\qty(-\frac{g}{4V}) \alpha_6 \alpha_3 \frac{(-p_y + ip_x) (-p''_y - ip''_x)}{pp''}\\
	\end{aligned}\\
	=&
	-\frac{g}{4V} \frac{(-p_y + ip_x) (-p'_y - ip'_x)}{pp'} \alpha_2 \alpha_3 T \sum_{\alpha_5 \alpha_6} \sum_{\varepsilon_{n''} \bm{p}''} \qty(-\frac{g}{4V}) \mathcal{G}_{\alpha_5} (\bm{p}'' + \bm{q}, i\varepsilon_{n''} + i\Omega_k) \mathcal{G}_{\alpha_6} (-\bm{p}'', -i\varepsilon_{n''}).
\end{align}
This expression does not depend on the internal wave numbers and spins for summation over loops because those dependencies are canceled. Similar cancellation occurs even in the diagrams containing more than two $H_{\mathrm{int}}$, and those diagrams depend only on the left-most and right-most wavenumber and spins. Consequently, the summation over the series of diagrams can be carried out in a way similar to that for superconductors without spin-orbit interaction shown in Fig.~\ref{fluctuation-fig1}, i.e.
\begin{align}
	&
	L_{\alpha_2\alpha_3} (\bm{q}, \bm{p}, \bm{p}', i\Omega_k)\\
	=&
	-\frac{g}{4V} \frac{(-p_y + ip_x) (-p'_y - ip'_x)}{pp'} \alpha_2 \alpha_3 \\
	&\times
	\begin{aligned}[t]
	\Biggl\{
	&
	1 
	-
	T \sum_{\alpha_5 \alpha_6} \sum_{\varepsilon_{n''} \bm{p}''} \qty(-\frac{g}{4V}) \mathcal{G}_{\alpha_5} (\bm{p}'' + \bm{q}, i\varepsilon_{n''} + i\Omega_k) \mathcal{G}_{\alpha_6} (-\bm{p}'', -i\varepsilon_{n''}) \\
	&+
	\Biggl[T \sum_{\alpha_5 \alpha_6} \sum_{\varepsilon_{n''} \bm{p}''} \qty(-\frac{g}{4V}) \mathcal{G}_{\alpha_5} (\bm{p}'' + \bm{q}, i\varepsilon_{n''} + i\Omega_k) \mathcal{G}_{\alpha_6} (-\bm{p}'', -i\varepsilon_{n''}) \Biggr]^2 \cdots \Biggr\}
	\end{aligned}\\
	=&
	-\frac{1}{4V} \alpha_2\alpha_3 \frac{(-p_y + ip_x) (-p'_y-ip'_x)}{pp'} \qty[g^{-1} - \Pi(\bm{q}, \Omega_k)]^{-1},
\end{align}
where we introduce the notations
\begin{align}
	\Pi(\bm{q}, \Omega_k)
	=&
	\frac{1}{4} \sum_{\alpha \beta} T \sum_{\varepsilon_n} P_{\alpha \beta} (\bm{q}, \varepsilon_n + \Omega_k, -\varepsilon_n)\\
	P_{\alpha \beta} (\bm{q}, \varepsilon_1, \varepsilon_2)
	=&
	\frac{1}{V} \sum_{\bm{p}} \mathcal{G}_\alpha (\bm{p}+\bm{q}, i\varepsilon_1) \mathcal{G}_\beta (-\bm{p}, i\varepsilon_2).
\end{align}
We find that $P_{\alpha\beta}(\bm{q}, \varepsilon_1, \varepsilon_2)$ is written as
\begin{align}
	P_{\alpha \beta} (\bm{q}, \varepsilon_1, \varepsilon_2)
	=&
	\int \frac{d^2 \bm{p}}{(2\pi)^2} \frac{1}{i\varepsilon_1 - \qty(\frac{\hbar^2}{2m} (\bm{p}+\bm{q})^2 + \alpha \lambda \norm{\bm{p}+\bm{q}}) + \mu}
	\frac{1}{i\varepsilon_2 - \qty(\frac{\hbar^2}{2m} \bm{p}^2 + \beta \lambda \norm{\bm{p}}) + \mu}\\
	\approx&
	\int \frac{d^2 \bm{p}}{(2\pi)^2} 
	\frac{1}{i\varepsilon_1 - \qty(\frac{\hbar}{2m} p^2 + \alpha \lambda p - \mu) - \qty(\frac{\hbar^2}{m} + \alpha \frac{\lambda}{p} )\bm{p}\vdot \bm{q}}
	\frac{1}{i\varepsilon_2 - \qty(\frac{\hbar^2}{2m} p^2 + \beta \lambda p - \mu)},
\end{align}
which becomes when $\lambda p_{\mathrm{F}} \ll \mu$
as
\begin{align}
	&
	P_{\alpha \beta} (\bm{q}, \varepsilon_1, \varepsilon_2)\\
	\approx&
	N(0) \int \frac{d\Omega_p}{2\pi} \int d\xi \; \frac{1}{i\varepsilon_1 - \xi - \alpha \lambda p_{\mathrm{F}} - \qty(\frac{\hbar^2}{m} + \alpha \frac{\lambda}{p_{\mathrm{F}}})\bm{p}_{\mathrm{F}} \vdot \bm{q}} \frac{1}{i\varepsilon_2 - \xi - \beta \lambda p_{\mathrm{F}}}\\
	=&
	2\pi N(0) \theta(-\varepsilon_1 \varepsilon_2) \frac{1}{\abs{\varepsilon_1 - \varepsilon_2}} \qty{1 - \frac{i(\alpha - \beta)\lambda p_{\mathrm{F}} \sign(\varepsilon_1)}{\abs{\varepsilon_1 - \varepsilon_2}} - \frac{\qty[(\alpha - \beta) \lambda p_{\mathrm{F}}]^2 + \qty(\frac{\hbar^2}{m} + \alpha \frac{\lambda}{p_{\mathrm{F}}})^2 \expval{(\bm{p}_{\mathrm{F}} \vdot \bm{q})^2}_{\mathrm{F.S.}}}{\abs{\varepsilon_1 - \varepsilon_2}^2}}.
\end{align}
We thus see that 
\begin{align}
	\frac{1}{4} \sum_{\alpha \beta} P_{\alpha \beta}(\bm{q}, \varepsilon_1, \varepsilon_2)
	=&
	2\pi N(0) \theta(-\varepsilon_1\varepsilon_2) \frac{1}{\abs{\varepsilon_1 - \varepsilon_2}} \qty(1 - \frac{A(\bm{q})}{\abs{\varepsilon_1 - \varepsilon_2}^2})
\end{align}
with  
\begin{align}
	A(\bm{q})
	=&
	2(\lambda p_{\mathrm{F}})^2 + \qty[\qty(\frac{\hbar^2}{m})^2 + \qty(\frac{\lambda}{p_{\mathrm{F}}})^2] \expval{(\bm{p}_{\mathrm{F}} \vdot \bm{q})^2}_{\mathrm{F.S.}}.
\end{align}
With use of this expression, 
$	\Pi(\bm{q}, \Omega_k)
$ becomes
\begin{align}
	\Pi(\bm{q}, \Omega_k)
	=&
	T\sum_{\varepsilon_n} \frac{1}{4} \sum_{\alpha \beta} P_{\alpha \beta} (\bm{q}, \varepsilon_n + \Omega_k, -\varepsilon_n)\\
	=&
	N(0) \qty[\psi\qty(\frac{1}{2} + \frac{\abs{\Omega_k}}{4\pi T} + \frac{\hbar \omega_{\mathrm{D}}}{2\pi T})  - \psi\qty(\frac{1}{2} + \frac{\abs{\Omega_k}}{4\pi T}) + \frac{A(\bm{q})}{2(4\pi T)^2} \psi^{(2)}\qty(\frac{1}{2} + \frac{\abs{\Omega_k}}{4\pi T})].
\end{align}
The transition temperature $ T_{\mathrm{c}}$ is determined by the condition that $L(\bm{q}=0, i\Omega_k = 0)$ diverges at $T\to T_{\mathrm{c}}$, i.e.,
\begin{align}
	0
	=
	g^{-1} - \Pi(0, 0) 
	=&
	g^{-1} - N(0) \qty[\psi\qty(\frac{1}{2} + \frac{\hbar \omega_{\mathrm{D}}}{2\pi T_{\mathrm{c}}}) - \psi\qty(\frac{1}{2}) + \frac{A(0)}{2(4\pi T)^2}\psi^{(2)}\qty(\frac{1}{2})]. \label{intrinsic-eq2}
\end{align}
With use of the relation 
\begin{align}
	\epsilon
	=
	\ln \frac{T}{T_{\mathrm{c}}}
	=
	\ln\qty(\frac{\hbar \omega_{\mathrm{D}}}{2\pi T_{\mathrm{c}}}) - \ln\qty(\frac{\hbar \omega_{\mathrm{D}}}{2\pi T}) 
	\approx
	\frac{1}{N(0)} \qty[g^{-1} - \Pi(\bm{q}, \Omega_k)] + \psi\qty(\frac{1}{2}) - \psi\qty(\frac{1}{2} + \frac{\abs{\Omega_k}}{4\pi T}) - \clength^2 q^2,\\
\end{align}
we obtain 
\begin{equation}
    g^{-1} - \Pi(\bm{q}, \Omega_k) 
	=
	N(0) \qty[\epsilon + \psi\qty(\frac{1}{2} + \frac{\abs{\Omega_k}}{4\pi T}) - \psi\qty(\frac{1}{2}) + \clength^2 q^2].
	\label{eq: g-Pi}
\end{equation}	
Here we introduce  the coherence length in the presence of intrinsic spin-orbit interaction through the relation 
\begin{align}
	\clength^2 q^2
	\equiv&
	- \frac{A(\bm{q}) - A(0)}{2(4\pi T)^2} \psi^{(2)}\qty(\frac{1}{2})\\
	=&
	-\frac{\qty(\frac{\hbar^2}{m})^2 + \qty(\frac{\lambda}{k_{\mathrm{F}}})^2}{2(4\pi T)^2} \psi^{(2)}\qty(\frac{1}{2}) \expval{(\bm{k}_{\mathrm{F}}\vdot \bm{q})^2}_{\mathrm{F.S.}}\\
	=&
	-\psi^{(2)}\qty(\frac{1}{2}) \frac{\qty(\frac{\hbar^2}{m})^2 + \qty(\frac{\lambda}{k_{\mathrm{F}}})^2}{4(4\pi T)^2}k_{\mathrm{F}}^2 q^2,
	\end{align}
	which becomes 
\begin{equation}
    \clength^2
	\approx
	14 \zeta(3) \frac{\qty(\frac{\hbar^2}{m})^2 k_{\mathrm{F}}^2}{4(4\pi T)^2}
	=
	\frac{7\zeta(3)}{32\pi^2} \frac{\hbar^2 v_{\mathrm{F}}^2}{T^2}
\end{equation}
because we assume that $\mu \gg \lambda k_{\mathrm{F}}$. With use of Eq.~\eqref{eq: g-Pi} the fluctuation propagator in the presence of spin-orbit interaction is given by
\begin{align}
	L_{\alpha_1\alpha_2\alpha_3\alpha_4} (\bm{q}, \bm{p}, \bm{p}', i\Omega_k)
	=&
	\qty(\frac{\alpha_2}{2} \frac{-p'_y - ip'_x}{p}) \qty(\frac{\alpha_3}{2} \frac{-p_y + ip_x}{p}) L(\bm{q}, i\Omega_k)\\
	L(\bm{q}, i\Omega_k)
	\equiv&
	-\frac{1}{N(0)V} \frac{1}{\epsilon + \psi\qty(\frac{1}{2} + \frac{\abs{\Omega_k}}{4\pi T}) - \psi\qty(\frac{1}{2}) + \clength^2 \bm{q}^2}
\end{align}
where $L(\bm{q}, i\Omega_k)$ is the fluctuation propagator without the spin-orbit interaction, which coincides with Eq.~\eqref{fluctuation-eq10}. 
In the end of this subsection, we consider $\lambda$ dependence of $T_{\mathrm{c}}$. 
From Eq.~\eqref{intrinsic-eq2}, we obtain
\begin{align}
    \frac{T_{\mathrm{c}}}{T_{\mathrm{c}0}}
	=&
	\exp[-14 \zeta(3)\qty(\frac{\lambda k_{\mathrm{F}}}{4\pi T_{\mathrm{c}}})^2] \label{intrinsic-eq3}, 
\end{align}
where $T_{\mathrm{c}0}$ is the transition temperature for  $\lambda=0$. 
From Eq.~\eqref{intrinsic-eq3}, we find that the transition to superconducting state occurs at a finite temperature when
\begin{align}
	\frac{\lambda k_{\mathrm{F}}}{T_{\mathrm{c}0}}
	<&
	\frac{2\pi}{\sqrt{7e\zeta(3)}}
	\approx
	1.313\cdots
\end{align}
are satisfied.
Figure~\ref{intrinsic-fig3} shows $\lambda$-dependence of  $T_{\mathrm{c}}$.
\subsection{Aslamazov-Larkin term} \label{intrinsic-section2}
\begin{figure}[tbp]
	\centering
	\includegraphics[width=0.7\textwidth, page=3]{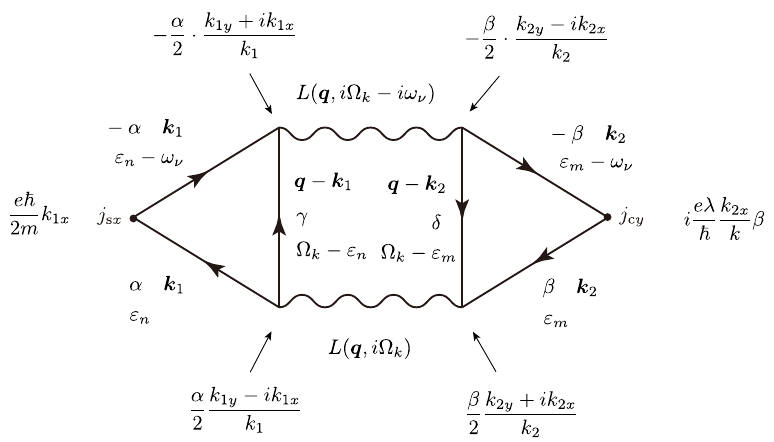}
	\caption{Feynman diagram for the AL term for intrinsic spin Hall effect.}
	\label{intrinsic-fig5}
\end{figure}

The diagram for AL terms is shown in Fig.~\ref{intrinsic-fig5}. 
The outline of the procedure to calculate the AL terms for intrinsic spin Hall effect is the same as that for the AL terms for extrinsic  spin Hall effect. The details of calculation along these procedures are given in Supplemental materials. 
The resultant expression for the AL term of the intrinsic spin Hall conductivity is given by
\begin{align}
	\sigma_{xy}^{\mathrm{AL}}
	\approx&
		\frac{e^2 \hbar^5}{2^{13} \pi^3 \clength^4 T \lambda^2 m^3} \qty[\Im \psi^{(1)}\qty(\frac{1}{2} + \frac{i\lambda k_{\mathrm{F}}}{2\pi T})]^2 \ln \frac{1}{\epsilon},
\end{align}
which reduces, when $\lambda k_{\mathrm{F}} \ll T$, to 
\begin{align}
	\sigma_{xy}^{\mathrm{AL}}
	\approx&
	\frac{e^2}{16\pi \hbar} \frac{T}{\varepsilon_{\mathrm{F}}} \ln \frac{1}{\epsilon}.
\end{align}
\subsection{DOS and Maki-Thompson terms}

\begin{figure}[tbp]
	\centering
	\includegraphics[width=0.6\textwidth,page=4]{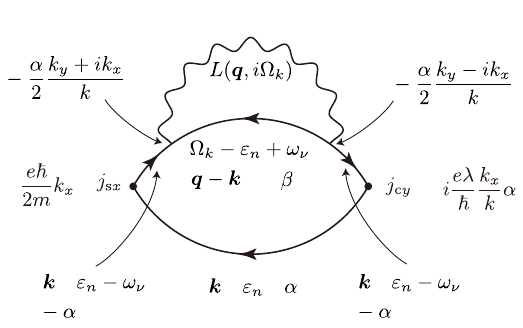}
	\caption{Feynman diagram of DOS term in intrinsic spin Hall conductivity.}
	\label{intrinsic-fig6}
\end{figure}

\begin{figure}[tbp]
	\centering
	\includegraphics[page=3]{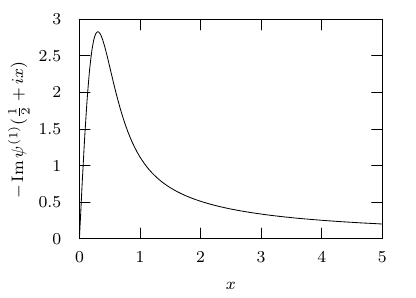}
	\caption{Profile of $-\Im \psi^{(1)}(1 / 2 + ix)$.}
	\label{intrinsic-fig7}
\end{figure}

The DOS term and MT terms in intrinsic spin Hall effects are calculated in a way similar to those in extrinsic spin Hall effect, but there are no anomalous terms in MT terms for intrinsic spin Hall effect, and thus the cutoff is not necessary to be introduced. See the subsections~\ref{subsec: Dos-Extrinsic} and \ref{subsec: Maki-Thompson}, where the procedures to calculate the DOS and MT terms are given. 
The diagram for the DOS term in intrinsic spin Hall effect is shown in Fig.~\ref{intrinsic-fig6}. 
The DOS term for intrinsic spin Hall conductivity is given by
\begin{align}
	\sigma_{xy}^{\mathrm{DOS}}
	=&
	-\frac{e^2 \hbar}{512 \pi^2 m \clength^2 \lambda k_{\mathrm{F}}} 
	\qty[-\Im \psi^{(1)}\qty(\frac{1}{2} + \frac{i \lambda k_{\mathrm{F}}}{2\pi T})]
	\ln \frac{1}{\epsilon}.
\end{align}
Figure~\ref{intrinsic-fig7} shows the profile of $-\Im \psi^{(1)}(1 / 2 + i x)$.
With use of $-\Im \psi^{(1)}(1 / 2 + i x) \approx 14\zeta(3)x$
for $x \ll 1$, we obtain
\begin{align}
    \sigma_{xy}^{\mathrm{DOS}}
    =&
    -\frac{e^2}{32\pi \hbar} \frac{T}{\varepsilon_{\mathrm{F}}} \ln \frac{1}{\epsilon}
\end{align}
for $\lambda k_{\mathrm{F}} \ll T$.

\begin{figure}[tbp]
	\centering
	\includegraphics[width=0.6\textwidth,page=5]{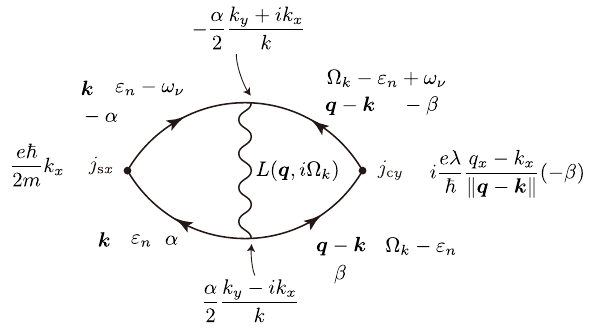}
	\caption{Feynman diagram of MT term in intrinsic Hall conductivity.}
	\label{intrinsic-fig8}
\end{figure}

\begin{figure}[tbp]
	\centering
    \includegraphics[page=4]{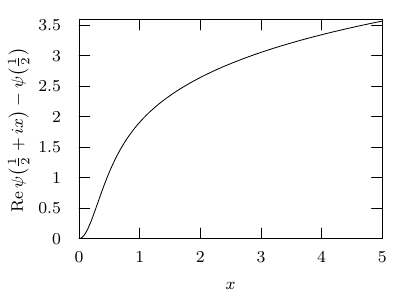}
	\caption{The profile of $\Re \psi\qty(1 / 2 + i x)  - \psi\qty(1 / 2)$.}
	\label{intrinsic-fig9}
\end{figure}

The diagram for the MT term in intrinsic spin Hall effect is shown in Fig.~\ref{intrinsic-fig8}. 
The MT term for intrinsic spin Hall conductivity is given by
\begin{align}
	\sigma_{xy}^{\mathrm{MT}}
	=&
	\frac{e^2 \hbar T}{64\pi m \clength^2 (\lambda k_{\mathrm{F}})^2} \qty[\Re \psi\qty(\frac{1}{2} + \frac{i\lambda k_{\mathrm{F}}}{2\pi T}) - \psi\qty(\frac{1}{2})] \ln \frac{1}{\epsilon}.
\end{align}
Figure \ref{intrinsic-fig9} shows the profile of 
$\Re \psi(1 / 2 + i x) - \psi(1 / 2)$. 
With use of $\Re \psi(1/2 + ix) - \psi(1/2) \approx 7\zeta(3)x^2$
for $x\ll 1$, we obtain 
\begin{align}
	\sigma_{xy}^{\mathrm{MT}}
	=&
	\frac{e^2}{16\pi \hbar} \frac{T}{\varepsilon_{\mathrm{F}}} \ln \frac{1}{\epsilon}
\end{align}
for $\lambda k_{\mathrm{F}} \ll T$. 
\section{Discussion} \label{discussion-section0}

\subsection{\texorpdfstring{Summary of the results for $D=2$}{Summary of the resuls for D = 2}}\label{discussion-section3}
We summarize the results for $D=2$ case, where the spin Hall conductivity diverges in the limit $\epsilon\rightarrow +0$.  
\begin{table}[tbp]
	\centering
	\caption{Extrinsic spin Hall conductivity for $D=2$ in the dirty limit. 
	The results are normalized by the spin Hall conductivity in the normal state ;$\sigma_{xy}^{\mathrm{SJ(normal)}}$ or   $\sigma_{xy}^{\mathrm{SS(normal)}}$. }
	\label{discussion-table1}
		\begin{tabular}{crr}
		\hline\hline
		 & side jump & skew scattering\\ 
		 \hline
		 \tspace AL / normal & $2 \dfrac{T}{\varepsilon_{\mathrm{F}}}\ln \dfrac{1}{\epsilon}$ & $4 \dfrac{T}{\varepsilon_{\mathrm{F}}} \ln \dfrac{1}{\epsilon}$ \\
		 \tspace DOS / normal  & $-\dfrac{7\zeta(3)}{\pi^3} \dfrac{\hbar / \tau}{\varepsilon_{\mathrm{F}}} \ln \dfrac{1}{\epsilon}$ & $-\dfrac{14\zeta(3)}{\pi^3} \dfrac{\hbar / \tau}{\varepsilon_{\mathrm{F}}} \ln \dfrac{1}{\epsilon}$\\
		 \tspace MT (reg) / normal  & $0$ & $\dfrac{7\zeta(3)}{\pi^3} \dfrac{\hbar / \tau}{\varepsilon_{\mathrm{F}}} \ln \dfrac{1}{\epsilon}$  \\
		 \tspace MT (an) / normal  & $0$ & $-\dfrac{\pi}{8} \dfrac{\hbar / \tau}{\varepsilon_{\mathrm{F}}} \dfrac{1}{\epsilon - \gamma_\varphi} \ln \dfrac{\epsilon}{\gamma_\varphi}$  \\
		 \hline\hline
		\end{tabular}
\end{table}
\begin{table}[tbp]
	\centering
	\caption{Extrinsic spin Hall conductivity for $D=2$ in the clean limit. 
	The results are normalized by the spin Hall conductivity in the normal state $\sigma_{xy}^{\mathrm{SJ(normal)}}$ or $\sigma_{xy}^{\mathrm{SS(normal)}}$. }
	\label{discussion-table2}
	\begin{tabular}{crr}
	\hline\hline
	 & side jump & skew scattering\\ 
	 \hline
	 \tspace AL / normal & $2 \dfrac{T}{\varepsilon_{\mathrm{F}}} \ln \dfrac{1}{\epsilon}$ & $2 \dfrac{T}{\varepsilon_{\mathrm{F}}}\ln \dfrac{1}{\epsilon}$\\
	 \tspace DOS / normal  & $ -\dfrac{2\pi^3}{7\zeta(3)} \dfrac{T}{\varepsilon_{\mathrm{F}}} \dfrac{T \tau}{\hbar} \ln \dfrac{1}{\epsilon} $ & $-\dfrac{4\pi^3}{7\zeta(3)} \dfrac{T}{\varepsilon_{\mathrm{F}}} \dfrac{T \tau}{\hbar} \ln \dfrac{1}{\epsilon}$\\
	 \tspace MT (reg) / normal & $0$ & $2 \dfrac{T}{\varepsilon_{\mathrm{F}}} \ln \dfrac{1}{\epsilon}$  \\
	 \tspace MT (an) / normal & $0$ & $-\dfrac{\pi^4}{28\zeta(3)} \dfrac{T}{\varepsilon_{\mathrm{F}}} \dfrac{1}{\epsilon - \gamma_{\varphi}} \ln \dfrac{\epsilon}{\gamma_{\varphi}}$  \\
	 \hline\hline
	\end{tabular}
\end{table}

\begin{table}[tbp]
	\centering
	\caption{Intrinsic spin Hall conductivity via Rashba-type spin-orbit interaction for $\lambda k_{\mathrm{F}} \ll T$. The results are normalized by that in the normal state; $\sigma_{xy}^{\mathrm{(normal)}}=e^2 / 8\pi\hbar$ (Eq.~\eqref{sinova-eq4}).}
\label{discussion-table3}
	\begin{tabular}{cr}
	\hline\hline
	\tspace AL / normal & $\dfrac{1}{2} \dfrac{T}{\varepsilon_{\mathrm{F}}} \ln \dfrac{1}{\epsilon}
	$\\
	\tspace DOS / normal  & $-\dfrac{1}{4} \dfrac{T}{\varepsilon_{\mathrm{F}}} \ln \dfrac{1}{\epsilon}$\\
	\tspace MT / normal  & $\dfrac{1}{2} \dfrac{T}{\varepsilon_{\mathrm{F}}} \ln \dfrac{1}{\epsilon}$\\
	\hline\hline
	\end{tabular}
\end{table}
In tables \ref{discussion-table1}, \ref{discussion-table2}, \ref{discussion-table3}, we note two properties common in extrinsic and intrinsic effects. One is that the singularity in the AL terms is $\ln (1/\epsilon)$, which is weaker than the power-law singularity in the AL terms in electric conductivity. The power-counting argument is given in \ref{discussion-section1}. As another point, we notice that all contributions contain the factor $1/\varepsilon_{\mathrm{F}}$, which weakens the fluctuation effect. The origin of this factor is discussed in \ref{discussion-section2}.

\begin{table}[tbp]
	\centering
	\caption{Dominant contribution in extrinsic spin Hall conductivity in the dirty limit.}
	\label{discussion-table4}
	\begin{tabular}{ccc}
	\hline\hline
	 & $\sigma_{xy}^{\mathrm{SJ(normal)}}      \gg \sigma_{xy}^{\mathrm{SS(normal)}}$ 
& $\sigma_{xy}^{\mathrm{SJ(normal)}}
          \ll 
   \sigma_{xy}^{\mathrm{SS(normal)}}$                  \\
   \hline
		$\gamma_{\varphi} \ll 1$ 
		& DOS-SJ 
		& MT(an)-SS\\
		$\gamma_{\varphi} \gg 1$ 
		& DOS-SJ 
		& DOS-SS + MT(reg) -SS\\
		\hline\hline
	\end{tabular}
\end{table}

\begin{table}[tbp]
	\centering
	\caption{Dominant contribution in extrinsic spin Hall conductivity in the clean limit.}
	\label{discussion-table5}
	\begin{tabular}{ccc}
	\hline\hline
	 & $\sigma_{xy}^{\mathrm{SJ(normal)}} \gg \sigma_{xy}^{\mathrm{SS(normal)}}$ 
	 & $\sigma_{xy}^{\mathrm{SJ(normal)}} \ll \sigma_{xy}^{\mathrm{SS(normal)}}$\\ 
	 \hline
		\tspace $\dfrac{T\tau}{\hbar} \ll \dfrac{1}{\epsilon - \gamma_{\varphi}}$ 
		& DOS-SJ 
		& MT(an)-SS\\
		\tspace $\dfrac{T\tau}{\hbar} \gg \dfrac{1}{\epsilon - \gamma_{\varphi}}$ 
		& DOS-SJ 
		& DOS-SS\\
		\hline\hline
	\end{tabular}
\end{table}

First, we discuss the extrinsic case. 
In tables~\ref{discussion-table1} and \ref{discussion-table2}, we see that there are three energy scales; $\varepsilon_{\mathrm{F}}$, $T$, $\hbar / \tau$. 
We summarize the dominant contribution in tables \ref{discussion-table4} and \ref{discussion-table5}. 
In the dirty limit, the DOS terms with the side jump process is dominant when $\sigma_{xy}^{\mathrm{SJ(normal)}} \gg \sigma_{xy}^{\mathrm{SS(normal)}}$.
When $\sigma_{xy}^{\mathrm{SJ(normal)}} \ll \sigma_{xy}^{\mathrm{SS(normal)}}$, 
either anomalous MT terms or the sum of the DOS term and the regular part of the MT terms is dominant, depending on the magnitude of $\gamma_{\varphi}$. 
In the clean limit, the DOS terms with side jump process is dominant when  $\sigma_{xy}^{\mathrm{SJ(normal)}} \gg \sigma_{xy}^{\mathrm{SS(normal)}}$.
When  $\sigma_{xy}^{\mathrm{SJ(normal)}} \ll \sigma_{xy}^{\mathrm{SS(normal)}}$,
either anomalous part of the MT term or the DOS term is dominant depending on relative magnitude of $T \tau / \hbar$ and $1 / (\epsilon - \gamma_{\varphi})$. 
Both in the dirty and clean limits, dominant contributions have signs opposite to that in the normal state. 

\begin{figure}[tbp]
	\begin{minipage}[t]{0.4\linewidth}
		\centering
        \includegraphics[page=1]{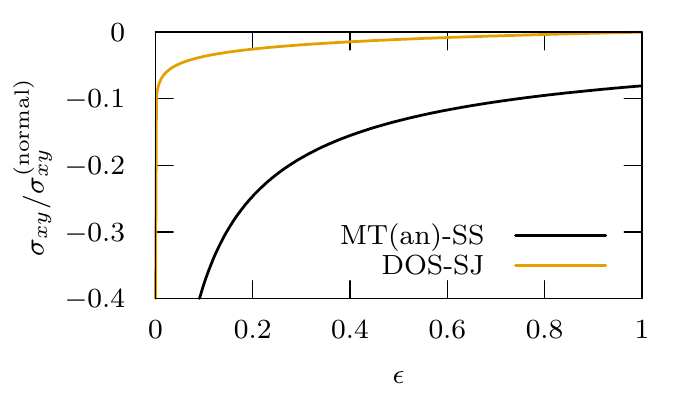}
		\caption{
			Estimate of fluctuation effect based on the parameters for Nb. 
		}
		\label{discussion-fig1}
	\end{minipage}
	\hspace{1cm}
	\begin{minipage}[t]{0.4\linewidth}
		\centering
        \includegraphics[page=2]{graphs.pdf}
		\caption{
			Estimate of fluctuation effect based on the parameters for clean Al, where skew scattering is considered to be dominant. 
		}
		\label{discussion-fig2}
	\end{minipage}
\end{figure}
We give an estimate of dominant contribution in fluctuation effect based on the parameters for Nb and clean Al when $\gamma_{\varphi} \ll 1$ in figures \ref{discussion-fig1} and \ref{discussion-fig2}. We have assumed in these  estimations $\gamma_{\varphi}$ independent of temperature but in reality importance of temperature-dependence in $\gamma_{\varphi} \ll 1$ has been pointed out \cite{Mori1990,Craven1973}. 

Next, we discuss the intrinsic case.
All terms are independent of $\lambda k_{\mathrm{F}}$ as in the normal state when $\lambda k_{\mathrm{F}} \ll T$, and thus a tiny Rashba-type spin-orbit interaction makes contribution of fluctuations finite.

Fluctuation effects on intrinsic spin Hall conductivity except for DOS term has the same sign as that in the normal state, in contrast to the extrinsic case. In ref.~\cite{Dimitrova2005}, the lowest order correction due to two-body {\it repulsive} interactions was found to suppress the intrinsic spin Hall conductivity in the two-dimensional Rashba model. The enhancement of spin Hall conductivity due to two-body attraction in the present paper and the suppression due to repulsion in ref.~\cite{Dimitrova2005} seem consistent with each other.  

\subsection{\texorpdfstring{Power-counting of singularity in the limit $\epsilon \to 0$}{Power-counting of singularity in the limit epsilon to 0}}
\label{discussion-section1}
We first consider the origin of the singularity near $\epsilon\rightarrow +0$ by power counting argument. After that, we discuss the physical implication of this result. 

Before considering the singularity of AL terms in the spin Hall conductivity, we first review the origin of the singularity in those terms in electric conductivity, which is given by
\begin{align}
	\sigma_{xx}^{\mathrm{AL}}
	\propto&
	\frac{1}{i\omega} \qty[\sum_{\bm{q}} \bm{q}^2 T \sum_{\Omega_k} L(\bm{q}, i\Omega_k) L(\bm{q}, i\Omega_k + i\omega_\nu) B_{\mathrm{c}}(i\Omega_k, i\omega_\nu)^2 ]_{i\omega_\nu \to \hbar \omega \approx 0}.
\end{align}
In contrast to spin Hall conductivity, the spin current vertex $B_{\mathrm{s}}$ does not appear. 
The charge current vertex in the zero frequency limit $B_{\mathrm{c}}(0, 0)$ is nonzero. We can thus replace $B_{\mathrm{c}}(i\Omega_k, i\omega_\nu)$ by $B_{\mathrm{c}}(0,0)$ 
and obtain
\begin{align}
	\sigma_{xx}^{\mathrm{AL}}
	\propto&
	\frac{1}{i\omega} B_{\mathrm{c}}(0,0)^2 \qty[\sum_{\bm{q}} \bm{q}^2 T \sum_{\Omega_k} L(\bm{q}, i\Omega_k) L(\bm{q}, i\Omega_k + i\omega_\nu) ]_{i\omega_\nu \to \hbar \omega \approx 0}\\
	\propto&
	B_{\mathrm{c}}(0,0)^2 \underbrace{\sum_{\bm{q}} \bm{q}^2}_{\epsilon^{(D+2)/2}} \int \underbrace{d\Omega}_{\epsilon} \underbrace{\coth(\frac{\Omega}{2T})}_{\epsilon^{-1}} \underbrace{\dv{\Omega}}_{\epsilon^{-1}} [\Im \underbrace{L^R(\bm{q}, \Omega)}_{\epsilon^{-1}}]^2
	\label{eq: sigmaxx-AL}
\end{align}
in order to extract the most diverging contribution $\epsilon \to 0$. 
Note that $\omega$-linear term comes from the fluctuation part, and $\Omega$-derivative of the fluctuation part appears in the last line.  

We count the power of $\epsilon$ in Eq.~\eqref{eq: sigmaxx-AL}. From the form of $L$, 
each quantity scales as $\Omega\propto \epsilon$ and $q\propto \epsilon^{1/2}$ and accordingly $\dv{\Omega}\propto\epsilon^{-1}$, $L\propto \epsilon^{-1}$, $dq \; q^{D+1}\propto \epsilon^{(D+2) / 2}$. The fluctuation propagator $L$ is appreciable when $\Omega / T \ll 1$, where we can replace $\coth (\Omega / 2T) \approx (\Omega / 2T)^{-1}$ thus $\coth(\Omega / 2T)$ yields a factor of  $\epsilon^{-1}$. Consequently, we see that $\sigma_{xx}^{\mathrm{AL}}
\propto \epsilon^{D / 2 - 2}$. 

We turn to the AL terms in spin Hall conductivity. In the zero frequency limit, $B_{\mathrm{s}}(0,0)=0$ and the dominant contribution comes from the $\omega$-linear in $B_{\mathrm{s}}(i\Omega_k, i\omega_\nu)$ and we obtain 
\begin{align}
	\sigma_{xy}^{\mathrm{AL}}
	\propto&
	\underbrace{\sum_{\bm{q}} \bm{q}^2}_{\epsilon^{(D+2)/2}} \underbrace{L(\bm{q}, 0)^2}_{\epsilon^{-2}} B_{\mathrm{c}2}(0, 0) \dv{x}\qty[B_{\mathrm{s}2}(-x, x) + B_{\mathrm{s}2} (0, x)]_{x=0},
\end{align}
where the derivative of $L$ does not appear but that of $B_{\mathrm{s}}$ does. $B_{\mathrm{s}}$ is regular in $\omega$ and thus derivative $B_{\mathrm{s}}$ does not yield any power of $\epsilon^{-1}$. As a result, $\sigma_{xy}^{\mathrm{AL}}\propto \epsilon^{0}$. The power-counting argument does not distinguish $\ln\epsilon$ from $\epsilon^{0}$ and thus this argument correctly accounts for singularity of $\sigma_{xy}^{\mathrm{AL}}$. 

Next, we discuss the power of $\epsilon$ in the DOS terms and the MT terms. It suffices to consider the contribution from $\Omega_k=0$ in electric conductivity and spin Hall conductivity. Consequently, the singularities in both quantities are the same. 

The power of $\epsilon$ comes only from $\sum_{\bm{q}} L(\bm{q}, 0)$ in the DOS terms(for extrinsic and intrinsic cases) and the MT terms for intrinsic case and the regular part of the MT terms for extrinsic case.  $L\propto \epsilon^{-1}$, $\sum_{\bm{q}} \propto \int dq q^{D-1}\propto\epsilon^{D / 2}$ and thus $\sigma_{xy}^{\mathrm{DOS}},\sigma_{xy}^{\mathrm{MT(reg)}}\propto \epsilon^{0}$, which is consistent with $\sigma_{xy}^{\mathrm{DOS}},\sigma_{xy}^{\mathrm{MT(reg)}}\propto \ln(1/\epsilon)$. 

The power of $\epsilon$ comes only from $\sum_{\bm{q}} L(\bm{q}, 0) (\gamma_{\varphi} - \clength^2 \bm{q}^2)^{-1}$ in anomalous part of the MT terms for extrinsic case. 
$L\propto \epsilon^{-1}$, $\sum_{\bm{q}} \propto \int dq q^{D-1}\propto\epsilon^{D / 2}$, $\clength^2 q^2 \propto \epsilon$ and thus $\sigma_{xy}^{\mathrm{MT(an)}} \propto (\gamma_\varphi - \epsilon)^{-1}$, which is consistent with $\sigma_{xy}^{\mathrm{MT(an)}} \propto (\gamma_\varphi - \epsilon)^{-1}\ln \epsilon$.

We have seen that the AL terms in spin Hall conductivity have weaker singularity than the AL terms in electric conductivity. The AL terms in electric conductivity represent the effect of transport carried by the dynamically fluctuating Cooper-pairs\cite{Larkinlate2005}; the effects of those terms can also be described by the time-dependent-Ginzburg-Landau theory, which is the effective theory for boson(Cooper pairs) obtained by integrating out the fermionic degrees of freedom. In the case of s-wave superconductors, however, the Cooper pairs carry electric charges but do not spin. Accordingly, the AL terms in the spin Hall conductivity represent a different physical process from that in electric conductivity. 
The $\omega$-linear term in the response function comes from the fluctuation propagator in the case of the AL terms in electric conductivity, while the $\omega$-linear term comes from the spin current vertex in the spin Hall conductivity. We could thus say that the AL terms in electric conductivity describe the dynamical effect of Cooper-pairs. In contrast, the AL terms in spin Hall conductivity come from the dynamical part of the spin current vertex with the static effect of fluctuating Cooper-pairs. This kind of dynamical aspect of the spin current vertex in the spin Hall effect has been pointed out in an earlier study\cite{Taira2021}, where the vortex spin Hall effect in the presence of magnetic field and spin accumulation is discussed.  

Singularity in the DOS terms in spin Hall conductivity is the same as that in electric conductivity. This can be understood by recalling that the DOS terms represent the quasiparticle contribution. In this process, spin/charge is carried by quasiparticles. The presence of the fluctuating Cooper-pairs suppresses the density of states of the quasiparticles above the transition temperature\cite{DiCastro1990, Abrahams1970}.  
Thereby, electric conductivity is suppressed. The quasiparticles carry spin as well as charge and thus those terms for the spin Hall conductivity diverges in a way similar to electric conductivity.  

\subsection{\texorpdfstring{Magnitude of spin Hall conductivity; $\epsilon$-independent factors }{Magnitude of spin Hall conductivity; epsilon-independent factors }}\label{discussion-section2}

In this subsection, we discuss the other factor than 
$\epsilon$-dependence in spin Hall conductivity. In tables~\ref{discussion-table1}, \ref{discussion-table2}, and \ref{discussion-table3}, we  notice the factor $\varepsilon_{\mathrm{F}}^{-1}$ in all cases. This factor reduces the effects of fluctuation on spin Hall conductivity. We will inspect the origin of this factor by counting the power of $k_{\mathrm{F}}$ in the expressions for spin Hall conductivity normalized by that in the normal state.

The charge current vertex and spin current vertex together yield $k_{\mathrm{F}}^2$ in the AL terms in the presence of side jump, and they do $k_{\mathrm{F}}^4$ in the AL terms in the presence of skew scattering. Expansion of Green function with respect to $\bm{q}$ gives $k_{\mathrm{F}}^2$. The sum of the product of the fluctuation propagators yields $\sum_{\bm{q}} \bm{q}^2 L^2(\bm{q}, 0) \propto \clength^{-(D+2)} \propto k_{\mathrm{F}}^{-(D+2)}$. Those factors amount to $k_{\mathrm{F}}^{0}$ in $D=2$ for side jump and $k_{\mathrm{F}}^2$ for skew scattering. 
In the normal state, neither the factor $k_{\mathrm{F}}^2$ stemming from expansion of green function nor $k_{\mathrm{F}}^{-(D+2)}$ coming from the sum with respect to $\bm{q}$ exist. Consequently, the spin Hall conductivity in the normal state has one power of $\varepsilon_{\mathrm{F}}$ larger than the fluctuation conductivity in the AL terms in $D=2$. 
In the DOS terms and the MT terms, we do not expand the green function concerning $\bm{q}$ and thus the power of $k_{\mathrm{F}}$ coming from the spin current vertex and charge current vertex factor is the same as that in the normal state. In the DOS terms and the MT terms, the sum of fluctuation propagator yields $\sum_{\bm{q}} L(\bm{q}, 0) \propto \clength^{-D} \propto k_{\mathrm{F}}^{-D}$. By the last factor, the spin Hall conductivity in the normal state has a multiplicative factor $\varepsilon_{\mathrm{F}}$, compared to the fluctuation conductivity in the DOS terms and MT terms. 

Namely, the integral $\sum_{\norm{\bm{q}} < \clength^{-1}} \bm{q}^2 L^2(\bm{q}, 0)$ or $\sum_{\norm{\bm{q}} < \clength^{-1}} L(\bm{q}, 0)$ yields factor $1/\clength$ and reduces magnitude of the spin Hall conductivity. The small phase volume of $\bm{q}$ restricted by the condition $\norm{\bm{q}} <\clength^{-1}$ or the {\it support} of $L(\bm{q},0)$ implies that a limited number of electrons can contribute to the fluctuation part of the spin Hall conductivity. This fact reflects in additional factor $\varepsilon_{\mathrm{F}}^{-1}$ in fluctuation conductivity, compared to spin Hall conductivity in the normal state. 
%
%
\subsection{Relation to anomalous Hall effect}
As mentioned in Sec.~\ref{introduction-section1}, it is known that there is connection between extrinsic spin Hall effect and extrinsic anomalous Hall effect\cite{Tse2006,Sinova2015,Nagaosa2010}. In this subsection, we discuss the relation to the reference \cite{Li2020a}, where the superconducting fluctuation on anomalous Hall effect was addressed.

The uniform component of the spin and charge current density operator can be written as Eqs.~\eqref{eq: spin_current} and \eqref{eq: charge_current}, respectively. 
These equations are identical except that (i) $\bm{j}_{\mathrm{s}}$ contains the factor $1 / 2$ (ii)$\sigma_{\alpha \beta}^z$ and $\delta_{\alpha \beta}$ are swapped.
Therefore, the Feynman diagrams of the spin Hall effect and anomalous Hall effect become very similar. One of the differences is that diagrams of anomalous Hall effect contain an odd number of $\sigma_{\alpha \beta}^z$. In a ferromagnetic metal, physical quantities of up spin electron and down spin electron, such as density of states, have different values. Thus, we can incorporate the difference of the quantities into the coupling constant of spin-orbit interaction $\alpha_{\mathrm{so}}$ by taking an average of spin direction (see Eq.~(2.7) in \cite{Li2020a}).

From the above discussion, we can rewrite the results in this paper to the results in \cite{Li2020a} by replacing the strength of spin-orbit interaction $\lambda_0^2$ to $8 \alpha_{\mathrm{so}} / p_{\mathrm{F}}^2$. However, the procedures in this paper for extracting the most diverging term slightly differ from that in \cite{Li2020a}. Because of this, the results in \cite{Li2020a} are different from our results by a numerical factor.

Besides, diagram containing more fluctuation propagators have more factor of $\varepsilon_{\mathrm{F}}^{-1}$ as mentioned in the last of \ref{discussion-section2}. Li and Levchenko calculated the diagrams that contain more fluctuation propagators than diagrams in this paper and showed that these contributions have the factor of $\varepsilon_{\mathrm{F}}^{-2}$ (see Table 1 in \cite{Li2020a}). 
The nonlinear fluctuation effects are more singular with $\epsilon$ than the lowest order contributions of the fluctuation effects and they are dominant when $ \hbar /(\varepsilon_{\mathrm{F}}\tau)\ll \epsilon \ll \sqrt{\hbar /(\varepsilon_{\mathrm{F}}\tau)}$ in dirty 2D superconductors\cite{Li2020a}. For simplicity, we restrict the the lowest order contributions of the fluctuation effects. This treatment is valid when $ \sqrt{\hbar /(\varepsilon_{\mathrm{F}}\tau)}\ll \epsilon \ll 1$ in 2D dirty superconductors.

\subsection{Future Issues}
As we are motivated in the present study, the experiments by Jeon et al.\cite{Jeon2020} imply important roles of superconducting fluctuations in spin injection into superconductors or spin-charge conversion in superconductors above $T_{\rm c}$. As future issues, fluctuation effects on spin-pumping, spin-Seebeck, and charge-imbalance related to spin injection into superconductors are worthwhile to address. 

Spin injection from magnets to metals can be driven by electromagnetic field (spin pumping) or thermal gradient at the interface (spin-Seebeck effect). While both subjects for superconductors have been addressed within the mean field theory\cite{Inoue2017, Kato2019}, fluctuation effects on these effects have yet to be considered. As developed in \cite{Inoue2017, Kato2019}, the spin currents injected via spin-pumping and spin-Seebeck effect depend on 
local magnetic susceptibility
$\chi_{\mathrm{loc}}^R(\omega)$. In the limit $\omega \to 0$, the AL process vanishes as it occurs in spin Hall conductivity. When the dephasing is weak or moderate, the MT term becomes dominant and is proportional to $(\epsilon - \gamma_{\varphi})^{-1} \ln (\epsilon / \gamma_{\varphi})$ when $T \tau  / \hbar \ll 1$ or $1 \ll T \tau / \hbar \ll 1 / \sqrt{\epsilon}$, and $\epsilon^{-1 / 2}\ln(T \tau \sqrt{\epsilon})$ when $1 \ll 1/\sqrt{\epsilon} \ll T \tau / \hbar$\cite{Randeria1994}. 
Consideration of fluctuation effects on $\chi_{\mathrm{loc}}^R(\omega)$ for finite $\omega \neq 0$ will reveal fluctuation effects on spin-Seebeck and spin-pumping effects. 

Charge imbalance is another issue to be addressed. In nonequilibrium steady states in superconductors, quasiparticle density can deviate from equilibrium value, and excess or depletion of quasiparticle density is compensated by that of Cooper-pairs by charge neutrality condition. Spatial variation of Cooper-pair density (and hence that of the chemical potential of Cooper-pairs) induces the electric field so that electrochemical potential for Cooper-pairs is spatially uniform. The inverse spin Hall voltage measured in experiments in \cite{Jeon2020} is considered to be a consequence of this charge imbalance caused by a spin-charge conversion of quasiparticles in the superconductor. Charge-imbalance has been discussed theoretically in the Boltzmann-type transport theory. It is appropriate to deal with the charge imbalance within the Green function formalism, to incorporate superconducting fluctuations. 

The spin Hall effect in the normal state in Nb has been attributed to intrinsic effect \cite{Morota2011, Tanaka2008} based on a semi-quantitative model reflecting the multi-orbital electronic band structure. For a quantitative account for the experiments by Jeon et al. \cite{Jeon2020}, a theoretical study on the fluctuation effects based on a realistic model is desirable. In future research developed in this direction, the fluctuation effects on spin transport in the simple models used in the present paper will serve as a basis for understanding the results of realistic models and experiments. 

\section{Conclusion} \label{conclusion-section1}
In this paper, we theoretically study the effects of superconducting fluctuations on extrinsic spin Hall effects in two- and three-dimensional electron gas and intrinsic spin Hall effects in the two-dimensional  Rashba model. The AL, DOS, MT terms have logarithmic divergence $\ln\epsilon$ in the limit $\epsilon=(T - T_{\mathrm{c}})/T_{\mathrm{c}} \rightarrow +0$ in two-dimensional systems for both extrinsic and intrinsic spin Hall effects except the MT terms in extrinsic effect, which are proportional to $(\epsilon-\gamma_\varphi)^{-1}\ln\epsilon$ with a cutoff $\gamma_\varphi$ in two-dimensional systems. The fluctuation correction to the extrinsic spin Hall effect has an opposite sign to that in the normal state and suppresses the spin Hall effect. The correction to the intrinsic spin Hall effect has the same sign as that in the normal state and thus enhances the spin Hall effect.
The study of fluctuation effects on spin injection to superconductors based on more realistic models as well as the simple models is an important issue in the future.
\section{Acknowledgements}
This work was supported by JSPS KAKENHI Grant Number 19K05253 and 20K20891. AW and YK thank Yuta Suzuki for his comments on the intrinsic spin Hall effect in the Rashba model. 
\appendix
\section*{Appendix: List of Symbols}
\begin{description}
	\item[$T$] temperature, which has the dimension of energy in the present paper because we set $k_{\mathrm{B}}=1$. 
	\item[$\beta$] $1/T$ inverse temperature. 
	\item[$T_{\mathrm{c}}$] transition temperature. 
	\item[$\mu$] chemical potential unless it is used as the superscript/subscript.
	\item[$\varepsilon_{\bm{k}}$] $\varepsilon_{\bm{k}} \equiv \hbar^2 \bm{k}^2 / 2m$. 
	\item[$V$] volume of system. 
	\item[$\mathcal{V}(\bm{r})$] potential of impurities. 
	 $\mathcal{V}(\bm{r}) = \sum_{i} \mathcal{V}_{\mathrm{single}}(\bm{r} - \bm{R}_i)$, where $\mathcal{V}_{\mathrm{single}}(\bm{r})$ is potential of single impurity and $\bm{R}_{i}$ is position of impurities.
	\item[$v_0$] $v_0 \equiv \int d \bm{r} \mathcal{V}_{\mathrm{single}}(\bm{r})$. 
	\item[$n_{\mathrm{i}}$] density of impurities. 
	\item[$\bm{A}$] vector potential. 
	\item[$e$] electric charge unit, $e>0$. 
	\item[$\bm{\sigma}$] the Pauli matrices. $\bm{\sigma}={}^t(\sigma^x, \sigma^y, \sigma^z)$. 
	\item[$\lambda_0$] coupling constant of spin-orbit interaction in extrinsic spin Hall effect.
	\item[$\lambda$] coupling constant of spin-orbit interaction in intrinsic spin Hall effect. $\lambda$ is different from $\lambda_0$ for extrinsic spin Hall effect.
	\item[$m$] electron mass. 
	\item[$\acomm{\cdots}{\cdots}$] anticommutator. 
	\item[$\hat{\bm{z}}$] unit vector along $z$-direction. 
	\item[$\psi_\sigma(\bm{r})$] annihilation operator of electron with $z$-component of spin $\sigma$. 
	\item[$\psi_{\sigma \bm{k}}$] Fourier transform of $\psi_\sigma(\bm{r})$. 
	\item[$c_{\bm{k}\alpha}$] annihilation operator of one-particle state with wavenumber $\bm{k}$ in the band $\alpha$. It is defined by Eq.~\eqref{sinova-eq2}. 
\item[$j_{\mathrm{s}}$] spin current density operator or the uniform component of Fourier transform of the spin current density operator. 
	\item[$j_{\mathrm{c}}$] charge current density operator or the uniform component of Fourier transform of the charge current density operator.
\item[$\varepsilon_n$] Fermionic Matsubara frequency. $\varepsilon_n \equiv (2n+1)\pi T \quad (n \in \mathbb{Z})$. 
	\item[$\omega_n$] Bosonic Matsubara frequency (external frequency). $\omega_n \equiv 2n\pi T \quad (n \in \mathbb{Z})$. 
	\item[$\Omega_n$] Bosonic Matsubara frequency (internal frequency). $\Omega_n \equiv 2n\pi T \quad (n \in \mathbb{Z})$. 
	\item[$\widetilde{\varepsilon}_n$] $\widetilde{\varepsilon}_n \equiv \varepsilon_n + (\hbar / 2\tau) \sign(\varepsilon_n)$. 
	\item[$\Phi_{\mu\nu}(\bm{q}, i\omega_\nu)$] spin current-charge current response function with the wavevector $\bm{q}$ and frequency, $i\omega_\nu$ carried by an external field. 
	\item[$\mathcal{G}_\alpha(\bm{k},i\varepsilon_n)$] green function, which is defined by Eq.~\eqref{das-sarma-eq10} for extrinsic spin Hall effect and by Eq.~\eqref{sinova-eq3} for intrinsic spin Hall effect. 
	\item[$\tau$] impurity scattering time. 
	\item[$N(0)$] density of states at the Fermi surface. 
	\item[$D$] spatial dimension. $D=2$ or $D=3$. 
	\item[$\varepsilon_{\mathrm{F}}$] Fermi energy. 
	\item[$k_{\mathrm{F}}$] Fermi wavevector. 
	\item[$v_{\mathrm{F}}$] Fermi velocity. 
	\item[$F(z)$] $F(z) \equiv 1 / (e^{\beta z} + 1)$. 
	\item[$L(\bm{q}, i\Omega_k)$] propagator of superconducting fluctuation, fluctuation propagator. 
	\item[$C(\bm{q}, \varepsilon_1, \varepsilon_2)$] Cooperon. 
	\item[$\omega_{\mathrm{D}}$] Debye frequency. 
	\item[$\expval{\cdots}_{\mathrm{F.S.}}$] angular average over the fermi surface. 
	\item[$\psi(z)$] diGamma function. $\psi(z) \equiv \dv{z} \ln \Gamma(z) = \lim_{n_{\mathrm{c}} \to \infty} [- \sum_{n=0}^{n_{\mathrm{c}}-1} (n+z)^{-1} + \ln n_{\mathrm{c}}]$. The $n$-th order derivative of $\psi(z)$ is denoted by $\psi^{(n)}(z)$, which is called polyGamma function.   
	\item[$C_{\mathrm{Euler}}$] Euler-Mascheroni constant. $C_{\mathrm{Euler}}=0.577\cdots$. 
	\item[$\gamma_{\mathrm{E}}$] $\gamma_{\mathrm{E}} \equiv e^{C_{\mathrm{Euler}}}$. 
	\item[$\epsilon$] $\epsilon \equiv \ln (T / T_{\mathrm{c}})$, which becomes $\epsilon \approx (T - T_{\mathrm{c}}) / T_{\mathrm{c}}$ near $T\approx T_{\mathrm{c}}$. 
	\item[$\clength$] $\xi(T)\sqrt{\epsilon}$ with the coherence length $\xi(T)$ in Ginzburg-Landau theory. In the present paper, we call $\clength$ coherence length. It is defined by Eq.~\eqref{fluctuation-eq12}. 
	\item[$\mathcal{D}$] diffusion constant defined by Eq.~\eqref{fluctuation-eq12}. 
	\item[$\mathcal{B}(\bm{q}, i\Omega_k, i\omega_\nu)$] triangular part containing spin current/charge current vertex. \item[$\tau_{\varphi}$] phase-breaking time, which is necessary to introduce as a cutoff for extrinsic spin Hall effect in two-dimension systems as well as for electric conductivity. %
	
		\item[$\gamma_{\varphi}$] dimensionless parameter for phase breaking $\gamma_{\varphi} \equiv \pi \hbar / 8 T \tau_\varphi$. 
	\item[$\zeta(x)$] zeta function. 
	$\zeta(3) = 1.202\cdots$. 
\item[$\zeta_{\rm s}$] $\frac{e\hbar\lambda_0^2 n_{\mathrm{i}} v_0^2}{4m} \frac{k_{\mathrm{F}}^2}{D} \qty(N(0))^2$.
\item[$\zeta'_{\rm s}$] $\frac{e\hbar^3\lambda_0^2 n_{\mathrm{i}} v_0^3}{4m^2} \frac{k_{\mathrm{F}}^4}{D^2} \qty(N(0))^3$. 
\item[$\zeta_{\rm c}$] $\frac{2e\hbar^3}{m^2} \frac{k_{\mathrm{F}}^2}{D} N(0)$. 
\end{description}
\bibliography{reference.bib}
\end{document}


\title{Supplementary material for ``Fluctuation contribution to Spin Hall Effect in Superconductors"}
\author{Akimitsu Watanabe}
  \email{watanabeakimitsu@alumni.u-tokyo.ac.jp}
 \affiliation{
 Department of Physics, The University of Tokyo, Bunkyo, Tokyo 113-0033, Japan}
\author{Hiroto Adachi}
\email{hiroto.adachi@okayama-u.ac.jp}
\affiliation{Research Institute for Interdisciplinary Science, Okayama University, Okayama 700-8530, Japan}
\author{Yusuke~Kato}
 \email{yusuke@phys.c.u-tokyo.ac.jp}
\affiliation{Department of Basic Science, The University of Tokyo, Meguro-ku, Tokyo 153-8902, Japan}
%
\date{\today}
%
%
%
\maketitle
\section{Extrinsic spin Hall effect}
\subsection{Aslamazov-Larkin terms with side jump}
In this subsection, we derive the expression for the spin Hall conductivity following the procedure outlined in the main text. 
\subsubsection{Step 1: Feynman diagrams for Response function}
\begin{figure}[tbp]
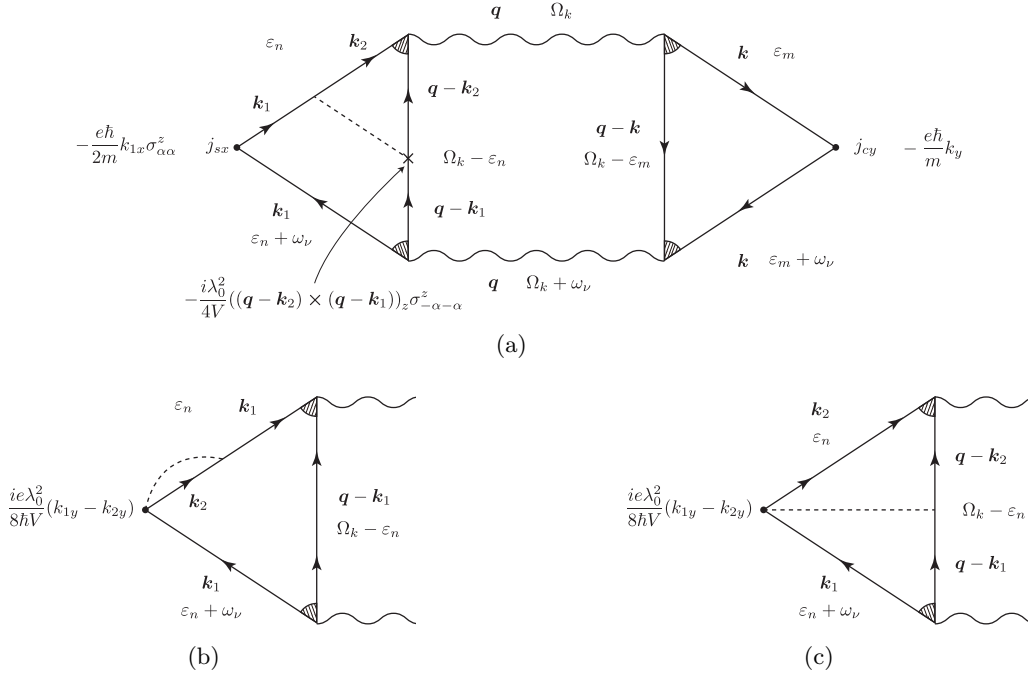

	\begin{minipage}[b]{0.9\linewidth}
		\centering
		\includegraphics[scale=0.7, page=10]{extrinsic_AL.pdf}
		\\(a)
		\label{al-fig4a}
	\end{minipage}\\
	\begin{minipage}[b]{0.45\linewidth}
		\centering
		\includegraphics[scale=0.7, page=8]{extrinsic_AL.pdf}
		\\(b)
		\label{al-fig4b}
	\end{minipage}
	\begin{minipage}[b]{0.45\linewidth}
		\centering
		\includegraphics[scale=0.7, page=9]{extrinsic_AL.pdf}
		\\(c)
		\label{al-fig4c}
	\end{minipage}
	\caption{Diagrams for AL terms with the side jump process}
	\label{al-fig4}
\end{figure}

Figure~\ref{al-fig4} shows the AL terms with the side jump process. Figures~\ref{al-fig4}(b) and (c) show only the triangular part of the spin current vertex because other parts are the same as that in (a). 
The diagram reflected with respect to the horizontal axis (i.e., the diagram "upside-down") contributes the spin Hall conductivity equally to the original diagram. Thus the contributions of (a) and (b) are doubled while that of (c) is not.  

Let $\mathcal{B}_{\rm s}^{\rm a,b,c}(\bm{q},i\Omega_k, i\omega_\nu)$, respectively, be the triangular parts in the spin current side in Figs.~\ref{al-fig4}(a), (b) and (c) and 
$\mathcal{B}_{\rm s}(\bm{q},i\Omega_k, i\omega_\nu)$ be the weighted sum of them, i. e.,
$$\mathcal{B}_{\rm s}(\bm{q},i\Omega_k, i\omega_\nu)=2\mathcal{B}_{\rm s}^{\rm a}(\bm{q},i\Omega_k, i\omega_\nu)+2\mathcal{B}_{\rm s}^{\rm b}(\bm{q},i\Omega_k, i\omega_\nu)+\mathcal{B}_{\rm s}^{\rm c}(\bm{q},i\Omega_k, i\omega_\nu).$$ 
We also denote the triangular part in the charge current side by $\mathcal{B}_{\rm c}(\bm{q},i\Omega_k, i\omega_\nu)$, which is common to Figure~\ref{al-fig4}(a), (b) and (c). In terms of $\mathcal{B}_{\rm s}(\bm{q},i\Omega_k, i\omega_\nu)$,  
$\mathcal{B}_{\rm c}(\bm{q},i\Omega_k, i\omega_\nu)$, and the fluctuation propagator, the response function is written as
\begin{align}
	\Phi_{xy}(0, i\omega_\nu)
	=&
	\frac{1}{V}\sum_{\bm{q}} T \sum_{\Omega_k} L(\bm{q}, i\Omega_k) L(\bm{q}, i\Omega_k + i\omega_\nu)
	\mathcal{B}_{\mathrm{s}}(\bm{q},i\Omega_k, i\omega_\nu) \mathcal{B}_{\mathrm{c}}(\bm{q},i\Omega_k, i\omega_\nu).
	\label{eq: al-eq9-}
\end{align}
The expressions for $\mathcal{B}_{\rm s}^{\rm a,b,c}(\bm{q},i\Omega_k, i\omega_\nu)$ and $\mathcal{B}_{\rm c}(\bm{q},i\Omega_k, i\omega_\nu)$ are given by
 \begin{align}
 \mathcal{B}_{\rm s}^{\rm a}(\bm{q},i\Omega_k, i\omega_\nu)
	=&
	\qty[-\frac{i\lambda_0^2}{4V} \qty((\bm{q} - \bm{k}_2) \cp (\bm{q}-\bm{k}_1))_z \sigma_{-\alpha -\alpha}^z] n_{\mathrm{i}}v_0^2 V \cdot \frac{1}{V} \\
	&\times
	\mathcal{G}(\bm{k}_1, i\varepsilon_n) \mathcal{G}(\bm{k}_1, i\varepsilon_n + i\omega_\nu) \mathcal{G}(\bm{q}-\bm{k}_1, i\Omega_k - i\varepsilon_n) 
	\mathcal{G}(\bm{k}_2, i\varepsilon_n) \mathcal{G}(\bm{q}- \bm{k}_2, i\Omega_k - i\varepsilon_n)\\
	&\times
	C(\bm{q}, \varepsilon_n, \varepsilon_{k-n}) C(\bm{q}, \varepsilon_{n+\nu}, \varepsilon_{k-n}),\\
 \mathcal{B}_{\rm s}^{\rm b}(\bm{q},i\Omega_k, i\omega_\nu)
	=&
	2T \sum_{\varepsilon_n} \sum_{\bm{k}_1, \bm{k}_2} \frac{ie \lambda_0^2}{8\hbar V} (k_{1y}- k_{2y}) n_{\mathrm{i}} v_0^2V \frac{1}{V}
	\mathcal{G}(\bm{k}_1, i\varepsilon_n) \mathcal{G}(\bm{k}_1, i\varepsilon_{n+\nu}) \mathcal{G}(\bm{q}-\bm{k}_1, i\varepsilon_{k-n}) \mathcal{G}(\bm{k}_2, i\varepsilon_n)\\
	&\times
	C(\bm{q}, \varepsilon_n, \varepsilon_{k-n}) C(\bm{q}, \varepsilon_{n+\nu}, \varepsilon_{k-n}),\\
\mathcal{B}_{\rm s}^{\rm c}(\bm{q},i\Omega_k, i\omega_\nu)
	=&
	2T \sum_{\varepsilon_n} \sum_{\bm{k}_1, \bm{k}_2} \frac{ie\lambda_0^2}{8\hbar V} (k_{1y} - k_{2y}) n_{\mathrm{i}} v_0^2 V \cdot \frac{1}{V} \\
	&\times
	\mathcal{G}(\bm{k}_1, i\varepsilon_{n+\nu}) \mathcal{G}(\bm{q}-\bm{k}_1, i\varepsilon_{k-n}) \mathcal{G}(\bm{k}_2, i\varepsilon_n) \mathcal{G}(\bm{q}-\bm{k}_2, i\varepsilon_{k-n})\\
	&\times
	C(\bm{q}, \varepsilon_n, \varepsilon_{k-n}) C(\bm{q}, \varepsilon_{n+\nu}, \varepsilon_{k-n}),\\
\end{align}
and 
\begin{align}
 \mathcal{B}_{\rm c}(\bm{q},i\Omega_k, i\omega_\nu)=&
	2T \sum_{\varepsilon_m} C(\bm{q}, \varepsilon_m, \Omega_k - \varepsilon_m) C(\bm{q}, \varepsilon_m + \omega_\nu, \Omega_k - \varepsilon_m) 
	\sum_{\bm{k}} \qty(-\frac{e\hbar}{m} k_{y}) \\
	&\times
	\mathcal{G}(\bm{k},i\varepsilon_m) \mathcal{G}(\bm{k}, i\varepsilon_m + i\omega_\nu) \mathcal{G}(\bm{q}-\bm{k}, i\Omega_k - i\varepsilon_m).\label{eq: Bc-bold} 
\end{align}

\subsubsection{\texorpdfstring{Step2: Expansion of triangular parts with $\bm{q}$}{Step2: Expansion of triangular parts with q}}

As in the case of calculation of electric conductivity, we expand the Green function in the side of spin current vertex as $\mathcal{G}(\bm{q}-\bm{k}_2, i\Omega_k - i\varepsilon_n) \approx \mathcal{G}(\bm{k}_2, i\Omega_k - i\varepsilon_n) - (\hbar^2 / m) (\bm{k}_2 \vdot \bm{q}) \mathcal{G}(\bm{k}_2, i\Omega_k - i\varepsilon_n)^2$ so that the wavevector $\bm{k}_2$ stemming from the external product in the $H_{\mathrm{SO}}$ is squared and the integral with $\bm{k}_2$ does not vanish. Similarly, we expand $\mathcal{G}(\bm{q}-\bm{k},i\Omega_k - i\varepsilon_m)$ in the side of charge current vertex so that the wavevector $k_y$ is squared. 
With this consideration, we expand $\mathcal{B}_{\rm s}^{\rm a,b,c}$ and $\mathcal{B}_{\rm c}$ with $\bm{q}$ up to the first order as 
\begin{align}
 \mathcal{B}_{\rm s}^{\rm a,b,c}(\bm{q},i\Omega_k, i\omega_\nu)\approx -i q_y \zeta_{\rm s}V B_{\rm s}^{\rm a,b,c}(i\Omega_k, i\omega_\nu), \quad
  \mathcal{B}_{\rm c}(\bm{q},i\Omega_k, i\omega_\nu)\approx  q_y \zeta_{\rm c}V B_{\rm c}(i\Omega_k, i\omega_\nu),   
\end{align}
with 
\begin{equation}
    \zeta_{\rm s}=\frac{e\hbar\lambda_0^2 n_{\mathrm{i}} v_0^2}{4m} \frac{k_{\mathrm{F}}^2}{D} \qty(N(0))^2,\quad
\zeta_{\rm c}=\frac{2e\hbar^3}{m^2} \frac{k_{\mathrm{F}}^2}{D} N(0) \label{eq: zeta-c-def}
\end{equation}
and 
\begin{align}
	B_{\mathrm{s}}^{\rm a,b,c}(i\Omega_k, i\omega_\nu)
	=&
	T\sum_{\varepsilon_n} C(0, \varepsilon_n, \varepsilon_{k-n}) C(0, \varepsilon_{n+\nu}, \varepsilon_{k-n})
	I^{\mathrm{a,b,c}}(i\varepsilon_n, i\Omega_k, i\omega_\nu),
	\\
	B_{\mathrm{c}}(i\Omega_k, i\omega_\nu)
	=&
	T\sum_{\varepsilon_n} C(0, \varepsilon_n, \varepsilon_{k-n}) C(0, \varepsilon_{n+\nu}, \varepsilon_{k-n}) I(i\varepsilon_n, i\Omega_k, i\omega_\nu) \label{al-eq2}.
\end{align}	
Here we introduce the following notations:  
\begin{align}
    I^{\mathrm{a}}(i\varepsilon_n , i\Omega_k, i\omega_\nu)
	=&
	\int d\xi_1 \mathcal{G}(\bm{k}_1, i\varepsilon_n) \mathcal{G}(\bm{k}_1, i\varepsilon_{n+\nu}) \mathcal{G}(\bm{k}_1, i\varepsilon_{k-n}) 
	\int d\xi_2 \mathcal{G}(\bm{k}_2, i\varepsilon_n) \mathcal{G}(\bm{k}_2, i\varepsilon_{k-n}),\\
	I^{\mathrm{b}}(i\varepsilon_n , i\Omega_k, i\omega_\nu)
	=&
	\int d\xi_1 \mathcal{G}(\bm{k}_1, i\varepsilon_n) \mathcal{G}(\bm{k}_1, i\varepsilon_{n+\nu}) \mathcal{G}(\bm{k}_1, i\varepsilon_{k-n})^2 
	\int d\xi_2 \mathcal{G}(\bm{k}_2, i\varepsilon_n),\\
	I^{\mathrm{c}}(i\varepsilon_n , i\Omega_k, i\omega_\nu)
	=&
	\int d\xi_1 \mathcal{G}(\bm{k}_1, i\varepsilon_{n+\nu}) \mathcal{G}(\bm{k}_1, i\varepsilon_{k-n})^2
	\int d\xi_2 \mathcal{G}(\bm{k}_2, i\varepsilon_n) \mathcal{G}(\bm{k}_2, i\varepsilon_{k-n})\\
	-&
	\int d\xi_1 \mathcal{G}(\bm{k}_1, i\varepsilon_{n+\nu}) \mathcal{G}(\bm{k}_1, i\varepsilon_{k-n})
	\int d\xi_2 \mathcal{G}(\bm{k}_2, i\varepsilon_n) \mathcal{G}(\bm{k}_2, i\varepsilon_{k-n})^2\\
    I(i\varepsilon_n , i\Omega_k, i\omega_\nu)
	=&
	\int d\xi\; \mathcal{G}(\bm{k}, i\varepsilon_n) \mathcal{G}(\bm{k}, i\varepsilon_{n+\nu}) \mathcal{G}(\bm{k}, i\varepsilon_{k-n})^2. \label{al-eq3}
\end{align}
We then rewrite the response function as
\begin{align}
	\Phi_{xy}(0, i\omega_\nu)
		\approx&
	-i\zeta_{\rm s}\zeta_{\rm c}
	V\sum_{\bm{q}} \frac{\bm{q}^2}{D} T \sum_{\Omega_k} L(\bm{q}, i\Omega_k) L(\bm{q}, i\Omega_k + i\omega_\nu)
	B_{\mathrm{s}}(i\Omega_k, i\omega_\nu) B_{\mathrm{c}}(i\Omega_k, i\omega_\nu),
	\label{al-eq9}
\end{align}
with 
\begin{align}
	B_{\mathrm{s}}(i\Omega_k, i\omega_\nu)
	=&
	2B_{\mathrm{s}}^{\rm a}(i\Omega_k, i\omega_\nu)+2B_{\mathrm{s}}^{\rm b}(i\Omega_k, i\omega_\nu)+B_{\mathrm{s}}^{\rm c}(i\Omega_k, i\omega_\nu).
	\end{align}
%
	\subsubsection{\texorpdfstring{Step3: Integral in $B_{\rm s}(i\Omega_k, i\omega_\nu), B_{\rm c}(i\Omega_k, i\omega_\nu)$ with wavevectors}{Step3: Integral in B\_s(iΩ\_k, iω\_ν), B\_c(iΩ\_k , iω\_ν) with wavevectors}}
We first perform the $\xi$ integral for $I$ and then will perform the $\xi_1,\xi_2$ integrals for $I^{\rm a,b,c}$. 

\begin{figure}[tbp]
	\centering
	\includegraphics[width=\textwidth, page=5]{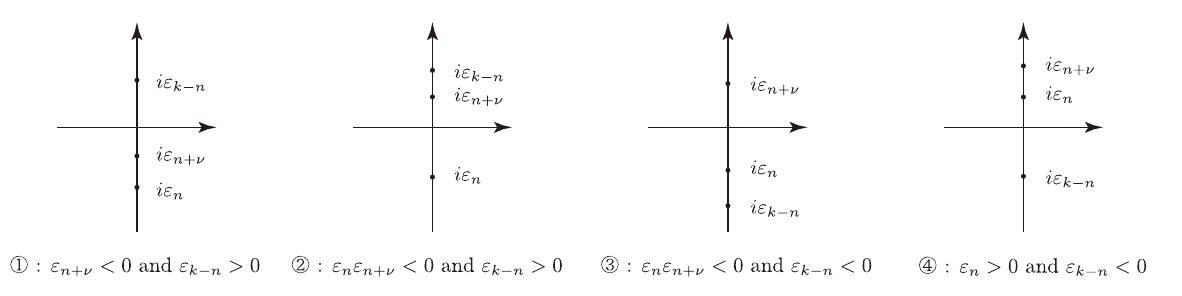}
	\caption{Locations of the poles for $\xi$ integral in $I$.}
	\label{al-fig2}
\end{figure}

The $\xi$ integral for $I$ is considered separately according to the distribution of the poles, which is shown in Fig.~\ref{al-fig2}. For each case \ctext{1} $\sim$ \ctext{4}, the ranges for $\varepsilon_n$ and $\Omega_k$ are given in Table \ref{table1}. 
\begin{table}[tbp]
	\centering
	\caption{The range of $\varepsilon_n, \Omega_k$}
	\label{table1}
	\scalebox{0.9}
	{
	\begin{tabular}{l|cccc}
		& \ctext{1}: \begin{tabular}{c} $\varepsilon_{n+\nu} < 0$  and \\$ \varepsilon_{k-n}>0$\end{tabular} 
		& \ctext{2}: \begin{tabular}{c} $\varepsilon_n \varepsilon_{n+\nu} < 0$ and \\$\varepsilon_{k-n} > 0$  \end{tabular}
		& \ctext{3}: \begin{tabular}{c} $\varepsilon_n \varepsilon_{n+\nu} < 0$ and \\$\varepsilon_{k - n} < 0$  \end{tabular}
		& \ctext{4}: \begin{tabular}{c} $\varepsilon_{n} > 0$  and \\ $\varepsilon_{k-n} < 0$ \end{tabular}\\ \hline
		\fbox{1} : $\Omega_k \leq -\omega_\nu$ & $\varepsilon_n < \Omega_k$ & NA & $-\omega_\nu < \varepsilon_n < 0$ & $0 < \varepsilon_n$\\
		\fbox{2} : $-\omega_\nu \leq \Omega_k \leq 0$ & $\varepsilon_n < -\omega_\nu$ & $-\omega_\nu < \varepsilon_n < \Omega_k$ & $\Omega_k < \varepsilon_n < 0$ & $0 < \varepsilon_n$\\
		\fbox{3} : $0 \leq \Omega_k$ & $\varepsilon_n < -\omega_\nu$ & $-\omega_\nu < \varepsilon_n < 0$ & NA & $\Omega_k < \varepsilon_n$
	\end{tabular}
	}
\end{table}
Performing the $\xi$ integral, we can obtain
\begin{align}
	I(i\varepsilon_n , i\Omega_k, i\omega_\nu)
    =&
	\int d\xi \frac{1}{\xi-i\widetilde{\varepsilon}_n} \frac{1}{\xi - i\widetilde{\varepsilon}_{n+\nu}} \frac{1}{(\xi - i\widetilde{\varepsilon}_{k-n})^2}\\
    =&
	-\frac{2\pi}{\widetilde{\varepsilon}_{n+\nu} - \widetilde{\varepsilon}_n} \qty[\frac{\ctext{1} + \ctext{2} - \ctext{4}}{(\widetilde{\varepsilon}_n - \widetilde{\varepsilon}_{k-n})^2} + \frac{-\ctext{1} + \ctext{3} + \ctext{4}}{(\widetilde{\varepsilon}_{n+\nu} - \widetilde{\varepsilon}_{k-n})^2}].
\end{align}
Here the symbol $\ctext{i}$ in the equation is unity when $\varepsilon_n\in$ $\ctext{i}$ and zero otherwise. For example, when \fbox{1} ($\Omega_k \leq -\omega_\nu$), \ctext{1} = $\theta(\Omega_k - \varepsilon_n)$.

In terms of this notation, $B_{\mathrm{c}}(i\Omega_k, i\omega_\nu)$ can be written as
\begin{align}
	B_{\mathrm{c}}(i\Omega_k, i\omega_\nu)
	=&
	-\frac{2\pi T}{\widetilde{\varepsilon}_{n+\nu} - \widetilde{\varepsilon}_n} \sum_{\varepsilon_n} \frac{\widetilde{\varepsilon}_n - \widetilde{\varepsilon}_{k-n}}{\varepsilon_n - \varepsilon_{k-n}} \frac{\widetilde{\varepsilon}_{n+\nu} - \widetilde{\varepsilon}_{k-n}}{\varepsilon_{n+\nu} - \varepsilon_{k-n}}
	\qty[\frac{\ctext{1} + \ctext{2} - \ctext{4}}{(\widetilde{\varepsilon}_n - \widetilde{\varepsilon}_{k-n})^2} + \frac{-\ctext{1} + \ctext{3} + \ctext{4}}{(\widetilde{\varepsilon}_{n+\nu} - \widetilde{\varepsilon}_{k-n})^2}]\\
	=&
	-2\pi T \sum_{\varepsilon_n} \frac{1}{(2\varepsilon_n - \Omega_k) (2\varepsilon_n - \Omega_k + \omega_\nu)} 
	\begin{aligned}[t]
		\Biggl[&
		\qty(\frac{1}{2\varepsilon_n - \Omega_k - \hbar / \tau} + \frac{1}{2\varepsilon_n - \Omega_k + \omega_\nu - \hbar / \tau}) \ctext{1}\\
		-&
		\qty(\frac{1}{2\varepsilon_n - \Omega_k + \hbar / \tau} + \frac{1}{2\varepsilon_n - \Omega_k + \omega_\nu + \hbar / \tau}) \ctext{4}
		\Biggr]
	\end{aligned}\\
	&-
	\frac{2\pi T}{\omega_\nu + \hbar / \tau} \sum_{\varepsilon_n} 
	\qty[\frac{\ctext{2}}{(2\varepsilon_n - \Omega_k) (2\varepsilon_n - \Omega_k - \hbar / \tau)} + \frac{\ctext{3}}{(2\varepsilon_n - \Omega_k + \omega_\nu) (2\varepsilon_n - \Omega_k + \omega_\nu + \hbar / \tau)}].
\end{align}
When \fbox{1} ($\Omega_k \leq -\omega_\nu$), $	B_{\mathrm{c}}$ becomes
\begin{align}
	&
	B_{\mathrm{c}1}(i\Omega_k, i\omega_\nu)\\
	=&
	-2\pi T 
	\begin{aligned}[t]
		\Biggl[&
			\sum_{\varepsilon_n<0} \frac{1}{(2\varepsilon_n + \Omega_k)(2\varepsilon_n + \Omega_k + \omega_\nu)} \qty(\frac{1}{2\varepsilon_n + \Omega_k - \hbar / \tau} + \frac{1}{2\varepsilon_n + \Omega_k + \omega_\nu - \hbar /\tau})\\
			-&
			\sum_{\varepsilon_n > 0} \frac{1}{(2\varepsilon_n - \Omega_k) (2\varepsilon_n - \Omega_k + \omega_\nu)} \qty(\frac{1}{2\varepsilon_n - \Omega_k + \hbar / \tau} + \frac{1}{2\varepsilon_n - \Omega_k + \omega_\nu + \hbar / \tau})
		\Biggr]
	\end{aligned}\\
	-&
	\frac{2\pi T}{\omega_\nu + \hbar / \tau}
	\sum_{\varepsilon_n > 0}
	\begin{aligned}[t]
		\Biggl[&
		\frac{1}{(2\varepsilon_n - \Omega_k - \omega_\nu)(2\varepsilon_n - \Omega_k - \omega_\nu + \hbar / \tau)} \\
		-&
		\frac{1}{(2\varepsilon_n - \Omega_k + \omega_\nu) (2\varepsilon_n - \Omega_k + \omega_\nu + \hbar / \tau)}
		\Biggr].
	\end{aligned}\label{al-eq10}
\end{align}
After the analytic continuation $i\Omega_k \to \Omega$, 	$B_{\mathrm{c}1}$ becomes a regular function of $\Omega$ for $\Im[\Omega] \leq -\omega_\nu$.  

When \fbox{2} ($-\omega_\nu \leq \Omega_k\leq 0$)$,	B_{\mathrm{c}}$ becomes
\begin{align}
	&
	B_{\mathrm{c}2}(i\Omega_k, i\omega_\nu)\\
	=&
	-2\pi T \sum_{\varepsilon_n < 0} \frac{1}{(2\varepsilon_n - \Omega_k - 2\omega_\nu) (2\varepsilon_n - \Omega_k -\omega_\nu)} \qty(\frac{1}{2\varepsilon_n - \Omega_k - 2\omega_\nu - \hbar / \tau} + \frac{1}{2\varepsilon_n - \Omega_k - \omega_\nu - \hbar / \tau})\\
	&+
	2\pi T \sum_{\varepsilon_n > 0} \frac{1}{(2\varepsilon_n - \Omega_k)(2\varepsilon_n - \Omega_k + \omega_\nu)} \qty(\frac{1}{2\varepsilon_n - \Omega_k + \hbar / \tau} + \frac{1}{2\varepsilon_n - \Omega_k + \omega_\nu + \hbar / \tau })\\
	&-
	\frac{2\pi T}{\omega_\nu + \hbar / \tau} 
	\begin{aligned}[t]
		\Biggl[&
			\sum_{\varepsilon_n < 0}
			\frac{1}{(2\varepsilon_n + \Omega_k) (2\varepsilon_n + \Omega_k - \hbar / \tau)} - \frac{1}{(2\varepsilon_n - \Omega_k - 2\omega_\nu)(2\varepsilon_n - \Omega_k - 2\omega_\nu - \hbar / \tau)}\\
			+&
			\sum_{\varepsilon_n > 0} 
			\frac{1}{(2\varepsilon_n + \Omega_k + \omega_\nu)(2\varepsilon_n + \Omega_k + \omega_\nu + \hbar / \tau)} - \frac{1}{(2\varepsilon_n - \Omega_k + \omega_\nu) (2\varepsilon_n - \Omega_k + \omega_\nu + \hbar / \tau)}
		\Biggr]. \label{al-eq11}
	\end{aligned}
\end{align}
This becomes regular function for $-\omega_\nu \leq \Im [\Omega] \leq 0$ after analytic continuation $i\Omega_k \to \Omega$.

When \fbox{3} ($\Omega_k \geq 0$), $B_{\mathrm{c}}$ becomes
\begin{align}
	&
	B_{\mathrm{c}3}(i\Omega_k, i\omega_\nu)\\
	=&
	-2\pi T
	\begin{aligned}[t]
		\Biggl[&
			\sum_{\varepsilon_n < 0} \frac{1}{(2\varepsilon_n - \Omega_k - 2\omega_\nu) (2\varepsilon_n - \Omega_k - \omega_\nu)} \qty(\frac{1}{2\varepsilon_n - \Omega_k - 2\omega_\nu - \hbar / \tau} + \frac{1}{2\varepsilon_n - \Omega_k - \omega_\nu - \hbar / \tau})\\
			-&
			\sum_{\varepsilon_n > 0} \frac{1}{(2\varepsilon_n + \Omega_k) (2\varepsilon_n + \Omega_k + \omega_\nu)} \qty(\frac{1}{2\varepsilon_n + \Omega_k + \hbar / \tau} + \frac{1}{2\varepsilon_n + \Omega_k + \omega_\nu + \hbar / \tau})
		\Biggr]
	\end{aligned}\\
	&-
	\frac{2\pi T}{\omega_\nu + \hbar / \tau} \sum_{\varepsilon_n < 0}
	\qty[\frac{1}{(2\varepsilon_n - \Omega_k) (2\varepsilon_n - \Omega_k - \hbar / \tau)} - \frac{1}{(2\varepsilon_n - \Omega_k - 2\omega_\nu) (2\varepsilon_n - \Omega_k - 2\omega_\nu - \hbar / \tau)}] \label{al-eq12}.
\end{align}
This is a regular function for $\Im[\Omega] \geq 0$ after analytic continuation $i\Omega_k \to \Omega$. 
%
%

We next calculate $I^{\mathrm{a,b,c}}$. 
After the integrals with $\xi_1$ and $\xi_2$, they become
\begin{align}
	I^{\mathrm{a}}
	=&
	4\pi^2 i \frac{1}{(\widetilde{\varepsilon}_{n+\nu} - \widetilde{\varepsilon}_{n}) (\widetilde{\varepsilon}_n - \widetilde{\varepsilon}_{k-n})} \qty(-\frac{\ctext{1} + \ctext{2} + \ctext{4}}{\widetilde{\varepsilon}_n - \widetilde{\varepsilon}_{k-n}} + \frac{\ctext{1} + \ctext{4}}{\widetilde{\varepsilon}_{n+\nu} - \widetilde{\varepsilon}_{k-n}})\\
	I^{\mathrm{b}}
	=&
	-2\pi^2 i \frac{1}{\widetilde{\varepsilon}_{n+\nu} - \widetilde{\varepsilon}_n} \qty[\frac{\ctext{1} + \ctext{2} + \ctext{4}}{(\widetilde{\varepsilon}_n - \widetilde{\varepsilon}_{k-n})^2} + \frac{-\ctext{1} + \ctext{3} - \ctext{4}}{(\widetilde{\varepsilon}_{n+\nu} - \widetilde{\varepsilon}_{k-n})^2}]\\
	I^{\mathrm{c}}
	=&
	\begin{aligned}[t]
		-4\pi^2 i \frac{(\widetilde{\varepsilon}_{n+\nu} - \widetilde{\varepsilon}_n) (\ctext{1} + \ctext{4})}{(\widetilde{\varepsilon}_{n} - \widetilde{\varepsilon}_{k-n})^2 (\widetilde{\varepsilon}_{n+\nu} - \widetilde{\varepsilon}_{k-n})^2}
	\end{aligned}.
\end{align}
With use of this result, we obtain
$B_{\mathrm{s}}^{\mathrm{a}}$ becomes
\begin{align}
	B_{\mathrm{s}}^{\mathrm{a}}(i\Omega_k, i\omega_\nu)
	=&
	T\sum_{\varepsilon_n} C(0, \varepsilon_n, \varepsilon_{k-n}) C(0, \varepsilon_{n+\nu}, \varepsilon_{k-n}) I^{\mathrm{a}}(i\varepsilon_n, i\Omega_k, i\omega_\nu)\\
		=&
	-4\pi^2 i T \sum_{\varepsilon_n}\qty[\frac{\ctext{1} + \ctext{4}}{(\varepsilon_n - \varepsilon_{k-n}) (\varepsilon_{n+\nu} - \varepsilon_{k-n}) (\widetilde{\varepsilon}_n - \widetilde{\varepsilon}_{k-n})} + \frac{1}{\omega_\nu + \hbar / \tau} \frac{\ctext{2}}{(\varepsilon_n - \varepsilon_{k-n}) (\widetilde{\varepsilon}_{n} - \widetilde{\varepsilon}_{k-n})}]\\
	=&
	-4\pi^2 i T \sum_{\varepsilon_n}
	\begin{aligned}[t]
		\Biggl[&
			\frac{\ctext{1}}{(2\varepsilon_n - \Omega_k) (2\varepsilon_n - \Omega_k + \omega_\nu) (2\varepsilon_n - \Omega_k - \hbar / \tau)}\\
			+&
			\frac{\ctext{4}}{(2\varepsilon_n - \Omega_k) (2\varepsilon_n - \Omega_k + \omega_\nu) (2\varepsilon_n - \Omega_k + \hbar / \tau)}\\
			+&
			\frac{1}{\omega_\nu + \hbar / \tau} \frac{\ctext{2}}{(2\varepsilon_n - \Omega_k) (2\varepsilon_n - \Omega_k - \hbar / \tau)}
		\Biggr].
	\end{aligned}
\end{align}
It then follows that 
\begin{align}
	B_{\mathrm{s}1}^{\mathrm{a}}(i\Omega_k, i\omega_\nu)
	=&
	-4\pi^2 i T
	\begin{aligned}[t]
		\Biggl[&
			\sum_{\varepsilon_n < 0} \frac{1}{(2\varepsilon_n + \Omega_k)(2\varepsilon_n + \Omega_k + \omega_\nu) (2\varepsilon_n + \Omega_k - \hbar /\tau)}\\
			+&
			\sum_{\varepsilon_n > 0} \frac{1}{(2\varepsilon_n - \Omega_k) (2\varepsilon_n - \Omega_k + \omega_\nu) (2\varepsilon_n - \Omega_k + \hbar / \tau)}
		\Biggr],
	\end{aligned}\\
	B_{\mathrm{s}2}^{\mathrm{a}} (i\Omega_k, i\omega_\nu)
	=&
	-4\pi^2 i T
	\begin{aligned}[t]
		\Biggl[&
			\sum_{\varepsilon_n < 0} \frac{1}{(2\varepsilon_n - \Omega_k - 2\omega_\nu) (2\varepsilon_n - \Omega_k - \omega_\nu) (2\varepsilon_n - \Omega_k - 2\omega_\nu - \hbar / \tau)}\\
			+&
			\sum_{\varepsilon_n > 0} \frac{1}{(2\varepsilon_n - \Omega_k) (2\varepsilon_n - \Omega_k + \omega_\nu) (2\varepsilon_n - \Omega_k + \hbar / \tau)}
		\Biggr]
	\end{aligned}\\
	&-
	\frac{4\pi^2 i T}{\omega_\nu + \hbar / \tau}
	\qty[\sum_{\varepsilon_n < 0} \frac{1}{(2\varepsilon_n + \Omega_k) (2\varepsilon_n + \Omega_k - \hbar / \tau)} - \frac{1}{(2\varepsilon_n - \Omega_k - 2\omega_\nu) (2\varepsilon_n - \Omega_k - 2\omega_\nu - \hbar / \tau)}],\\
	B_{\mathrm{s}3}^{\mathrm{a}}(i\Omega_k, i\omega_\nu)
	=&
	-4\pi^2 i T
	\begin{aligned}[t]
		\Biggl[&
		\sum_{\varepsilon_n< 0}\frac{1}{(2\varepsilon_n - \Omega_k - 2\omega_\nu) (2\varepsilon_n - \Omega_k - \omega_\nu) (2\varepsilon_n - \Omega_k - 2\omega_\nu - \hbar / \tau)}\\
		+&
		\sum_{\varepsilon_n > 0} \frac{1}{(2\varepsilon_n + \Omega_k) (2\varepsilon_n + \Omega_k + \omega_\nu) (2\varepsilon_n + \Omega_k + \hbar / \tau)}
		\Biggr]
	\end{aligned}\\
	&-
	\frac{4\pi^2 i T}{\omega_\nu + \hbar / \tau} \sum_{\varepsilon_n < 0}
	\qty[\frac{1}{(2\varepsilon_n - \Omega_k) (2\varepsilon_n - \Omega_k - \hbar / \tau)} - \frac{1}{(2\varepsilon_n - \Omega_k - 2\omega_\nu) (2\varepsilon_n - \Omega_k - 2\omega_\nu - \hbar / \tau)}].
\end{align}
Next we consider $B_{\mathrm{s}}^{\mathrm{b}}$
\begin{align}
	B_{\mathrm{s}}^{\mathrm{b}} (i\Omega_k, i\omega_\nu)
	=&
	T \sum_{\varepsilon_n} C(0, \varepsilon_n, \varepsilon_{k-n}) C(0, \varepsilon_{n+\nu}, \varepsilon_{k-n}) I^{\mathrm{b}}(i\varepsilon_n, i\Omega_k, i\omega_\nu)\\
	=&
	-2\pi^2 i T	\sum_{\varepsilon_n}
	\begin{aligned}[t]
		\Biggl[&
			\frac{1}{(2\varepsilon_n - \Omega_k) (2\varepsilon_n - \Omega_k + \omega_\nu)} \qty(\frac{1}{2\varepsilon_n - \Omega_k - \hbar / \tau} + \frac{1}{2\varepsilon_n - \Omega_k + \omega_\nu - \hbar / \tau}) \ctext{1}\\
			+&
			\frac{1}{(2\varepsilon_n - \Omega_k) (2\varepsilon_n - \Omega_k + \omega_\nu)} \qty(\frac{1}{2\varepsilon_n - \Omega_k + \hbar / \tau} + \frac{1}{2\varepsilon_n - \Omega_k + \omega_\nu + \hbar / \tau}) \ctext{4}
		\Biggr]
	\end{aligned}\\
	&-
	\frac{2\pi^2 i T}{\omega_\nu + \hbar / \tau} \sum_{\varepsilon_n}
	\qty[\frac{\ctext{2}}{(2\varepsilon_n - \Omega_k) (2\varepsilon_n - \Omega_k - \hbar /\tau)} + \frac{\ctext{3}}{(2\varepsilon_n - \Omega_k + \omega_\nu) (2\varepsilon_n - \Omega_k + \omega_\nu + \hbar / \tau)}],
\end{align}
from which it follows that 
\begin{align}
	&
	B_{\mathrm{s}1}^{\mathrm{b}} (i\Omega_k, i\omega_\nu)\\
	=&
	-2\pi^2 i T
	\begin{aligned}[t]
		\Biggl[&
			\sum_{\varepsilon_n < 0} \frac{1}{(2\varepsilon_n + \Omega_k) (2\varepsilon_n + \Omega_k + \omega_\nu)} \qty(\frac{1}{2\varepsilon_n + \Omega_k - \hbar / \tau} + \frac{1}{2\varepsilon_n + \Omega_k + \omega_\nu - \hbar / \tau}) \\
			+&
			\sum_{\varepsilon_n > 0} \frac{1}{(2\varepsilon_n - \Omega_k) (2\varepsilon_n - \Omega_k + \omega_\nu)} \qty(\frac{1}{2\varepsilon_n - \Omega_k + \hbar / \tau} + \frac{1}{2\varepsilon_n - \Omega_k + \omega_\nu + \hbar / \tau})
		\Biggr]
	\end{aligned}\\
	&-
	\frac{2\pi^2 i T}{\omega_\nu + \hbar / \tau} \sum_{\varepsilon_n > 0} 
	\qty[\frac{1}{(2\varepsilon_n - \Omega_k - \omega_\nu) (2\varepsilon_n - \Omega_k + \omega_\nu - \hbar / \tau)} - \frac{1}{(2\varepsilon_n - \Omega_k + \omega_\nu) (2\varepsilon_n - \Omega_k + \omega_\nu + \hbar / \tau)}]\\
	 &
	B_{\mathrm{s}2}^{\mathrm{b}}(i\Omega_k, i\omega_\nu)\\
	=&
	-2\pi^2 i T
	\begin{aligned}[t]
		\Biggl[&
			\sum_{\varepsilon_n < 0} \frac{1}{(2\varepsilon_n - \Omega_k - 2\omega_\nu) (2\varepsilon_n - \Omega_k - \omega_\nu)} \qty(\frac{1}{2\varepsilon_n - \Omega_k - 2\omega_\nu - \hbar / \tau} + \frac{1}{2\varepsilon_n - \Omega_k - \omega_\nu - \hbar / \tau})\\
			+&
			\sum_{\varepsilon_n > 0} \frac{1}{(2\varepsilon_n - \Omega_k) (2\varepsilon_n - \Omega_k + \omega_\nu)} \qty(\frac{1}{2\varepsilon_n - \Omega_k + \hbar / \tau} + \frac{1}{2\varepsilon_n - \Omega_k + \omega_\nu + \hbar / \tau})
		\Biggr]
	\end{aligned}\\
	&-
	\frac{2\pi^2 i T}{\omega_\nu + \hbar / \tau}
	\begin{aligned}[t]
		\Biggl\{&
			\sum_{\varepsilon_n < 0} \qty[\frac{1}{(2\varepsilon_n + \Omega_k) (2\varepsilon_n + \Omega_k - \hbar / \tau)} - \frac{1}{(2\varepsilon_n - \Omega_k - 2\omega_\nu) (2\varepsilon_n - \Omega_k - 2\omega_\nu - \hbar / \tau)}]\\
			+&
			\sum_{\varepsilon_n > 0} \qty[\frac{1}{(2\varepsilon_n + \Omega_k + \omega_\nu)  (2\varepsilon_n + \Omega_k + \omega_\nu + \hbar / \tau)} - \frac{1}{(2\varepsilon_n - \Omega_k + \omega_\nu) (2\varepsilon_n - \Omega_k + \omega_\nu + \hbar / \tau)}]
		\Biggr\}
	\end{aligned}\\
	 &
	 B_{\mathrm{s}3}^{\mathrm{b}}(i\Omega_k, i\omega_\nu)\\
	=&
	-2\pi^2 i T
	\begin{aligned}[t]
		\Biggl[&
			\sum_{\varepsilon_n < 0} \frac{1}{(2\varepsilon_n - \Omega_k - 2\omega_\nu) (2\varepsilon_n - \Omega_k - \omega_\nu)} \qty(\frac{1}{2\varepsilon_n - \Omega_k - 2\omega_\nu - \hbar / \tau} + \frac{1}{2\varepsilon_n - \Omega_k - \omega_\nu - \hbar / \tau}) \\
			+&
			\sum_{\varepsilon_n > 0} \frac{1}{(2\varepsilon_n + \Omega_k) (2\varepsilon_n + \Omega_k + \omega_\nu)} \qty(\frac{1}{2\varepsilon_ n+ \Omega_k + \hbar / \tau} + \frac{1}{2\varepsilon_n + \Omega_k + \omega_\nu + \hbar / \tau})
		\Biggr]
	\end{aligned}\\
	&-
	\frac{2\pi^2 i T}{\omega_\nu + \hbar / \tau} \sum_{\varepsilon_n < 0}
	\qty[\frac{1}{(2\varepsilon_n - \Omega_k) (2\varepsilon_n - \Omega_k - \hbar / \tau)} - \frac{1}{(2\varepsilon_n - \Omega_k - 2\omega_\nu) (2\varepsilon_n - \Omega_k - 2\omega_\nu - \hbar / \tau)}].
\end{align}
Finally we consider	$B_{\mathrm{s}}^{\mathrm{c}}(i\Omega_k, i\omega_\nu)$
\begin{align}
	B_{\mathrm{s}}^{\mathrm{c}}(i\Omega_k, i\omega_\nu)
	=&
	T \sum_{\varepsilon_n} C(0, \varepsilon_n, \varepsilon_{k-n}) C(0, \varepsilon_{n+\nu}, \varepsilon_{k-n}) I^{\mathrm{c}}(i\varepsilon_n, i\Omega_k, i\omega_\nu)\\
	=&
	-4\pi^2 i T \omega_\nu \sum_{\varepsilon_n}
	\begin{aligned}[t]
		\Biggl[&
		\frac{\ctext{1}}{(2\varepsilon_n - \Omega_k) (2\varepsilon_n - \Omega_k + \omega_\nu) (2\varepsilon_n - \Omega_k - \hbar / \tau) (2\varepsilon_n - \Omega_k + \omega_\nu - \hbar / \tau)} \\
		+&
		\frac{\ctext{4}}{(2\varepsilon_n - \Omega_k) (2\varepsilon_n - \Omega_k + \omega_\nu) (2\varepsilon_n - \Omega_k + \hbar / \tau) (2\varepsilon_n - \Omega_k + \omega_\nu + \hbar / \tau)}
		\Biggr],
	\end{aligned}
\end{align}
from which it follows that
\begin{align}
	&
	B_{\mathrm{s}1}^{\mathrm{c}}(i\Omega_k, i\omega_\nu)\\
	=&
	-4\pi^2 i T \omega_\nu 
	\begin{aligned}[t]
		\Biggl[&
		\sum_{\varepsilon_n < 0} \frac{1}{(2\varepsilon_n + \Omega_k) (2\varepsilon_n + \Omega_k + \omega_\nu) (2\varepsilon_n + \Omega_k - \hbar /\tau) (2\varepsilon_n + \Omega_k + \omega_\nu - \hbar / \tau)}\\
		+&
		\sum_{\varepsilon_n > 0} \frac{1}{(2\varepsilon_n - \Omega_k) (2\varepsilon_n - \Omega_k + \omega_\nu) (2\varepsilon_n - \Omega_k + \hbar / \tau) (2\varepsilon_n - \Omega_k + \omega_\nu + \hbar / \tau)}
		\Biggr],
	\end{aligned}\\
	 &
	B_{\mathrm{s}2}^{\mathrm{c}}(i\Omega_k, i\omega_\nu)\\
	=&
	-4\pi^2 i T \omega_\nu
	\begin{aligned}[t]
		\Biggl[&
			\sum_{\varepsilon_n < 0} \frac{1}{(2\varepsilon_n - \Omega_k - 2\omega_\nu) (2\varepsilon_n - \Omega_k - \omega_\nu) (2\varepsilon_n - \Omega_k - 2\omega_\nu - \hbar / \tau) (2\varepsilon_n - \Omega_k - \omega_\nu - \hbar / \tau)}\\
			+&
			\sum_{\varepsilon_n > 0} \frac{1}{(2\varepsilon_n - \Omega_k) (2\varepsilon_n - \Omega_k + \omega_\nu) (2\varepsilon_n - \Omega_k + \hbar / \tau) (2\varepsilon_n - \Omega_k + \omega_\nu + \hbar / \tau)}
		\Biggr],
	\end{aligned}\\
	 &
	 B_{\mathrm{s}3}^{\mathrm{c}}(i\Omega_k, i\omega_\nu)\\
	=&
	-4\pi^2 i T \omega_\nu
	\begin{aligned}[t]
		\Biggl[&
		\sum_{\varepsilon_n < 0} \frac{1}{(2\varepsilon_n - \Omega_k - 2\omega_\nu) (2\varepsilon_n - \Omega_k - \omega_\nu) (2\varepsilon_n - \Omega_k - 2\omega_\nu - \hbar / \tau) (2\varepsilon_n - \Omega_k - \omega_\nu - \hbar / \tau)}\\
		+&
		\sum_{\varepsilon_n > 0} \frac{1}{(2\varepsilon_n + \Omega_k) (2\varepsilon_n + \Omega_k + \omega_\nu) (2\varepsilon_n + \Omega_k + \hbar / \tau) (2\varepsilon_n + \Omega_k + \omega_\nu + \hbar / \tau)}
		\Biggr].
	\end{aligned}
\end{align}
From these expressions, we see that $B_{\mathrm{s}}^{\mathrm{a,b,c}}$ become regular functions of $\Omega$ for the regions \fbox{1}, \fbox{2}, \fbox{3}, respectively, after analytic continuation  
$i\Omega_k \to \Omega$. 
We immediately notice from these expressions that 
\begin{align}
	B_{\mathrm{s}1} (\Omega, 0) = B_{\mathrm{s}2}(0, 0) = B_{\mathrm{s}3}(\Omega, 0) = 0 \label{al-eq1}.
\end{align}
\subsubsection{\texorpdfstring{Step4: Sum over $\Omega_k$}{Step4: Sum over Ω\_k}}
\begin{figure}[tbp]
	\centering
	\includegraphics[page=6]{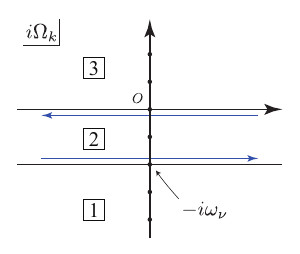}
	\caption{The path for integral.}
	\label{al-fig3}
\end{figure}

With the use of results in the previous two subsections, we can single out the term diverging in the limit of $\epsilon \to 0$ in Eq.\eqref{al-eq9} by summing over $\Omega_k$.
For the region \fbox{2}, we transform the sum to the integral along the path shown in Fig.~\ref{al-fig3} and obtain
\begin{align}
	&
	T \sum_{\Omega_k} L(\bm{q}, i\Omega_k) L(\bm{q}, i\Omega_k + i\omega_\nu) B_{\mathrm{c}}(i\Omega_k , i\omega_\nu) B_{\mathrm{s}}(i\Omega_k, i\omega_\nu)\label{eq: al-eq1+}\\
	=&
	\frac{T}{2} L^A(\bm{q}, -i\omega_\nu) L(\bm{q}, 0) B_{\mathrm{c}2}(-i\omega_\nu, i\omega_\nu) B_{\mathrm{s}2}(-i\omega_\nu, i\omega_\nu)\\
	&+
	\frac{T}{2} L(\bm{q}, 0) L^R(\bm{q}, i\omega_\nu) B_{\mathrm{c}2}(0, i\omega_\nu) B_{\mathrm{s}2}(0, i\omega_\nu)\\
	&+
	\pv \int_{-\infty}^{\infty} \frac{d\Omega}{4\pi i} \coth \frac{\Omega}{2T}
	\begin{aligned}[t]
		\Bigl[&
			L^A(\bm{q}, \Omega - i\omega_\nu) L^R(\bm{q}, \Omega) B_{\mathrm{c}2}(\Omega - i\omega_\nu, i\omega_\nu) B_{\mathrm{s}2}(\Omega - i\omega_\nu, i\omega_\nu)\\
			-&
			L^A(\bm{q}, \Omega) L^R(\bm{q}, \Omega + i\omega_\nu) B_{\mathrm{c}2}(\Omega, i\omega_\nu) B_{\mathrm{s}2}(\Omega, i\omega_\nu)
		\Bigr]
	\end{aligned}\\
	&+
	T \sum_{\Omega_k < 0} L^A(\bm{q}, i\Omega_k - i\omega_\nu) L^A(\bm{q}, i\Omega_k) B_{\mathrm{c}1}(i\Omega_k - i\omega_\nu, i\omega_\nu) B_{\mathrm{s}1}(i\Omega_k - i\omega_\nu, i\omega_\nu)\\
	&+
	T \sum_{\Omega_k > 0} L^R(\bm{q}, i\Omega_k) L^R(\bm{q}, i\Omega_k + i\omega_\nu) B_{\mathrm{c}3}(i\Omega_k, i\omega_\nu)B_{\mathrm{s}3}(i\Omega_k, i\omega_\nu).
\end{align}

\subsubsection{\texorpdfstring{Step5: Analytic continuation and expansion with $\omega$}{Step5: Analytic continuation and expansion with ω}}
We then expand it $\omega$ after analytical continuation $i\omega_\nu \to \hbar \omega \approx 0$. From \eqref{al-eq1}, we see that the terms containing no derivative of $B_{\mathrm{s}}$ vanish. As a result, the expression in Eq.~\eqref{eq: al-eq1+} becomes
\begin{align}
	&
	\qty[T \sum_{\Omega_k} L(\bm{q}, i\Omega_k) L(\bm{q}, i\Omega_k + i\omega_\nu) B_{\mathrm{c}}(i\Omega_k, i\omega_\nu) B_{\mathrm{s}}(i\Omega_k, i\omega_\nu)]_{i\omega_\nu \to \hbar \omega \approx 0}\\
	=&
	\hbar \omega \cdot \frac{T}{2} L(\bm{q}, 0)^2 B_{\mathrm{c}2}(0, 0) \dv{x} \qty[B_{\mathrm{s}2}(-x, x) + B_{\mathrm{s}2}(0, x)]_{x=0} \label{al-eq3+}\\
	&-
	\hbar \omega \cdot \pv \int_{-\infty}^{\infty} \frac{d\Omega}{4\pi i} \coth \frac{\Omega}{2T} \dv{\Omega} \qty[L^A(\bm{q}, \Omega) L^R(\bm{q}, \Omega) B_{\mathrm{c}2}(\Omega, 0) B_{\mathrm{s}2}(\Omega, 0)] \label{al-eq4}\\
	&+
	\hbar \omega \cdot T \sum_{\Omega_k < 0} L^A(\bm{q}, i\Omega_k)^2 B_{\mathrm{c}1} (i\Omega_k, 0) \qty[\pdv{x} B_{\mathrm{s}1}(i\Omega_k, x)]_{x=0} \label{al-eq5}\\
	&+
	\hbar \omega \cdot T \sum_{\Omega_k > 0} L^R(\bm{q}, i\Omega_k)^2 B_{\mathrm{c}3} (i\Omega_k, 0) \qty[\pdv{x} B_{\mathrm{s}3}(i\Omega_k, x)]_{x=0} \label{al-eq6}\\
	&+
	\order{\omega^2}.
\end{align}
\subsubsection{\texorpdfstring{Step6: Retaining the diverging term in the limit $\epsilon\rightarrow +0$}{Step6: Retaining the diverging term in the limit ϵ→+0}}
We find that Eqs.~\eqref{al-eq4}, \eqref{al-eq5}, \eqref{al-eq6} do not contribute the terms diverging in the limit $\epsilon\rightarrow +0$ as shown in sec.~\ref{subsec: appendixB}. It thus suffices to consider Eq.~\eqref{al-eq3+}. We then obtain 
\begin{align}
	\sigma_{xy}^{\mathrm{AL-SJ}}
	\approx&
	-\frac{1}{i\omega} \qty(\frac{i\zeta_{\rm s}\zeta_{\rm c}}{D})
	\qty[V \sum_{\bm{q}} \bm{q}^2 T \sum_{\Omega_k} L(\bm{q}, i\Omega_k) L(\bm{q}, i\Omega_k + i\omega_\nu) B_{\mathrm{s}}(i\Omega_k, i\omega_\nu) B_{\mathrm{c}}(i\Omega_k, i\omega_\nu)]_{i\omega_\nu \to \hbar \omega \approx 0}\\
	\approx&
	-\frac{1}{i\omega} \cdot 
	 \qty(\frac{i\zeta_{\rm s}\zeta_{\rm c}}{D})
     V\sum_{\bm{q}} \qty(\bm{q}^2 L(\bm{q}, 0)^2)
	\hbar \omega \frac{T}{2} B_{\mathrm{c}2}(0, 0) \dv{x}\qty[B_{\mathrm{s}2}(-x, x) + B_{\mathrm{s}2} (0, x)]_{x=0}. \label{eq: al-eq13-}
\end{align}
Note that the divergence in the limit $\epsilon \to 0$ comes from the points $\Omega_k = 0$ and $\Omega_k = -\omega_\nu$.
In the case of AL term of electric conductivity, these contributions are completely canceled out by taking an integration path properly (refer to the derivation of Eq.~(16) in Ref.~\cite{[{}][{ [Sov. Phys. JETP $\bm{14}$ 633 (1962)].}]Perel1962}) for this kind of details.)
In this paper, on the other hand, the sums in the ranges $\Omega_k>0$ and $\Omega_k < -\omega_\nu$ are evaluated directly without using contour integration. Hence, one-half of the contribution from $\Omega_k=0$ and $\Omega_k = -\omega_\nu$ remains.

\subsubsection{\texorpdfstring{Step7: Sum over $\varepsilon_n$}{Step7: Sum over ε\_n}}
Performing the sum over $\varepsilon_n$ in $B_{\mathrm{s}2}$ and $B_{\mathrm{c}2}$, 
we can express Eq.~\eqref{al-eq3+} more explicitly in terms of the polyGamma functions.
A direct calculation shows that 
\begin{align}
	&
	\dv{x} \qty[B_{\mathrm{s}2}^{\mathrm{a}}(-x, x) + B_{\mathrm{s}2}^{\mathrm{a}}(0,x)]_{x=0}\\
	=&
	-4\pi \qty(\frac{\tau}{\hbar})^3 \qty[\frac{\hbar}{4\pi T \tau} \psi^{(1)}\qty(\frac{1}{2}) + \psi\qty(\frac{1}{2}) - \psi\qty(\frac{1}{2} + \frac{\hbar}{4\pi T \tau})]\\
	&
	\dv{x} \qty[B_{\mathrm{s}2}^{\mathrm{b}}(-x, x) + B_{\mathrm{s}2}^{\mathrm{b}}(0, x)]_{x=0}\\
	=&
	-2\pi \qty(\frac{\tau}{\hbar})^3 \frac{\hbar}{4\pi T \tau} \qty[\psi^{(1)}\qty(\frac{1}{2}) - \psi^{(1)}\qty(\frac{1}{2} + \frac{\hbar}{4\pi T \tau})]\\
	&
	\dv{x} \qty[B_{\mathrm{s}2}^{\mathrm{c}} (-x, x) + B_{\mathrm{s}2}^{\mathrm{c}} (0, x)]_{x=0}\\
	=&
	-4\pi \qty(\frac{\tau}{\hbar})^3 \qty[\frac{\hbar}{4\pi T \tau} \psi^{(1)}\qty(\frac{1}{2}) + \frac{\hbar}{4\pi T \tau} \psi^{(1)}\qty(\frac{1}{2} + \frac{\hbar}{4\pi T \tau}) + 2\psi\qty(\frac{1}{2}) - 2\psi\qty(\frac{1}{2} + \frac{\hbar}{4\pi T \tau})],
\end{align}
from which we obtain 
\begin{align}
	&
	\dv{x} \qty[B_{\mathrm{s}2}(-x, x) + B_{\mathrm{s}2} (0, x)]_{x=0}\\
    =&
	-16\pi\qty(\frac{\tau}{\hbar})^3 \qty[\frac{\hbar}{4\pi T \tau} \psi^{(1)}\qty(\frac{1}{2}) + \psi\qty(\frac{1}{2}) - \psi\qty(\frac{1}{2} + \frac{\hbar}{4\pi T \tau})].\label{eq: Bs2-AL-SJ}
\end{align}
$B_{\mathrm{c}2}(0,0)$ can be reduced from Eq.~\eqref{al-eq11} to 
\begin{align}
	B_{\mathrm{c}2}(0, 0)
	=&
	2\pi T \qty[-\sum_{\varepsilon_n < 0} \frac{1}{(2\varepsilon_n)^2} \frac{2}{2\varepsilon_n - \hbar / \tau} + \sum_{\varepsilon_n > 0} \frac{1}{(2\varepsilon_n )^2} \frac{2}{2\varepsilon_n + \hbar / \tau}]\\
	=&
	2\qty(\frac{\tau}{\hbar})^2 \qty[\frac{\hbar}{4\pi T \tau}\psi^{(1)}\qty(\frac{1}{2}) + \psi\qty(\frac{1}{2}) - \psi\qty(\frac{1}{2} + \frac{\hbar}{4\pi T \tau})].\label{eq: Bc2-AL-SJ}
\end{align}
\subsubsection{\texorpdfstring{Step8: Sum over $\bm{q}$}{Step8: Sum over q}}
The remaining part to be calculated in Eq.~\eqref{eq: al-eq13-} is $V \sum_{\bm{q}} \bm{q}^2 L(\bm{q}, 0)^2$, which becomes
\begin{align}
	V \sum_{\bm{q}} \bm{q}^2 L(\bm{q}, 0)^2
	=&
	\frac{1}{N(0)^2} \frac{1}{V} \sum_{\norm{\bm{q}} < \clength^{-1}}  \frac{\bm{q}^2}{(\epsilon + \clength^2 \bm{q}^2)^2}
	=
	\frac{1}{N(0)^2} \frac{1}{(2\pi)^2} \int_0^{\clength^{-1}} dq\; 2\pi q \cdot q^2 \frac{1}{(\epsilon  +\clength^2 q^2)^2}\\
	=&
	\frac{1}{2\pi N(0)^2} \frac{1}{2\clength^4} \int_0^1 dx \frac{x}{(\epsilon + x)^2}
	=
	\frac{1}{4\pi N(0)^2 \clength^4} \int_0^1 dx \qty[\frac{1}{\epsilon + x} - \frac{\epsilon}{(\epsilon + x)^2}]\\
	=&
	\frac{1}{4\pi N(0)^2 \clength^4} \qty[\ln (\epsilon + x) + \frac{\epsilon}{\epsilon + x}]_0^1
	=
	\frac{1}{4\pi N(0)^2 \clength^4} \qty[\ln \frac{1}{\epsilon} + \ln (1+\epsilon) + \frac{\epsilon}{\epsilon + 1} - 1]\\
	\approx&
	\frac{1}{4\pi N(0)^2 \clength^4} \ln \frac{1}{\epsilon} \label{al-eq14}
\end{align}
for $D=2$ and
\begin{align}
	V\sum_{\bm{q}}\bm{q}^2 L(\bm{q}, 0)^2
	=&
	\frac{1}{N(0)^2} \frac{1}{V} \sum_{\norm{\bm{q}} < \clength^{-1}} \frac{\bm{q}^2}{(\epsilon + \clength^2 \bm{q}^2)^2}
	=
	\frac{1}{N(0)^2} \frac{1}{(2\pi)^3} \int_0^{\clength^{-1}} dq \cdot 4\pi q^2 \frac{q^2}{(\epsilon + \clength^2 q^2)^2}\\
	=&
	\frac{1}{2\pi^2 N(0)^2} \int_0^{\clength^{-1}}dq \frac{q^4}{(\epsilon + \clength^2 q^2)^2}
	=
	\frac{1}{2\pi^2 N(0)^2\clength^5} \int_0^1 dx \frac{x^4}{(\epsilon + x^2)^2}\\
	=&
	\frac{1}{2\pi^2 N(0)^2 \clength^5} \frac{2 + 3\epsilon - 3\sqrt{\epsilon}(1+\epsilon) \arccot(\sqrt{\epsilon})}{2(1+\epsilon)}\\
	\approx&
	\frac{1}{2\pi^2 N(0)^2 \clength^5}\label{eq: al-eq14++}
\end{align}
for $D=3$.

Here the cutoff $\norm{\bm{q}} < \clength^{-1} $ is introduced. 
\subsubsection{Step9: Expression for spin Hall conductivity}
Substituting Eqs.~\eqref{eq: Bs2-AL-SJ}, \eqref{eq: Bc2-AL-SJ}, \eqref{al-eq14}, \eqref{eq: al-eq14++}, into Eq.~\eqref{eq: al-eq13-}, we arrive at the expression for spin Hall conductivity
\begin{align}
	\sigma_{xy}^{\mathrm{AL-SJ}}
		\approx&
	\frac{16\pi \zeta_{\rm s}\zeta_{\rm c}\hbar T}{D N(0)^2} \qty(\frac{\tau}{\hbar})^5
	\qty[\frac{\hbar}{4\pi T \tau} \psi^{(1)}\qty(\frac{1}{2}) + \psi\qty(\frac{1}{2}) - \psi\qty(\frac{1}{2} + \frac{\hbar}{4\pi T \tau})]^2
	\begin{dcases}
		\frac{1}{4\pi\clength^4} \ln \frac{1}{\epsilon} & (D=2)\\
		\frac{1}{2\pi^2 \clength^5} & (D=3)
	\end{dcases}\\
=&
	\frac{8\pi e^2 \hbar^5 \lambda_0^2 n_{\mathrm{i}} v_0^2 k_{\mathrm{F}}^4 N(0) T}{D^3 m^3} \qty(\frac{\tau}{\hbar})^5
	\qty[\frac{\hbar}{4\pi T \tau} \psi^{(1)}\qty(\frac{1}{2}) + \psi\qty(\frac{1}{2}) - \psi\qty(\frac{1}{2} + \frac{\hbar}{4\pi T \tau})]^2\\
	&\times
	\begin{dcases}
		\frac{1}{4\pi\clength^4} \ln \frac{1}{\epsilon} & (D=2)\\
		\frac{1}{2\pi^2 \clength^5} & (D=3)
	\end{dcases}\\
=&	
	\frac{2e^2 \hbar}{Dm} \lambda_0^2 k_{\mathrm{F}}^2 \frac{T}{\varepsilon_{\mathrm{F}}} N(0)
	\begin{dcases}
		\frac{1}{2} \ln \frac{1}{\epsilon} & (D=2)\\
		\frac{1}{\pi \clength} & (D=3). 
	\end{dcases}
\end{align}
\subsubsection{\texorpdfstring{Behaviours of Eqs.~(\ref{al-eq4}), (\ref{al-eq5}), (\ref{al-eq6}) in the limit $\epsilon\rightarrow +0$}{Behaviours of Eqs.~(\ref{al-eq4}), (\ref{al-eq5}), (\ref{al-eq6}) in the limit ϵ→+0}}\label{subsec: appendixB}
This subsection supplements the argument in Step 6, where we assume that  Eqs.~\eqref{al-eq4}$\sim$ \eqref{al-eq6} do not contribute to the terms diverging in the limit $\epsilon\rightarrow +0$. 

We first consider Eq.~\eqref{al-eq4}, which is rewritten as
\begin{align}
	\hbar \omega \cdot \pv \int_{-\infty}^{\infty} \frac{d\Omega}{4\pi i} \qty(\dv{\Omega} \coth \frac{\Omega}{2T}) L^A(\bm{q}, \Omega) L^R(\bm{q}, \Omega) B_{\mathrm{c}2} (\Omega, 0) B_{\mathrm{s}2} (\Omega, 0)
\end{align}
with use of partial integral. 
We will see, immediately below, that $B_{\mathrm{c}2} (\Omega, 0)$ is an even function of $\Omega$ while $B_{\mathrm{s}2} (\Omega, 0)$ 
is odd.
We see that $B_{\mathrm{c}2} (\Omega, 0)$ is an even function of $\Omega$ from the last line of the following expressions: 
\begin{align}
	B_{\mathrm{c}2}(\Omega, 0)
	=&
	-2\pi T \sum_{\varepsilon_n < 0} \frac{1}{(2\varepsilon_n + i\Omega)^2} \frac{2}{2\varepsilon_n + i\Omega - \hbar /\tau} + 2\pi T \sum_{\varepsilon_n > 0} \frac{1}{(2\varepsilon_n + i\Omega)^2} \frac{2}{2\varepsilon_n + i\Omega + \hbar / \tau}\\
	-&
	\frac{2\pi T}{\hbar / \tau}
	\begin{aligned}[t]
		\Biggl\{&
			\sum_{\varepsilon_n < 0} \qty[\frac{1}{(2\varepsilon_n - i\Omega) (2\varepsilon_n - i\Omega - \hbar / \tau)} - \frac{1}{(2\varepsilon_n + i\Omega) (2\varepsilon_n + i\Omega - \hbar / \tau)}]\\
			+&
			\sum_{\varepsilon_n > 0} \qty[\frac{1}{(2\varepsilon_n - i\Omega) (2\varepsilon_n - i\Omega + \hbar / \tau)} - \frac{1}{(2\varepsilon_n + i\Omega) (2\varepsilon_n + i\Omega + \hbar / \tau)}]
		\Biggr\}
	\end{aligned}\\
	=&
	4\pi T \sum_{\varepsilon_n > 0} \qty[\frac{1}{(2\varepsilon_n - i\Omega)^2 (2\varepsilon_n - i\Omega + \hbar / \tau)} + \frac{1}{(2\varepsilon_n + i\Omega)^2 (2\varepsilon_n + i\Omega + \hbar / \tau)}].
\end{align}
Next we see that $B_{\mathrm{s}2}(\Omega, 0)$  
is odd from the following expressions: 
\begin{align}
	&
	B_{\mathrm{s}2}(\Omega, 0)\\
	=&
	-8\pi^2 i T \sum_{\varepsilon_n > 0} \qty[-\frac{1}{(2\varepsilon_n - i\Omega)^2 (2\varepsilon_n - i\Omega + \hbar / \tau)} + \frac{1}{(2\varepsilon_n + i\Omega)^2 (2\varepsilon_n + i\Omega + \hbar / \tau)}]\\
	-&
	\frac{4\pi^2 i T}{\hbar / \tau} \sum_{\varepsilon_n < 0} \qty[\frac{1}{(2\varepsilon_n - i\Omega) (2\varepsilon_n - i\Omega - \hbar / \tau)} - \frac{1}{(2\varepsilon_n + i\Omega) (2\varepsilon_n + i\Omega - \hbar / \tau)}]\\
	-&
	4\pi^2 i T \sum_{\varepsilon_n > 0} \qty[-\frac{1}{(2\varepsilon_n - i\Omega)^2} \frac{2}{2\varepsilon_n - i\Omega + \hbar / \tau} + \frac{1}{(2\varepsilon_n + i\Omega)^2} \frac{2}{2\varepsilon_n + i\Omega + \hbar / \tau}]\\
	-&
	\frac{4\pi^2 i T}{\hbar / \tau} 
	\begin{aligned}[t]
		\Biggl\{&
			\sum_{\varepsilon_n < 0} \qty[\frac{1}{(2\varepsilon_n - i\Omega) (2\varepsilon_n - i\Omega - \hbar / \tau)} - \frac{1}{(2\varepsilon_n + i\Omega) (2\varepsilon_n + i\Omega - \hbar / \tau)}]\\
			+&
			\sum_{\varepsilon_n > 0} \qty[\frac{1}{(2\varepsilon_n - i\Omega) (2\varepsilon_n - i\Omega + \hbar / \tau)} - \frac{1}{(2\varepsilon_n + i\Omega) (2\varepsilon_n + i\Omega + \hbar / \tau)}]
		\Biggr\}.
	\end{aligned}
\end{align}
Further we note that both $L^{\rm A}(\bm{q}, \Omega) L^{\rm R}(\bm{q}, \Omega)$ and $\dv{\Omega} \coth (\Omega / 2T)$ are even  functions of $\Omega$ and thus the integrand in Eq.~\eqref{al-eq4} is odd and the integral vanishes.

Next we consider Eqs.~\eqref{al-eq5} and \eqref{al-eq6}.
We can show that these two do not diverge in the limit $\epsilon \to 0$ by showing that  
\begin{align}
	&
	\hbar \omega \cdot T \sum_{\Omega_k<0} L^{\rm A}(\bm{q}, i\Omega_k)^2 B_{\mathrm{c}1}(i\Omega_k, 0) \qty[\pdv{x} B_{\mathrm{s}1}(i\Omega_k, x)]_{x=0}\\
	+&
	\hbar \omega \cdot T \sum_{\Omega_k > 0} L^{\rm R}(\bm{q}, i\Omega_k)^2 B_{\mathrm{c}3}(i\Omega_k, 0) \qty[\pdv{x} B_{\mathrm{s}3}(i\Omega_k, x)]_{x=0}\\
\end{align}
converges when $\epsilon = 0$. 
In the expressions for $B_{\mathrm{c}}$ and $B_{\mathrm{s}}$ in the limit of $i\omega_\nu \to 0$, $\Omega_k$ dependence enters only through the form of $2\varepsilon_n + \sign(\varepsilon_n) \abs{\Omega_k}$ or  $2\varepsilon_n + \sign(\varepsilon_n) (\abs{\Omega_k} + \hbar / \tau)$. We sum them with $\varepsilon_n$ after the partial fraction decomposition and find that the resultant expression is linear combination of $\psi^{(m)} (\frac{1}{2} + \abs{\Omega_k} / 4\pi T)$ and $\psi^{(m)} (\frac{1}{2} + \abs{\Omega_k} / 4\pi T + \hbar / 4\pi T \tau)$ ($m=0, 1,2 \cdots$). For $m=0$, the $\Omega_k$-dependences in all partial fractions enter through the form of $\psi(\frac{1}{2} + \abs{\Omega_k} / 4\pi T + \hbar / 4\pi T \tau) - \psi(\frac{1}{2} + \abs{\Omega_k} / 4\pi T)$, which is bounded from above as
\begin{align}
	\psi\qty(\frac{1}{2} + \frac{\abs{\Omega_k}}{4\pi T} + \frac{\hbar}{4\pi T \tau}) - \psi\qty(\frac{1}{2} + \frac{\abs{\Omega_k}}{4\pi T})
	=&
	\sum_{n=0}^{\infty}
	\frac{1}{n + \frac{1}{2} + \frac{\abs{\Omega_k}}{4\pi T}} - \frac{1}{n + \frac{1}{2} + \frac{\abs{\Omega_k}}{4\pi T} + \frac{\hbar}{4\pi T \tau}}\\
	=&
	\frac{\hbar}{4\pi T \tau} \sum_{n=0}^{\infty} \frac{1}{\qty(n + \frac{1}{2} + \frac{\abs{\Omega_k}}{4\pi T}) \qty(n + \frac{1}{2} + \frac{\abs{\Omega_k}}{4\pi T} + \frac{\hbar}{4\pi T \tau})}\\
	<&
	\frac{\hbar}{4\pi T \tau} \sum_{n=0}^{\infty} \frac{1}{\qty(n + \frac{1}{2} + \frac{\abs{\Omega_k}}{4\pi T})^2}\\
	=&
	\frac{\hbar}{4\pi T} \psi^{(1)} \qty(\frac{1}{2} + \frac{\abs{\Omega_k}}{4\pi T})
\end{align}
and it thus suffices to find the upper bound for $m \geq 1$. 

We have the upper bound for $x \geq \frac{1}{2}, a \geq 0, m \geq 1$ by 
\begin{align}
	\abs{\psi^{(m)}\qty(x+a)}
	=&
	\abs{(-1)^{m+1} m! \sum_{n=0}^{\infty} \frac{1}{(n+x+a)^{m+1}}}\\
	\leq&
	m! \sum_{n=0}^{\infty} \frac{1}{(n+x)^{m+1}}\\
	<&
	m! \qty(x^{-(m+1)} + \int_{x}^{\infty} dz \frac{1}{z^{m+1}})\\
	=&
	m!\qty(x^{-(m+1)} + \qty[-\frac{1}{m} z^{-m}]_x^{\infty})\\
	=&
	m! \qty(\frac{1}{x^{m+1}} + \frac{1}{m} \frac{1}{x^m})\\
	=&
	m! \qty(\frac{1}{m} + \frac{1}{x}) \frac{1}{x^m}\\
	\leq&
	m! \qty(1 + \frac{1}{1 / 2}) \frac{1}{x^m}\\
	=&
	3(m!) \frac{1}{x^m}.
\end{align}
With use of it, we find that the expression for 
$m , l \geq 1,\; a, b = 0 \text{ or } \frac{\hbar}{4\pi T \tau}$ does not diverge when  $\epsilon =0$ in the following calculation:
\begin{align}
	&
	\abs{\sum_{\bm{q}}\bm{q}^2 \sum_{\Omega_k > 0} L(\bm{q}, i\Omega_k)^2 \psi^{(m)}\qty(\frac{1}{2} + \frac{\Omega_k}{4\pi T} + a) \psi^{(l)}\qty(\frac{1}{2} + \frac{\Omega_k}{4\pi T} + b)}\\
	<&
	3(m!) \cdot 3(l!) \abs{\sum_{\norm{\bm{q}}< \clength^{-1}} \bm{q}^2 \sum_{\Omega_k} \frac{1}{N(0)^2} \frac{1}{\qty(0 + \psi\qty(\frac{1}{2} + \frac{\Omega_k}{4\pi T}) - \psi\qty(\frac{1}{2}) + \clength^2 q^2)^2} \frac{1}{(\Omega_k)^{m+l}}}\\
	\leq&
	\frac{3(m!) \cdot 3(l!)}{N(0)^2} \abs{\sum_{\norm{\bm{q}}< \clength^{-1}} \bm{q}^2 \frac{1}{\qty(\psi(1) - \psi(1 / 2) + \clength^2 \bm{q}^2)^2} \sum_{\Omega_k} \frac{1}{(\Omega_k)^{m+l}}}\\
	=&
	\frac{3(m!) \cdot 3(l!)}{N(0)^2} \qty(\sum_{\Omega_k} \frac{1}{(\Omega_k)^{m+l}}) \frac{V}{(2\pi)^2} \int_0^{\clength^{-1}} 2\pi q dq\; q^2 \frac{1}{(\psi(1) - \psi(1 / 2) + \clength^2 \bm{q}^2)}\\
	<&
	\frac{3(m!) \cdot 3(l!)}{N(0)^2} \qty(\sum_{\Omega_k} \frac{1}{(\Omega_k)^{m+l}}) \frac{V}{2\pi} \frac{1}{\qty(\psi(1) - \psi(1 / 2))^2} \int_0^{\clength^{-1}} q^3 dq
	< \infty.
\end{align}
The quantities $B_{\mathrm{s}3}, B_{\mathrm{c}3}$ are the sum of terms such as $\psi^{(m)}(\frac{1}{2} + \abs{\Omega_k} / 4\pi T  + a)$ and thus Eq.~\eqref{al-eq6} do not diverge when $\epsilon =0$. Similarly, we can find that Eq.~\eqref{al-eq5} does not diverge either.  
\subsection{Aslamazov-Larkin terms with skew scattering}
The expression for the spin Hall conductivity via the AL terms with a skew scattering is similar to that via the AL terms with a side jump. We thus briefly explain the derivation in the present case, emphasizing the difference from the previous case. 
\subsubsection{Step 1: Feynman diagrams for Response function}
Figure \ref{al-fig1} shows the diagrams for the AL terms with the skew scattering process. 
%
Among them, only the diagram shown in Fig.~\ref{al-fig1}(d) has $\sigma_{-\alpha -\alpha}^z$ and hence it has the opposite sign to other diagrams (a), (b), and (c). 
The diagram reflected with respect to the horizontal axis (i.e., the diagram "upside-down") contributes the spin Hall conductivity equally to the original diagram. Thus the contributions of (a), (b), (c), and (d) should be doubled.  
The response function can be written in the form of 
\begin{align}
	\Phi_{xy}(0, i\omega_\nu)
	=&
	\frac{1}{V}\sum_{\bm{q}} T \sum_{\Omega_k} L(\bm{q}, i\Omega_k) L(\bm{q}, i\Omega_k + i\omega_\nu)
	\mathcal{B}'_{\mathrm{s}}(\bm{q},i\Omega_k, i\omega_\nu) \mathcal{B}_{\mathrm{c}}(\bm{q},i\Omega_k, i\omega_\nu).
	\label{eq: al-eq9-ss}
\end{align}
Here $\mathcal{B}_{\rm c}(\bm{q},i\Omega_k, i\omega_\nu)$ has been defined by Eq.~\eqref{eq: Bc-bold} in the previous subsection.  The triangular part in the side of spin current $\mathcal{B}'_{\rm s}(\bm{q},i\Omega_k, i\omega_\nu)$ in the present case is different from $\mathcal{B}_{\rm s}(\bm{q},i\Omega_k, i\omega_\nu)$ in the previous subsection.  
The contribution of each diagram to $\mathcal{B}'_{\rm s}(\bm{q},i\Omega_k, i\omega_\nu)$ has the form of 
\begin{align}
	&2T \sum_{\varepsilon_n} 
	C(\bm{q}, \varepsilon_n, \Omega_k - \varepsilon_n) C(\bm{q}, \varepsilon_n + \omega_\nu, \Omega_k - \varepsilon_n)
	\sum_{\bm{k}_1 \bm{k}_2 \bm{k}_3} \qty(-\frac{e\hbar}{2m} k_{1x}) \qty(- \frac{i\lambda_0^2}{4V} (k_{2x}k_{1y} - k_{2y} k_{1x})) n_{\mathrm{i}} v_0^3 V \cdot \frac{1}{V^2}\\
	&\times
	\text{(Product of the Green functions)}.\\
\end{align}
%
\subsubsection{\texorpdfstring{Step2: Expansion of triangular parts with $\bm{q}$}{Step2: Expansion of triangular parts with q}}
 As in the AL terms with side jump process, nonzero contribution with the lowest order of $\bm{q}$ comes from expansion of the Green function $\mathcal{G}(\bm{q}-\bm{k}_2, i\Omega_k - i\varepsilon_n) \approx \mathcal{G}(\bm{k}_2, i\Omega_k - i\varepsilon_n) - (\hbar^2 / m) (\bm{k}_2 \vdot \bm{q}) \mathcal{G}(\bm{k}_2, i\Omega_k - i\varepsilon_n)^2$ in the side of the spin current vertex so that $\bm{k}_2$ coming from the external product in $H_{\mathrm{SO}}$ is to be squared with $\bm{k}_2 \vdot \bm{q}$. 
 
 With this consideration, we expand $\mathcal{B}_{\rm s}'$ with $\bm{q}$ up to the first order as 
\begin{align}
 \mathcal{B}_{\rm s}'(\bm{q},i\Omega_k, i\omega_\nu)\approx i 2q_y \zeta'_{\rm s}V B'_{\rm s}(i\Omega_k, i\omega_\nu), 
\end{align}
with 
\begin{equation}
    \zeta'_{\rm s}=\frac{e\hbar^3\lambda_0^2 n_{\mathrm{i}} v_0^3}{4m^2} \frac{k_{\mathrm{F}}^4}{D^2} \qty(N(0))^3. 
\end{equation}
The response function is then written as
\begin{align}
	\Phi_{xy}(0, i\omega_\nu)
    \approx&
    i 2\zeta_{\rm s}'\zeta_{\rm c}
		V\sum_{\bm{q}} \frac{\bm{q}^2}{D} T \sum_{\Omega_k} L(\bm{q}, i\Omega_k) L(\bm{q}, i\Omega_k + i\omega_\nu) B'_{\mathrm{s}}(i\Omega_k , i\omega_\nu) B_{\mathrm{c}}(i\Omega_k , i\omega_\nu).\label{eq: Phi-al-ss-BB}
\end{align}
The quantity $B_{\mathrm{c}}(i\Omega_k, i\omega_\nu)$ has been defined in Eq.~\eqref{al-eq2} with Eq.~\eqref{al-eq3} and $\zeta_{\rm c}$ has been defined by Eq.~\eqref{eq: zeta-c-def}.  The expression for $B'_{\mathrm{s}}(i\Omega_k, i\omega_\nu)$ is given by the sum of the contributions from each diagram 
\begin{align}
	B'_{\mathrm{s}}(i\Omega_k, i\omega_\nu)
	=&
	B_{\mathrm{s}}^{'\rm a}(i\Omega_k, i\omega_\nu)+B_{\mathrm{s}}^{'\rm b}(i\Omega_k, i\omega_\nu)+B_{\mathrm{s}}^{'\rm c}(i\Omega_k, i\omega_\nu)+B_{\mathrm{s}}^{'\rm d}(i\Omega_k, i\omega_\nu),
	\end{align}
where 
\begin{align}
	B_{\mathrm{s}}^{'\rm a,b,c}(i\Omega_k, i\omega_\nu)
	=&
	T\sum_{\varepsilon_n} C(0, \varepsilon_n, \varepsilon_{k-n}) C(0, \varepsilon_{n+\nu}, \varepsilon_{k-n})
	I'^{\mathrm{a,b,c}}(i\varepsilon_n, i\Omega_k, i\omega_\nu)
	\label{al-eq15}\\
	B_{\mathrm{s}}^{'\rm d}(i\Omega_k, i\omega_\nu)
	=&
	-T\sum_{\varepsilon_n} C(0, \varepsilon_n, \varepsilon_{k-n}) C(0, \varepsilon_{n+\nu}, \varepsilon_{k-n})
	I'^{\mathrm{d}}(i\varepsilon_n, i\Omega_k, i\omega_\nu)
	\label{al-eq15-d}
\end{align}	
represent the contribution from the diagram (a), (b), (c) and (d). 
In Eqs.~\eqref{al-eq15} and \eqref{al-eq15-d}, we have introduced  
\begin{align}
	I^{'\mathrm{a}}(i\varepsilon_n, i\Omega_k, i\omega_\nu)
	= &
	\int d\xi_1\; \mathcal{G}(\bm{k}_1, i\varepsilon_n) \mathcal{G}(\bm{k}_1, i\varepsilon_{n+\nu})
	\int d\xi_2\; \mathcal{G}(\bm{k}_2, i\varepsilon_n) \mathcal{G}(\bm{k}_2, i\varepsilon_{k-n})^2\\
	&\times
	\int d\xi_3\; \mathcal{G}(\bm{k}_3, i\varepsilon_{n+\nu}) \mathcal{G}(\bm{k}_3, i\varepsilon_{k-n}),\\
	I^{'\mathrm{b}}(i\varepsilon_n, i\Omega_k, i\omega_\nu)
	=&
	\int d\xi_1\; \mathcal{G}(\bm{k}_1, i\varepsilon_n) \mathcal{G}(\bm{k}_1, i\varepsilon_{n+\nu}) \mathcal{G}(\bm{k}_1, i\varepsilon_{k-n})
	\int d\xi_2\; \mathcal{G}(\bm{k}_2, i\varepsilon_n) \mathcal{G}(\bm{k}_2, i\varepsilon_{k-n})^2\\
	&\times
	\int d\xi_3\; \mathcal{G}(\bm{k}_3, i\varepsilon_{k-n}),\\
	I^{'\mathrm{c}}(i\varepsilon_n, i\Omega_k, i\omega_\nu)
	=&
	\int d\xi_1\; \mathcal{G}(\bm{k}_1, i\varepsilon_n) \mathcal{G}(\bm{k}_1, i\varepsilon_{n+\nu})
	\int d\xi_2\; \mathcal{G}(\bm{k}_2, i\varepsilon_n) \mathcal{G}(\bm{k}_2, i\varepsilon_{k-n})^2 \mathcal{G}(\bm{k}_2, i\varepsilon_{n+\nu})\\
	&\times
	\int d\xi_3\; \mathcal{G}(\bm{k}_3, i\varepsilon_{n+\nu}),\\
	I^{'\mathrm{d}}(i\varepsilon_n, i\Omega_k, i\omega_\nu)
	=&
	\int d\xi_1\; \mathcal{G}(\bm{k}_1, i\varepsilon_n) \mathcal{G}(\bm{k}_1, i\varepsilon_{n+\nu}) \mathcal{G}(\bm{k}_1, i\varepsilon_{k-n})
	\int d\xi_2\; \mathcal{G}(\bm{k}_2, i\varepsilon_n) \mathcal{G}(\bm{k}_2, i\varepsilon_{k-n})^2\\
	&\times
	\int d\xi_3\; \mathcal{G}(\bm{k}_3, i\varepsilon_n).
\end{align}
\begin{figure}[tbp]
	\begin{minipage}[b]{0.8\linewidth}
		\centering
		\includegraphics[width=\textwidth, page=1]{extrinsic_AL.pdf}
		\\(a)
		\label{al-fig1a}
	\end{minipage}\\
	\begin{minipage}[b]{0.4\linewidth}
		\centering
		\includegraphics[scale=0.7, page=2]{extrinsic_AL.pdf}
		\\(b)
		\label{al-fig1b}
	\end{minipage}
	\begin{minipage}[b]{0.4\linewidth}
		\centering
		\includegraphics[scale=0.7, page=3]{extrinsic_AL.pdf}
		\\(c)
		\label{al-fig1c}
	\end{minipage}\\
	\begin{minipage}[b]{0.4\linewidth}
		\centering
		\includegraphics[scale=0.7, page=4]{extrinsic_AL.pdf}
		\\(d)
		\label{al-fig1d}
	\end{minipage}
	\caption{Diagrams for the AL terms with skew scattering process}
	\label{al-fig1}
\end{figure}

\subsubsection{\texorpdfstring{Step3: Integral in $B'_{\rm s}(i\Omega_k, i\omega_\nu)$ with wavevectors}{Step3: Integral in B'\_s(iΩ\_k, iω\_ν) with wavevectors}}
First we note that $I^{'\mathrm{a}}=0$ because at least one integral among the $\int d\xi_1$, $\int d\xi_2$, and $\int d\xi_3$ vanishes for all regions \ctext{1} $\sim$ \ctext{4}; 
$\int d\xi_1\cdots=0$ in \ctext{1}, $\int d\xi_3\cdots=0$ in \ctext{2}, $\int d\xi_2\cdots=0$ in \ctext{3}, and $\int d\xi_1\cdots=0$ in \ctext{4}. 

Performing the integrals, $I^{'\mathrm{b}}$ can be written as
\begin{align}
	I^{'\mathrm{b}}
    =&
	-4\pi^3 i \frac{1}{(\widetilde{\varepsilon}_{n+\nu} - \widetilde{\varepsilon}_n)(\widetilde{\varepsilon}_n - \widetilde{\varepsilon}_{k-n})^2}
	\qty(\frac{\ctext{1} + \ctext{2} - \ctext{4}}{\widetilde{\varepsilon}_n - \widetilde{\varepsilon}_{k-n}} + \frac{-\ctext{1} + \ctext{4}}{\widetilde{\varepsilon}_{n+\nu} - \widetilde{\varepsilon}_{k-n}}),\\
\end{align}
with use of 
\begin{align}
	&\int d\xi_1\; \mathcal{G}(\bm{k}_1, i\varepsilon_n) \mathcal{G}(\bm{k}_1, i\varepsilon_{n+\nu}) \mathcal{G}(\bm{k}_1, i\varepsilon_{k-n})
    =
    2\pi i \frac{1}{\widetilde{\varepsilon}_{n+\nu} - \widetilde{\varepsilon}_n} \qty(\frac{\ctext{1} + \ctext{2} - \ctext{4}}{\widetilde{\varepsilon}_n - \widetilde{\varepsilon}_{k-n}} + \frac{-\ctext{1} + \ctext{3} + \ctext{4}}{\widetilde{\varepsilon}_{n+\nu} - \widetilde{\varepsilon}_{k-n}}),\\
	&\int d\xi_2 \mathcal{G}(\bm{k}_2, i\varepsilon_n) \mathcal{G}(\bm{k}_2, i\varepsilon_{k-n})^2
	=
	-2\pi i \frac{\ctext{1}+\ctext{2}-\ctext{4}}{(\widetilde{\varepsilon}_n - \widetilde{\varepsilon}_{k-n})^2},\\
	&\int d\xi_3 \mathcal{G}(\bm{k}_3, i\varepsilon_{k-n})
	=
	-i\pi(\ctext{1} + \ctext{2} - \ctext{3} - \ctext{4}).
\end{align}
The quantity $I^{'\mathrm{c}}$ becomes
\begin{align}
	I^{'\mathrm{c}}
	=&
	4i\pi^3 \frac{1}{(\widetilde{\varepsilon}_{n+\nu} - \widetilde{\varepsilon}_n)^2} \qty[\frac{\ctext{2}}{(\widetilde{\varepsilon}_n - \widetilde{\varepsilon}_{k-n})^2} + \frac{\ctext{3}}{(\widetilde{\varepsilon}_{n+\nu} - \widetilde{\varepsilon}_{k-n})^2}]
\end{align}
with use of 
\begin{align}
	&\int d\xi_1 \mathcal{G}(\bm{k}_1, i\varepsilon_n) \mathcal{G}(\bm{k}_1, i\varepsilon_{n+\nu})
	=
	2\pi \frac{\ctext{2} + \ctext{3}}{\widetilde{\varepsilon}_{n+\nu} - \widetilde{\varepsilon}_{n}}\\
	 &\int d\xi_2 \mathcal{G}(\bm{k}_2, i\varepsilon_n) \mathcal{G}(\bm{k}_2, i\varepsilon_{k-n})^2 \mathcal{G}(\bm{k}_2, i\varepsilon_{n+\nu})=
	-\frac{2\pi}{\widetilde{\varepsilon}_{n+\nu} - \widetilde{\varepsilon}_n} \qty[\frac{\ctext{1} + \ctext{2} - \ctext{4}}{(\widetilde{\varepsilon}_n - \widetilde{\varepsilon}_{k-n})^2} + \frac{-\ctext{1} + \ctext{3} + \ctext{4}}{(\widetilde{\varepsilon}_{n+\nu} - \widetilde{\varepsilon}_{k-n})^2}]\\
	 &\int d\xi_3 \mathcal{G}(\bm{k}_3, i\varepsilon_{n+\nu})
	=
	-i\pi \sign(\varepsilon_{n+\nu})
	= 
	-i\pi(-\ctext{1} + \ctext{2} + \ctext{3} + \ctext{4}).
\end{align}
The $\xi_1$ integral and the $\xi_2$ integral in $I^{'\mathrm{d}}$ are the same as those in $I^{'\mathrm{b}}$. The $\xi_3$ integral is given by
\begin{align}
	\int d\xi_3 \mathcal{G}(\bm{k}_3, i\varepsilon_n)
	=&
	-i\pi \sign(\varepsilon_n)
	=
	-i\pi(-\ctext{1} - \ctext{2} - \ctext{3} + \ctext{4}),
\end{align}
from which,
\begin{align}
	I^{'\mathrm{d}}
	=&
	-4i\pi^3 \frac{1}{(\widetilde{\varepsilon}_{n+\nu} - \widetilde{\varepsilon}_n) (\widetilde{\varepsilon}_n - \widetilde{\varepsilon}_{k-n})^2} \qty(\frac{-\ctext{1} - \ctext{2} + \ctext{4}}{\widetilde{\varepsilon}_n - \widetilde{\varepsilon}_{k-n}} + \frac{\ctext{1} - \ctext{4}}{\widetilde{\varepsilon}_{n+\nu} - \widetilde{\varepsilon}_{k-n}})
	=
	-I^{'\mathrm{b}}
\end{align}
follows. From the relation $I^{'\mathrm{a}} + I^{'\mathrm{b}} + I^{'\mathrm{c}} - I^{'\mathrm{d}} = 2I^{'\mathrm{b}} + I^{'\mathrm{c}}$, we see that  
$B'_{\mathrm{s}}(i\Omega_k, i\omega_\nu) = 2B_{\mathrm{s}}^{'\mathrm{b}}(i\Omega_k, i\omega_\nu) + B_{\mathrm{s}}^{'\mathrm{c}}(i\Omega_k, i\omega_\nu)$. 

We see that 
\begin{align}
	B_{\mathrm{s}}^{'\mathrm{b}} (i\Omega_k, i\omega_\nu)
	=&
	T \sum_{\varepsilon_n} C(0, \varepsilon_n, \varepsilon_{k-n}) C(0, \varepsilon_{n+\nu}, \varepsilon_{k-n}) I^{\mathrm{b}}(i\varepsilon_n, i\Omega_k, i\omega_\nu)\\
	=&
	-4\pi^3 i T \sum_{\varepsilon_n}
	\begin{aligned}[t]
		\Biggl[&
			\frac{\ctext{1}}{(2\varepsilon_n - \Omega_k) (2\varepsilon_n - \Omega_k + \omega_\nu) (2\varepsilon_n - \Omega_k - \hbar / \tau)^2}\\
			-&
			\frac{\ctext{4}}{(2\varepsilon_n - \Omega_k) (2\varepsilon_n - \Omega_k + \omega_\nu) (2\varepsilon_n - \Omega_k + \hbar  / \tau)^2}\\
			+&
			\frac{1}{\omega_\nu + \hbar / \tau} \frac{\ctext{2}}{(2\varepsilon_n - \Omega_k) (2\varepsilon_n - \Omega_k - \hbar / \tau)^2}
		\Biggr],
	\end{aligned}
\end{align}
from which it follows that for \fbox{1}, \fbox{2}, and \fbox{3}, 
\begin{align}
	&B_{\mathrm{s}1}^{'\mathrm{b}} (i\Omega_k, i\omega_\nu)\\
	=&
	-4\pi^3 i T 
	\begin{aligned}[t]
		\Biggl[&
		\sum_{\varepsilon_n < 0} \frac{1}{(2\varepsilon_n + \Omega_k) (2\varepsilon_n + \Omega_k + \omega_\nu) (2\varepsilon_n + \Omega_k - \hbar / \tau)^2} \\
		-&
		\sum_{\varepsilon_n > 0} \frac{1}{(2\varepsilon_n - \Omega_k) (2\varepsilon_n - \Omega_k + \omega_\nu) (2\varepsilon_n - \Omega_k + \hbar / \tau)^2}
		\Biggr],
	\end{aligned}\\
	 &B_{\mathrm{s}2}^{'\mathrm{b}} (i\Omega_k, i\omega_\nu)\\
	=&
	-4\pi^3 i T
	\begin{aligned}[t]
		\Biggl[&
			\sum_{\varepsilon_n < 0} \frac{1}{(2\varepsilon_n - \Omega_k - 2\omega_\nu) (2\varepsilon_n - \Omega_k - \omega_\nu) (2\varepsilon_n - \Omega_k - 2\omega_\nu - \hbar / \tau)^2}\\
			-&
			\sum_{\varepsilon_n > 0} \frac{1}{(2\varepsilon_n - \Omega_k) (2\varepsilon_n - \Omega_k + \omega_\nu) (2\varepsilon_n - \Omega_k + \hbar / \tau)^2}
		\Biggr],\label{eq:  Bs2b}
	\end{aligned}\\
	-&
	\frac{4\pi^3 iT}{\omega_\nu + \hbar / \tau} \sum_{\varepsilon_n < 0} \qty[\frac{1}{(2\varepsilon_n + \Omega_k) (2\varepsilon_n + \Omega_k - \hbar / \tau)^2} - \frac{1}{(2\varepsilon_n - \Omega_k - 2\omega_\nu) (2\varepsilon_n - \Omega_k - 2\omega_\nu - \hbar / \tau)^2}]\\
	 &B_{\mathrm{s}3}^{'\mathrm{b}}(i\Omega_k, i\omega_\nu)\\
	=&
	-4\pi^3 i T
	\begin{aligned}[t]
		\Biggl[&
			\sum_{\varepsilon_n < 0} \frac{1}{(2\varepsilon_n - \Omega_k - 2\omega_\nu) (2\varepsilon_n - \Omega_k - \omega_\nu) (2\varepsilon_n - \Omega_k - 2\omega_\nu -\hbar / \tau)^2}\\
			-&
			\sum_{\varepsilon_n > 0} \frac{1}{(2\varepsilon_n + \Omega_k) (2\varepsilon_n + \Omega_k + \omega_\nu) (2\varepsilon_n + \Omega_k + \hbar / \tau)^2}
		\Biggr]
	\end{aligned}\\
	-&
	\frac{4\pi^3 iT}{\omega_\nu + \hbar / \tau} \sum_{\varepsilon_n< 0} \qty[\frac{1}{(2\varepsilon_n - \Omega_k) (2\varepsilon_n - \Omega_k - \hbar / \tau)^2} - \frac{1}{(2\varepsilon_n - \Omega_k - 2\omega_\nu) (2\varepsilon_n - \Omega_k - 2\omega_\nu - \hbar / \tau)^2}].
\end{align}
Next we consider
$B_{\mathrm{s}}^{'\mathrm{c}} (i\Omega_k, i\omega_\nu)$, which is written as
\begin{align}
	B_{\mathrm{s}}^{'\mathrm{c}} (i\Omega_k, i\omega_\nu)
	=&
	\frac{4i\pi^3 T}{(\omega_\nu + \hbar / \tau)^2} \sum_{\varepsilon_n}
	\qty[\frac{\ctext{2}}{(2\varepsilon_n - \Omega_k) (2\varepsilon_n - \Omega_k - \hbar / \tau)} + \frac{\ctext{3}}{(2\varepsilon_n - \Omega_k + \omega_\nu) (2\varepsilon_n - \Omega_k + \omega_\nu + \hbar / \tau)}],
\end{align}
which we write down for \fbox{1}, \fbox{2}, and \fbox{3} as
\begin{align}
	&
	B_{\mathrm{s}1}^{'\mathrm{c}} (i\Omega_k, i\omega_\nu)\\
	=&
	\frac{4i\pi^3 T}{(\omega_\nu + \hbar / \tau)^2}
	\sum_{\varepsilon_n > 0} \qty[\frac{1}{(2\varepsilon_n - \Omega_k - \omega_\nu) (2\varepsilon_n - \Omega_k - \omega_\nu + \hbar / \tau)} - \frac{1}{(2\varepsilon_n - \Omega_k + \omega_\nu) ( 2\varepsilon_n - \Omega_k + \omega_\nu + \hbar / \tau)}]\\
	 &
	B_{\mathrm{s}2}^{'\mathrm{c}} (i\Omega_k, i\omega_\nu)\\
	=&
	\frac{4i\pi^3 T}{(\omega_\nu + \hbar / \tau)^2}
	\begin{aligned}[t]
		\Biggl\{&
			\sum_{\varepsilon_n < 0} \qty[\frac{1}{(2\varepsilon_n + \Omega_k) (2\varepsilon_n + \Omega_k - \hbar /\tau)} - \frac{1}{(2\varepsilon_n - \Omega_k - 2\omega_\nu) (2\varepsilon_n - \Omega_k - 2\omega_\nu - \hbar / \tau)}]\\
			+&
			\sum_{\varepsilon_n > 0} \qty[\frac{1}{(2\varepsilon_n + \Omega_k + \omega_\nu) (2\varepsilon_n + \Omega_k + \omega_\nu + \hbar / \tau)} - \frac{1}{(2\varepsilon_n - \Omega_k + \omega_\nu) (2\varepsilon_n - \Omega_k + \omega_\nu + \hbar / \tau)}]
		\Biggr\}
	\end{aligned}\label{eq:  Bs2c}\\
	 &
	B_{\mathrm{s}3}^{'\mathrm{c}} (i\Omega_k, i\omega_\nu)\\
	=&
	\frac{4i\pi^3 T}{(\omega_\nu + \hbar / \tau)^2} \sum_{\varepsilon_n < 0} \qty[\frac{1}{(2\varepsilon_n - \Omega_k) (2\varepsilon_n - \Omega_k - \hbar / \tau)} - \frac{1}{(2\varepsilon_n - \Omega_k - 2\omega_\nu) (2\varepsilon_n - \Omega_k - 2\omega_\nu - \hbar / \tau)}].
\end{align}
Both $B_{\mathrm{s}}^{'\mathrm{b}}$ and $B_{\mathrm{s}}^{'\mathrm{c}}$ become analytic functions for the regions \fbox{1}, \fbox{2}, \fbox{3} after analytical continuation $i\Omega_k \to \Omega$. 
In the present case, the relation \eqref{al-eq1}, i. e,
$B'_{\mathrm{s}1} (\Omega, 0) = B'_{\mathrm{s}2}(0, 0) = B'_{\mathrm{s}3}(\Omega, 0) = 0$ 
holds as \eqref{al-eq1} in the case of the AL terms with side jump process.
\subsubsection{\texorpdfstring{Step4 $\sim$ Step6}{Step4 ∼ Step6}}
We can proceed from step 4 to step 6 in the same way as those in the case of the AL terms with side jump process. 
We then obtain the expression corresponding to \eqref{eq: al-eq13-} as
\begin{align}
	\sigma_{xy}^{\mathrm{AL-SS}}
	\approx&
	\frac{1}{i\omega} \cdot 
	 \qty(\frac{i2\zeta'_{\rm s}\zeta_{\rm c}}{D})
     V\sum_{\bm{q}} \qty(\bm{q}^2 L(\bm{q}, 0)^2)
	\hbar \omega \frac{T}{2} B_{\mathrm{c}2}(0, 0) \dv{x}\qty[B'_{\mathrm{s}2}(-x, x) + B'_{\mathrm{s}2} (0, x)]_{x=0}. \label{eq: al-eq13-SS}
\end{align}
\subsubsection{\texorpdfstring{Step7: Sum over $\varepsilon_n$ and Step8: Sum over $\bm{q}$}{Step7: Sum over ε\_n and Step8: Sum over q}}
Performing summation \eqref{eq:  Bs2b} and \eqref{eq:  Bs2c} over $\varepsilon_n$, we find that 
\begin{align}
	&
	\dv{x} \qty[B_{\mathrm{s}2}^{'\mathrm{b}} (-x, x) + B_{\mathrm{s}2}^{'\mathrm{b}} (0, x)]_{x=0}\\
	 =&
	 6\pi^2 \qty(\frac{\tau}{\hbar})^4 \qty[\frac{\hbar}{4\pi T \tau} \psi^{(1)}\qty(\frac{1}{2}) + \frac{\hbar}{4\pi T \tau} \psi^{(1)}\qty(\frac{1}{2} + \frac{\hbar}{4\pi T \tau}) + 2\psi\qty(\frac{1}{2}) - 2\psi\qty(\frac{1}{2} + \frac{\hbar}{4\pi T \tau})],\\
	&
	\dv{x} \qty[B_{\mathrm{s}2}^{'\mathrm{c}}(-x, x) + B_{\mathrm{s}2}^{'\mathrm{c}}(0,x)]_{x=0}\\
=&
	4\pi^2 \qty(\frac{\tau}{\hbar})^4 \frac{\hbar}{4\pi T \tau} \qty[\psi^{(1)}\qty(\frac{1}{2}) - \psi^{(1)}\qty(\frac{1}{2} + \frac{\hbar}{4\pi T \tau})],
\end{align}
from which, we rewrite a factor in Eq.~\eqref{eq: al-eq13-SS} as
\begin{align}
	&
	\dv{x} \qty[B'_{\mathrm{s}2}(-x, x) + B'_{\mathrm{s}2}(0, x)]_{x=0}\\
   =&
	8\pi^2 \qty(\frac{\tau}{\hbar})^4 \qty[2 \frac{\hbar}{4\pi T \tau} \psi^{(1)}\qty(\frac{1}{2}) + \frac{\hbar}{4\pi T \tau} \psi^{(1)}\qty(\frac{1}{2} + \frac{\hbar}{4\pi T \tau}) + 3 \psi\qty(\frac{1}{2}) - 3\psi\qty(\frac{1}{2} + \frac{\hbar}{4\pi T \tau})].\label{eq:  Bs2prime-derivative}
\end{align}

The sum over $\bm{q}$ in Eq.~\eqref{eq: al-eq13-SS} is given by Eq.~\eqref{al-eq14} for $D=2$ or Eq.~\eqref{eq: al-eq14++} for $D=3$. 
\subsubsection{Step9: Expression for spin Hall conductivity}
Now we are ready to write down the expression for the spin Hall conductivity. Substituting Eqs.~\eqref{eq: Bc2-AL-SJ}, \eqref{al-eq14},  \eqref{eq: al-eq14++},  and \eqref{eq:  Bs2prime-derivative},   into Eq.~\eqref{eq: al-eq13-SS},  we obtain
\begin{align}
	\sigma_{xy}^{\mathrm{AL-SS}}
	\approx&
	\frac{16\pi^2 \zeta'_{\rm s}\zeta_{\rm c}\hbar  T}{D N(0)^2} \qty(\frac{\tau}{\hbar})^6 \\
	&\times
	\qty[\frac{\hbar}{4\pi T \tau} \psi^{(1)}\qty(\frac{1}{2}) + \psi\qty(\frac{1}{2}) - \psi\qty(\frac{1}{2} + \frac{\hbar}{4\pi T \tau})] \\
	&\times
	\qty[2 \frac{\hbar}{4\pi T \tau}\psi^{(1)}\qty(\frac{1}{2}) + \frac{\hbar}{4\pi T \tau} \psi^{(1)}\qty(\frac{1}{2} + \frac{\hbar}{4\pi T \tau}) + 3 \psi\qty(\frac{1}{2}) - 3\psi\qty(\frac{1}{2} + \frac{\hbar}{4\pi T \tau})]\\
	&\times
	\begin{dcases}
		\frac{1}{4\pi \clength^4} \ln \frac{1}{\epsilon} & (D=2)\\
		\frac{1}{2\pi^2 \clength^5} & (D=3)
	\end{dcases}\\
    =&
	\frac{8\pi^2 e^2 \hbar^7 \lambda_0^2 n_{\mathrm{i}} v_0^3 k_{\mathrm{F}}^6 N(0)^2 T}{D^4 m^4} \qty(\frac{\tau}{\hbar})^6 \\
	&\times
	\qty[\frac{\hbar}{4\pi T \tau} \psi^{(1)}\qty(\frac{1}{2}) + \psi\qty(\frac{1}{2}) - \psi\qty(\frac{1}{2} + \frac{\hbar}{4\pi T \tau})] \\
	&\times
	\qty[2 \frac{\hbar}{4\pi T \tau}\psi^{(1)}\qty(\frac{1}{2}) + \frac{\hbar}{4\pi T \tau} \psi^{(1)}\qty(\frac{1}{2} + \frac{\hbar}{4\pi T \tau}) + 3 \psi\qty(\frac{1}{2}) - 3\psi\qty(\frac{1}{2} + \frac{\hbar}{4\pi T \tau})]\\
	&\times
	\begin{dcases}
		\frac{1}{4\pi \clength^4} \ln \frac{1}{\epsilon} & (D=2)\\
		\frac{1}{2\pi^2 \clength^5} & (D=3)
	\end{dcases}.
\end{align}

\subsection{DOS terms with side jump}
\subsubsection{Step1: Feynman diagrams for Response function} 
\begin{figure}[tbp]
	\begin{minipage}[b]{0.85\linewidth}
		\begin{minipage}[b]{0.49\linewidth}
			\centering
			\includegraphics[width=\linewidth,page=1]{extrinsic_DOS_sj.pdf}
		\end{minipage}
		\begin{minipage}[b]{0.49\linewidth}
			\centering
			\includegraphics[width=\linewidth, page=2]{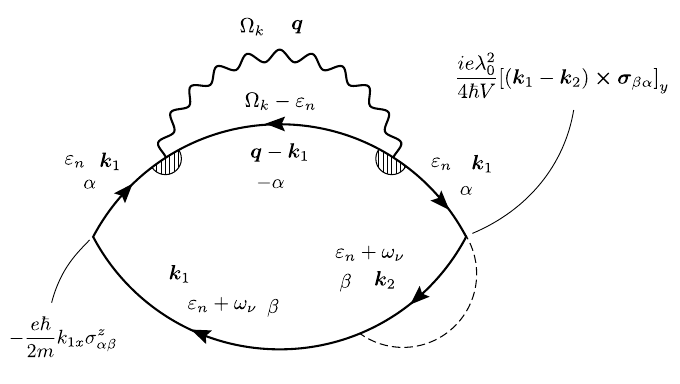}
		\end{minipage}
		\\(a)
	\label{figa}
	\end{minipage}\\

	\begin{minipage}[b]{0.85\linewidth}
		\begin{minipage}[b]{0.49\linewidth}
			\centering
			\includegraphics[width=\linewidth,page=3]{extrinsic_DOS_sj.pdf}
		\end{minipage}
		\begin{minipage}[b]{0.49\linewidth}
			\centering
			\includegraphics[width=\linewidth, page=4]{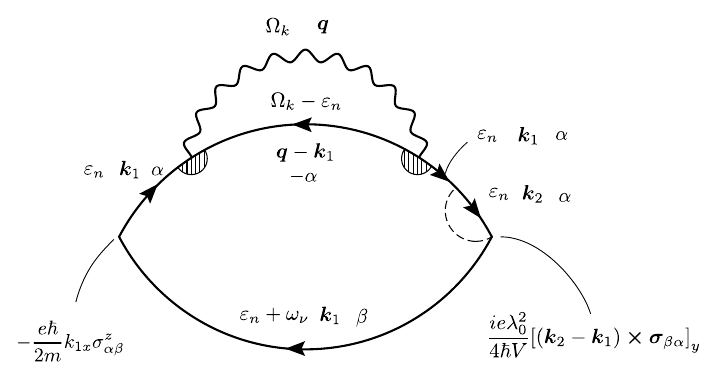}
		\end{minipage}
		\\(b)
	\label{figb}
	\end{minipage}\\

	\begin{minipage}[b]{0.85\linewidth}
		\begin{minipage}[b]{0.49\linewidth}
			\centering
			\includegraphics[width=\linewidth,page=5]{extrinsic_DOS_sj.pdf}
		\end{minipage}
		\begin{minipage}[b]{0.49\linewidth}
			\centering
			\includegraphics[width=\linewidth, page=6]{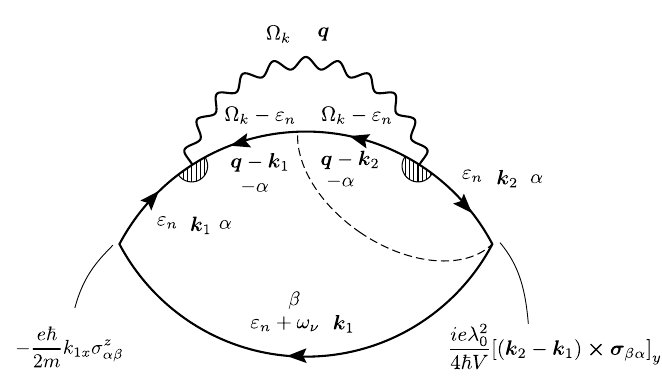}
		\end{minipage}
		\\(c)
	\label{figc}
	\end{minipage}\\

	\begin{minipage}[b]{0.85\linewidth}
		\begin{minipage}[b]{0.49\linewidth}
			\centering
			\includegraphics[width=\linewidth,page=7]{extrinsic_DOS_sj.pdf}
		\end{minipage}
		\begin{minipage}[b]{0.49\linewidth}
			\centering
			\includegraphics[width=\linewidth, page=8]{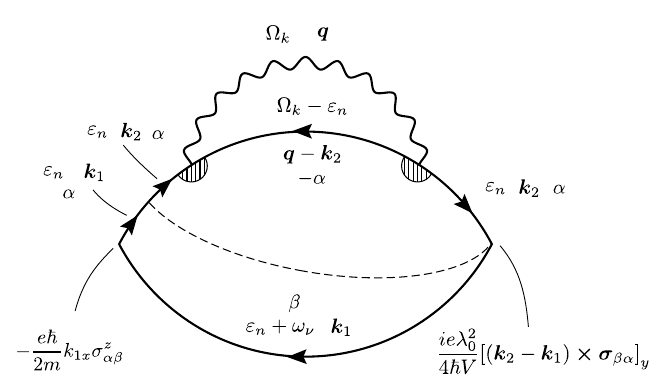}
		\end{minipage}
		\\(d)
	\label{figd}
	\end{minipage}
	\caption{Diagrams for DOS term in the presence of side jump process}
	\label{DOS-sj-figures}
\end{figure}

The response function $\Phi_{xy}$ corresponding to Fig.~\ref{DOS-sj-figures} is given by
\begin{align}
	&
	\Phi_{xy}(0,i\omega_\nu)\\
	=&
	-\frac{2 T^2}{V} \sum_{\Omega_k \varepsilon_n} \sum_{\bm{k}_1\bm{k}_2 \bm{q}} \qty[\qty(\frac{ie\lambda^2_0}{8\hbar V} \qty(k_{1y} - k_{2y})) \qty(- \frac{e\hbar}{m} k_{1y}) + \qty(-\frac{e\hbar}{2m} k_{1x}) \qty(-\frac{ie\lambda_0^2}{4\hbar V} \qty(k_{2x}-k_{1x}))] \frac{1}{V} \cdot n_{\mathrm{i}} v_0^2 V \\
	 &\times
	 (\text{the product of the Green function contained in the diagram })\\
	 &\times
	 C(\bm{q}, \varepsilon_n, \Omega_k - \varepsilon_n)^2 L(\bm{q}, i\Omega_k)\\
	\approx&
	\frac{i e^2 \lambda_0^2 n_{\mathrm{i}} v_0^2}{2m} \frac{k_F^2}{D} N(0)^2
	T \sum_{\Omega_k}\sum_{\bm{q}} L(\bm{q}, i\Omega_k) \Sigma^{\mathrm{DOS-SJ}}_{xy} (\bm{q}, \Omega_k, \omega_\nu),\label{DOS-sj-eq4}
\end{align}
where the notations 
\begin{align}
	\Sigma^{\mathrm{DOS-SJ}}_{xy}(\bm{q}, \Omega_k, \omega_\nu) \coloneqq& T\sum_{\varepsilon_n} C(\bm{q}, \varepsilon_n, \Omega_k-\varepsilon_n)^2 (-I^{\mathrm{a}} + I^{\mathrm{b}} + I^{\mathrm{c}} + I^{\mathrm{d}}) \label{DOS-sj-eq3}\\
	I^{\rm a} (\bm{q}, \varepsilon_n, \Omega_k, \omega_\nu) \coloneqq& 
	\int d\xi_1 d\xi_2 \;
	 \mathcal{G}(\bm{k}_1, i\varepsilon_n)^2 \mathcal{G}(\bm{k}_1, i\varepsilon_n + i\omega_\nu) \mathcal{G}(\bm{k}_2, i\varepsilon_n + i\omega_\nu) \mathcal{G}(\bm{q}-\bm{k}_1, i\Omega_k - i\varepsilon_n) \\
	I^{\rm b} (\bm{q}, \varepsilon_n, \Omega_k, \omega_\nu) \coloneqq& 
	\int d\xi_1 d\xi_2 \;
	 \mathcal{G}(\bm{k}_1, i\varepsilon_n)^2 \mathcal{G}(\bm{k}_1, i\varepsilon_n + i\omega_\nu) \mathcal{G}(\bm{k}_2, i\varepsilon_n) \mathcal{G}(\bm{q}-\bm{k}_1, i\Omega_k - i\varepsilon_n) \\
	I^{\rm c} (\bm{q}, \varepsilon_n, \Omega_k, \omega_\nu) \coloneqq& 
	\int d\xi_1 d\xi_2 \;
	 \mathcal{G}(\bm{k}_1, i\varepsilon_n) \mathcal{G}(\bm{k}_2, i\varepsilon_n) \mathcal{G}(\bm{k}_1, i\varepsilon_n + i\omega_\nu) \mathcal{G}(\bm{q}-\bm{k}_1, i\Omega_k-i\varepsilon_n) \mathcal{G}(\bm{q}-\bm{k}_2, i\Omega_k - i\varepsilon_n)\\
	I^{\rm d} (\bm{q}, \varepsilon_n, \Omega_k, \omega_\nu) \coloneqq& 
	\int d\xi_1 d\xi_2 \;
	 \mathcal{G}(\bm{k}_1, i\varepsilon_n) \mathcal{G}(\bm{k}_2, i\varepsilon_n)^2 \mathcal{G}(\bm{k}_1, i\varepsilon_n + i\omega_\nu) \mathcal{G}(\bm{q}-\bm{k}_2, i\Omega_k - i\varepsilon_n)
\end{align}
are introduced. 
\subsubsection{\texorpdfstring{Step2: Setting  $\Omega_k = 0$ in all quantities and $\bm{q}=0$ in all but $L(\bm{q},i\Omega_k)$}{Step2: Setting Ω\_k = 0 in all quantities and q = 0 in all but L(q, iΩ\_k)}} 
In the following, we retain the term with $\Omega_k = 0$, which is most diverging with respect to $\epsilon$. We also set $\bm{q}=0$ in $I^{\rm a,b,c,d}$ because only the $\bm{q}$-dependence in  $L(\bm{q},i\Omega_k)$ is important. 
\subsubsection{Step3: Integral over wavevectors} 
\begin{figure}[tbp]
	\centering
	\includegraphics[width=0.8\textwidth]{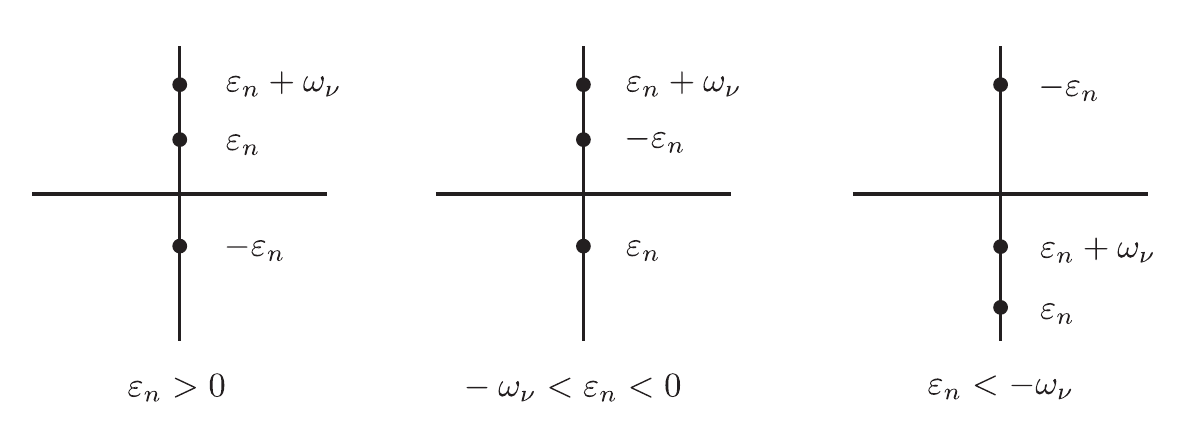}
	\caption{The locations of the poles.}
	\label{pole}
\end{figure}
The expressions for $I^{\rm a,b,c,d}$ depend on the location of $\Omega_k$ and $\omega_\nu$, as shown in Fig.~\ref{pole}. Performing integral with $\xi_1$ and $\xi_2$, we obtain%
\begin{align}
	&I^{\rm a}(0, \varepsilon_n, 0, \omega_\nu) =
	 2i\pi^2 \frac{1}{\widetilde{\varepsilon}_n + \widetilde{\varepsilon}_{n+\nu}}
	 \qty[\frac{\theta(\varepsilon_n \varepsilon_{n+\nu})}{(2\widetilde{\varepsilon}_n)^2} + \frac{\theta(-\varepsilon_n \varepsilon_{n+\nu})}{(\omega_\nu + \hbar / \tau)^2} - \frac{\theta(-\varepsilon_n \varepsilon_{n+\nu})}{(2\widetilde{\varepsilon}_n)^2}],\\
	&I^{\rm b}(0, \varepsilon_n, 0, \omega_\nu) 
	=
	 2i\pi^2 \frac{1}{\widetilde{\varepsilon}_n + \widetilde{\varepsilon}_{n+\nu}}
	 \qty[\frac{\theta(\varepsilon_n \varepsilon_{n+\nu})}{(2\widetilde{\varepsilon}_n)^2} - \frac{\theta(-\varepsilon_n \varepsilon_{n+\nu})}{(\omega_\nu + \hbar / \tau)^2} + \frac{\theta(-\varepsilon_n \varepsilon_{n+\nu})}{(2\widetilde{\varepsilon}_n)^2}],\\
	&I^{\rm c}(0, \varepsilon_n, 0, \omega_\nu) 
	=
	-i\pi^2 \frac{1}{\abs{\widetilde{\varepsilon}_n}^2} 
	\qty[\frac{\theta(\varepsilon_n) + \theta(-\varepsilon_{n+\nu})}{\widetilde{\varepsilon}_n + \widetilde{\varepsilon}_{n+\nu}} + \frac{\theta(-\varepsilon_n \varepsilon_{n+\nu})}{\omega_\nu + \hbar / \tau}],\\
	&I^{\rm d}(0, \varepsilon_n, 0, \omega_\nu) 
	= 
	i\pi^2 \frac{1}{\abs{\widetilde{\varepsilon}_n}^2} \frac{ \theta(-\varepsilon_n \varepsilon_{n+\nu})}{\omega_\nu + \hbar / \tau}. 
\end{align}
\subsubsection{\texorpdfstring{Step4: Sum over $\varepsilon_n$}{Step4: Sum over ε\_n}} 
Substituting these expressions into Eq.~\eqref{DOS-sj-eq3} and rewriting the sum over $\varepsilon_n$ with the polygamma function, we obtain   
\begin{align}
	&\Sigma^{\mathrm{DOS-SJ}}_{xy}(0, 0, \omega_\nu)\\
	=&
	\frac{i\pi}{(\omega_\nu + \hbar / \tau)^2} 
	\qty{\psi \qty(\frac{1}{2} + \frac{\omega_\nu}{2 \pi T})- \psi \qty(\frac{1}{2})
	-(\omega_\nu + 2 \hbar / \tau)\frac{1}{4\pi T}\qty[\psi^{(1)} \qty(\frac{1}{2} + \frac{\omega_\nu}{2 \pi T}) - \psi^{(1)} \qty(\frac{1}{2})]}\\
	 &-
	\frac{i\pi}{ (\omega_\nu-\hbar / \tau )^2}\qty[\psi \qty(\frac{1}{2} + \frac{\omega_\nu}{4 \pi T} + \frac{\hbar}{4 \pi T \tau}) - \psi \qty(\frac{1}{2} + \frac{\omega_\nu}{2 \pi T})]\\
		 &+
	\frac{i\pi}{ (\omega_\nu+\hbar / \tau )^2}\qty[\psi \qty(\frac{1}{2} + \frac{\omega_\nu}{4 \pi T} + \frac{\hbar}{4 \pi T \tau}) - \psi \qty(\frac{1}{2})]\\
	 &-
	\frac{i\pi}{ (\omega_\nu-\hbar / \tau )}\frac{1}{4 \pi T}\psi^{(1)} \qty(\frac{1}{2} + \frac{\omega_\nu}{2 \pi T})\\
	 &-
	\frac{i\pi}{ (\omega_\nu+\hbar / \tau )}\frac{1}{4 \pi T}\psi^{(1)}\qty(\frac{1}{2}).
\label{eq: sigma-DOS-SJ}
\end{align}
\subsubsection{\texorpdfstring{Step5: Analytic continuation and expansion with $\omega$}{Step5: Analytic continuation and expansion with ω}}
Analytic continuation $i\omega_\nu \rightarrow \hbar\omega$ and expansion with $\omega$ up to the first order are followed by 
\begin{align}
	\Sigma^{\mathrm{DOS-SJ}}_{xy} (0, 0, \omega)
	=&
	\hbar \omega \cdot 2\pi \qty(\frac{\tau}{\hbar})^3 
	\qty[
	2\psi\qty(\frac{1}{2}) - 2\psi \qty(\frac{1}{2} + \frac{\hbar}{4\pi T \tau})
		+ 3\frac{\hbar}{4\pi T \tau} \psi^{(1)} \qty(\frac{1}{2}) 
	-\qty(\frac{\hbar}{4\pi T \tau})^2 \psi^{(2)} \qty(\frac{1}{2} )
	]\\ 
	&+
	\order{\omega^2}.\label{DOS-sj-eq2}
\end{align}
\subsubsection{\texorpdfstring{Step6: Sum over $\bm{q}$}{Step6: Sum over q}}
The remaining factor to be considered is 
$\sum_{\bm{q}} L(\bm{q}, 0)$, which becomes 
\begin{align}
	\sum_{\bm{q}} L(\bm{q}, 0)
	=&
	-\frac{1}{V N(0)} \sum_{\norm{\bm{q}} < \clength^{-1}} \frac{1}{\epsilon + \clength^2 \bm{q}^2}
	=
	-\frac{1}{N(0)} \int_{\norm{\bm{q}} < \clength^{-1}} \frac{d^D \bm{q}}{(2\pi)^D} \frac{1}{\epsilon + \clength^2 \bm{q}^2}\\
	=&
	-\frac{1}{(2\pi)^D N(0)} \frac{1}{\clength^D} \int_0^1 dx \frac{x^{D-1}}{\epsilon + x^2}
	\begin{dcases}
		2\pi & (D=2)\\
		4\pi & (D=3)
	\end{dcases}\\
	\approx&
	-\frac{1}{N(0) \clength^D} 
	\begin{dcases}
		\frac{1}{4\pi} \ln \frac{1}{\epsilon} & (D=2)\\
		\frac{1}{2\pi^2} & (D=3)
	\end{dcases}.\label{eq: Lq-sum}
\end{align}

\subsubsection{Step7: Expression for spin Hall conductivity}
Putting Eqs.~\eqref{DOS-sj-eq2} and \eqref{eq: Lq-sum} into Eq.~\eqref{DOS-sj-eq4} and multiplying by the factor of two, coming from the contribution of the diagrams reflected vertically,   
we arrive at 
\begin{align}
	\sigma_{xy}^{\mathrm{DOS-SJ}}
	=&
	-\frac{2\pi e^2 \lambda_0^2n_{\mathrm{i}} v_0^2 k_F^2 N(0) T \hbar}{Dm \clength^D} \qty(\frac{\tau}{\hbar})^3\\
	&\times
	\qty[
	2\psi\qty(\frac{1}{2}) - 2\psi \qty(\frac{1}{2} + \frac{\hbar}{4\pi T \tau})
		+ 3\frac{\hbar}{4\pi T \tau} \psi^{(1)} \qty(\frac{1}{2}) 
	-\qty(\frac{\hbar}{4\pi T \tau})^2 \psi^{(2)} \qty(\frac{1}{2} )
	]\\ 
	&\times
	\begin{dcases}
		\frac{1}{4\pi} \ln \frac{1}{\epsilon} & (D=2)\\
		\frac{1}{2\pi^2} & (D=3)
	\end{dcases} \label{DOS-sj-eq8}.
\end{align}
%
%
%
\subsection{DOS terms with skew scattering}
\subsubsection{Step1: Feynman diagrams for Response function} 
\begin{figure}[tbp]
	\centering
	\begin{tabular}{ccc}
		\begin{minipage}[b]{0.4\textwidth}
			\centering
			\includegraphics[width=\textwidth]{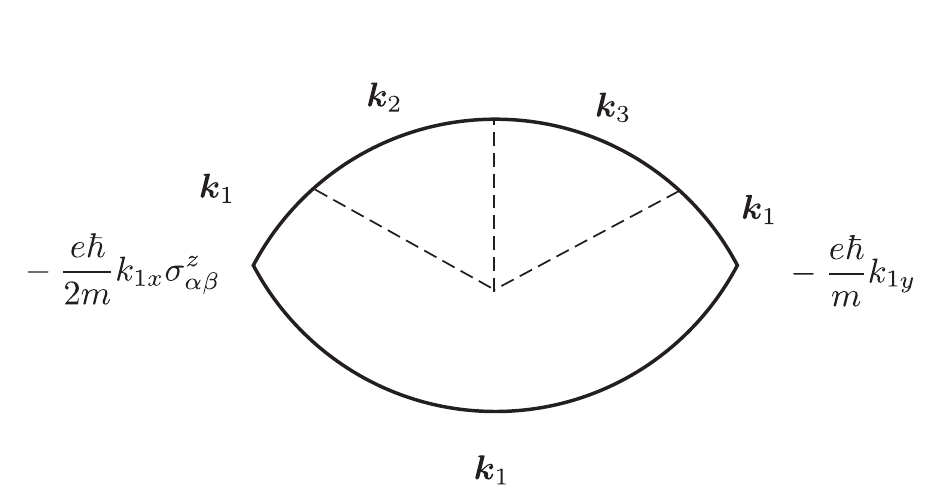}
			\caption{Diagrams with impurity vertices only in one-side of the lines.}
			\label{DOS-ss-katagawa}
		\end{minipage}
		&
		\hspace{1cm}
		&
		\begin{minipage}[b]{0.4\textwidth}
			\centering
			\includegraphics[width=\textwidth]{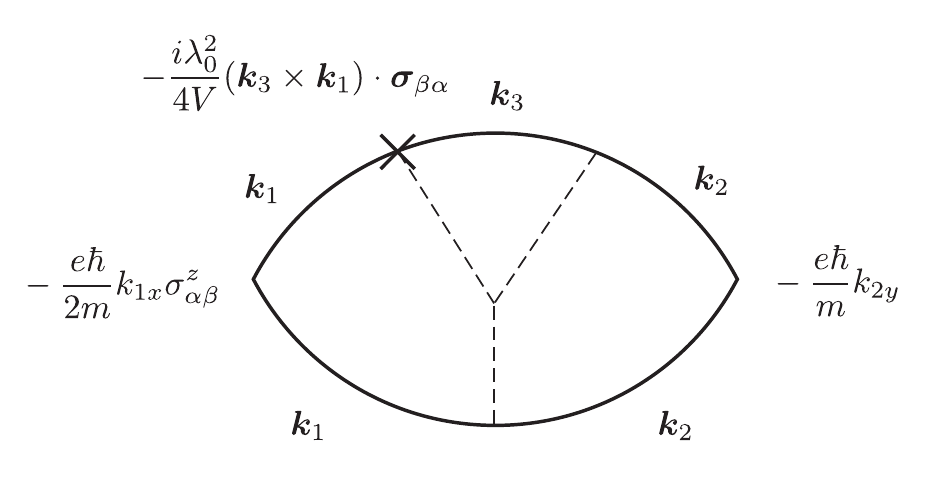}
			\caption{Diagrams with $H_{\mathrm{SO}}$ on the side with another impurity }
			\label{DOS-ss-koritu}
		\end{minipage}
	\end{tabular}
\end{figure}

\begin{figure}[tbp]
	\centering
	\includegraphics[width=0.7\textwidth]{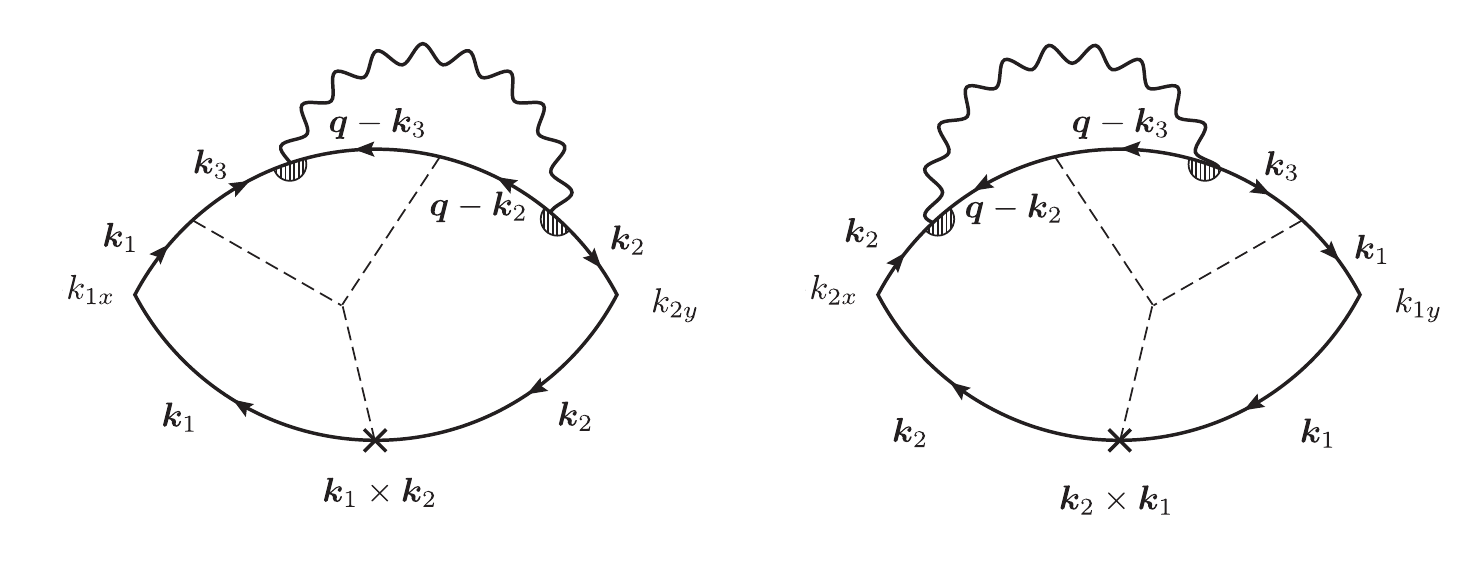}
	\caption{A diagram (left) and its reflection (right) with respect to the vertical axis}
	\label{DOS-ss-LR}
\end{figure}

It is convenient to see which diagrams really contribute as the DOS terms with skew-scattering process before explicit calculation. This is because there are so many diagrams of this type. 
\begin{itemize}
    \item We first note that the diagrams with impurity vertex only in one-side of the particle line (e.g. Fig.~\ref{DOS-ss-katagawa}) do not contribute when we retain the lowest order contribution with respect to  $\bm{q}$. In those diagrams, $k_{1x}, k_{1y}$ stem from $j_{\mathrm{s}}, j_{\mathrm{c}}$ while the external product in $H_{\mathrm{SO}}$ yields either $\pm \bm{k}_1\times \bm{k}_2$, $\pm\bm{k}_1 \times \bm{k}_3$, or $\pm \bm{k}_2 \times \bm{k}_3$. Thus only $k_{1x}^2 k_{1y}$ or $k_{1x} k_{1y}^2$ appears and the integrals with  $k_{1x}$ or $k_{1y}$ vanish. 
    \item We note that diagrams contribute only when $H_{\mathrm{SO}}$ is on the line with one impurity vertex in the lowest order contribution with respect to $\bm{q}$ (see Fig.~\ref{DOS-ss-koritu}). In order for the diagram to contribute, $k_{1x}, k_{1y}$ stemming from $j_{\mathrm{s}}, j_{\mathrm{c}}$ have to be squared with  $k_{1x}, k_{1y}$ coming from $H_{\mathrm{SO}}$. It is possible when $H_{\mathrm{SO}}$ is on the side with no other impurity vertex  but not when  $H_{\mathrm{SO}}$ are on the side with another impurity vertex.
    \item The diagram {\it reflected with vertical line} contributes equally to the original diagram (See. Fig.~\ref{DOS-ss-LR}). Note that the left end point and the right end point, respectively, represent the spin current vertex and charge current vertex both for the original and reflected diagrams. When we assign the wavenumbers of the Green function in the diagrams that the parts corresponding with each other in both diagram have the same wave number, as shown in Fig.~\ref{DOS-ss-LR}. Then the original diagram have $\bm{k}_1 \times \bm{k}_2$ while the reflected diagram $\bm{k}_2 \times \bm{k}_1$. In the former, $k_{1x}, k_{2y}$ come from $j_{\mathrm{s}}, j_{\mathrm{c}}$ while $k_{2x}, k_{1y}$ come from $j_{\mathrm{s}}, j_{\mathrm{c}}$ in the latter. Thus $k_{1x}k_{2y}(k_{1x}k_{2y} - k_{1y}k_{2x})$ in the original diagram corresponds to $k_{2x}k_{1y}(k_{2x}k_{1y} - k_{2y}k_{1x})$ in the reflected diagram. These two equally yield $(k_F^2/ D)^2$ after the integral with $\xi_1$ and $\xi_2$.
\end{itemize}

\begin{figure}[tbp]
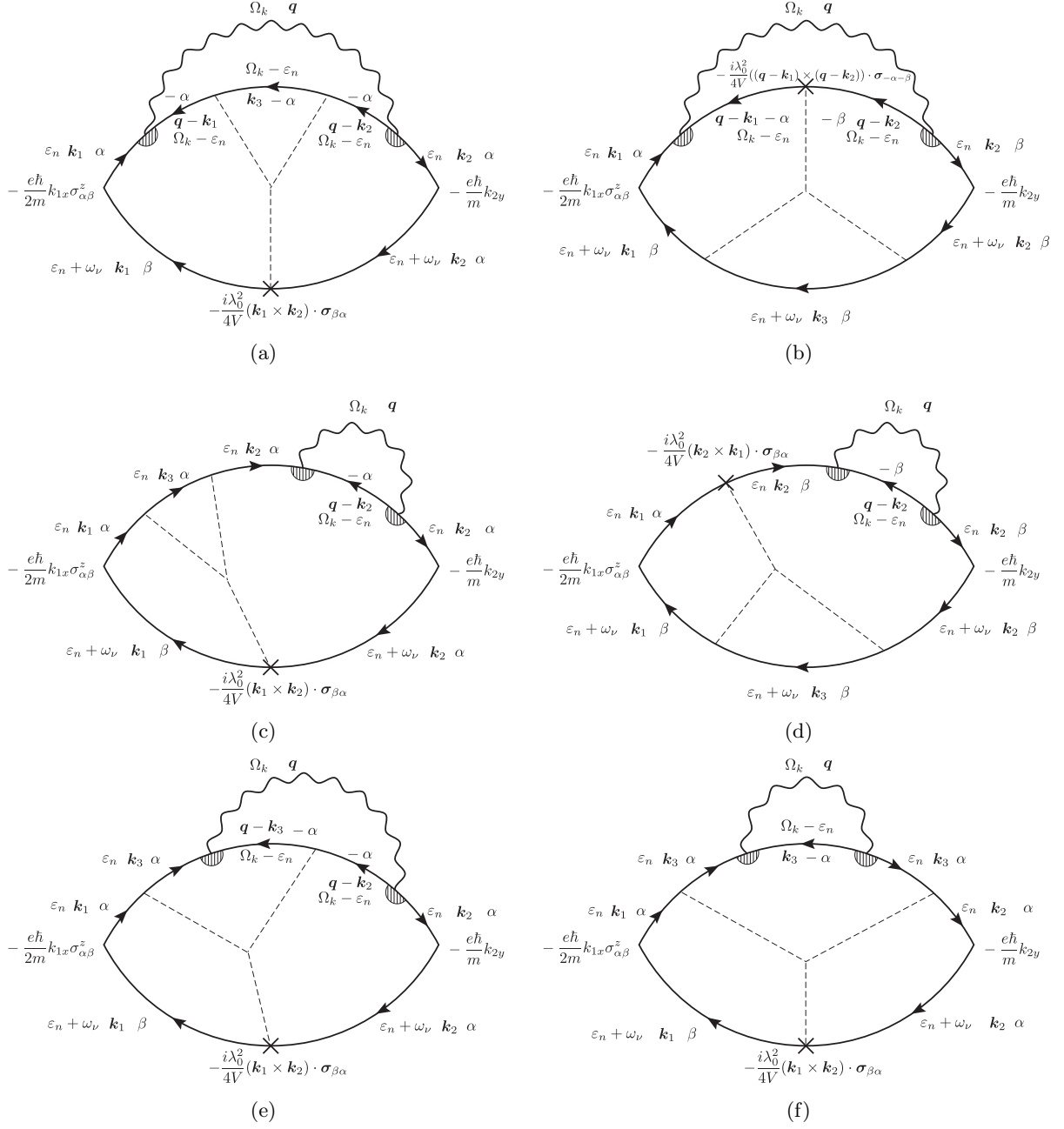

	\begin{minipage}[b]{0.45\linewidth}
		\centering
		\includegraphics[width=\textwidth]{DOS_ss_a.pdf}
		\\(a)
	\end{minipage}
	\begin{minipage}[b]{0.45\linewidth}
		\centering
		\includegraphics[width=\textwidth]{DOS_ss_b.pdf}
		\\(b)
	\end{minipage}\\
	\begin{minipage}[b]{0.45\linewidth}
		\centering
		\includegraphics[width=\textwidth]{DOS_ss_c.pdf}
		\\(c)
	\end{minipage}
	\begin{minipage}[b]{0.45\linewidth}
		\centering
		\includegraphics[width=\textwidth]{DOS_ss_d.pdf}
		\\(d)
	\end{minipage}\\
	\begin{minipage}[b]{0.45\linewidth}
		\centering
		\includegraphics[width=\textwidth]{DOS_ss_e.pdf}
		\\(e)
	\end{minipage}
	\begin{minipage}[b]{0.45\linewidth}
		\centering
		\includegraphics[width=\textwidth]{DOS_ss_f.pdf}
		\\(f)
	\end{minipage}
	\caption{Six types of diagrams contributing to the spin Hall conductivity as the DOS terms with the skew-scattering process.}
	\label{DOS-ss-figs}
\end{figure}

Considering the above points, we focus on the six types of the diagram shown in Fig.~\ref{DOS-ss-figs}. 

Next, we consider the coefficient coming from the other parts than the Green function, Cooperon, and fluctuation propagator.
In the diagrams shown in Fig.\ref{DOS-ss-figs}, the number of impurities and the number of wavevectors are common, and hence the coefficients in the diagram are an integer multiple of that in (a), which turns out to be $(ie^2\hbar^2 \lambda_0^2 n_{\mathrm{i}} v_0^3 V /4m^2) (k_{\mathrm{F}} /D)^2 N(0)^3$ as shown later. 

Let $k_1$ and $k_2$ be wavevectors coming from $j_{\mathrm{s}}$ and $j_{\mathrm{c}}$, respectively. Then every diagram have common factor $k_{1x}k_{2y}\sigma_{\alpha \beta}^z$ while $H_{\mathrm{SO}}$ yields \begin{align}
	-\frac{i\lambda_0^2}{4V} (\bm{k}_1 \times \bm{k}_2)\cdot \bm{\sigma}_{\beta \alpha}
\end{align}
in (a), (c), (e), (f) and 
\begin{align}
	-\frac{i\lambda_0^2}{4V} ((\bm{q}- \bm{k}_1 )\times (\bm{q}- \bm{k}_2))\cdot \bm{\sigma}_{-\alpha -\beta} \approx -\frac{i\lambda_0^2}{4V} (\bm{k}_1 \times \bm{k}_2)\cdot \bm{\sigma}_{-\alpha -\beta}
\end{align}
in (b)
and
\begin{align}
	-\frac{i\lambda_0^2}{4V} (\bm{k}_2 \times \bm{k}_1)\cdot \bm{\sigma}_{\beta \alpha}
\end{align}
in (d). Summation with $\sigma$ yields 
\begin{align}
	\sum_{\alpha \beta} \sigma_{\alpha \beta}^z \sigma_{\beta \alpha}^\mu 
	=&
	\delta_{z\mu} \sum_{\alpha \beta} \sigma_{\alpha \beta}^z \sigma_{\beta \alpha}^z 
	=
	\delta_{z \mu} \sum_{\alpha} (\sigma_{\alpha \alpha}^z)^2 
	=
	2\delta_{z\mu} \\
	\sum_{\alpha \beta} \sigma_{\alpha \beta}^z \sigma_{-\alpha -\beta}^\mu 
	=&
	\delta_{z\mu} \sum_{\alpha} \sigma_{\alpha \alpha}^z\sigma_{-\alpha -\alpha}^z 
	=
	-2\delta_{z\mu}.
\end{align}
Thus, diagrams (c), (e), and (f) have the same coefficient as (a) while those (b) and (d) have minus the coefficient in (a).
The DOS terms with skew scattering process for $\sigma^{\mathrm{DOS-SS}}_{xy}$ is thus given by
\begin{align}
	\Phi^{\mathrm{DOS-SS}}_{xy}(0,i\omega_\nu)
	=&
	-\frac{1}{V} \cdot 2T^2 \sum_{\Omega_k, \varepsilon_n} \sum_{\bm{k}_1, \bm{k}_2, \bm{k}_3, \bm{q}} 
	\qty(-\frac{e\hbar}{2m} k_{1x}) \qty(-\frac{i\lambda_0^2}{4V} (k_{1x}k_{2y} - k_{1y}k_{2x})) \qty(-\frac{e\hbar}{m}k_{2y}) 
	\frac{n_{\mathrm{i}} v_0^3 V}{V^2}\\
	 &\times
	 C(\bm{q}, \varepsilon_n, \Omega_k-\varepsilon_n)^2 L(\bm{q}, i\Omega_k)
	 \times \text{(product of the Green functions)}
	\label{DOS-ss-eq3-}
\end{align}
\subsubsection{\texorpdfstring{Step2: Setting  $\Omega_k = 0$ in all quantities and $\bm{q}=0$ in all but $L(\bm{q},i\Omega_k)$}{Step2: Setting Ω\_k = 0 in all quantities and q = 0 in all but L(q, iΩ\_k)}}
Setting  $\Omega_k = 0$ in all quantities and $\bm{q}=0$ in all but $L(\bm{q},i\Omega_k)$, Eq.~\eqref{DOS-ss-eq3-} reduces to 
\begin{align}
	\Phi^{\mathrm{DOS-SS}}_{xy}(0,i\omega_\nu)
	\approx &
	\underbrace{2}_{\mathclap{\text{diagram with up-side-down}}} \times  \frac{1}{V} \frac{ie^2 \hbar^2 \lambda_0^2 n_{\mathrm{i}} v_0^3 V}{4m^2} \qty(\frac{k_F^2}{D})^2 N(0)^3 T \sum_{\bm{q}} L(\bm{q}, 0)\\
	 &\times
	 T \sum_{\varepsilon_n}C(0, \varepsilon_n, -\varepsilon_n)^2 
	 (I^{\rm a} - I^{\rm b} + 2I^{\rm c} - 2I^{\rm d} + 2I^{\rm e} + I^{\rm f}). \label{DOS-ss-eq3}
\end{align}
Here we introduce $I^{\rm a,b,c,d,e,f}$ as
\begin{align}
	I^{\rm a}(\varepsilon_n, \omega_\nu) 
	=&
	\begin{aligned}[t]
		&\int d\xi_1 \;\mathcal{G}(\bm{k}_1, i\varepsilon_n) \mathcal{G}(\bm{k}_1, i\varepsilon_n+i\omega_\nu) \mathcal{G}(\bm{k}_1, -i\varepsilon_n)\\
		\times
		&\int d\xi_2\; \mathcal{G}(\bm{k}_2, i\varepsilon_n) \mathcal{G}(\bm{k}_2, i\varepsilon_n+i\omega_\nu) \mathcal{G}(\bm{k}_2, -i\varepsilon_n)\\
		\times
		&\int d\xi_3\; \mathcal{G}(\bm{k}_3, -i\varepsilon_n),
	\end{aligned}\\
	I^{\rm b}(\varepsilon_n, \omega_\nu) 
	=&
	\begin{aligned}[t]
		&\int d\xi_1\; \mathcal{G}(\bm{k}_1, i\varepsilon_n) \mathcal{G}(\bm{k}_1, i\varepsilon_n+i\omega_\nu) \mathcal{G}(\bm{k}_1, -i\varepsilon_n)\\
		\times
		&\int d\xi_2 \;\mathcal{G}(\bm{k}_2, i\varepsilon_n) \mathcal{G}(\bm{k}_2, i\varepsilon_n+i\omega_\nu) \mathcal{G}(\bm{k}_2, -i\varepsilon_n)\\
		\times
		&\int d\xi_3 \;\mathcal{G}(\bm{k}_3, i\varepsilon_n + i\omega_\nu),
	\end{aligned}\\
	I^{\rm c}(\varepsilon_n, \omega_\nu) 
	=&
	\begin{aligned}[t]
		&\int d\xi_1 \;\mathcal{G}(\bm{k}_1, i\varepsilon_n) \mathcal{G}(\bm{k}_1, i\varepsilon_n+i\omega_\nu) \\
		\times
		&\int d\xi_2 \;\mathcal{G}(\bm{k}_2, i\varepsilon_n)^2 \mathcal{G}(\bm{k}_2, i\varepsilon_n+i\omega_\nu) \mathcal{G}(\bm{k}_2, -i\varepsilon_n)\\
		\times
		&\int d\xi_3 \;\mathcal{G}(\bm{k}_3, i\varepsilon_n),
	\end{aligned}\\
	I^{\rm d}(\varepsilon_n, \omega_\nu) 
	=&
	\begin{aligned}[t]
		&\int d\xi_1\; \mathcal{G}(\bm{k}_1, i\varepsilon_n) \mathcal{G}(\bm{k}_1, i\varepsilon_n+i\omega_\nu) \\
		\times
		&\int d\xi_2\; \mathcal{G}(\bm{k}_2, i\varepsilon_n)^2 \mathcal{G}(\bm{k}_2, i\varepsilon_n+i\omega_\nu) \mathcal{G}(\bm{k}_2, -i\varepsilon_n)\\
		\times
		&\int d\xi_3\; \mathcal{G}(\bm{k}_3, i\varepsilon_n + i\omega_\nu),
	\end{aligned}\\
	I^{\rm e}(\varepsilon_n, \omega_\nu) 
	=&
	\begin{aligned}[t]
		&\int d\xi_1\; \mathcal{G}(\bm{k}_1, i\varepsilon_n) \mathcal{G}(\bm{k}_1, i\varepsilon_n+i\omega_\nu) \\
		\times
		&\int d\xi_2\; \mathcal{G}(\bm{k}_2, i\varepsilon_n) \mathcal{G}(\bm{k}_2, i\varepsilon_n+i\omega_\nu) \mathcal{G}(\bm{k}_2, -i\varepsilon_n)\\
		\times
		&\int d\xi_3\; \mathcal{G}(\bm{k}_3, i\varepsilon_n ) \mathcal{G}(\bm{k}_3, -i\varepsilon_n),
	\end{aligned}\\
	I^{\rm f}(\varepsilon_n, \omega_\nu) 
	=&
	\begin{aligned}[t]
		&\int d\xi_1\; \mathcal{G}(\bm{k}_1, i\varepsilon_n) \mathcal{G}(\bm{k}_1, i\varepsilon_n+i\omega_\nu) \\
		\times
		&\int d\xi_2\; \mathcal{G}(\bm{k}_2, i\varepsilon_n) \mathcal{G}(\bm{k}_2, i\varepsilon_n+i\omega_\nu) \\
		\times
		&\int d\xi_3\; \mathcal{G}(\bm{k}_3, i\varepsilon_n )^2 \mathcal{G}(\bm{k}_3, -i\varepsilon_n).
	\end{aligned}\\
\end{align}
\subsubsection{Step3: Integral over wavevectors} 
Performing the integral with $\xi_1,\xi_2,\xi_3$, we obtain  
\begin{align}
	I^{\rm a}(\varepsilon_n, \omega_\nu)-I^{\mathrm{b}}(\varepsilon_n, \omega_\nu) 
	=&
	-8i \pi^3 \frac{\sign(\varepsilon_n)}{\abs{2\widetilde{\varepsilon}_n}^2 (\widetilde{\varepsilon}_n + \widetilde{\varepsilon}_{n+\nu})^2} \theta(\varepsilon_n \varepsilon_{n+\nu}),\\
	I^{\rm c}(\varepsilon_n, \omega_\nu) - I^{\mathrm{d}}(\varepsilon_n, \omega_\nu)
	=&
	-8i\pi^3 \frac{\theta(-\varepsilon_n \varepsilon_{n+\nu})}{(\omega_\nu + \hbar / \tau) (\widetilde{\varepsilon}_n + \widetilde{\varepsilon}_{n+\nu})} \qty[\frac{1}{(\omega_\nu + \hbar / \tau)^2} - \frac{1}{(2\widetilde{\varepsilon}_n)^2}],\\
	I^{\rm e}(\varepsilon_n, \omega_\nu) 
	=&
	-8i \pi^3 \frac{\theta(-\varepsilon_n \varepsilon_{n+\nu})}{(\omega_\nu + \hbar / \tau)^2 \abs{2\widetilde{\varepsilon}_n}^2},\\
	I^{\rm f}(\varepsilon_n, \omega_\nu) 
	=&
	8i\pi^3 \frac{\theta(-\varepsilon_n \varepsilon_{n+\nu})}{(\omega_\nu + \hbar / \tau)^2 \abs{2\widetilde{\varepsilon}_n}^2}.
\end{align}
\subsubsection{\texorpdfstring{Step4: Sum over $\varepsilon_n$}{Step4: Sum over ε\_n}}
Performing the sum over $\varepsilon_n$, we obtain 
\begin{align}
   &T \sum_{\varepsilon_n} C(0, \varepsilon_n, -\varepsilon_n)^2 (I^{\rm a} - I^{\rm b} + 2I^{\rm c} - 2I^{\rm d} + 2I^{\rm e} + I^{\rm f})\\
 =&
	-2i \pi^2 \Biggl\{
	\begin{aligned}[t]
		&\frac{2}{(-\omega_\nu + \hbar / \tau )^3} 
		\qty[\psi\qty(\frac{1}{2} + \frac{\omega_\nu}{4\pi T} + \frac{\hbar}{4 \pi T \tau}) - \psi\qty(\frac{1}{2} + \frac{\omega_\nu}{2\pi T})]\\
		-&
		\frac{2}{(\omega_\nu + \hbar / \tau )^3} 
		\qty[\psi\qty(\frac{1}{2} + \frac{\omega_\nu}{4\pi T} + \frac{\hbar}{4 \pi T \tau}) - \psi\qty(\frac{1}{2})]\\
		-&
		\frac{1}{4\pi T(-\omega_\nu + \hbar / \tau )^2} 
		\qty[\psi^{(1)} \qty(\frac{1}{2} + \frac{\omega_\nu}{4\pi T} + \frac{\hbar}{4 \pi T \tau}) + \psi^{(1)} \qty(\frac{1}{2} + \frac{\omega_\nu}{2\pi T})]\\
		+&
		\frac{1}{4\pi T(\omega_\nu + \hbar / \tau )^2} 
		\qty[\psi^{(1)} \qty(\frac{1}{2} + \frac{\omega_\nu}{4\pi T} + \frac{\hbar}{4 \pi T \tau}) + \psi^{(1)} \qty(\frac{1}{2})] \Biggr\}
	\end{aligned}\\
    &+
	\frac{4i \pi^2}{(\omega_\nu + \hbar / \tau)^3} 
	\qty[\psi\qty(\frac{1}{2} + \frac{\omega_\nu}{2 \pi T}) - \psi \qty(\frac{1}{2})-\frac{(\omega_\nu + 2 \hbar / \tau )}{4\pi T} 
	\qty[\psi^{(1)}\qty(\frac{1}{2} + \frac{\omega_\nu}{2 \pi T}) 
	     -\psi^{(1)}\qty(\frac{1}{2})]]\\
%
	&-\frac{i\pi}{2(\omega_\nu + \hbar / \tau)^2 T} \qty[\psi^{(1)} \qty(\frac{1}{2}) - \psi^{(1)} \qty(\frac{1}{2} + \frac{\omega_\nu}{2\pi T})].
\end{align}
\subsubsection{\texorpdfstring{Step5: Analytic continuation and expansion with $\omega$}{Step5: Analytic continuation and expansion with ω}}
After analytic continuation $i\omega_\nu \to \hbar \omega \approx 0$ and expansion with $\omega$, we obtain
\begin{align}
	&T \sum_{\varepsilon_n} C(0, \varepsilon_n, -\varepsilon_n)^2 (I^{\rm a} - I^{\rm b} + 2I^{\rm  c} - 2I^{\rm d} + 2I^{\rm e} + I^{\rm f})\\
	&\xrightarrow{\omega_\nu \to -i\hbar \omega \approx 0}\\
	&
	\hbar \omega\cdot 8\pi^2 \qty(\frac{\tau}{\hbar})^4 \Biggl\{
		\begin{aligned}[t]
			-&
			3\qty[\psi\qty(\frac{1}{2} + \frac{\hbar}{4 \pi T \tau}) - \psi\qty(\frac{1}{2})]
			+
			\frac{\hbar}{4\pi T \tau} \qty[\psi^{(1)}\qty(\frac{1}{2} + \frac{\hbar}{4 \pi T \tau}) + 3\psi^{(1)}\qty(\frac{1}{2})]\\
			-&
			\qty(\frac{\hbar}{4\pi T \tau})^2 \psi^{(2)}\qty(\frac{1}{2})
			\Biggr\} + \order{\omega^2}.
		\end{aligned} \label{eq: TsumC}
\end{align}
\subsubsection{\texorpdfstring{Step6: Sum over $\bm{q}$ and Step7: Expression for spin Hall conductivity}{Step6: Sum over q and Step7: Expression for spin Hall conductivity}}
With use of Eqs.\eqref{DOS-ss-eq3}, \eqref{eq: TsumC}, and \eqref{eq: Lq-sum}, we obtain
\begin{align}
	\sigma^{\mathrm{DOS-SS}}_{xy}
=&
	-\frac{e^2 \hbar^3 \lambda_0^2 n_{\mathrm{i}} v_0^3 N(0)^2 T}{2m^2 \clength^D} \qty(\frac{k_{\mathrm{F}}^2}{D})^2 
	\\
	&\times
	8\pi^2 \qty(\frac{\tau}{\hbar})^4 \Biggl\{
	   \begin{aligned}[t]
		   -&
			3\qty[\psi\qty(\frac{1}{2} + \frac{\hbar}{4 \pi T \tau}) - \psi\qty(\frac{1}{2})]
			+
			\frac{\hbar}{4\pi T \tau} \qty[\psi^{(1)}\qty(\frac{1}{2} + \frac{\hbar}{4 \pi T \tau}) + 3\psi^{(1)}\qty(\frac{1}{2})]\\
			-&
			\qty(\frac{\hbar}{4\pi T \tau})^2 \psi^{(2)}\qty(\frac{1}{2})
			\Biggr\}
		\end{aligned}\\
	&\times
	\begin{dcases}
		\frac{1}{4\pi} \ln \frac{1}{\epsilon} & (D=2)\\
		\frac{1}{2\pi^2} & (D=3).
	\end{dcases}
\end{align}
\subsection{Maki-Thompson terms with side jump}
\subsubsection{Step1: Feynman diagrams for Response function} 
\begin{figure}[tbp]
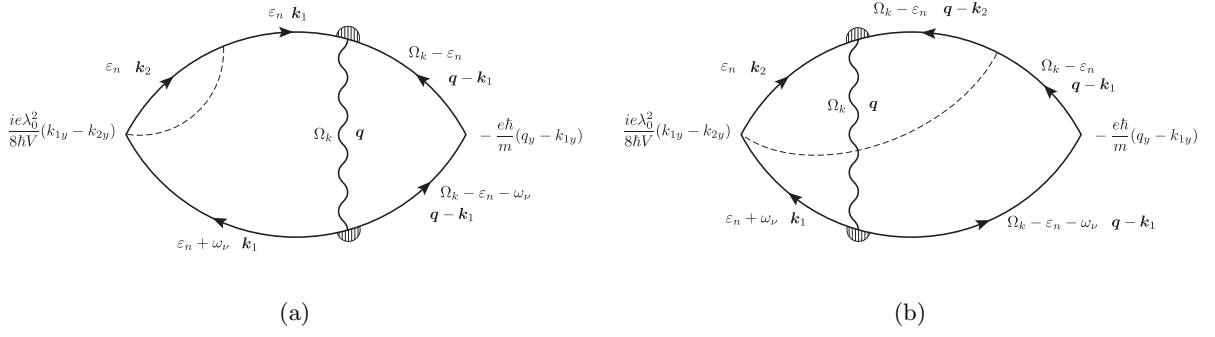

	\begin{minipage}[b]{0.45\linewidth}
		\centering
		\includegraphics[width=\textwidth]{mt_sj_a.pdf}
		\\(a)
	\end{minipage}
	\begin{minipage}[b]{0.45\linewidth}
		\centering
		\includegraphics[width=\textwidth]{mt_sj_b.pdf}
		\\(b)
	\end{minipage}\\
	\caption{Diagram for the MT terms with side jump process.}
	\label{mt-sj-figs}
\end{figure}
We show that the sum of the MT terms with the side jump process, which is shown in Fig.~\ref{mt-sj-figs}, vanishes when we set $\Omega_k = 0$. 
Similar to the DOS terms, the diagram reflected with respect to the horizontal axis contributes equally to the original diagram. The response functions $\Phi_{xy}$ for each diagram are given by
\begin{align}
	\Phi_{xy}^{\mathrm{MT-SJ-a}} 
	=&
	-\frac{2T^2}{V} \sum_{\Omega_k \varepsilon_n} \sum_{\bm{k}_1\bm{k}_2\bm{q}} \frac{ie\lambda_0^2}{8\hbar V}(k_{1y} - k_{2y}) \qty(-\frac{e\hbar}{m}) (q_y - k_{1y}) \frac{n_{\mathrm{i}} v_0^2 V}{V} \\
	&\times
	\mathcal{G}(\bm{k}_1, i\varepsilon_n) \mathcal{G}(\bm{k}_1, i\varepsilon_n + i\omega_\nu) \mathcal{G}(\bm{q}-\bm{k}_1, i\Omega_k-i\varepsilon_n) 
	\mathcal{G}(\bm{q}-\bm{k}_1, i\Omega_k - i\varepsilon_n - i\omega_\nu) \\
	&\times
	\mathcal{G}(\bm{k}_2, i\varepsilon_n) \\
	&\times
	C(\bm{q}, \varepsilon_n, \Omega_k - \varepsilon_n) C(\bm{q}, \varepsilon_n + \omega_\nu, \Omega_k-\varepsilon_n - \omega_\nu) L(\bm{q}, i\Omega_k),\\
	\Phi_{xy}^{\mathrm{MT-SJ-b}} 
	=&
	-\frac{2T^2}{V} \sum_{\Omega_k \varepsilon_n} \sum_{\bm{k}_1\bm{k}_2\bm{q}} \frac{ie\lambda_0^2}{8\hbar V} (k_{1y} - k_{2y}) \qty(-\frac{e\hbar}{m}) (q_y - k_{1y}) \frac{n_{\mathrm{i}} v_0^2 V}{V} \\
	&\times
	\mathcal{G}(\bm{k}_1, i\varepsilon_n + i\omega_\nu) \mathcal{G}(\bm{q}-\bm{k}_1, i\Omega_k-i\varepsilon_n) 
	\mathcal{G}(\bm{q}-\bm{k}_1, i\Omega_k - i\varepsilon_n - i\omega_\nu) \\
	&\times
	\mathcal{G}(\bm{k}_2, i\varepsilon_n) \mathcal{G}(\bm{q}-\bm{k}_2, i\Omega_k - i\varepsilon_n)\\
	&\times
	C(\bm{q}, \varepsilon_n, \Omega_k - \varepsilon_n) C(\bm{q}, \varepsilon_n + \omega_\nu, \Omega_k-\varepsilon_n - \omega_\nu) L(\bm{q}, i\Omega_k).\\
\end{align}
\subsubsection{\texorpdfstring{Step2: Setting  $\Omega_k = 0$ in all quantities and $\bm{q}=0$ in Green functions}{Step2: Setting Ω\_k = 0 in all quantities and q = 0 in Green functions}}
We set $\Omega_k = 0$ for all quantities and set $\bm{q}=0$ for Green functions, $\Phi_{xy}^{\mathrm{MT-SJ-a}}$ and $\Phi_{xy}^{\mathrm{MT-SJ-b}}$ reduce to 
\begin{align}
	\Phi_{xy}^{\mathrm{MT-SJ-a}} 
	\approx&
	-2T^2\frac{k_{\mathrm{F}}^2}{D}\frac{ie^2\lambda_0^2}{8}\frac{n_{\mathrm{i}} v_0^2 }{m} N(0)^2 \sum_{q} L(\bm{q},0) \sum_{\varepsilon_n}  
		C(\bm{q}, \varepsilon_n,  - \varepsilon_n) C(\bm{q}, \varepsilon_n + \omega_\nu, -\varepsilon_n - \omega_\nu)I^{\rm a}(\varepsilon_n,\omega_\nu),\\
	\Phi_{xy}^{\mathrm{MT-SJ-b}} 
	\approx&
	-2T^2\frac{k_{\mathrm{F}}^2}{D}\frac{ie^2\lambda_0^2}{8}\frac{n_{\mathrm{i}} v_0^2 }{m} N(0)^2 
	\sum_{q} L(\bm{q},0) \sum_{\varepsilon_n}  
		C(\bm{q}, \varepsilon_n,  - \varepsilon_n) C(\bm{q}, \varepsilon_n + \omega_\nu, -\varepsilon_n - \omega_\nu)I^{\rm b}(\varepsilon_n,\omega_\nu),\\
\end{align}
with 
\begin{align}
	I^{\rm a}  
	=&
	\int d\xi_1\;
	\mathcal{G}(\bm{k}_1, i\varepsilon_n) \mathcal{G}(\bm{k}_1, i\varepsilon_n + i\omega_\nu) \mathcal{G}(\bm{k}_1, -i\varepsilon_n) \mathcal{G}(\bm{k}_1, -i\varepsilon_n - i\omega_\nu)
	\int d\xi_2
	\mathcal{G}(\bm{k}_2, i\varepsilon_n) ,\\
	I^{\rm b}
	=&
	\int d\xi_1\;
	\mathcal{G}(\bm{k}_1, i\varepsilon_n) \mathcal{G}(\bm{k}_1, -i\varepsilon_n) 
	\mathcal{G}(\bm{k}_1, - i\varepsilon_n - i\omega_\nu) 
	\int d\xi_2\;
	\mathcal{G}(\bm{k}_2, i\varepsilon_n) \mathcal{G}(\bm{k}_2, - i\varepsilon_n).
\end{align}
\subsubsection{Step3: Integral over wavevectors} 
\begin{figure}[tbp]
	\centering
	\includegraphics[width=0.8\textwidth]{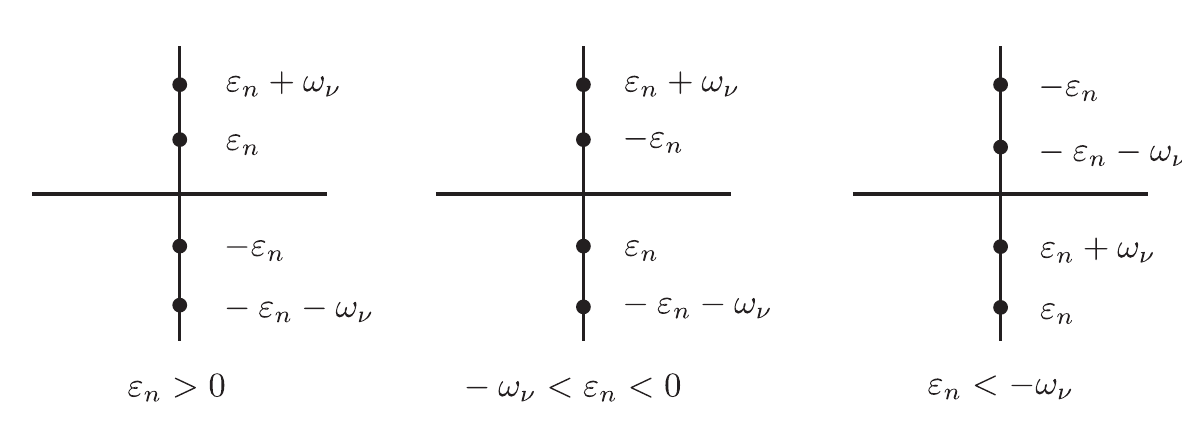}
	\caption{Location of poles for $\xi_1$ integral in $I^{\rm a}$.}
	\label{mt-sj-a-pole}
\end{figure}

\begin{figure}[tbp]
	\centering
	\includegraphics[width=0.8\textwidth]{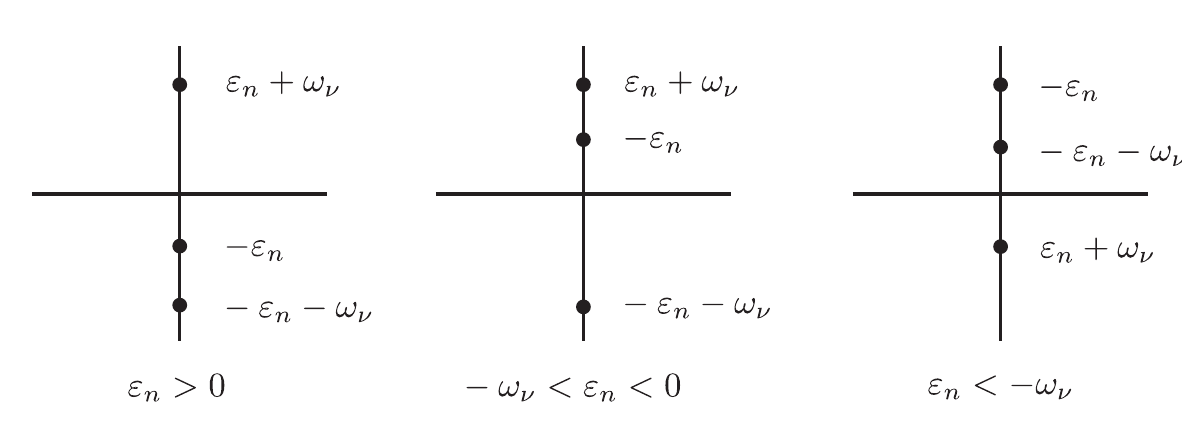}
	\caption{Location of poles for $\xi_1$ integral in $I^{\rm b}$.}
	\label{mt-sj-b-pole}
\end{figure}

The $\xi_1$ integrals for $I^{\rm a}$ and $I^{\rm b}$ are performed separately for the three cases of distribution of the poles shown in Fig.~\ref{mt-sj-a-pole} or Fig.~\ref{mt-sj-b-pole}. Performing integral with $\xi_1$ and $\xi_2$, we obtain 
\begin{align}
	&I^{\rm a} = -i\pi^2 \sign(\varepsilon_n)
	\frac{1}{\abs{\te_{n+\nu}} + \abs{\te_n}} \frac{1}{\abs{\te_{n+\nu}} \abs{\te_n}},\\
	&I^{\rm b} = i \pi^2 \sign(\varepsilon_n) \frac{1}{\abs{\te_{n+\nu}} + \abs{\te_n}} \frac{1}{\abs{\te_{n+\nu}}  \abs{\te_n}},
\end{align}
i.e. $I^{\rm a} + I^{\rm b} = 0$. We thus see that the MT terms with side jump process vanish in total.
\subsection{Maki-Thompson terms with skew scattering}
%
%
\subsubsection{\texorpdfstring{Step1: Feynman diagrams for Response function and Step2: Put $\Omega_k$ = 0 in all quantities}{Step1: Feynman diagrams for Response function and Step2: Put Ω\_k = 0 in all quantities}}
\begin{figure}[tbp]
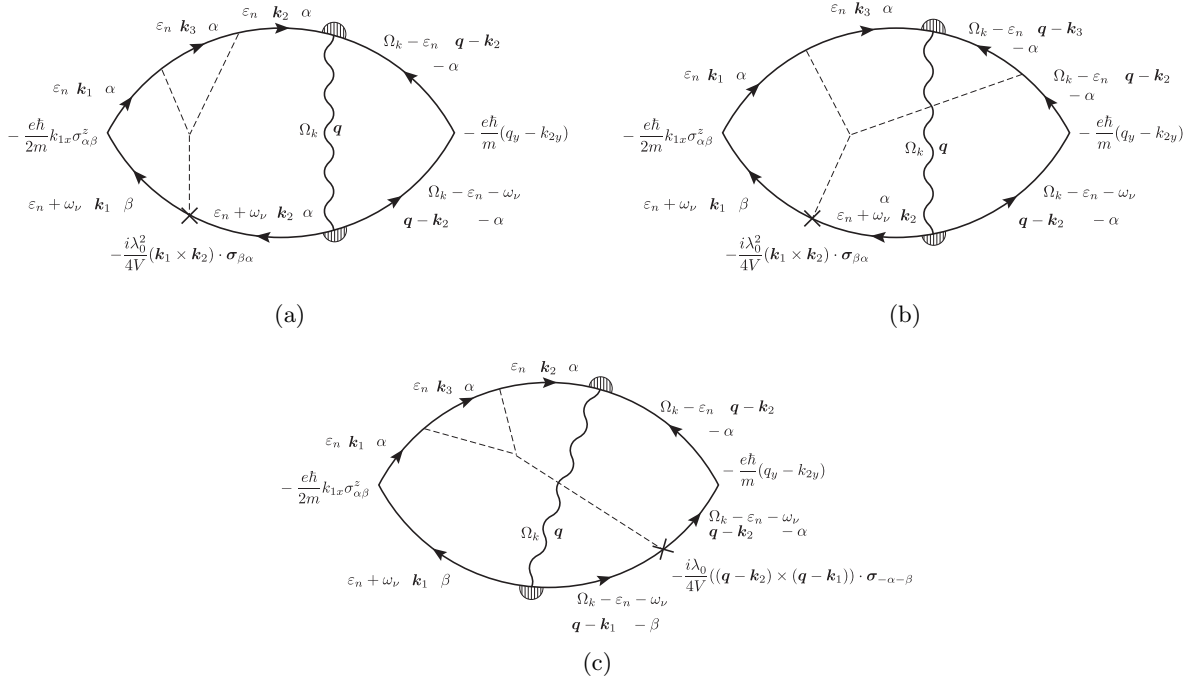

	\begin{minipage}[b]{0.45\linewidth}
		\centering
		\includegraphics[width=\textwidth]{mt_ss_a.pdf}
		\\(a)
	\end{minipage}
	\begin{minipage}[b]{0.45\linewidth}
		\centering
		\includegraphics[width=\textwidth]{mt_ss_b.pdf}
		\\(b)
	\end{minipage}\\
	\begin{minipage}[b]{0.5\linewidth}
		\centering
		\includegraphics[width=\textwidth]{mt_ss_c.pdf}
		\\(c)
	\end{minipage}
	\caption{Diagrams for the MT terms with skew scattering process.}
	\label{mt-ss-figs}
\end{figure}
Figure \ref{mt-ss-figs} shows the diagrams for the MT terms with the skew scattering process. The vertically or horizontally reflected diagrams contribute equally to the original diagram and thus, we focus on the diagrams shown in Fig.~\ref{mt-ss-figs} and multiply the resultant expressions by four.  

We see that (b) is different from (a) only in the arguments of Green functions. 
In the diagram (c), the vector product part $\bm{k}_2\times \bm{k}_1$ is minus that in (a). However, the summation over spin index recovers the same coefficient as (a), including the sign. Thus, we first perform the integral of the product of the Green function with $\xi_1$ and $\xi_2$ for (a), (b), and (c), and then sum them up. 

The resultant expression for $\Phi_{xy}^{\mathrm{MT-SS}}$ is given by 
\begin{align}
	\Phi_{xy}^{\mathrm{MT-SS}} 
	=&
	-\frac{2T^2}{V} \sum_{\Omega_k \varepsilon_n} \sum_{\bm{k}_1 \bm{k}_2 \bm{k}_3\bm{q}} \qty(-\frac{e\hbar}{2m}k_{1x}) \qty(-\frac{i\lambda_0^2}{4V}) (k_{1x}k_{2y} - k_{1y}k_{2x}) 
	\qty(-\frac{e\hbar}{m}) (q_y - k_{2y}) \frac{1}{V^2} n_{\mathrm{i}}v_0^3 V\\
	&\times 
	(\text{product of Green functions})\\
	&\times
	C(\bm{q}, \varepsilon_n, \Omega_k-\varepsilon_n) C(\bm{q}, \varepsilon_n + \omega_\nu, \Omega_k - \varepsilon_n-\omega_\nu) L(\bm{q}, i\Omega_k)\\
	\approx &
	-\underbrace{4}_{\mathclap{\text{reflected diagrams}}}\cdot \frac{ie^2\hbar^2 \lambda_0^2 n_{\mathrm{i}} v_0^3}{4m^2} \qty(\frac{k_{\mathrm{F}}^2}{D})^2 N(0)^3 \cdot
	 T \sum_{\Omega_k} \sum_{\bm{q}} L(\bm{q}, i\Omega_k) \Sigma^{\mathrm{MT-SS}} (\bm{q}, \Omega_k, \omega_\nu) \\
	\approx &
	-\frac{ie^2\hbar^2 \lambda_0^2 n_{\mathrm{i}} v_0^3 k_{\mathrm{F}}^4 N(0)^3 T}{D^2 m^2} \cdot
	  \sum_{\bm{q}} L(\bm{q}, 0) \Sigma^{\mathrm{MT-SS}} (\bm{q}, 0, \omega_\nu), 
\end{align}
where we introduce the notations
\begin{align}
	 \Sigma^{\mathrm{MT-SS}} (\bm{q}, \Omega_k, \omega_\nu)
	=&
	T\sum_{\varepsilon_n} C(\bm{q}, \varepsilon_n, \Omega_k-\varepsilon_n) C(\bm{q}, \varepsilon_n + \omega_\nu, \Omega_k - \varepsilon_n - \omega_\nu) (I^{\rm a} + I^{\rm b} + I^{\rm c}),\\
	I^{\rm a}
	=&
	\int d\xi_1\; \mathcal{G}(\bm{k}_1, i\varepsilon_n) \mathcal{G}(\bm{k}_1, i\varepsilon_n + i\omega_\nu) \\
	&\times
	\int d\xi_2\; \mathcal{G}(\bm{k}_2, i\varepsilon_n) \mathcal{G}(\bm{k}_2, i\varepsilon_n + i\omega_\nu) \mathcal{G}(\bm{k}_2, -i\varepsilon_n) \mathcal{G}(\bm{k}_2, -i\varepsilon_n - i\omega_\nu) \\
	&\times
	\int d\xi_3 \; \mathcal{G}(\bm{k}, i\varepsilon_n),\\
	I^{\rm b}
	=&
	\int d\xi_1\; \mathcal{G}(\bm{k}_1, i\varepsilon_n) \mathcal{G}(\bm{k}_1, i\varepsilon_n + i\omega_\nu) \\
	&\times
	\int d\xi_2\; \mathcal{G}(\bm{k}_2, i\varepsilon_n + i\omega_\nu) \mathcal{G}(\bm{k}_2, -i\varepsilon_n) \mathcal{G}(\bm{k}_2, -i\varepsilon_n - i\omega_\nu)\\
	&\times
	\int d\xi_3\; \mathcal{G}(\bm{k}_3, i\varepsilon_n) \mathcal{G}(\bm{k}_3, -i\varepsilon_n),\\
	I^{\rm c}
	=&
	\int d\xi_1\; \mathcal{G}(\bm{k}_1, i\varepsilon_n) \mathcal{G}(\bm{k}_1, i\varepsilon_n + i\omega_\nu) \mathcal{G}(\bm{k}_1, -i\varepsilon_n - i\omega_\nu)\\
	&\times
	\int d\xi_2\; \mathcal{G}(\bm{k}_2, i\varepsilon_n) \mathcal{G}(\bm{k}_2, -i\varepsilon_n) \mathcal{G}(\bm{k}_2, -i\varepsilon_n - i\omega_\nu)\\
	&\times
	\int d\xi_3\; \mathcal{G}(\bm{k}_3, i\varepsilon_n) .
\end{align}
\subsubsection{Step2: Integral over wavevectors} 
The $\xi_2$, $\xi_3$ integrals of $I^{\rm a} + I^{\rm b}$ in the present case are the same as the $\xi_1$, $\xi_2$ integrals of $I^{\rm a} + I^{\rm b}$ in the MT terms with side jump process and thus it vanishes. 

\begin{figure}[tbp]
	\centering
	\includegraphics[width=0.8\textwidth]{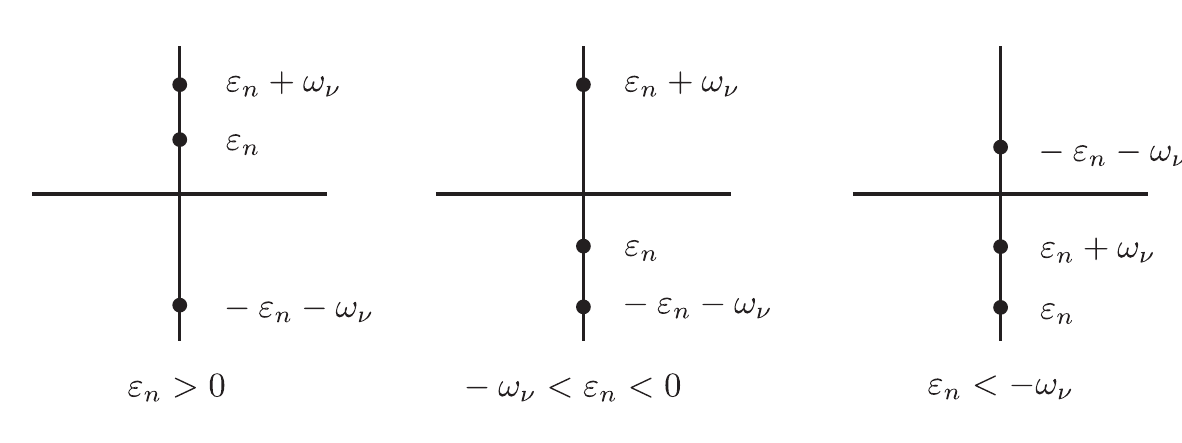}
	\caption{Location of poles for the $\xi_1$ integral in $I^{\rm c}$.}
	\label{mt-ss-c-1-pole}
\end{figure}

\begin{figure}[tbp]
	\centering
	\includegraphics[width=0.8\textwidth]{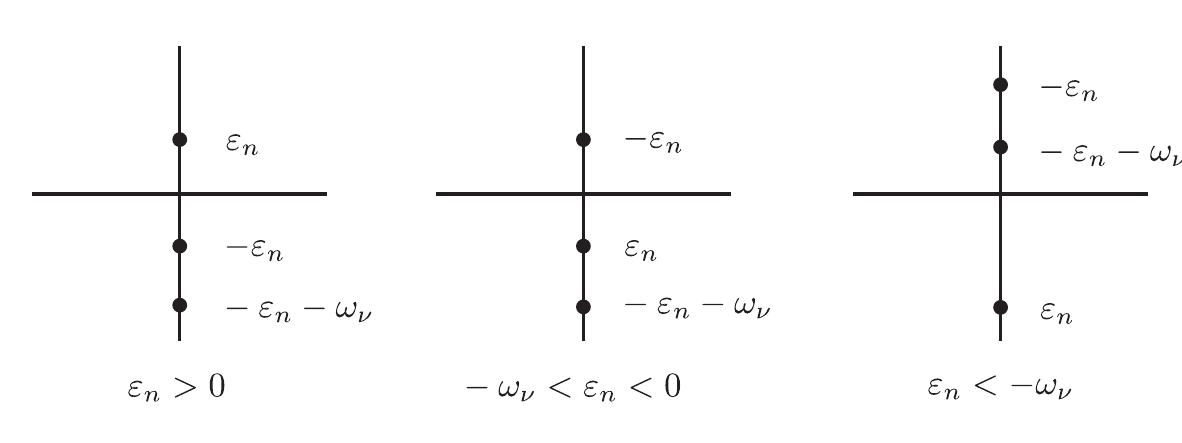}
	\caption{Location of poles for the $\xi_2$ integral in $I^{\rm c}$.}
	\label{mt-ss-c-2-pole}
\end{figure}

The $\xi_1$ integral and the $\xi_2$ integral in $I^{\rm c}$ are performed, considering the distribution of the poles, which are shown in Fig.~\ref{mt-ss-c-1-pole} and Fig.~\ref{mt-ss-c-2-pole}. 
Performing the integral with $\xi_1,\xi_2,\xi_3$, we obtain
\begin{align}
	I^{\rm c} = -4i \pi^3 \frac{\sign(\varepsilon_{n+\nu})}{\abs{2\te_n} \abs{2\te_{n+\nu}}(\abs{\te_{n+\nu}} + \abs{\te_n})^2}.
\end{align}
With use of it, we calculate $\Sigma^{\mathrm{MT-SS}}$
\begin{align}
	\Sigma^{\mathrm{MT-SS}}(\bm{q}, 0, \omega_\nu) 
	=&
	T\sum_{\varepsilon_n} C(\bm{q}, \varepsilon_n, -\varepsilon_n) C(\bm{q}, \varepsilon_n + \omega_\nu, -\varepsilon_n - \omega_\nu) I^{\rm c}\\
	=&
	-4i\pi^3 T \sum_{\varepsilon_n} \frac{\abs{2\te_{n+\nu}}}{\abs{2\varepsilon_{n+\nu}} + \Dq} \frac{\abs{2\te_n}}{\abs{2\varepsilon_n} + \Dq} 
	\frac{\sign(\varepsilon_{n+\nu})}{\abs{2\te_n} \abs{2\te_{n+\nu}}(\abs{\te_{n+\nu}} + \abs{\te_n})^2}\\
	=&
	-4i\pi^3 T \sum_{\varepsilon_n} \frac{1}{\abs{2\varepsilon_{n+\nu}} + \Dq} \frac{1}{\abs{2\varepsilon_n} + \Dq} 
	\frac{\sign(\varepsilon_{n+\nu})}{ (\abs{\te_{n+\nu}} + \abs{\te_n})^2}.
\end{align}
\subsubsection{\texorpdfstring{Step4: Sum over $\varepsilon_n$ }{Step4: Sum over ε\_n }}
We note that the sum over $n<-\nu$ and that over $n\geq 0$ cancel with each other; the sum over $n<-\nu$ becomes
\begin{align}
	&\sum_{n=-\infty}^{-\nu-1} 
	\frac{1}{\abs{2\varepsilon_{n+\nu}} + \Dq} \frac{1}{\abs{2\varepsilon_n} + \Dq} 
	\frac{\sign(\varepsilon_{n+\nu})}{ (\abs{\te_{n+\nu}} + \abs{\te_n})^2}\\
	=&
	-\sum_{n=-\infty}^{-\nu-1} 
	\frac{1}{-2\varepsilon_n -2\omega_\nu + \Dq} \frac{1}{-2\varepsilon_n + \Dq} \frac{1}{(-2\varepsilon_n - \omega_\nu + \hbar / \tau)^2}\\
	=&
	-\sum_{n=-\infty}^{-1} 
	\frac{1}{-2\varepsilon_n + \Dq} \frac{1}{-2\varepsilon_n + 2\omega_\nu + \Dq} \frac{1}{(-2\varepsilon_n + \omega_\nu + \hbar / \tau)^2} \\
	=&
	-\sum_{n=0}^{\infty} \frac{1}{2\varepsilon_n + \Dq} \frac{1}{2\varepsilon_n + 2\omega_\nu + \Dq} \frac{1}{(2\varepsilon_n + \omega_\nu + \hbar / \tau)^2} \\
	=&
	-\sum_{n=0}^{\infty} \frac{1}{\abs{2\varepsilon_n} + \Dq} \frac{1}{\abs{2\varepsilon_{n+\nu}} + \Dq} \frac{1}{(\abs{\te_n} + \abs{\te_{n+\nu}})^2},
\end{align}
which is minus the contribution from $n\geq 0$. 

The sum over $-\nu\leq n<0$ becomes
\begin{align}
	&\pi T \sum_{n=-\nu}^{-1} 
	\frac{1}{\abs{2\varepsilon_{n+\nu}} + \Dq} \frac{1}{\abs{2\varepsilon_n} + \Dq} 
	\frac{\sign(\varepsilon_{n+\nu})}{ (\abs{\te_{n+\nu}} + \abs{\te_n})^2}\\
	=&
	\pi T \sum_{n=-\nu}^{-1} \frac{1}{2\varepsilon_n + 2\omega_\nu + \Dq} \frac{1}{-2\varepsilon_n + \Dq} \frac{1}{(\varepsilon_n + \omega_\nu + \hbar / 2\tau - (\varepsilon_n- \hbar / 2\tau))^2}\\
	=&
	\frac{\pi T}{(\omega_\nu + \hbar / \tau)^2} \sum_{n=-\nu}^{-1} \frac{1}{2\varepsilon_n + 2\omega_\nu + \Dq} \frac{1}{-2\varepsilon_n + \Dq}\\
	=&
	\frac{1}{4(\omega_\nu + \hbar / \tau)^2} \frac{1}{\hbar \mathcal{D} \bm{q}^2 + \omega_\nu} \qty[\psi\qty(\frac{1}{2} + \frac{2\omega_\nu}{4\pi T} + \frac{\hbar \mathcal{D} \bm{q}^2}{4\pi T}) - \psi\qty(\frac{1}{2} + \frac{\hbar \mathcal{D} \bm{q}^2}{4\pi T})].
\end{align}

\subsubsection{\texorpdfstring{Step5: Analytic continuation and expansion with $\omega$}{Step5: Analytic continuation and expansion with ω}}
Analytic continuation $i\omega_\nu\rightarrow \omega$, the resultant expression is written in terms of polyGamma function as 
\begin{align}
	\Sigma^{\mathrm{MT-SS}} (\bm{q}, 0, \omega_\nu) 
	\xrightarrow{\omega_\nu\to -i\hbar \omega \approx 0} -4i\pi^2\cdot \frac{\tau}{\hbar}
	\qty[-\frac{i\pi \hbar \omega\tau}{16T\hbar} \frac{1}{-i\hbar \omega + \Dq} - \frac{i\hbar \omega\tau}{32 \pi^2 T^2\hbar} \psi^{(2)}\qty(\frac{1}{2})].
\end{align}
\subsubsection{Step6: Separation into regular and anomalous parts}

We introduce the notations
\begin{align}
	\Sigma^{\mathrm{MT-SS(reg)}} =& -\hbar \omega \frac{\tau^2}{8T^2 \hbar^2} \psi^{(2)} \qty(\frac{1}{2})\\
	\Sigma^{\mathrm{MT-SS(an)}} =& -\hbar \omega \frac{\pi^3\tau^2}{4T \hbar^2} \frac{1}{-i\hbar \omega + \Dq}
\end{align}
for the regular part and the anomalous part. 

In the response function 
\begin{align}
	\Phi_{xy}^{\mathrm{MT-SS}}
	\approx&
	-\frac{ie^2\hbar^2 \lambda_0^2 n_{\mathrm{i}} v_0^3}{m^2} \qty(\frac{k_{\mathrm{F}}^2}{D})^2 N(0)^3 \cdot
	T \sum_{\bm{q}} L(\bm{q}, 0)\qty(\Sigma^{\mathrm{MT-SS(reg)}} + \Sigma^{\mathrm{MT-SS(an)}}), 
\end{align}
we consider $\sum_{\bm{q}} L(\bm{q}, 0) \Sigma^{\mathrm{MT-SS(reg)}}$ and $\sum_{\bm{q}} L(\bm{q}, 0) \Sigma^{\mathrm{MT-SS(an)}}$, separately. 
\subsubsection{Step7: Regular part in MT term of  spin Hall conductivity}

In the regular part, $\Sigma^{\mathrm{MT-SS(reg)}}$ has no $q$-dependence and thus 
\begin{align}
	\sum_{\bm{q}} L(\bm{q}, 0) \Sigma^{\mathrm{MT-SS(reg)}} 
	=&
	\Sigma^{\mathrm{MT-SS(reg)}} \sum_{\bm{q}} L(\bm{q}, 0)\\
	=&
	\hbar \omega \frac{\tau^2}{8T^2 \hbar^2} \psi^{(2)} \qty(\frac{1}{2}) \cdot \frac{1}{N(0) \clength^D}
	\begin{dcases}
		\frac{1}{4\pi} \ln \frac{1}{\epsilon} & (D=2)\\
		\frac{1}{2\pi^2} & (D=3)
	\end{dcases}.
\end{align}
We then obtain
\begin{align}
	\sigma_{xy}^{\mathrm{MT-SS(reg)}}
=&
	-\frac{e^2 \hbar^3 \lambda_0^2 n_{\mathrm{i}} v_0^3 k_{\mathrm{F}}^4 N(0)^2 \tau^2}{8D^2 m^2 T \hbar^2 \clength^D} \psi^{(2)}\qty(\frac{1}{2})
	\begin{dcases}
		\frac{1}{4\pi} \ln \frac{1}{\epsilon} & (D=2)\\
		\frac{1}{2\pi^2} & (D=3)
	\end{dcases}.\\
\end{align}
\subsubsection{Step8: Anomalous part in MT term of  spin Hall conductivity}
In the anomalous part, we introduce the phase-breaking time $\tau_{\varphi}$ as a cutoff in $\Sigma^{\mathrm{MT-SS(an)}}$ and obtain 
\begin{align}
	&
	\sum_{\bm{q}} L(\bm{q}, 0) \Sigma^{\mathrm{MT-SS(an)}} \\
	=&
	\hbar \omega \frac{\pi^3\tau^2}{4T \hbar^2} \frac{1}{N(0)} \int_{\norm{\bm{q}}< \clength^{-1}} \frac{d^Dq}{(2\pi)^D} \frac{1}{\epsilon + \frac{\pi}{8T} \Dq} \frac{1}{\hbar / \tau_\varphi + \Dq} \\
	=&
	\hbar \omega \frac{1}{N(0)} \frac{\pi^4 \tau^2}{32 T^2 \hbar^2} \frac{1}{\clength^D}
	\frac{1}{(2\pi)^D} \int_0^1 dx\; x^{D-1} \frac{1}{\epsilon+ x^2} \frac{1}{\gamma_\varphi + x^2} 
	\begin{dcases}
		2\pi & (D=2)\\
		4\pi & (D=3)
	\end{dcases},
\end{align}
with $\gamma_\varphi \coloneqq \frac{\pi \hbar}{8T\tau_\varphi}$.

The integral becomes 
\begin{align}
	\int_0^1 dx\; x\frac{1}{\epsilon +x^2} \frac{1}{ \gamma_\varphi + x^2}
\approx&
	 \frac{1}{2 (\epsilon - \gamma_\varphi)} \ln \frac{\epsilon}{\gamma_\varphi}
\end{align}
for $D=2$ and 
\begin{align}
	\int_0^1 dx\; x^2 \frac{1}{\epsilon+ x^2} \frac{1}{\gamma_\varphi + x^2}
\approx&
\frac{\pi}{2\pi (\sqrt{\epsilon}+ \sqrt{\gamma_\varphi})} 
\end{align}
for $D=3$. 
With use of it, we arrive at
\begin{align}
	\sigma_{xy}^{\mathrm{MT-SS(an)}}
	=&
	-\frac{\pi^4 e^2 \hbar^3 \lambda_0^2 n_{\mathrm{i}} v_0^3 k_{\mathrm{F}}^4 N(0)^2 \tau^2}{32 D^2 m^2 T \hbar^2 \clength^D} 
	\begin{dcases}
		 \frac{1}{4\pi (\epsilon - \gamma_\varphi)} \ln \frac{\epsilon}{\gamma_\varphi} & (D=2)\\
		\frac{1}{4\pi (\sqrt{\epsilon}+ \sqrt{\gamma_\varphi})} & (D=3)
	\end{dcases}.
\end{align}
\section{Intrinsic spin Hall effect}
\subsection{Aslamazov-Larkin terms}
\subsubsection{Step 1: Feynman diagrams for Response function}
\begin{figure}[tbp]
	\centering
	\includegraphics[width=0.7\textwidth, page=3]{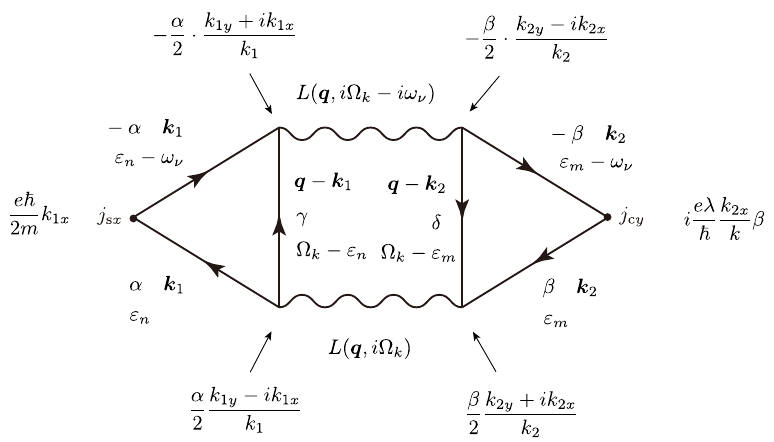}
	\caption{Diagrams for the AL terms for the intrinsic spin Hall effect.}
	\label{intrinsic-fig5}
\end{figure}

Figure \ref{intrinsic-fig5} shows the diagrams for the AL terms.
The response function $\Phi_{xy}^{\mathrm{AL}}(0, i\omega_\nu)$ is written as
\begin{align}
	&
	\Phi_{xy}^{\mathrm{AL}}(0, i\omega_\nu)\\
	=&
	\frac{1}{V} T\sum_{\bm{q}, \Omega_k} L(\bm{q}, i\Omega_k) L(\bm{q}, i\Omega_k - i\omega_\nu)\\
	&\times
	\sum_{\bm{k}_1} \qty(-\frac{1}{4}) \sum_{\alpha \gamma} T \sum_{\varepsilon_n} \frac{e\hbar}{2m} k_{1x} \mathcal{G}	_{\alpha} (\bm{k}_1, i\varepsilon_n) \mathcal{G}_{-\alpha} (\bm{k}_1, i\varepsilon_n - i\omega_\nu) \mathcal{G}_{\gamma} (\bm{q} - \bm{k}_1, i\Omega_k - i\varepsilon_n)\\
	&\times
	\sum_{\bm{k}_2} \qty(-\frac{1}{4}) \sum_{\beta \delta} T \sum_{\varepsilon_m} \frac{e\lambda}{\hbar} i \frac{k_{2x}}{k_2} \beta \mathcal{G}_{\beta} (\bm{k}_2, i\varepsilon_m) \mathcal{G}_{-\beta} (\bm{k}_2, i\varepsilon_m - i\omega_\nu) \mathcal{G}_{\delta}(\bm{q}-\bm{k}_2, i\Omega_k - i\varepsilon_m).
\end{align}
\subsubsection{\texorpdfstring{Step2: Expansion of triangular parts with $\bm{q}$}{Step2: Expansion of triangular parts with q}}
With use of $\norm{\bm{q}-\bm{k}} = \sqrt{\bm{k}^2 - 2\bm{k}\vdot \bm{q} + \bm{q}^2} \approx k\sqrt{1 - 2\bm{k}\vdot \bm{q} / k^2} \approx k(1 - \bm{k}\vdot \bm{q} / k^2) = k - \bm{k}\vdot \bm{q} / k$, we expand the Green function with $\bm{q}$, 
\begin{align}
	\mathcal{G}_{\alpha} (\bm{q}-\bm{k}, i\varepsilon_n) 
	=&
	\frac{1}{i\varepsilon_n - \qty[\frac{\hbar^2}{2m} (\bm{q}-\bm{k})^2 + \alpha \lambda \norm{\bm{q}-\bm{k}}] + \mu}\\
	\approx&
	\mathcal{G}_{\alpha}(\bm{k}, i\varepsilon_n) -\mathcal{G}_{\alpha}(\bm{k}, i\varepsilon_n)^2 \qty(\frac{\hbar^2}{m} + \frac{\alpha \lambda}{k}) \bm{k}\vdot \bm{q}.
\end{align}
We then reduce the expression for $\Phi_{xy}^{\mathrm{AL}}(0,i\omega_\nu)$ as
\begin{align}
	\Phi_{xy}^{\mathrm{AL}}(0,i\omega_\nu)
	=&
	V^2 \frac{e\hbar}{2m} \frac{e\lambda}{\hbar} i \frac{1}{V} T\sum_{\bm{q} \Omega_k} L(\bm{q}, i\Omega_k) L(\bm{q}, i\Omega_k-i\omega_\nu)\\
	&\times
	\frac{1}{V}\sum_{\bm{k}_1} k_{1x} \bm{k}_1\vdot \bm{q} \frac{1}{4} \sum_{\alpha \gamma} \qty(\frac{\hbar^2}{m} + \frac{\gamma \lambda}{k_1}) T\sum_{\varepsilon_n} \mathcal{G}_{\alpha} (\bm{k}_1, i\varepsilon_n) \mathcal{G}_{-\alpha} (\bm{k}_1, i\varepsilon_n - i\omega_\nu) \mathcal{G}_{\gamma}^2(\bm{k}_1, i\Omega_k - i\varepsilon_n)\\
	&\times
	\frac{1}{V} \sum_{\bm{k}_2} \frac{k_{2x}}{k_2} \bm{k}_2\vdot \bm{q}	\frac{1}{4} \sum_{\beta \delta} \qty(\frac{\hbar^2}{m} + \frac{\delta \lambda}{k_2}) \beta T\sum_{\varepsilon_m} \mathcal{G}_\beta(\bm{k}_2, i\varepsilon_m) \mathcal{G}_{-\beta}(\bm{k}_2, i\varepsilon_m - i\omega_\nu) \mathcal{G}_\delta^2 (\bm{k}_2, i\Omega_k - i\varepsilon_m)\\
	\approx&
	\frac{ie^2 \lambda}{16m} N(0)^2 k_{\mathrm{F}}^3 \cdot V \sum_{\bm{q}} \bm{q}^2	T \sum_{\Omega_k} L(\bm{q}, i\Omega_k) L(\bm{q}, i\Omega_k - i\omega_\nu)
	B_{\mathrm{c}}(i\Omega_k, i\omega_\nu) B_{\mathrm{s}}(i\Omega_k, i\omega_\nu)\label{eq: Phi-AL-intrinsic}
\end{align}
in terms of 
\begin{align}
	B_{\mathrm{c}}(i\Omega_k, i\omega_\nu)
	=&
	\frac{1}{4} \sum_{\alpha \beta} \qty(\frac{\hbar^2}{m} + \frac{\beta \lambda}{k_{\mathrm{F}}}) T \sum_{\varepsilon_n} I_{\alpha \beta}(i\varepsilon_n, i\Omega_k, i\omega_\nu), \label{intrinsic-eq13}\\
	B_{\mathrm{s}}(i\Omega_k, i\omega_\nu)
	=&
	\frac{1}{4} \sum_{\alpha \beta} \qty(\frac{\hbar^2}{m} + \frac{\beta \lambda}{k_{\mathrm{F}}}) \alpha  T \sum_{\varepsilon_n} I_{\alpha \beta}(i\varepsilon_n, i\Omega_k, i\omega_\nu) \label{intrinsic-eq14},\\
	I_{\alpha \beta}(i\varepsilon_n, i\Omega_k, i\omega_\nu)
	=&
	\int d\xi\; \mathcal{G}_{\alpha}(\bm{k}, i\varepsilon_n) \mathcal{G}_{-\alpha}(\bm{k}, i\varepsilon_n-i\omega_\nu) \mathcal{G}_{\beta}^2(\bm{k}, i\Omega_k - i\varepsilon_n).
\end{align}
\subsubsection{\texorpdfstring{Step3: Integral in $B_{\rm s}(i\Omega_k, i\omega_\nu), B_{\rm c}(i\Omega_k, i\omega_\nu)$ with wavevectors}{Step3: Integral in B\_s(iΩ\_k, iω\_ν), B\_c(iΩ\_k, iω\_ν) with wavevectors}}

\begin{figure}[tbp]
	\centering
	\includegraphics[width=\textwidth, page=6]{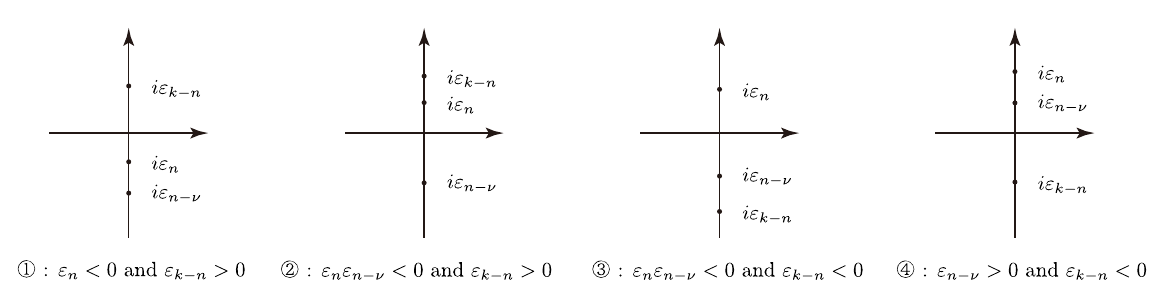}
	\caption{Locations of the poles for $\xi$ integral in $I_{\alpha\beta}$.}
	\label{intrinsic-fig10}
\end{figure}

We consider the $\xi$-integral in $I_{\alpha\beta}$
\begin{align}
	I_{\alpha \beta}(i\varepsilon_n, i\Omega_k, i\omega_\nu)
	\approx&
	\int d\xi \frac{1}{i\varepsilon_n - \xi - \alpha \lambda k_{\mathrm{F}}} \frac{1}{i\varepsilon_{n-\nu} - \xi + \alpha \lambda k_{\mathrm{F}}} \frac{1}{(i\varepsilon_{k-n} - \xi - \beta \lambda k_{\mathrm{F}})^2}\\
	=&
	\int d\xi \frac{1}{\xi- (i\varepsilon_n - \alpha \lambda k_{\mathrm{F}})} \frac{1}{\xi - (i\varepsilon_{n-\nu} + \alpha \lambda k_{\mathrm{F}})} \frac{1}{[\xi - (i\varepsilon_{k-n} - \beta \lambda k_{\mathrm{F}})]^2}.
\end{align}
The $\xi$-integral in $I_{\alpha\beta}$ is considered separately for the four cases of the distribution of the poles (see Fig.~\ref{intrinsic-fig10}). The ranges of $\varepsilon_n, \Omega_k$ are shown in Table~\ref{table2}.
\begin{table}[tbp]
	\centering
	\caption{The ranges of $\varepsilon_n, \Omega_k$.}
	\label{table2}
	\scalebox{0.9}
	{
	\begin{tabular}{l|cccc}
		&\ctext{1}\begin{tabular}{l} $\varepsilon_n < 0$ and\\$ \varepsilon_{k-n}>0$\end{tabular} 
		&\ctext{2}\begin{tabular}{l} $\varepsilon_n \varepsilon_{n-\nu} < 0$ and \\$\varepsilon_{k-n} > 0$  \end{tabular}
		&\ctext{3}\begin{tabular}{l} $\varepsilon_n \varepsilon_{n-\nu} < 0$ and \\$\varepsilon_{k - n} < 0$  \end{tabular}
		&\ctext{4}\begin{tabular}{l} $\varepsilon_{n-\nu} > 0$ and \\ $\varepsilon_{k-n} < 0$ \end{tabular}\\ \hline
		\fbox{1} : $\Omega_k \leq 0$ & $\varepsilon_n < \Omega_k$ & NA & $0 < \varepsilon_n < \omega_\nu$ & $\omega_\nu < \varepsilon_n$\\
		\fbox{2} : $0\leq \Omega_k \leq \omega_\nu$ & $\varepsilon_n < 0$ & $0 < \varepsilon_n < \Omega_k$ & $\Omega_k < \varepsilon_n < \omega_\nu$ & $\omega_\nu < \varepsilon_n$\\
		\fbox{3} : $\omega_\nu \leq \Omega_k$ & $\varepsilon_n < 0$ & $0 < \varepsilon_n < \omega_\nu$ & NA & $\Omega_k < \varepsilon_n$
	\end{tabular}
	}
\end{table}
Taking account of Fig.~\ref{intrinsic-fig10} and Table~\ref{table2}, we perform the $\xi$ integral as
\begin{align}
	&
	I_{\alpha \beta}(i\varepsilon_n, i\Omega_k, i\omega_\nu)\\
	=&
	\frac{2\pi i}{i\omega_\nu - 2\alpha \lambda k_{\mathrm{F}}} \qty[\frac{-\ctext{1}+\ctext{3}+\ctext{4}}{(2i\varepsilon_n - i\Omega_k + (-\alpha + \beta)\lambda k_{\mathrm{F}})^2} - \frac{-\ctext{1}-\ctext{2}+\ctext{4}}{(2i\varepsilon_n - i\Omega_k - i\omega_\nu + (\alpha + \beta)\lambda k_{\mathrm{F}})^2}].\label{eq: M4-6}
\end{align}
\subsubsection{\texorpdfstring{Step4: Sum over $\varepsilon_n$}{Step4: Sum over ε\_n}}
Further we sum over $\varepsilon_n$, which yields
the expression \begin{align}
	&
	T \sum_{\varepsilon_n} I_{\alpha \beta} (i\varepsilon_n, i\Omega_k, i\omega_\nu)\\
=&
	\frac{1}{(8i\pi T) (i\omega_\nu - 2\alpha \lambda k_{\mathrm{F}})}
	\begin{aligned}[t]
		\Biggl[&
			-
			\psi^{(1)}\qty(\frac{1}{2} + \frac{\abs{\Omega_k}}{4\pi T} + \frac{i(-\alpha + \beta)\lambda k_{\mathrm{F}}}{4\pi T})
			+
			\psi^{(1)}\qty(\frac{1}{2} + \frac{\abs{\Omega_k}}{4\pi T} - \frac{i(-\alpha + \beta)\lambda k_{\mathrm{F}}}{4\pi T})\\
		    &+
			\psi^{(1)}\qty(\frac{1}{2} + \frac{\abs{\Omega_k - \omega_\nu}}{4\pi T} + \frac{i(\alpha + \beta)\lambda k_{\mathrm{F}}}{4\pi T} )
			-
			\psi^{(1)}\qty(\frac{1}{2} + \frac{\abs{\Omega_k - \omega_\nu}}{4\pi T} - \frac{i(\alpha + \beta)\lambda k_{\mathrm{F}}}{4\pi T} )
		\Biggr] \label{intrinsic-eq7}
	\end{aligned}
\end{align}
in terms of the polygamma function. 
\subsubsection{\texorpdfstring{Step5: Sum over $\Omega_k$}{Step5: Sum over Ω\_k}}
\begin{figure}[tbp]
	\centering
	\includegraphics[page=7, scale=0.9]{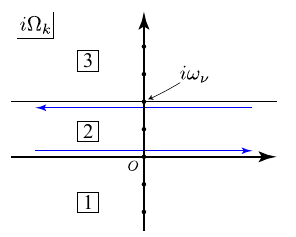}
	\caption{Path for $i\Omega_k$.}
	\label{intrinsic-fig11}
\end{figure}

Next we sum over $\Omega_k$ , separating the sum into three regions shown  in Fig.~\ref{intrinsic-fig11}. The function $\psi^{(1)}(z)$ is analytic in the region $\Re(z) > 0$ and hence Eq.~\eqref{intrinsic-eq7} is analytic  function of $i\Omega_k$ in each region.
Let $B_{\mathrm{c}i}, B_{\mathrm{s}i} \; (i = 1, 2,3)$ be analytic continuation of $B_{\mathrm{s}}$ in each region \fbox{1}, \fbox{2}, and \fbox{3}. 
We note that $B_{\mathrm{s}1}(i\Omega_k, 0)=B_{\mathrm{s}3}(i\Omega_k, 0)=0$ through explicit calculation
\begin{align}
	&
	B_{\mathrm{s}i}(i\Omega_k, 0)\\
	=&
	\frac{1}{4} \sum_{\alpha \beta} \qty(\frac{\hbar^2}{m} + \frac{\beta \lambda}{k_{\mathrm{F}}}) \alpha \cdot
	\frac{1}{(8i\pi T) (- 2\alpha \lambda k_{\mathrm{F}})}\\
	 \times&
	\begin{aligned}[t]
		\Biggl[&
			-
			\psi^{(1)}\qty(\frac{1}{2} \pm \frac{\Omega_k}{4\pi T} + \frac{i(-\alpha + \beta)\lambda k_{\mathrm{F}}}{4\pi T})
			+
			\psi^{(1)}\qty(\frac{1}{2} \pm \frac{\Omega_k}{4\pi T} - \frac{i(-\alpha + \beta)\lambda k_{\mathrm{F}}}{4\pi T})\\
		    &+
			\psi^{(1)}\qty(\frac{1}{2} \pm \frac{\Omega_k}{4\pi T} + \frac{i(\alpha + \beta)\lambda k_{\mathrm{F}}}{4\pi T} )
			-
			\psi^{(1)}\qty(\frac{1}{2} \pm \frac{\Omega_k}{4\pi T} - \frac{i(\alpha + \beta)\lambda k_{\mathrm{F}}}{4\pi T} )
		\Biggr]
	\end{aligned}\\
	=&
	\frac{1}{4} \sum_{\alpha \beta} \qty(\frac{\hbar^2}{m} + \frac{\beta \lambda}{k_{\mathrm{F}}}) \cdot
	\frac{1}{(8i\pi T) (- 2 \lambda k_{\mathrm{F}})}\\
	 \times&
	\begin{aligned}[t]
		\Biggl[&
			-
			\psi^{(1)}\qty(\frac{1}{2} \pm \frac{\Omega_k}{4\pi T} + \frac{i(\alpha + \beta)\lambda k_{\mathrm{F}}}{4\pi T})
			+
			\psi^{(1)}\qty(\frac{1}{2} \pm \frac{\Omega_k}{4\pi T} - \frac{i(\alpha + \beta)\lambda k_{\mathrm{F}}}{4\pi T})\\
		    &+
			\psi^{(1)}\qty(\frac{1}{2} \pm \frac{\Omega_k}{4\pi T} + \frac{i(\alpha + \beta)\lambda k_{\mathrm{F}}}{4\pi T} )
			-
			\psi^{(1)}\qty(\frac{1}{2} \pm \frac{\Omega_k}{4\pi T} - \frac{i(\alpha + \beta)\lambda k_{\mathrm{F}}}{4\pi T} )
		\Biggr] 
	\end{aligned}\\
	=&
	0,\label{intrinsic-eq12}
\end{align}
where the upper (lower) sign in the symbol $\pm$ is for $i=1$ ($i=3$). 
\subsubsection{\texorpdfstring{Step6: Analytic continuation and expansion with $\omega$}{Step6: Analytic continuation and expansion with ω}}
With this result and the fact that Eq.~\eqref{eq: M4-6} vanishes when $\Omega_k=\omega_\nu=0$, $B_{\mathrm{s}2}(0,0)=0$. 
Thus, $\order{\omega}$ terms in Eq.~\eqref{eq: Phi-AL-intrinsic} originate only from the terms containing the derivative of $B_{\mathrm{s}1,2,3}$, i.e.  
\begin{align}
	&
	T \sum_{\Omega_k} L(\bm{q}, i\Omega_k) L(\bm{q}, i\Omega_k-i\omega_\nu) B_{\mathrm{c}}(i\Omega_k, i\omega_\nu) B_{\mathrm{s}}(i\Omega_k, i\omega_\nu)\\
=&
	\frac{T}{2} L^R(\bm{q}, i\omega_\nu) L(\bm{q}, 0) B_{\mathrm{c}2}(i\omega_\nu, i\omega_\nu) B_{\mathrm{s}2}(i\omega_\nu, i\omega_\nu)\\
	&+
	\frac{T}{2} L(\bm{q}, 0) L^A(\bm{q}, -i\omega_\nu) B_{\mathrm{c}2}(0, i\omega_\nu) B_{\mathrm{s}2}(0, i\omega_\nu)\\
	&+
	\pv \int_{-\infty}^{\infty} \frac{d\Omega}{4\pi i} \coth \frac{\Omega}{2T} 
	\begin{aligned}[t]
		\Bigl[&
			L^R(\bm{q}, \Omega) L^A(\bm{q}, \Omega - i\omega_\nu) B_{\mathrm{c}2}(\Omega, i\omega_\nu) B_{\mathrm{s}2}(\Omega, i\omega_\nu)\\
			-&
			L^R(\bm{q}, \Omega + i\omega_\nu) L^A(\bm{q}, \Omega) B_{\mathrm{c}2}(\Omega+i\omega_\nu, i\omega_\nu) B_{\mathrm{s}2}(\Omega + i\omega_\nu, i\omega_\nu)
		\Bigr]
	\end{aligned}\\
	&+
	T \sum_{\Omega_k < 0} L^A(\bm{q}, i\Omega_k) L^A(\bm{q}, i\Omega_k - i\omega_\nu) B_{\mathrm{c}1}(i\Omega_k, i\omega_\nu) B_{\mathrm{s}1}(i\Omega_k, i\omega_\nu)\\
	&+
	T \sum_{\Omega_k > 0} L^R(\bm{q}, i\Omega_k + i\omega_\nu) L^R(\bm{q}, i\Omega_k) B_{\mathrm{c}3}(i\Omega_k + i\omega_\nu, i\omega_\nu) B_{\mathrm{s}3}(i\Omega_k + i\omega_\nu, i\omega_\nu)\\
	\xrightarrow{i\omega_\nu \to \hbar \omega \approx 0}&\\
	&
	\hbar \omega \cdot \frac{T}{2} L(\bm{q}, 0)^2 B_{\mathrm{c}2}(0, 0) \qty[\pdv{x} B_{\mathrm{s}2}(x,x)]_{x=0}\label{intrinsic-eq6+}\\
	+&
	\hbar \omega \cdot \frac{T}{2} L(\bm{q}, 0)^2 B_{\mathrm{c}2}(0, 0) \qty[\pdv{x} B_{\mathrm{s}2}(0, x)]_{x=0}\label{intrinsic-eq7+}\\
	-&
	\hbar \omega \cdot \pv \int_{-\infty}^{\infty} \frac{d\Omega}{4\pi i} \coth \frac{\Omega}{2T} \pdv{\Omega} 
	\qty[L^R(\bm{q}, \Omega) L^A(\bm{q}, \Omega) B_{\mathrm{c}2}(\Omega, 0) B_{\mathrm{s}2}(\Omega, 0)] \label{intrinsic-eq8}\\
	+&
	\hbar \omega \cdot T \sum_{\Omega_k<0} L^A(\bm{q}, i\Omega_k)^2 B_{\mathrm{c}1}(i\Omega_k, 0) \qty[\pdv{x} B_{\mathrm{s}1}(i\Omega_k, x)]_{x=0} \label{intrinsic-eq9}\\
	+&
	\hbar \omega \cdot T \sum_{\Omega_k > 0} L^R(\bm{q}, i\Omega_k)^2 B_{\mathrm{c}3}(i\Omega_k, 0) \qty[\pdv{x} B_{\mathrm{s}3}(i\Omega_k, x)]_{x=0} \label{intrinsic-eq10}\\
	+&
	\order{\omega^2}.
\end{align}
We note that Eqs.~\eqref{intrinsic-eq8}, \eqref{intrinsic-eq9}, and \eqref{intrinsic-eq10} do not yield singularity in the limit $\epsilon\to 0$ and thus ignore them. Consequently, we obtain
\begin{align}
	&
	T \sum_{\Omega_k} L(\bm{q}, i\Omega_k) L(\bm{q}, i\Omega_k-i\omega_\nu) B_{\mathrm{c}}(i\Omega_k, i\omega_\nu) B_{\mathrm{s}}(i\Omega_k, i\omega_\nu)\\
	\xrightarrow{i\omega_\nu \to \hbar \omega \approx 0}&
	\hbar \omega \cdot \frac{T}{2} L(\bm{q}, 0)^2 B_{\mathrm{c}2}(0, 0) \pdv{x}\qty[B_{\mathrm{s}2}(x, x) + B_{\mathrm{s}2}(0, x)]_{x=0}\label{eq: Oomega}\\
	+&
	\text{(regular terms in the limit $\epsilon \to 0$)} + \order{\omega^2}.
\end{align}
We calculate each factor in Eq.~\eqref{eq: Oomega} as 
\begin{align}
	&
	\pdv{x}\qty[B_{\mathrm{s}2}(x, x) + B_{\mathrm{s}2}(0, x)]_{x=0}
=
	-\frac{1}{8\pi T(\lambda k_{\mathrm{F}})^2} \frac{\hbar^2}{m} \Im \qty[\psi^{(1)}\qty(\frac{1}{2} + \frac{i\lambda k_{\mathrm{F}}}{2\pi T})],\label{eq: dBs2-AL-intrinsic}\\
	&
	B_{\mathrm{c}2}(0, 0)
=
	-\frac{1}{8\pi T \lambda k_{\mathrm{F}}} \frac{\hbar^2}{m} \Im \qty[\psi^{(1)}\qty(\frac{1}{2} + \frac{i\lambda k_{\mathrm{F}}}{2\pi T})]\label{eq: Bc2-L-intrinsic}.
\end{align}
\subsubsection{\texorpdfstring{Step8: Sum over $\bm{q}$}{Step8: Sum over q}}
The part related to $\bm{q}$ is written with use of  Eq.~\eqref{al-eq14}
 by
\begin{align}
	V \sum_{\bm{q}} \bm{q}^2 L(\bm{q}, 0)^2
	\approx&
	\frac{1}{4\pi N(0)^2 \clength^4} \ln \frac{1}{\epsilon}. \label{eq: Sum-q-L1-AL-intrinsic}
\end{align}
\subsubsection{Step9: Expression for spin Hall conductivity}
Putting Eqs.~\eqref{eq: Oomega}, \eqref{eq: dBs2-AL-intrinsic}, \eqref{eq: Bc2-L-intrinsic}, and \eqref{eq: Sum-q-L1-AL-intrinsic} into Eq.~\eqref{eq: Phi-AL-intrinsic}, we obtain
\begin{align}
	\sigma_{xy}^{\mathrm{AL}}
	=&
	\frac{\Phi_{xy}^{\mathrm{AL}}(0,i\omega_\nu \to \hbar \omega)}{i\omega}\\
	\approx&
	\frac{1}{i\omega} \qty[\frac{ie^2\lambda}{16m} N(0)^2 k_{\mathrm{F}}^3 V\sum_{\bm{q}} \bm{q}^2 T \sum_{\Omega_k} L(\bm{q}, i\Omega_k) L(\bm{q}, i\Omega_k - i\omega_\nu) B_{\mathrm{c}}(i\Omega_k, i\omega_\nu) B_{\mathrm{s}}(i\Omega_k, i\omega_\nu)]_{i\omega_\nu \to \hbar \omega}\\
	\approx&
	\frac{1}{i\omega}\cdot \frac{ie^2 \lambda}{16m} N(0)^2 k_{\mathrm{F}}^3 \cdot V\sum_{\bm{q}} \bm{q}^2 \cdot \frac{T}{2} \hbar \omega L(\bm{q}, 0)^2 B_{\mathrm{c}2}(0, 0) \pdv{x} \qty[B_{\mathrm{s}2}(x, x) + B_{\mathrm{s}2}(0, x)]_{x=0} \label{intrinsic-eq11}\\
    =&
	\frac{e^2 \hbar^5}{2^{13} \pi^3 \clength^4 T \lambda^2 m^3} \qty{\Im \qty[\psi^{(1)}\qty(\frac{1}{2} + \frac{i\lambda k_{\mathrm{F}}}{2\pi T})]}^2 \ln \frac{1}{\epsilon}.
\end{align}
\subsection{DOS terms}
\subsubsection{Step 1: Feynman diagrams for Response function}
\begin{figure}[tbp]
	\centering
	\includegraphics[width=0.6\textwidth,page=4]{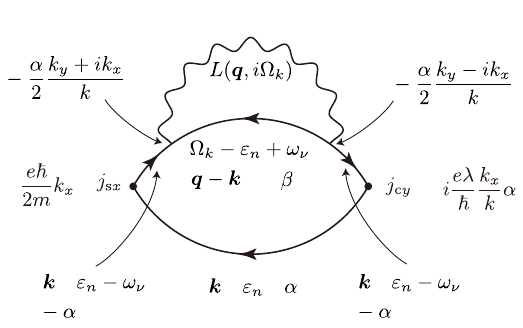}
	\caption{Diagrams of the DOS terms for intrinsic spin Hall effect.}
	\label{intrinsic-fig6}
\end{figure}

Figure \ref{intrinsic-fig6} shows the diagram of the DOS terms contributing to the intrinsic spin Hall effect.
The response function is written as
\begin{align}
	\Phi_{xy}^{\mathrm{DOS}}(0, i\omega_\nu)
	=&
	-\frac{1}{V} T\sum_{\bm{q}, \Omega_k} L(\bm{q}, i\Omega_k) \sum_{\bm{k}}\sum_{\alpha \beta} T\sum_{n} \frac{e\hbar}{2m} k_x \frac{e\lambda}{\hbar} i \frac{k_x}{k} \alpha \qty(-\frac{\alpha}{2})^2 \frac{k^2}{k^2}\\
	\times &
	\mathcal{G}_\alpha(\bm{k}, i\varepsilon_n) \mathcal{G}_{-\alpha}(\bm{k}, i\varepsilon_n - i\omega_\nu)^2 \mathcal{G}_\beta(\bm{q}-\bm{k}, i\Omega_k - i\varepsilon_n + i\omega_\nu).
\end{align}
\subsubsection{\texorpdfstring{Step2: Setting  $\Omega_k = 0$ in all quantities and $\bm{q}=0$ in all but $L(\bm{q},i\Omega_k)$}{Setting Ω\_k = 0 in all quantities and q = 0 in all but L(q, iΩ\_k)}}
Setting  $\Omega_k = 0$ in all quantities and $\bm{q}=0$ in all but $L(\bm{q},i\Omega_k)$, Eq.~\eqref{DOS-ss-eq3-} reduces to 
\begin{align}
	\Phi_{xy}^{\mathrm{DOS}}(0, i\omega_\nu)
	\approx&
	-T \cdot \frac{ie^2 \lambda k_{\mathrm{F}}}{16m} \cdot \sum_{\bm{q}} L(\bm{q}, 0) \cdot \sum_{\alpha \beta} \alpha A_{\alpha \beta} (i\omega_\nu, 0) \label{intrinsic-eq4}\\
	A_{\alpha \beta}(i\omega_\nu , i\Omega_k)
	\coloneqq&
	\frac{1}{V} \sum_{\bm{k}} T \sum_{\varepsilon_n} \mathcal{G}_\alpha(\bm{k}, i\varepsilon_n) \mathcal{G}_{-\alpha} (\bm{k}, i\varepsilon_n - i\omega_\nu)^2 \mathcal{G}_\beta(\bm{k}, i\Omega_k - i\varepsilon_n + i\omega_\nu)
\end{align}

\subsubsection{Step3: Integral over wavevectors} 
We assume that $\abs{\lambda k} \ll \varepsilon_{\mathrm{F}}$ and can then approximate as $1 / V \sum_{\bm{k}} = N(0) \int d\xi, \quad \xi = \hbar^2 \bm{k}^2 / 2m-\mu$. With use of this, $A_{\alpha \beta}(i\omega_\nu, 0)$ can be reduced to  
\begin{align}
	&
	A_{\alpha \beta}(i\omega_\nu, 0)\\
    \approx&
	T\sum_{\varepsilon_n} N(0) \int d\xi\; \frac{1}{\xi - (i\varepsilon_n - \alpha \lambda k_{\mathrm{F}})} \frac{1}{\qty(\xi - (i\varepsilon_{n-\nu} + \alpha \lambda k_{\mathrm{F}}))^2} \frac{1}{\xi- (-i\varepsilon_{n-\nu} - \beta \lambda k_{\mathrm{F}})}\\
	=&
		2\pi i N(0) T \sum_{\varepsilon_n}
	\begin{aligned}[t]
	\Biggl\{
	 &
		\sign(\varepsilon_n) \frac{1}{2i\varepsilon_n + i\omega_\nu - (\alpha - \beta) \lambda k_{\mathrm{F}}} \frac{1}{[ 2i\varepsilon_n + (\alpha + \beta) \lambda k_{\mathrm{F}} ]^2}\\
	&+
		\theta(-\varepsilon_n \varepsilon_{n-\nu}) \frac{1}{(i\omega_\nu - 2\alpha \lambda k_{\mathrm{F}})^2} \frac{1}{2i\varepsilon_n - i\omega_\nu - (\alpha - \beta) \lambda k_{\mathrm{F}}}
	\Biggr\}
	\end{aligned}
\end{align}
With use of the relation
\begin{align}
	&
	\frac{1}{x+a} \frac{1}{(x+b)^2} =
	\frac{1}{(b-a)^2} \qty(\frac{1}{x+a} - \frac{1}{x+b}) - \frac{1}{b-a} \frac{1}{(x+b)^2},
\end{align}
we can further rewrite $A_{\alpha \beta}(i\omega_\nu, 0)$ as
\begin{align}
	A_{\alpha \beta}(i\omega_\nu, 0)
	\approx&
	2\pi i N(0) T
	\begin{aligned}[t]
		\Biggl\{&
			\frac{1}{(i\omega_\nu - 2\alpha \lambda k_{\mathrm{F}})^2} \sum_{\varepsilon_n}\sign(\varepsilon_n) \qty[ \frac{1}{2i\varepsilon_n + i\omega_\nu - (\alpha - \beta) \lambda k_{\mathrm{F}}} - \frac{1}{2i\varepsilon_n + (\alpha + \beta) \lambda k_{\mathrm{F}}} ] \\
		&+
		\frac{1}{i\omega_\nu - 2\alpha \lambda k_{\mathrm{F}}} \sum_{\varepsilon_n} \sign(\varepsilon_n) \frac{1}{[2i\varepsilon_n + (\alpha + \beta) \lambda k_{\mathrm{F}}]^2} \\
		&+
		\frac{1}{(i\omega_\nu - 2\alpha \lambda k_{\mathrm{F}})^2} \sum_{\varepsilon_n} \theta(-\varepsilon_n \varepsilon_{n-\nu}) \frac{1}{2i\varepsilon_n - i\omega_\nu - (\alpha - \beta) \lambda k_{\mathrm{F}}}
		\Biggr\}
	\end{aligned}\label{eq: A-cal}
\end{align}

\subsubsection{\texorpdfstring{Step4: Sum over $\varepsilon_n$}{Step4: Sum over ε\_n}}
The first $\sum_{\varepsilon_n}$ in Eq.~\eqref{eq: A-cal} becomes
\begin{align}
	&
	\sum_{\varepsilon_n} \sign(\varepsilon_n) \qty[ \frac{1}{2i\varepsilon_n + i\omega_\nu - (\alpha - \beta) \lambda k_{\mathrm{F}}} - \frac{1}{2i\varepsilon_n + (\alpha + \beta) \lambda k_{\mathrm{F}}} ]\\
	=&
	\frac{1}{4\pi i T} 
	\begin{aligned}[t]
		\Biggl[
			&
			\psi\qty(\frac{1}{2} + \frac{i(\alpha + \beta) \lambda k_{\mathrm{F}}}{4\pi T}) + \psi\qty(\frac{1}{2} - \frac{i(\alpha + \beta) \lambda k_{\mathrm{F}}}{4\pi T}) \\
			&-
			\psi\qty(\frac{1}{2} + \frac{\omega_\nu}{4\pi T} + \frac{i(\alpha - \beta) \lambda k_{\mathrm{F}}}{4\pi T}) - \psi\qty(\frac{1}{2} - \frac{\omega_\nu}{4\pi T} - \frac{i(\alpha - \beta) \lambda k_{\mathrm{F}}}{4\pi T})
			\Biggr]
	\end{aligned}
\end{align}
and the second $\sum_{\varepsilon_n}$ in Eq.~\eqref{eq: A-cal} becomes
\begin{align}
	\sum_{\varepsilon_n} \sign(\varepsilon_n) \frac{1}{[2i\varepsilon_n + (\alpha + \beta) \lambda k_{\mathrm{F}}]^2}
	=&
	\frac{1}{(4\pi i T)^2} \qty[\psi^{(1)}\qty(\frac{1}{2} - \frac{i(\alpha + \beta) \lambda k_{\mathrm{F}}}{4\pi T}) - \psi^{(1)}\qty(\frac{1}{2} + \frac{i(\alpha + \beta) \lambda k_{\mathrm{F}}}{4\pi T})].
\end{align}
The third $\sum_{\varepsilon_n}$ becomes
\begin{align}
	&
	\sum_{\varepsilon_n} \theta(-\varepsilon_n \varepsilon_{n-\nu}) \frac{1}{2i\varepsilon_n - i\omega_\nu - (\alpha - \beta) \lambda k_{\mathrm{F}}}\\
	=&
	\frac{1}{4\pi i T} \qty[\psi\qty(\frac{1}{2} + \frac{\omega_\nu}{4\pi T} + \frac{i(\alpha - \beta) \lambda k_{\mathrm{F}}}{4\pi T}) - \psi\qty(\frac{1}{2} - \frac{\omega_\nu}{4\pi T} + \frac{i(\alpha - \beta) \lambda k_{\mathrm{F}}}{4\pi T})].
\end{align}
Putting them together, we can obtain
\begin{align}
	&
	A_{\alpha \beta}(i\omega_\nu, 0)\\
	=&
	\frac{N(0)}{2}
	\begin{aligned}[t]
		\Biggl\{&
			\frac{1}{(i\omega_\nu - 2\alpha \lambda k_{\mathrm{F}})^2}
			\begin{aligned}[t]
				\Biggl[&
					\psi\qty(\frac{1}{2} + \frac{i(\alpha + \beta) \lambda k_{\mathrm{F}}}{4\pi T})
					+
					\psi\qty(\frac{1}{2} - \frac{i(\alpha + \beta) \lambda k_{\mathrm{F}}}{4\pi T})\\
					-&
					\psi\qty(\frac{1}{2} - \frac{i\omega_\nu}{4\pi iT} + \frac{i(\alpha - \beta) \lambda k_{\mathrm{F}}}{4\pi T})
					- 
					\psi\qty(\frac{1}{2} - \frac{i\omega_\nu}{4\pi iT} - \frac{i(\alpha - \beta) \lambda k_{\mathrm{F}}}{4\pi T})
				\Biggr]
			\end{aligned}\\
			+&
			\frac{1}{i\omega_\nu - 2\alpha \lambda k_{\mathrm{F}}} \cdot \frac{1}{4\pi i T}
			\qty[\psi^{(1)}\qty(\frac{1}{2} - \frac{i(\alpha + \beta) \lambda k_{\mathrm{F}}}{4\pi T})
			-
			\psi^{(1)}\qty(\frac{1}{2} + \frac{i(\alpha + \beta) \lambda k_{\mathrm{F}}}{4\pi T})]
		\Biggr\}.
	\end{aligned}
\end{align}
\subsubsection{\texorpdfstring{Step5: Analytic continuation and expansion with $\omega$}{Step5: Analytic continuation and expansion with ω}}
After analytic continuation $i\omega_\nu \to \hbar \omega \approx 0$, we expand the resultant expression with $\omega$ as
\begin{align}
	&
	\sum_{\alpha \beta} \alpha A_{\alpha \beta}(i\omega_\nu \to \hbar \omega, 0)\\
	=&
	\frac{N(0)}{2}\sum_{\alpha \beta} \alpha 
	\begin{aligned}[t]
		\Biggl\{&
			\frac{1}{(- 2\alpha \lambda k_{\mathrm{F}})^2}
			\begin{aligned}[t]
				\Biggl[&
					\psi\qty(\frac{1}{2} + \frac{i(\alpha + \beta) \lambda k_{\mathrm{F}}}{4\pi T})
					+
					\psi\qty(\frac{1}{2} - \frac{i(\alpha + \beta) \lambda k_{\mathrm{F}}}{4\pi T})\\
					-&
					\psi\qty(\frac{1}{2}  + \frac{i(\alpha - \beta) \lambda k_{\mathrm{F}}}{4\pi T})
					- 
					\psi\qty(\frac{1}{2} - \frac{i(\alpha - \beta) \lambda k_{\mathrm{F}}}{4\pi T})
				\Biggr]
			\end{aligned}\\
			+&
			\frac{1}{- 2\alpha \lambda k_{\mathrm{F}}} \cdot \frac{1}{4\pi i T}
			\qty[\psi^{(1)}\qty(\frac{1}{2} - \frac{i(\alpha + \beta) \lambda k_{\mathrm{F}}}{4\pi T})
			-
			\psi^{(1)}\qty(\frac{1}{2} + \frac{i(\alpha + \beta) \lambda k_{\mathrm{F}}}{4\pi T})]
		\Biggr\}
	\end{aligned}\\
	&+
	\frac{N(0)}{2}\sum_{\alpha \beta} \alpha 
	\begin{aligned}[t]
		\Biggl\{&
			\frac{2\hbar \omega}{(2\alpha \lambda k_{\mathrm{F}})^3}
			\begin{aligned}[t]
				\Biggl[&
					\psi\qty(\frac{1}{2} + \frac{i(\alpha + \beta) \lambda k_{\mathrm{F}}}{4\pi T})
					+
					\psi\qty(\frac{1}{2} - \frac{i(\alpha + \beta) \lambda k_{\mathrm{F}}}{4\pi T})\\
					-&
					\psi\qty(\frac{1}{2}  + \frac{i(\alpha - \beta) \lambda k_{\mathrm{F}}}{4\pi T})
					- 
					\psi\qty(\frac{1}{2} - \frac{i(\alpha - \beta) \lambda k_{\mathrm{F}}}{4\pi T})
				\Biggr]
			\end{aligned}\\
			+&
			\frac{1}{(- 2\alpha \lambda k_{\mathrm{F}})^2} \cdot \frac{\hbar \omega}{4\pi i T}
			\qty[\psi^{(1)}\qty(\frac{1}{2} + \frac{i(\alpha - \beta) \lambda k_{\mathrm{F}}}{4\pi T})
			+
			\psi^{(1)}\qty(\frac{1}{2} - \frac{i(\alpha - \beta) \lambda k_{\mathrm{F}}}{4\pi T})]\\
			+&
			\frac{-\hbar \omega}{(-2\alpha \lambda k_{\mathrm{F}})^2} \cdot \frac{1}{4\pi i T}
			\qty[\psi^{(1)}\qty(\frac{1}{2} - \frac{i(\alpha + \beta) \lambda k_{\mathrm{F}}}{4\pi T})
			-
			\psi^{(1)}\qty(\frac{1}{2} + \frac{i(\alpha + \beta) \lambda k_{\mathrm{F}}}{4\pi T})]
		\Biggr\}
	\end{aligned}\\
	&+
	\order{\omega^2}\\
	=&
	\omega \cdot \frac{\hbar N(0)}{8\pi T (\lambda k_{\mathrm{F}})^2} \Im \psi^{(1)}\qty(\frac{1}{2} + \frac{i\lambda k_{\mathrm{F}}}{2\pi T}) + \order{\omega^2}.
\end{align}
\subsubsection{\texorpdfstring{Step6: Sum over $\bm{q}$ and Step7: Expression for spin Hall conductivity}{Step6: Sum over q and Step7: Expression for spin Hall conductivity}}
With use of 
\begin{align}
	\sum_{\bm{q}} L(\bm{q}, 0)
	\approx
	-\frac{1}{4\pi \clength^2 N(0)} \ln \frac{1}{\epsilon},
\end{align}
the spin Hall conductivity is written as
\begin{align}
	\sigma_{xy}^{\mathrm{DOS}}
	=&
	\frac{\Phi_{xy}^{\mathrm{DOS}}(0, \hbar\omega) - \Phi_{xy}^{\mathrm{DOS}}(0,0)}{i\omega} 
	=
	\frac{\Phi_{xy}^{\mathrm{DOS}}(0, \hbar\omega)}{i\omega} \\
	\overset{\text{\eqref{intrinsic-eq4}}}\approx&
	\frac{1}{i\omega} (-T) \cdot \frac{i e^2 \lambda k_{\mathrm{F}}}{16m} \cdot \sum_{\bm{q}} L(\bm{q}, 0) \cdot \sum_{\alpha \beta} \alpha A_{\alpha \beta}(i\omega_\nu \to \hbar \omega, 0)\\
	\approx&
	-\frac{e^2 \hbar}{512 \pi^2 m \clength^2 \lambda k_{\mathrm{F}}} \qty[- \Im \psi^{(1)}\qty(\frac{1}{2} + \frac{i\lambda k_{\mathrm{F}}}{2\pi T})] \ln \frac{1}{\epsilon}.
\end{align}
\subsection{Maki-Thompson terms }
\subsubsection{Step1: Feynman diagrams for Response function} 
\begin{figure}[tbp]
	\centering
	\includegraphics[width=0.6\textwidth,page=5]{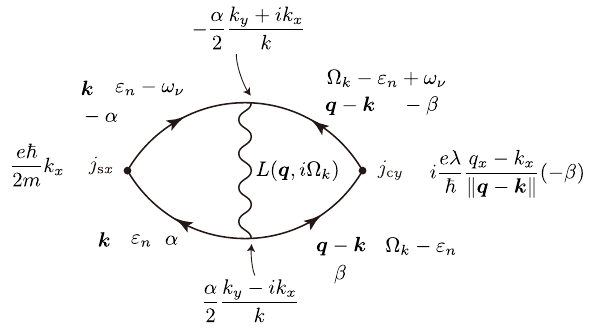}
	\caption{Diagram for the MT terms for the intrinsic spin Hall effect.}
	\label{intrinsic-fig8}
\end{figure}

Figure~\ref{intrinsic-fig8} shows the diagram of the MT terms for the intrinsic spin Hall effect. 
The response function is written as
\begin{align}
	\Phi_{xy}^{\mathrm{MT}}(0, i\omega_\nu)
	=&
	-\frac{1}{V} T \sum_{\bm{q}\Omega_k} L(\bm{q}, i\Omega_k) \sum_{\bm{k}}\sum_{\alpha \beta} T\sum_{\varepsilon_n} \frac{e\hbar}{2m} k_{x} \cdot i \frac{e\lambda}{\hbar} \frac{q_x - k_x}{\norm{\bm{q}-\bm{k}}} (-\beta)\cdot
	\qty(-\frac{\alpha}{2} \frac{k_y + ik_x}{k}) \cdot \frac{\alpha}{2} \frac{k_y - ik_x}{k}\\
	&\times
	\mathcal{G}_{\alpha}(\bm{k}, i\varepsilon_n) \mathcal{G}_{-\alpha}(\bm{k}, i\varepsilon_n - i\omega_\nu)
	\mathcal{G}_{\beta}(\bm{q}-\bm{k}, i\Omega_k - i\varepsilon_n) \mathcal{G}_{-\beta}(\bm{q}-\bm{k}, i\Omega_k - i\varepsilon_n + i\omega_\nu). \label{eq: Phi-MT-intrinsic}
\end{align}
\subsubsection{\texorpdfstring{Step2: Setting  $\Omega_k = 0$ in all quantities and $\bm{q}=0$ in all but $L(\bm{q},i\Omega_k)$}{Step2: Setting Ω\_k = 0 in all quantities and q = 0 in all but L(q, iΩ\_k)}}
In the following, we set $\Omega_k = 0$ and retain the  $\bm{q}$-dependence only in $L(\bm{q}, 0)$. The response function \eqref{eq: Phi-MT-intrinsic} then becomes
\begin{align}
	\Phi_{xy}^{\mathrm{MT}}(0, i\omega_\nu)
	\approx&
	i \frac{e^2 \lambda k_{\mathrm{F}} T N(0)}{4m} \sum_{\bm{q}} L(\bm{q}, 0) \cdot \frac{1}{4} \sum_{\alpha \beta} \beta \cdot T \sum_{\varepsilon_n}\\
	&\times
	\int d\xi \; \mathcal{G}_{\alpha}(\bm{k}, i\varepsilon_n) \mathcal{G}_{-\alpha}(\bm{k}, i\varepsilon_n - i\omega_\nu) \mathcal{G}_{\beta}(\bm{k}, -i\varepsilon_n) \mathcal{G}_{-\beta}(\bm{k},-i\varepsilon_n + i\omega_\nu).
\end{align}
\subsubsection{Step3: Integral over wavevectors} 
With use of $\varepsilon_{\alpha} \approx \frac{\hbar^2 \bm{k}^2}{2m} + \alpha \lambda k_{\mathrm{F}}$,
the $\xi$ integral becomes
\begin{align}
	&
	\int d\xi \; \mathcal{G}_{\alpha}(\bm{k}, i\varepsilon_n) \mathcal{G}_{-\alpha}(\bm{k}, i\varepsilon_n - i\omega_\nu) \mathcal{G}_{\beta}(\bm{k}, -i\varepsilon_n) \mathcal{G}_{-\beta}(\bm{k},-i\varepsilon_n + i\omega_\nu)\\
	\approx&
	\int d\xi\; \frac{1}{\xi-(i\varepsilon_n - \alpha \lambda k_{\mathrm{F}})} \frac{1}{\xi - (i\varepsilon_{n-\nu} + \alpha \lambda k_{\mathrm{F}})} \frac{1}{\xi - (-i\varepsilon_n - \beta \lambda k_{\mathrm{F}})} \frac{1}{\xi - (-i\varepsilon_{n-\nu} + \beta \lambda k_{\mathrm{F}})}\\
	=&
	2\pi i \sign(\varepsilon_n) \frac{1}{i\omega_\nu - 2\alpha \lambda k_{\mathrm{F}}} \frac{1}{2i\varepsilon_n + (\beta - \alpha) \lambda k_{\mathrm{F}}} \frac{1}{2i\varepsilon_n - i\omega_\nu - (\alpha + \beta) \lambda k_{\mathrm{F}}}\\
	&+
	2\pi i\theta(-\varepsilon_n \varepsilon_{n-\nu}) \frac{1}{-2i\varepsilon_n + i\omega_\nu + (\alpha + \beta)\lambda k_{\mathrm{F}}} \frac{1}{-2i\varepsilon_n + 2i\omega_\nu + (\beta - \alpha) \lambda k_{\mathrm{F}}} \frac{1}{i\omega_\nu + 2\beta \lambda k_{\mathrm{F}}}\\
	&+
	2\pi i \theta(\varepsilon_n \varepsilon_{n-\nu}) \sign(\varepsilon_n)
	\frac{1}{-i\omega_\nu + 2\alpha \lambda k_{\mathrm{F}}} \frac{1}{2i\varepsilon_n - i\omega_\nu + (\alpha + \beta)\lambda k_{\mathrm{F}}} \frac{1}{2i\varepsilon_n - 2i\omega_\nu + (\alpha - \beta)\lambda k_{\mathrm{F}}}
	\Biggr].
\end{align}
\subsubsection{\texorpdfstring{Step4: Sum over $\varepsilon_n$}{Step4: Sum over ε\_n}}
With use of the relation
\begin{align}
	T \sum_{\varepsilon_n >0} \frac{1}{2i\varepsilon_n + a} \frac{1}{2i\varepsilon_n + b}
	=&
	T \sum_{\varepsilon_n > 0}\qty(\frac{1}{2i\varepsilon_n + b} - \frac{1}{2i\varepsilon_n + a}) \frac{1}{a - b}\\
	=&
	\frac{T}{a-b} \cdot \frac{1}{4\pi i T} \sum_{\varepsilon_n>0}\qty(\frac{1}{n + \frac{1}{2} - \frac{ib}{4\pi T}} - \frac{1}{n+\frac{1}{2} - \frac{ia}{4\pi T}})\\
	=&
	\frac{1}{4\pi i (a-b)} \qty[\psi\qty(\frac{1}{2} - \frac{ia}{4\pi T}) - \psi\qty(\frac{1}{2} - \frac{ib}{4\pi T})],
\end{align}
the sum $\varepsilon_n$ becomes
\begin{align}
	&
	T \sum_{\varepsilon_n} \int d\xi \; \mathcal{G}_{\alpha}(\bm{k}, i\varepsilon_n) \mathcal{G}_{-\alpha}(\bm{k}, i\varepsilon_n - i\omega_\nu) \mathcal{G}_{\beta}(\bm{k}, -i\varepsilon_n) \mathcal{G}_{-\beta}(\bm{k},-i\varepsilon_n + i\omega_\nu)\\
=&
	\frac{1}{2(i\omega_\nu - 2\alpha \lambda k_{\mathrm{F}}) (i\omega_\nu + 2\beta \lambda k_{\mathrm{F}})}
	\begin{aligned}[t]
		\Biggl[&
		\psi\qty(\frac{1}{2} + \frac{i(\alpha - \beta) \lambda k_{\mathrm{F}}}{4\pi T}) - \psi\qty(\frac{1}{2} + \frac{i(i\omega_\nu)}{4\pi T} + \frac{i(\alpha + \beta)\lambda k_{\mathrm{F}}}{4\pi T})\\
		+&
		2\psi\qty(\frac{1}{2} - \frac{i(\alpha - \beta)\lambda k_{\mathrm{F}}}{4\pi T}) - 2\psi\qty(\frac{1}{2} - \frac{i(i\omega_\nu)}{4\pi T} - \frac{i(\alpha + \beta)\lambda k_{\mathrm{F}}}{4\pi T})\\
		+&
		\psi\qty(\frac{1}{2} - \frac{2i(i\omega_\nu)}{4\pi T} + \frac{i(\alpha - \beta)\lambda k_{\mathrm{F}}}{4\pi T}) - \psi\qty(\frac{1}{2} - \frac{i(i\omega_\nu)}{4\pi T} + \frac{i(\alpha + \beta)\lambda k_{\mathrm{F}}}{4\pi T})
		\Biggr]
	\end{aligned}\\
	&+
	\frac{1}{2(i\omega_\nu + 2\beta \lambda k_{\mathrm{F}}) (i\omega_\nu - 2\beta \lambda k_{\mathrm{F}})}
	\begin{aligned}[t]
		\Biggl[&
		\psi\qty(\frac{1}{2} + \frac{i(i\omega_\nu)}{4\pi T} + \frac{i(\alpha + \beta)\lambda k_{\mathrm{F}}}{4\pi T}) - \psi\qty(\frac{1}{2} - \frac{i(i\omega_\nu)}{4\pi T} + \frac{i(\alpha + \beta) \lambda k_{\mathrm{F}}}{4\pi T})\\
		+&
		\psi\qty(\frac{1}{2} - \frac{i(\alpha - \beta)\lambda k_{\mathrm{F}}}{4\pi T}) - \psi\qty(\frac{1}{2} + \frac{2i (i\omega_\nu)}{4\pi T} - \frac{i(\alpha - \beta)\lambda k_{\mathrm{F}}}{4\pi T})
		\Biggr].
	\end{aligned}
\end{align}
\subsubsection{\texorpdfstring{Step5: Analytic continuation and expansion with $\omega$}{Step5: Analytic continuation and expansion with ω}}
We perform the sum over $\alpha, \beta$ after analytic continuation $i\omega_\nu \to \hbar \omega \approx 0$,
\begin{align}
	&
	\frac{1}{4} \sum_{\alpha \beta}\beta \qty[T \sum_{\varepsilon_n} \int d\xi \; \mathcal{G}_{\alpha}(\bm{k}, i\varepsilon_n) \mathcal{G}_{-\alpha}(\bm{k}, i\varepsilon_n - i\omega_\nu) \mathcal{G}_{\beta}(\bm{k}, -i\varepsilon_n) \mathcal{G}_{-\beta}(\bm{k},-i\varepsilon_n + i\omega_\nu)]_{i\omega_\nu \to \hbar \omega}\\
	=&
	-\frac{1}{4} \sum_{\alpha \beta}\beta\cdot \frac{1}{8\alpha \beta (\lambda k_{\mathrm{F}})^2}
	\begin{aligned}[t]
		\Biggl[&
		2\psi\qty(\frac{1}{2} + \frac{i(\alpha - \beta)\lambda k_{\mathrm{F}}}{4\pi T}) + 2\psi\qty(\frac{1}{2} - \frac{i(\alpha - \beta)\lambda k_{\mathrm{F}}}{4\pi T})\\
		-&
		2\psi\qty(\frac{1}{2} + \frac{i(\alpha + \beta)\lambda k_{\mathrm{F}}}{4\pi T}) - 2\psi\qty(\frac{1}{2} - \frac{i(\alpha + \beta)\lambda k_{\mathrm{F}}}{4\pi T})
		\Biggr]
	\end{aligned}\label{eq: omega-0th}\\
	&+
	\hbar \omega \cdot \frac{1}{4} \sum_{\alpha \beta}\beta 
	\begin{aligned}[t]
		\Biggl\{&
			\frac{1}{2}\frac{1}{2\alpha \lambda k_{\mathrm{F}}(2\beta \lambda k_{\mathrm{F}})^2} 
			\begin{aligned}[t]
				\Biggl[&
				2\psi\qty(\frac{1}{2} + \frac{i(\alpha - \beta)\lambda k_{\mathrm{F}}}{4\pi T}) + 2\psi\qty(\frac{1}{2} - \frac{i(\alpha - \beta)\lambda k_{\mathrm{F}}}{4\pi T}) \\
				-&
				2\psi\qty(\frac{1}{2} + \frac{i(\alpha + \beta)\lambda k_{\mathrm{F}}}{4\pi T}) - 2\psi\qty(\frac{1}{2} - \frac{i(\alpha + \beta)\lambda k_{\mathrm{F}}}{4\pi T})
				\Biggr]
			\end{aligned}\\
			-&
			\frac{1}{2}\frac{1}{(2\alpha \lambda k_{\mathrm{F}})^2 2\beta \lambda k_{\mathrm{F}}}
			\begin{aligned}[t]
				\Biggl[&
				2\psi\qty(\frac{1}{2} + \frac{i(\alpha - \beta)\lambda k_{\mathrm{F}}}{4\pi T}) + 2\psi\qty(\frac{1}{2} - \frac{i(\alpha - \beta)\lambda k_{\mathrm{F}}}{4\pi T}) \\
				-&
				2\psi\qty(\frac{1}{2} + \frac{i(\alpha + \beta)\lambda k_{\mathrm{F}}}{4\pi T}) - 2\psi\qty(\frac{1}{2} - \frac{i(\alpha + \beta)\lambda k_{\mathrm{F}}}{4\pi T})
				\Biggr]
			\end{aligned}\\
			-&
			\frac{1}{8\alpha \beta (\lambda k_{\mathrm{F}})^2} \qty[-\frac{2i}{4\pi T} \psi^{(1)}\qty(\frac{1}{2} + \frac{i(\alpha - \beta) \lambda k_{\mathrm{F}}}{4\pi T}) + \frac{2i}{4\pi T} \psi^{(1)}\qty(\frac{1}{2} - \frac{i(\alpha + \beta)\lambda k_{\mathrm{F}}}{4\pi T})]\\
			-&
			\frac{1}{8\alpha \beta (\lambda k_{\mathrm{F}})^2} \qty[\frac{2i}{4\pi T} \psi^{(1)}\qty(\frac{1}{2} + \frac{i(\alpha + \beta)\lambda k_{\mathrm{F}}}{4\pi T}) - \frac{2i}{4\pi T} \psi^{(1)}\qty(\frac{1}{2} - \frac{i(\alpha - \beta)\lambda k_{\mathrm{F}}}{4\pi T})]
		\Biggr\}
	\end{aligned}\\
	&+
	\order{\omega^2}.
\end{align}
We see that $\order{\omega^0}$ terms \eqref{eq: omega-0th} 
cancel by
replacing $\beta$ by $-\beta$ in the first two terms in $[\cdots]$ so that they become minus the last two terms in $[\cdots]$. 
Let us then see $\order{\omega^1}$ terms; Among the four $[\cdots]$ in $\{\cdots\}$, the second one vanishes. The third and forth terms $[\cdots]$ cancel with each other. 

Thus the first $[\cdots]$ in $\{\cdots\}$ yields the $\order{\omega^1}$ term, i,e,
\begin{align}
	&
	\frac{1}{4} \sum_{\alpha \beta}\beta \qty[T \sum_{\varepsilon_n} \int d\xi \; \mathcal{G}_{\alpha}(\bm{k}, i\varepsilon_n) \mathcal{G}_{-\alpha}(\bm{k}, i\varepsilon_n - i\omega_\nu) \mathcal{G}_{\beta}(\bm{k}, -i\varepsilon_n) \mathcal{G}_{-\beta}(\bm{k},-i\varepsilon_n + i\omega_\nu)]_{i\omega_\nu \to \hbar \omega}\\
	=&
	\hbar \omega \cdot \frac{1}{2} \cdot \frac{1}{4} \sum_{\alpha \beta} \beta \frac{1}{2\alpha \lambda k_{\mathrm{F}} (2\beta \lambda k_{\mathrm{F}})^2} \\
	\times&
	2\qty[\psi\qty(\frac{1}{2} + \frac{i(\alpha - \beta)\lambda k_{\mathrm{F}}}{4\pi T}) + \psi\qty(\frac{1}{2} - \frac{i(\alpha - \beta)\lambda k_{\mathrm{F}}}{4\pi T}) - \psi\qty(\frac{1}{2} + \frac{i(\alpha + \beta)\lambda k_{\mathrm{F}}}{4\pi T}) - \psi\qty(\frac{1}{2} - \frac{i(\alpha + \beta)\lambda k_{\mathrm{F}}}{4\pi T})]\\
=&
	-\hbar \omega \cdot \frac{1}{4(\lambda k_{\mathrm{F}})^3} \qty[\Re \psi\qty(\frac{1}{2} + \frac{i\lambda k_{\mathrm{F}}}{2\pi T}) - \psi\qty(\frac{1}{2})].
\end{align}
\subsubsection{\texorpdfstring{Step6: Sum over $\bm{q}$ and Step7: Expression for spin Hall conductivity}{Step6: Sum over q and Step7: Expression for spin Hall conductivity}}
With use of it and 
\begin{align}
	\sum_{\bm{q}} L(\bm{q}, 0)
	\approx&
	-\frac{1}{4\pi \clength^2 N(0)} \ln \frac{1}{\epsilon},
\end{align}
the spin Hall conductivity becomes
\begin{align}
	\sigma_{xy}^{\mathrm{MT}}
	=&
	\frac{\Phi_{xy}^{\mathrm{MT}}(0, \hbar \omega) - \Phi_{xy}^{\mathrm{MT}}(0,0)}{i\omega}
	=
	\frac{\Phi_{xy}^{\mathrm{MT}}(0, \hbar \omega)}{i\omega}\\
	=&
	\frac{1}{i\omega}\cdot i \frac{e^2 \lambda k_{\mathrm{F}} T N(0)}{4m} \cdot \qty(-\frac{1}{4\pi \clength^2 N(0)} \ln \frac{1}{\epsilon}) \cdot \qty(-\hbar \omega \frac{1}{4(\lambda k_{\mathrm{F}})^3}) \qty[\Re \psi\qty(\frac{1}{2} + \frac{i\lambda k_{\mathrm{F}}}{2\pi T}) - \psi\qty(\frac{1}{2})]\\
	=&
	\frac{e^2 \hbar T}{64\pi m \clength^2 (\lambda k_{\mathrm{F}})^2} \qty[\Re \psi\qty(\frac{1}{2} + \frac{i\lambda k_{\mathrm{F}}}{2\pi T}) - \psi\qty(\frac{1}{2})] \cdot \ln \frac{1}{\epsilon}.
\end{align}

%
%
%
%
%
%
%
\bibliography{reference.bib}